\documentclass[10pt]{article}

\usepackage{amsmath}
\usepackage{amssymb}
\usepackage{mathrsfs}
\usepackage{graphics}

\usepackage{duotenzor}

\usepackage{makeidx}

\makeindex

\title{\textbf{Reformulating and Reconstructing Quantum Theory}}
\author{Lucien Hardy\\
\textit{Perimeter Institute,}\\
\textit{31 Caroline Street North,}\\
\textit{Waterloo, Ontario N2L 2Y5, Canada}}
\date{}

\newcounter{thm}

\newcounter{thmindex} \addtocounter{thmindex}{1000}

\newenvironment{T}{\begin{quote} \refstepcounter{thmindex}\refstepcounter{thm}\index{theorems!T\arabic{thmindex}@\textbf{T\arabic{thm}}} {\bf T\arabic{thm}}  } {\end{quote}}

\newcommand{\negs }{\hspace{-1pt}}

\newcommand{\ident}{\ensuremath{1 \hspace{-1.0mm} {\bf l}}}

\begin{document}

\pagestyle{empty}

\begin{titlepage}

\maketitle

\vskip 3.2cm

\[
\begin{Diagram}[1.8]{0}{-3}
\begin{move}{-7,5}
\Duobox{A}{0,0}
\linkedprep[0.7]{A}{1}{A1}{3}{0}\duosymbol[-3,-6]{\scriptstyle a}
\linkedprep[0.7]{A}{2}{A2}{4.5}{0}\duosymbol[-25,-6]{\scriptstyle c}
\linkedprep[0.7]{A}{3}{A3}{6}{0}\duosymbol[-44,-6]{\scriptstyle a}
\end{move}
\begin{move}{10,15}
\Duobox[2]{B}{0,0}
\linkedeffect[0.7]{B}{1}{1B}{-4.5}{0}\duosymbol[25,-6]{\scriptstyle a}
\linkedeffect[0.7]{B}{2}{2B}{-3}{0}\duosymbol[3,-6]{\scriptstyle a}
\linkedprep[0.7]{B}{1.5}{B15}{3}{0} \duosymbol[-3,-3]{\scriptstyle b}
\end{move}
\begin{move}{4,8}
\Duobox[2]{C}{0,0}
\linkedeffect[0.7]{C}{1}{1C}{-4.5}{0}\duosymbol[25,-6]{\scriptstyle c}
\linkedeffect[0.7]{C}{2}{2C}{-3}{0}\duosymbol[3,-6]{\scriptstyle a}
\linkedprep[0.7]{C}{1}{C1}{3}{0} \duosymbol[-3,-6]{\scriptstyle a}
\linkedprep[0.7]{C}{2}{C2}{4.5}{0} \otherside\duosymbol[-25,3]{\scriptstyle d}
\end{move}
\begin{move}{19,18}
\Duobox[2]{D}{0,0}
\linkedeffect[0.7]{D}{1}{1D}{-4.5}{0}\duosymbol[25,-2]{\scriptstyle b}
\linkedeffect[0.7]{D}{2}{2D}{-3}{0}\otherside\duosymbol[3,3]{\scriptstyle d}
\end{move}
\wire{A1}{1B}{1}{1}\opsymbol{a} \wire[0.4]{A2}{1C}{1}{1}\opsymbol{c} \wire[0.4]{A3}{2C}{1}{1}\otherside\opsymbol{a}
\wire{C1}{2B}{1}{1}\opsymbol{a} \wire{C2}{2D}{1}{1}\otherside\opsymbol{d} \wire[0.4]{B15}{1D}{1}{1}\opsymbol[-3,0]{b}
\end{Diagram}
\]

\thispagestyle{empty}
\end{titlepage}

\newpage

%\begin{center} \large{\bf Abstract} \end{center}
\begin{abstract}
\noindent We provide a reformulation of finite dimensional quantum theory in the circuit framework in terms of mathematical axioms, and a reconstruction of quantum theory from operational postulates.

We consider operations, $\mathsf{A}_\mathsf{a_1}^\mathsf{b_2}$ (where $\mathsf{a_1}$ is the input  and $\mathsf{b_2}$ is the output).  We also consider operators, $\hat A_\mathsf{a_1}^\mathsf{b_2} \in {\cal V}_\mathsf{a_1}\otimes{\cal V}^\mathsf{b_2}$ where ${\cal V}_\mathsf{a_1}$  (${\cal V}^\mathsf{b_2}$) is the space of Hermitian operators on the complex Hilbert space ${\cal H}_\mathsf{a_1}$ (${\cal H}^\mathsf{b_2}$). We say operations \emph{correspond} to operators if the probability for a circuit is given by replacing operations with operators then taking the trace.  For example, $\text{Prob}(\mathsf{A}^\mathsf{a_2}\mathsf{B}_\mathsf{a_2}^\mathsf{b_2} \mathsf{C}_\mathsf{b_2}) = \hat{A}^\mathsf{a_2}\hat{B}_\mathsf{a_2}^\mathsf{b_2} \hat{C}_\mathsf{b_2}$ (trace is implicit for repeated labels).  The mathematical axioms for quantum theory are the following
\begin{description}
\item[Axiom 1] Operations correspond to operators.
\item[Axiom 2] Every complete set of physical operators corresponds to a complete set of operations.
\end{description}
Physical operators have the property that they are positive after taking the partial transpose over the input space.

We show that these mathematical axioms are equivalent to a set of postulates couched in operational terms.  A \emph{maximal set of distinguishable states} is any set containing the maximum number of states for which there exists some measurement, called a \emph{maximal measurement}, which can identify which state from the set we have in a single shot.  A \emph{maximal effect} is associated with each result of a maximal measurement.  States are represented by vectors whose entries are probabilities.
%A \emph{filter} is a transformation that passes unchanged those states which would give rise only to a given subset of outcomes of a given maximal measurement %and block states which would give rise only to the complement set.
A set of states is said to be \emph{non-flat} if it is a spanning subset of the full set of states that give rise only to some subset of outcomes of some maximal measurement.  We show that classical probability theory and quantum theory are the only two theories consistent with the following set of postulates.
\begin{description}
\item[P1] \emph{Sharpness.}
Associated with any given pure state is a unique maximal effect giving probability equal to one.  This maximal effect does not give probability equal to one for any other pure state.
\item[P2] \emph{Information locality.} A maximal measurement on a composite system is effected if we perform maximal measurements on each of the components.
\item[P3] \emph{Tomographic locality.} The state of a composite system can be determined from the statistics collected by making measurements on the components.
\item[P4$'$] \emph{Permutability.} There exists a reversible transformation on any system effecting any given permutation of any given maximal set of distinguishable states for that system.
\item[P5] \emph{Sturdiness.} Filters are non-flattening.
\end{description}
We single out quantum theory if we replace {\bf P4$'$} by
%One way to single out quantum theory is to add the word \lq\lq compound" to postulate {\bf P4$'$}:
\begin{description}
\item[P4] \emph{Compound permutability.} There exists a compound reversible transformation on any system effecting any given permutation of any given maximal set of distinguishable states for that system.
\end{description}
A compound transformation is one that can be made from two sequential transformations (neither equal to the identity).
\end{abstract}

%\end{document}

\newpage

\pagestyle{plain}

\pagenumbering{roman} % i, ii, iii, iv, ...
\setcounter{page}{1}

{~}

%~\vskip 8mm

\tableofcontents

\newpage

\pagenumbering{arabic} % 1, 2, 3, 4, ...
\setcounter{page}{1}

\section*{Prelude: non-flattening transformations}
\addcontentsline{toc}{section}{Prelude: non-flattening transformations}

In physics we need to get control of some part of the world when we do experiments.  One device we often use for this purpose is a filter.  A filter cuts away the part of the world we are not interested in leaving only that bit we wish to do experiments on.  There are many filtering devices in typical quantum experiments. For example, we may have pinholes or slits that allow only particles with a particular range of positions through.  We may use frequency filters which allow only particles with a particular range of frequencies through.  We have velocity selectors that narrow down the range of velocities passing through.   Filters are, effectively, what we use to define the system we are interested in.

To define what we mean by a filter in operational terms we need a few basic notions.  A \emph{maximal set of distinguishable states} for a system is any set of states containing the maximum number of states for which there exists some measurement, called a \emph{maximal measurement}, which can identify which state from the set we have in a single shot.  In quantum theory a maximal set of distinguishable states would be those corresponding to any orthonormal basis of the Hilbert space.  A maximal measurement corresponds to one that measures a non-degenerate operator (i.e.\ a projection valued measure consisting only of rank one projectors).

It is interesting to consider restricting to those states which only give rise to a subset of outcomes of a maximal measurement.  We define
\begin{quote}
{\bf An informational subset of states} is the full set of states which only give rise to some given subset of outcomes of a given maximal measurement (and give probability zero for the other outcomes).
\end{quote}
Systems whose state is restricted to belong to a given informational subset of states have their information carrying capacity constrained.

In general operational theories, states can be represented by vectors whose entries are probabilities or by objects that are given by a linear map acting on such vectors of probabilities. In quantum theory, states can be represented by density operators which are, indeed, linearly related to probabilities.   We wish to define an important kind of set of states.
\begin{quote}
{\bf Non-flat sets of states.}  A set of states is non-flat if it is a spanning subset of some informational subset of states.
\end{quote}
Any set of states that is not non-flat is said to be flat.  A non-flattening transformation is one that transforms any non-flat input set of states into a non-flat output set.

We define a filter with respect to a particular subset of outcomes associated with a particular maximal measurement.
\begin{quote}
{\bf A filter} is a transformation that passes unchanged those states which would give rise only to the given subset of outcomes of the given maximal measurement and block states which would give rise only to the complement set of outcomes.
\end{quote}
In quantum theory a filter corresponds to projecting the state onto a given subspace of the Hilbert space.  For example, we may start out with a system associated with a Hilbert space of dimension 5.  We can project onto some particular 3 dimensional subspace of this Hilbert space.   The system after such a filtering operation is associated with the 3 dimensional Hilbert space.

The fifth postulate of the reconstruction of quantum theory to be given in Part \ref{thereconstruction} is the following.
\begin{quote}\index{sturdiness}\index{postulates!\textbf{P5} \emph{sturdiness}}
{\bf P5} \emph{Sturdiness.} Filters are non-flattening.
\end{quote}
This basically says, in a certain sense, that filters do not destroy more information than necessary.  If the informational subset associated with a particular non-flat set of states is the same as the informational subset associated with the given filter then the states will pass through unchanged and hence remain non-flat. But we can also have situations where the states in the non-flat set get partially absorbed by the filter.  {\bf P5} implies that, even in this case, the out coming set of states will be non-flat.  This implies that sets of states are, in a certain sense, sturdy against a fairly dramatic transformation.  On the other hand, given that there is no need for filtering transformations to flatten sets of states, it is reasonable that such transformations be non-flattening.   Filters are, indeed, non-flattening in both classical probability theory and in quantum theory (see Appendix \ref{appendixflattening}).

In this prelude we will illustrate how this postulate works in the context of quantum theory with examples. Consider, for example, a four dimensional Hilbert space, ${\cal H}_4$, with orthonormal basis $|n\rangle$ where $n=1$ to $4$.  Define
\begin{equation}
|mnx\rangle = \frac{1}{\sqrt{2}}(|m\rangle+|n\rangle) ~~~~\text{and}~~~~
|mny\rangle = \frac{1}{\sqrt{2}}(|m\rangle+i|n\rangle)
\end{equation}
Define
\begin{equation}
\rho_{n} = |n\rangle\langle n|, ~~~~ \rho_{mnx} = |mnx\rangle\langle mnx|, ~~~~|mny\rangle\langle mny|
\end{equation}
Now consider the following three sets of states for this space.
\begin{eqnarray}
\text{Set}~A &=& \{\rho_1, \rho_2, \rho_4, \rho_{12x}, \rho_{12y}, \rho_{14x}, \rho_{14y},  \rho_{24x}, \rho_{24y}\}  \\
\text{Set}~B &=& \{\rho_1, \rho_2, \rho_4, \rho_{12x}, \rho_{24y}, \rho_{14y}\} \\
\text{Set}~C &=& \{\rho_1, \rho_2, \rho_{12x}, \rho_{12y} \}
\end{eqnarray}
Sets $A$ and $B$ contain states only having support on the three dimensional Hilbert space spanned by $\{|1\rangle, |2\rangle, |4\rangle \}$.  Further, there does not exist a Hilbert space of smaller dimension which supports the states in either of these sets.  In quantum theory the space of the positive operators acting on an $N$ dimensional Hilbert space is of dimension $N^2$ (this is the number of real parameters required to specify a density matrix having support on an $N$ dimensional Hilbert space).  The states in set $A$ span the space of operators acting on this three dimensional Hilbert space (there are $9=3^2$ linearly independent states in $A$) and hence constitute a non-flat set. The same is not true of the states in set $B$ and hence this set is flat.  \index{flat set of states!in quantum theory}.   Set $C$ has support on the two dimensional Hilbert space spanned by $|1\rangle$ and $|2\rangle$.  There are $4$ linearly independent states in this set and hence the set is non-flat.

Now consider sending the states in set $A$ through a filter that projects onto the two dimensional Hilbert space spanned by $|1\rangle$ and $|2\rangle$.  The projection operator associated with this filter is
\begin{equation}
\hat F=|1\rangle\langle 1| + |2\rangle\langle 2|
\end{equation}
By applying this projector to the states in set $A$ we obtain
\begin{equation}
\text{Set}~A_F = \{ \rho_1, \rho_2, 0, \rho_{12x}, \rho_{12y}, \frac{1}{2}\rho_1, \frac{1}{2}\rho_1, \frac{1}{2}\rho_2, \frac{1}{2}\rho_2 \}
\end{equation}
These states have support on two dimensional Hilbert space spanned by $|1\rangle$ and $|2\rangle$. Further, we see that the set of states is non-flat since these states span the space of operators acting on this two dimensional Hilbert space (there are $4=2^2$ linearly independent states amongst the states in $A_F$).  Hence, when we send in the non-flat set of states $A$, we get out a non-flat set of states $A_F$.

Consider sending the states in set $C$ through a filter that projects on to the two dimensional Hilbert space spanned by the orthonormal vectors $|1\rangle$ and $|23x\rangle$.  The projector associated with this filter is
\begin{equation}
\hat G = |1\rangle\langle 1| + |23x\rangle\langle 23x|
\end{equation}
The states in $C$ become
\begin{equation}
\text{Set}~ C_G = \{ \rho_1, \frac{1}{2} \rho_{23x}, |x'\rangle\langle x'|, |y'\rangle\langle y'| \}
\end{equation}
where
\begin{equation}
|x'\rangle = \frac{1}{\sqrt{2}}|1\rangle + \frac{1}{2} |23x\rangle~~~~ \text{and} ~~~~ |y'\rangle = \frac{1}{\sqrt{2}}|1\rangle + \frac{i}{2} |23x\rangle
\end{equation}
The states in set $C_G$ are clearly non-flat also.  We see that in this case, if we send the non-flat set of states, $C$, into this filter we get out a non-flat set of states.  Interestingly, in this case, the states get closer to the $|1\rangle \langle 1|$ state.  The version of this fact that holds for general bits will play a role in the reconstruction of quantum theory from operational postulates.

This are just a few examples.  It turns out that, in quantum theory, if we send any non-flat set of states into a filter then we get a non-flat set of states out.  For a proof of this see Appendix \ref{appendixflattening}.  \index{non-flattening transformations!in quantum theory}

In quantum theory filters also have the property that they send pure states to pure states (up to normalization).  We call transformations that do this \emph{non-mixing transformations}.  Interestingly, in quantum theory, it turns out that all non-mixing transformations are also non-flattening (see Appendix \ref{appendixflattening}).  This is not surprising.  In general probabilistic theories, it is quite hard to see how we could flatten sets of states in such a way that all pure states remain pure (up to normalization).  We conjecture in the postlude that {\bf P5} can be replaced by the postulate that filters are non-mixing.

%\end{document}

\newpage

\part{Introduction}

In this work we will give a reformulation of quantum theory terms of mathematical axioms (in Part \ref{theduotensorframework}) and a reconstruction of quantum theory from operational postulates (in Part \ref{thereconstruction}).  We show that the mathematical axioms are equivalent to the usual formulation of quantum theory in terms of density matrices, positive operator valued measures, and completely positive maps).  We consider only the case of finite dimensional Hilbert spaces.   We then show that the operational postulates are equivalent to the mathematical axioms.

\section{Background}

\subsection{Motivation}

In quantum theory (QT) states and measurement outcomes are represented by positive operators on a complex Hilbert
space.  The probability for any particular outcome is given by the trace rule (also known as Born's rule).  Evolution is given by completely
positive maps (examples being unitary operators and von-Neumann Projection).  This structure is rather abstract. Why
Hilbert space? Why complex Hilbert space? Why represent states and measurement outcomes in this way?  Why do the usual
postulates of quantum theory take the particular form they do? In physics we answer \lq\lq why" questions like this by finding simpler
more natural postulates, axioms, or laws. For example Kepler's three laws of planetary motion were empirically
adequate for predicting planetary motion at the time.  However they are rather ad hoc.  One could argue that they are explained by
Newton's three laws of motion plus his universal law of gravitation.  The Lorentz transformations are rather abstract
and not at all natural in and of themselves. However, they were accounted for in a natural way by Einstein's two
postulates (that the laws of physics are the same in every inertial frame and that the speed of light is independent
of the source).  Quantum theory is ad hoc and abstract in the same way that Kepler's laws and the Lorentz
transformations are.  What is needed is some more natural postulates from which QT follows.

To gain an idea of how such postulates might look we need to think about what kind of theory QT is. Quantum theory
applies to a wide range of physical phenomena - spin degrees of freedom, interferometers, tunneling particles, etc.
However, what all these applications have in common is that quantum theory is used to calculate probabilities. Quantum
theory is a probability calculus.  In this sense, its natural predecessor is not Newtonian physics, but
rather what might be called classical probability theory (CProbT).  CProbT is the calculus used to calculate
probabilities for classical situations like tossing coins, throwing dice, predicting the weather, and so on.  Like
quantum theory, it comprises a set of rules which apply to a wide range of physical phenomena.  In writing down
postulates for classical probability theory we had better not make them specific to one type of situation (e.g. dice)
in which the theory might be applied.  Likewise, natural postulates for quantum theory should be natural for any
situation in which quantum theory might be applied.  Consequently we must expect a certain level of
abstraction - the postulates cannot make necessary reference to particular physical quantities (for example, position and momentum) which are only defined for some types of situation where QT might be applied.  Such considerations limit how \lq\lq physical" the postulates should be.  To deal with this issue we will outline a rather general operational framework (the circuit framework) pertaining to a wide range of physical phenomena.

\subsection{Previous work}

There is a long tradition of thinking about deriving quantum theory from more reasonable axioms or postulates going back to von Neumann \cite{von1996mathematical} and Mackey \cite{mackey1963mathematical}.  Much of the early work was in the quantum logical tradition, such as the papers of Birkhoff and von-Neumann \cite{birkhoff1936logic}, Zierler \cite{zierler1975axioms}, and Piron \cite{piron1964axiomatique}.  The convex probabilities framework (basically this is the idea of representing states as vectors of probabilities) goes back to originally to Mackey and has been worked on by many others since including Ludwig \cite{ludwig1985axiomatic}, Davies and Lewis \cite{davies1970operational}, Gunson \cite{gunson1967algebraic}, Mielnik \cite{mielnik1969theory}, Araki \cite{araki1980characterization}, Gudder {\it et al.\ } \cite{gudder1999convex}, Foulis and Randall \cite{foulis1979empirical}, and Fivel \cite{fivel1994interference}.

In the past decade a number of papers have been written on the topic of reconstructing quantum theory \cite{hardy2001quantum, clifton2003characterizing, d2008probabilistic, wilce2009four, rau2009quantum, rau2010measurement, goyal2008information, dakic2009quantum, masanes2010derivation, goyal2010origin, helland2009steps, fuchs2010quantum, fivel2010derivation, chiribella2010informational}.
In 2009 a conference on reconstructing quantum theory was held at Perimeter Institute (the talks can be viewed on PIRSA \cite{goyal2009reconstructing}).
Many of these have been inspired by ideas coming from quantum information which, generally, consider finite dimensional Hilbert spaces. This program was very much inspired by Fuchs's suggestion that we need to find information-theoretic reasons for the quantum axioms (presented in a number of talks and written up in \cite{fuchs2002quantum}).

Much of this work is in the convex probabilities framework. Recent treatments and developments of this framework can be found in \cite{hardy2001quantum}, \cite{barrett2007information}, \cite{barnum2011information}, \cite{chiribella2010probabilistic, chiribella2010informational}, \cite{hardy2009foliable,  hardy2009operational, hardy2009operational2, hardy2010formalism}. In parallel with this work, Abramsky and Coecke developed a categorical approach to quantum theory \cite{abramsky2004categorical}.  One of the most salient features of the categorical approach is that it gives rise to a kind of pictorialism \cite{coecke2010quantum}.  These pictures are basically the circuits in the circuit model presented in Part \ref{thecircuitframework}.   Motivated in part by this pictorial approach, Chiribella, D'Ariano, and Perinotti \cite{chiribella2010probabilistic, chiribella2010informational} and the present author \cite{hardy2009foliable, hardy2009operational, hardy2009operational2, hardy2010formalism} put forward various frameworks which show how probabilities can be put on top of such pictures in accord with the convex probabilities framework.

In order to reconstruct quantum theory from operational postulates we need to specify what we mean by quantum theory.  To this end, we will provide a reformulation of quantum theory using the duotensor framework \cite{hardy2010formalism} in terms of two mathematical axioms.  Although this mathematical reformulation was developed to aid the operational reconstruction, it stands alone and may be more interesting to some readers than the operational reconstruction.  The key point of this reformulation is that, rather than associating a completely positive map with an operation, we associated a positive operator acting on the tensor product of the input and output Hilbert spaces.

The two mathematical axioms presented here are part of a development of ideas by the author beginning with \cite{hardy2001quantum} in which a framework for general probabilistic theories was given.  This framework requires fixed causal structure. In a theory of quantum gravity we do not expect fixed causal structure.  To address this, the causaloid framework \cite{hardy2005probability} was developed for theories in which we do not need to have fixed causal structure.  We cannot assume an evolving state in such a situation.  Hence, in the causaloid framework, mathematical objects apply to arbitrary regions of spacetime.  We use the \emph{causaloid product} to combine such objects for non-overlapping regions.  Quantum theory was formulated in the causaloid framework.  In the case of quantum theory we do have fixed causal structure so the full machinery of the causaloid approach may be more than is necessary.  It is instructive, then, to apply the kind of thinking in the causaloid formalism where we consider arbitrary regions of spacetime to the situation in which we do have definite causal structure.  This was done in the duotensor framework \cite{hardy2010formalism}.  Motivated by the work of Abramsky and Coecke \cite{abramsky2004categorical}, pictorial techniques are used.   In the present work, we use the duotensor framework to associate operators with fragments of a circuit (fragments are the analogue of an arbitrary region of spacetime in the circuit framework).  Such operators can be combined with the \emph{circuit trace} to obtain the operator for a composite fragment.   It turns out that this approach is very similar to the quantum combs framework of Chiribella, D'Ariano, and Perinotti (CDP) \cite{chiribella2009theoretical}.  They associate  Cho-Jamio\l kowski operators with fragments and provide the \emph{link product} for combining them.  The basic equations of  CDP are related to the basic equations in the reformulation given here by appropriate insertion of partial transposes. The formula for the circuit trace is a little simpler than the link product, but the idea is similar. The circuit trace is an example of the causaloid product, as, effectively, is the link product.   Related ideas appear in the work of Aharonov, Popescu, Tollaksen, and Vaidman \cite{aharonov2009multiple} who consider multiple-time states and Oeckl \cite{oeckl2003general} who has developed a general boundary formalism for quantum theory.  The work of Aharonov {\it et al} and of Oeckl apply to the pure state case whereas the quantum combs framework and the framework to be presented here apply to the general mixed state case.   There are many other related approaches in which particular attention is given to issues concerning causality.  Sorkin has developed the causal set approach to quantum gravity \cite{sorkin1991spacetime}. Markopoulou developed the quantum causal histories approach \cite{markopoulou2000quantum}, and a dual point of view to this was provided by Blute, Ivanov, and Panangaden \cite{blute2003discrete}.  Leifer \cite{leifer2006quantum} has also done interesting work concerning the evolution of quantum systems on a causal circuit.

The project of reconstructing quantum theory from operational postulates presented here is a continuation of a project initiated by the author ten years ago \cite{hardy2001quantum}.  There a small number of operational axioms were given from which quantum theory can be reconstructed (see also Sec.\ \ref{previouswork}).  One of the axioms given in \cite{hardy2001quantum} is not particularly compelling.  This is the \emph{simplicity axiom} which says, basically, that states are specified by the smallest
number of probabilities consistent with the other axioms.  This forces us to take the second simplest case in the Wootters hierarchy \cite{wootters1990local, wootters1986quantum} of theories (see Sec.\ \ref{woottershierachy}).

The problem of replacing the simplicity axiom with more a compelling axiom was left open for a long time.  Then, in 2009, two papers appeared which addressed the problem.  First, Chiribella, D'Ariano, and Perinotti \cite{chiribella2010probabilistic} showed that the success probability for probabilistic teleportation is bounded by the inverse of the number of probabilities required to specify a state (see Lemma 22 in \cite{chiribella2010probabilistic}). In a subsequent paper they used this in a full derivation of quantum theory from operational axioms with no need for a simplicity axiom \cite{chiribella2010informational}.   Second, Daki\'c and Brukner \cite{dakic2009quantum} gave an argument to get rid of the simplicity axiom in a derivation of quantum theory based on the ideas in \cite{hardy2001quantum}.  They argued for a bound on the number of probabilities required to specify the state coming from considering entangled states for two generalized bits. Their argument was sharpened by Masanes and M\"uller \cite{masanes2010derivation} in another reconstruction of quantum theory.

In the this work we present a new set of postulates. These are mostly different from the axioms in \cite{hardy2001quantum}.   We adopt the technique of Chiribella, D'Ariano, and Perinotti \cite{chiribella2010probabilistic, chiribella2010informational} to avoid the need for a simplicity axiom though the approach of Daki\'c and Brukner \cite{dakic2009quantum} with the improvements due to Masanes and M\"uller \cite{masanes2010derivation} may also be applicable here.

To guide the readers intuition, sometimes remarks will be included in square parenthesis [like this] that discuss how things look in classical probability theory or quantum theory.

All figures in this paper were drawn using version 1.1 of the duotenzor package (see Appendix \ref{duotenzor}).

\section{Main results}

\subsection{The circuit framework}

In Part \ref{thecircuitframework} we will show how to describe certain types of experiment in operational terms within what we will call \emph{the circuit framework}.  In this model, an experiment consists of a bunch of apparatuses placed next to each other so that apertures on the apparatuses are aligned with one another. Each apparatus may have outcomes on it (as read off meters or detector clicks for example).  Each apparatus use will be associated with an operation (represented by a box in the graphical representation of a circuit).  An operation is associated with an outcome set, this being a subset of the possible outcomes on the apparatus.  An operation has a bunch of inputs and outputs and can be represented as
\begin{equation}
\begin{Diagram}{0}{-0.80}
\Opbox{A}{3,3}
\inwire[-5]{A}{1}\Opsymbol{a}
\inwire{A}{2}\Opsymbol{b}
\inwire[5]{A}{3}\Opsymbol{b}
\outwire[-5]{A}{1.5}\Opsymbol{b}
\outwire[5]{A}{2.5}\Opsymbol{c}
\end{Diagram}
~~~~~~~~~~~~\Longleftrightarrow ~~~~~~~~~~~~~\mathsf{A_{a_1b_2b_3}^{b_4c_5}}
\end{equation}
for example. The labels $\mathsf{a}$, $\mathsf{b}$, etc.\ correspond to different system types. The inputs enter the box at the bottom and are represented by subscripts in the symbolic notation. Outputs leave the box at the top and are represented by superscripts in the symbolic notation. The alignment of apertures is represented by a wire.  A circuit consists of a bunch of operations wired together so that there are no inputs or outputs left over. For example,
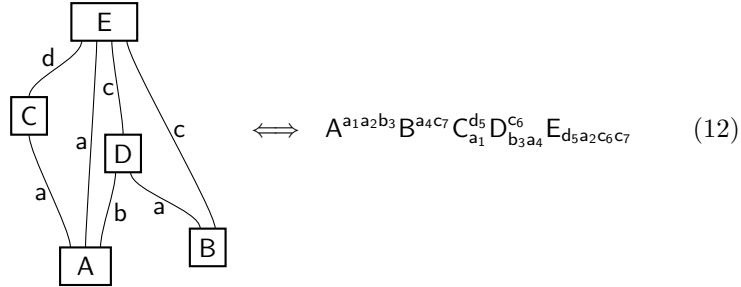
\begin{equation}
\begin{Diagram}{0}{-1.4}
%\constructiongrid{-2,-4}{19,15}
\Opbox{A}{8,-2}\Opbox[2]{B}{14.5,-1}\Opbox[2]{C}{5,6}\Opbox[2]{D}{10,4}\Opbox[4]{E}{9,11}
\wire{A}{C}{1}{1.5}\opsymbol{a}\wire{A}{E}{2}{2}\opsymbol{a}\wire{A}{D}{3}{1}\otherside\opsymbol[5,0]{b}
\wire{B}{D}{1}{2}\opsymbol{a}\wire{B}{E}{2}{4}\otherside\opsymbol{c}\wire{D}{E}{1.5}{3}\opsymbol{c}\wire{C}{E}{1.5}{1}\opsymbol[0,8]{d}
\end{Diagram}
~~
\Longleftrightarrow ~~  \mathsf{   A^{a_1a_2b_3} B^{a_4c_7} C_{a_1}^{d_5} D_{b_3a_4}^{c_6} E_{d_5a_2c_6c_7}  }
\end{equation}
In the symbolic notation the repeated integer label indices correspond to the placement of the wires.
Such circuits are to be understood graphically.  It is common to think of the vertical axis as corresponding to a background Newtonian time in interpreting circuit diagrams.   In this case it would matter how high up the page a box is placed.  We do not think in this way here.  There is absolutely no significance to the vertical position of the boxes on the page in the diagrams in this work.  The boxes can be moved to any position.  As long they maintain their orientation (so inputs remain as inputs and outputs remain as outputs) and the wires (which can be stretched) continue to be connected to the boxes in the same way, the diagram does not change its meaning.   Thinking about experiments as circuits provides a deeper foundation for understanding, in operational terms, notions like {\it system} and {\it state} and will form the backdrop against which the postulates are set.   We make three assumptions as part of the circuit model which are regarded as being too basic to be part of the postulate set.  The first two are ({\bf Assump 1}) that we can associate a probability with a circuit which depends only on the description of that circuit, and ({\bf Assump 2}) that non-trivial finite systems exist.  The third assumption says, basically, that hypothetical states  which are operationally indiscernible from some existing state to any accuracy actually exist (in fact the assumption is a bit more general than this). This assumption is one of mathematical convenience and allows us to deduce that the sets of states are closed.  It is not possible to operationally distinguish the case where we make this third assumption form the case where we do not as we cannot make arbitrarily accurate measurements.   Although very reasonable, these basic assumptions may have to be modified in a theory of quantum gravity.

\subsection{The reformulation}

In Part \ref{theduotensorframework} we will provide a reformulation of quantum theory in which objects called \emph{duotensors} mediate between circuits composed of operations (as in the circuit framework) and circuits composed of operators (these are mathematical objects acting on complex Hilbert spaces).

A duotensor \cite{hardy2010formalism} is like a tensor but with a bit more structure.  In Sec.\ \ref{operationsandduotensors} we will show how to associate duotensors with operations if operations have a certain property - that they are fully decomposable. It turns out that postulate {\bf P3} is equivalent to full decomposability of operations.   By associating operations with duotensors we can convert a circuit into a duotensor calculation for the probability associated with that circuit.  The ideas in this section were first presented in \cite{hardy2010formalism}.

In Sec.\ \ref{operatorsandduotensors} we will show how to associate operators with duotensors.  We consider the space, ${\cal V}_\mathsf{a}$, of Hermitian operators on a Hilbert space ${\cal H}_\mathsf{a}$ of dimension $N_\mathsf{a}$.  We also consider the space, ${\cal V}^\mathsf{a}$, of Hermitian operators on a Hilbert space ${\cal H}^\mathsf{a}$ also of dimension $N_\mathsf{a}$.   In general we are interested in operators in the space
\begin{equation}
{\cal V}_\mathsf{a_1b_2\dots c_3}^\mathsf{d_4e_5\dots f_6} := {\cal V}_\mathsf{a_1} \otimes {\cal V}_\mathsf{b_2} \otimes \dots \otimes {\cal V}_\mathsf{c_3}
\otimes {\cal V}^\mathsf{d_4} \otimes {\cal V}^\mathsf{e_5} \otimes \dots \otimes {\cal V}^\mathsf{f_6}
\end{equation}
We represent an operator in this space as
\begin{equation}
\begin{Diagram}[1.4]{0}{0}
\Dopbox[5]{A}{0,0}
\inwire{A}{1}\Opsymbol{a} \inwire{A}{2.2}\Opsymbol{b} \putlatex[30,15]{\ensuremath{\dots}} \inwire{A}{5}\Opsymbol{c}
\outwire{A}{1}\Opsymbol{d}\outwire{A}{2.2}\Opsymbol{e}\putlatex[30,-50]{\ensuremath{\dots}}   \outwire{A}{5}\Opsymbol{f}
\end{Diagram}
~~~~~\Leftrightarrow~~~~~\hat A_\mathsf{a_1b_2\dots c_3}^\mathsf{d_4e_5\dots f_6}
\end{equation}
It turns out that operators have the property that they are fully decomposable.  This means that we can associate a duotensor with every operator.
We can wire these operators together to form circuits.  The wires tell us how to match up the different parts of the tensor product space.

We can place wires between operators (denoted by a repeated label in symbolic notation).  A wire (or repeated label) indicates that we are taking the trace.  For example,
\begin{equation}
\begin{Diagram}{0}{-0.7}
\Dopbox[2]{A}{0,0} \Dopbox[2]{B}{2,5} \wire{A}{B}{1.5}{1.5} \opsymbol[-1,0]{a}
\end{Diagram}
~~~~~ \Longleftrightarrow ~~~~~
\hat A^\mathsf{a_1} \hat B_\mathsf{a_1}
\end{equation}
This is equal to the trace of the product of $\hat A^\mathsf{a_1}\in {\cal V}^\mathsf{a_1}$ and $\hat B_\mathsf{a_1}\in{\cal V}_\mathsf{a_1}$.   More generally we have expressions such as $\hat{A}^\mathsf{a_1b_2}\hat{B}_\mathsf{b_2}^\mathsf{c_3a_4}\hat{C}_\mathsf{a_5c_3a_4}^\mathsf{b_6}$ where we have more than one wire.  In this case the wire (or repeated label) indicates that we take the partial trace over the corresponding spaces.   This means that, for example,
\begin{equation}
\hat{A}^\mathsf{a_1b_2}\hat{B}_\mathsf{b_2}^\mathsf{c_3a_4}\hat{C}_\mathsf{a_5c_3a_4}^\mathsf{b_6} \in {\cal V}_\mathsf{a_5}^\mathsf{a_1b_6}
\end{equation}
This is similar to Einstein's summation convention.  Here the partial trace is implicit where ever we have a repeated index (or wire).

In Sec.\ \ref{operationsandoperators} we will bring the duotensor treatment of operations and operators together.  We will show that if a certain condition holds we can associate an operator, $\hat A_\mathsf{a_1b_2\dots c_3}^\mathsf{d_4e_5\dots f_6}$, with every operation, $\mathsf{A}_\mathsf{a_1b_2\dots c_3}^\mathsf{d_4e_5\dots f_6}$ such the probability for a circuit formed from operations is given by the trace of the corresponding operator expression.  For example,
\begin{equation}\label{mainresultsprobtrace}
\text{Prob}(\mathsf{A}^\mathsf{a_1b_2}\mathsf{B}_\mathsf{b_2}^\mathsf{c_3a_4}\mathsf{C}_\mathsf{a_1c_3a_4})
=\hat{A}^\mathsf{a_1b_2}\hat{B}_\mathsf{b_2}^\mathsf{c_3a_4}\hat{C}_\mathsf{a_1c_3a_4}
\end{equation}
This same example in diagrammatic form
\begin{equation}\label{mainresultsprobtracediagram}
\text{Prob}\left(~
\begin{Diagram}{0}{-1.4}
\Opbox[2]{A}{0,0} \Opbox[2]{B}{2,4} \Opbox{C}{-1,10}
\wire{A}{C}{1}{1} \opsymbol{a} \wire{A}{B}{2}{1.5}\opsymbol[-3,4]{b} \wire{B}{C}{1}{2} \opsymbol{c} \wire{B}{C}{2}{3}\otherside\opsymbol{a}
\end{Diagram}
~\right)
~=~
\begin{Diagram}{0}{-1.4}
\Dopbox[2]{A}{0,0} \Dopbox[2]{B}{2,4} \Dopbox{C}{-1,10}
\wire{A}{C}{1}{1} \opsymbol{a} \wire{A}{B}{2}{1.5}\opsymbol[-3,4]{b} \wire{B}{C}{1}{2} \opsymbol{c} \wire{B}{C}{2}{3}\otherside\opsymbol{a}
\end{Diagram}
\end{equation}
If we can calculate the probability for a circuit from the trace under such a mapping from operations to operators in this way, we will say that operations \emph{correspond} to operators. Note that this is different from the usual formulation in which completely positive maps are associated with operations.

We will see that operators, such as $\hat B_\mathsf{a_1}^\mathsf{b_2}$, can sensibly be associated with operations if, after taking the input transpose (this is the partial transpose over the input part of the space), we get a positive operator.   A complete set of physical operators, $\{\hat B_\mathsf{a_1}^\mathsf{b_2}[l]: l=1 ~\text{to} ~ L\}$, has the property that every operator in the set has positive input transpose and, further,
\begin{equation}\label{mainresultssumtoI}
\sum_{l=1}^{L} \hat B_\mathsf{a_1}^\mathsf{b_2}[l] \hat I_\mathsf{b_2} = \hat I_\mathsf{a_1}
\end{equation}
where $\hat I_\mathsf{a_1}$ is the identity operator acting on ${\cal H}_\mathsf{a_1}$.

We will show that quantum theory for finite dimensional Hilbert spaces can be formulated rather succinctly with the following two axioms.
\begin{quote}
\begin{description}
\item[Axiom 1] Operations correspond to operators.
\item[Axiom 2] Every complete set of physical operators corresponds to a complete set of operations.
\end{description}
\end{quote}
The operators here are understood to act on a complex Hilbert space.  A complete set of operations is a set of operations corresponding to the disjoint outcome sets of the same apparatus use.  A complete set of operations is a set of operations corresponding to the same setup (same apparatus with the same settings) where the associated outcome sets are disjoint and have union equal to the full set of outcomes for this setup.

Axiom 1 tells us that we can calculate probabilities for circuits by corresponding operator expressions as in (\ref{mainresultsprobtrace},\ref{mainresultsprobtracediagram}).  Axiom 2 guarantees that probabilities are greater than zero (the requirement that the operators have positive input transpose imposes this) and that the sum of the probabilities over all outcomes adds up to one as required (the condition in (\ref{mainresultssumtoI}) imposes this).

\subsection{The reconstruction}

In Part \ref{thereconstruction} we will show that classical probability theory and quantum theory are the only two theories consistent with the following postulates within the circuit framework.
\begin{description}
\item[P1] \emph{Sharpness.} Associated with any given pure state is a unique maximal effect giving probability equal to one.  This maximal effect does not give probability equal to one for any other pure state.
\item[P2] \emph{Information locality.} A maximal measurement on a composite system is effected if we perform maximal measurements on each of the components.
\item[P3] \emph{Tomographic locality.} The state of a composite system can be determined from the statistics collected by making measurements on the components.
\item[P4$'$] \emph{Permutability.} There exists a reversible transformation on any system effecting any given permutation of any given maximal set of distinguishable states for that system.
\item[P5] \emph{Sturdiness.} Filters are non-flattening.
\end{description}
We can single out quantum theory by adding anything that is inconsistent with classical probability theory yet consistent with quantum theory. One way to do this is to add the word \lq\lq compound" to postulate {\bf P4$'$}:
\begin{description}
\item[P4] \emph{Compound permutability.} There exists a compound reversible transformation on any system effecting any given permutation of any given maximal set of distinguishable states for that system.
\end{description}
A compound transformation is one that can be made from two sequential transformations (neither equal to the identity).  {\bf P4} fails, in particular, for a classical bit.  We will discuss these postulates in Sec.\ \ref{thepostulates}.  In the three sections after that we will show in detail how to reconstruct quantum theory (as given by the above two mathematical axioms) from these postulates.   Sec.\ \ref{filtersandsystemssection} and Sec.\ \ref{gebitssection} can be read without reading the sections on duotensors (Part \ref{theduotensorframework}).  Sec.\ \ref{partIII} requires the techniques developed in Part \ref{theduotensorframework}.

In Sec.\ \ref{filtersandsystemssection} we prove numerous results using only {\bf P1, P2, P3} and {\bf P4}$'$ (i.e.\ without using the assumption that filters are non-flattening of {\bf P5}).

We will show that {\bf P1} implies causality - namely that the future cannot influence the past.  This means we do not need this as a separate assumption.

We will show that there exists a reversible transformation between any pair of pure states.  The proof of this uses the following construction
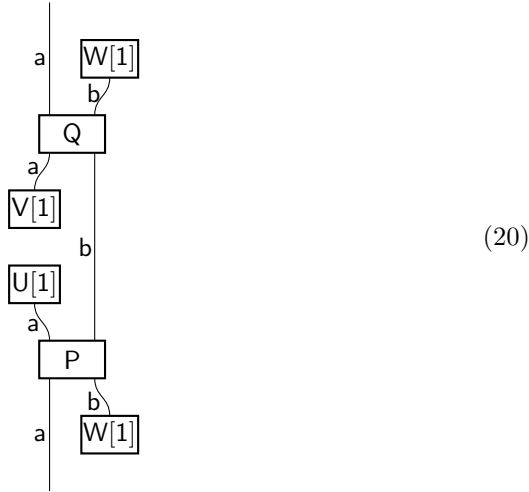
\begin{equation}\label{exampletransformation}
\begin{Diagram}{0}{-1.7}
\Opbox[4]{P}{0,0}
\opbox[4]{Q}{0,12} \opsymbol{Q}
\thispoint{IN}{-1.2, -7}   \wire{IN}{P}{1}{1} \opsymbol{a}
\opbox[3.4]{B1}{2,-4} \opsymbol{W[1]}
\wire{B1}{P}{2.2}{4} \opsymbol{b}
\opbox[3]{C1}{-2,4} \opsymbol{U[1]}
\wire{P}{C1}{1}{2} \opsymbol{a}
\wire{P}{Q}{4}{4} \opsymbol{b}
\opbox[3]{A1}{-2,8} \opsymbol{V[1]}
\wire{A1}{Q}{2}{1}  \opsymbol{a}
\opbox[3.4]{D1}{2,16} \opsymbol{W[1]}
\wire{Q}{D1}{4}{2.2} \opsymbol{b}
\thispoint{OUT}{-1.2, 19} \wire{Q}{OUT}{1}{1} \opsymbol{a}
\end{Diagram}
\end{equation}
where $\mathsf P$ and $\mathsf Q$ are transformations that permute a maximal sets of distinguishable states for the composite system entering them. By choosing appropriate permutations the incoming state is mapped on to the $\mathsf b$ system then mapped back onto the $\mathsf a$ system.

We define systems as being the thing we have after a filter.  By showing that two filters acting in parallel act as a filter on the composite, we are able to show that the composite of two systems is a system itself.   We show how to construct arbitrary filters using a transformation similar to (\ref{exampletransformation}) but where $\mathsf Q$ is the inverse of $\mathsf P$. Since we have arbitrary filters we have arbitrary systems.

Let $N_\mathsf{a}$ be the maximum number of distinguishable states for systems of type $\mathsf a$.  We show that if $N_\mathsf{a}=N_\mathsf{b}$ then systems of type $\mathsf a$ and $\mathsf b$ have the same properties.  We also show that the $K_\mathsf{a}=N_\mathsf{a}^r$ where $K_\mathsf{a}$ is the number of probabilities required to specify the state and $r$ is an integer greater than or equal to one.

In Sec.\ \ref{gebitssection} we consider systems having $N_\mathsf{a}=2$.  We call such systems generalized bits (or \emph{gebits} for short).  First we show that pure states for a gebit are represented by points on a hypersphere.  In the next step we use {\bf P5} (that filters are non-flattening) for the first time. We use this to show that all points on the hypersphere represent pure states for a gebit.  We do this by considering filtering on a getrit (a system having $N_\mathsf{b}=3$) to put constraints on a gebit.  Finally we show that the hypersphere must, in fact, be a 2-sphere.  This means it corresponds to the standard Bloch sphere of quantum theory.  This proof uses an ingenious technique developed by Chiribella, D'Ariano, and Perinotti \cite{chiribella2010probabilistic, chiribella2010informational} based on teleportation.  This implies that $r=2$, or $K_\mathsf{a}=N_\mathsf{a}^2$.  It is by means that the simplicity axiom of \cite{hardy2001quantum} is eliminated.

In Sec.\ \ref{partIII} we finally reconstruct quantum theory for systems of arbitrary $N_\mathsf{a}$ by establishing that there is a correspondence between operations and operators and showing that we can construct a complete set of operations corresponding to every complete set of physical operators.  Indeed, we show that the set of operations
\begin{equation}
\begin{Diagram}{0}{-2}
\opbox[1]{zero1}{0,-4} \opsymbol{0} \opbox[1]{zero2}{4,-4} \opsymbol{0} \opbox[1]{zero3}{10,-4} \opsymbol{0}
\Opbox[30]{P}{0,0}
\opbox[4]{U1}{-7,5} \opsymbol{U[1]}
\opbox[4]{phi1}{0,8} \opsymbol{\mathnormal{\phi[1]}}
\opbox[4]{phi2}{4,8} \opsymbol{\mathnormal{\phi[2]}}
\opbox[4]{phi3}{10,8} \opsymbol{\mathnormal{\phi[N_\mathsf{a}]}}
\opbox[13]{V1}{-8.4,11} \opsymbol{V[1]}
\opbox[2]{aux1t}{-8.4,20} \opsymbol{\mathnormal{l}}
\opbox[2]{aux2t}{-4.4,20} \opsymbol{T}
\Opbox[33]{Q}{-1.2,16}
\opbox[1]{zero1t}{0,20} \opsymbol{0} \opbox[1]{zero2t}{4,20} \opsymbol{0} \opbox[1]{zero3t}{10,20} \opsymbol{0}
\thispoint{in}{-7,-7} \thispoint{out}{-12.4,23}
\wire{in}{P}{1}{6.75} \opsymbol{a}
\wire{Q}{out}{3}{1}\opsymbol{b}
\wire{P}{U1}{6.75}{2.5} \opsymbol{a}
\wire{V1}{Q}{2}{3} \opsymbol{b}
\wire{V1}{Q}{7}{8} \opsymbol{c}
\wire{V1}{Q}{12}{13} \opsymbol{d}
\wire{Q}{aux1t}{8}{1.5} \opsymbol{c}
\wire{Q}{aux2t}{13}{1.5} \opsymbol{d}
\wire{zero1}{P}{1}{15.5} \wire{zero2}{P}{1}{20.5} \wire{zero3}{P}{1}{28}
\wire{Q}{zero1t}{18.5}{1} \wire{Q}{zero2t}{23.5}{1} \wire{Q}{zero3t}{31}{1}
\wire{P}{phi1}{15.5}{2.5} \wire{P}{phi2}{20.5}{2.5} \wire{P}{phi3}{28}{2.5}
\wire{phi1}{Q}{2.5}{18.5} \wire{phi2}{Q}{2.5}{23.5} \wire{phi3}{Q}{2.5}{31}
\placelatex{7,-4}{\dots}  \placelatex{7,8}{\dots}\placelatex{7,20}{\dots}
\end{Diagram}
\end{equation}
can be set to correspond to any complete set of positive operators.  Note the set is generated by considering different outcomes $l$.

\newpage

\part{The circuit framework}\label{thecircuitframework}

\section{Operational description in the circuit framework}

In this section and the following two sections we will describe the circuit framework to which the postulates will be applied.  In this section we will deal with the operational description of circuits.

\subsection{Apparatuses}

Physically we perform experiments by placing apparatuses\index{apparatuses} (such as lasers, beamsplitters, lenses, \dots) next to each other in an appropriate way.
\begin{description}
\item[Apparatuses:] An apparatus, $\mathcal A$, is a physical device having: (i) a means for determining what constitutes a single use of this apparatus (this could be given by gating the use of the apparatus with respect to an external clock); (ii) apertures which can be placed next to apertures on other apparatuses; (iii) settings (fixed, for example, by setting some knobs); and (iiii) outcomes, denoted by $\mathsf x_{\mathcal A}$, (read off a meter for example).
\item[Apertures:] An aperture\index{apertures} is a hole or some other such like that allows one use of an apparatus to be connected to another use of an apparatus (in such a way that we can imagine a system passing from one apparatus use to next apparatus use).  In the case that the two apparatus uses are sequential uses of the same apparatus then we can think of the aperture as simply corresponding to the spacetime region which interfaces these two uses.
\end{description}
In this paper we are considering a limited class of experiments, namely those corresponding to linking up apparatuses using apertures.
This is sufficient for the purposes of describing the kinds of experiments which are done to test classical probability theory and quantum theory.  However, we can imagine more general notions of apparatus that can be linked up in other ways \cite{hardy2009operational}. It seems likely that we need a more general notion of apparatuses to give an operational account of general relativity for example.

\subsection{Operations}\label{operations}

\index{operations}We can extract another notion, that of the operation, which we obtain by adding some structure to the notion of an apparatus use.  The point of this extra structure is, as we will see below, to prescribe the ways in which we use the apparatuses.
\begin{description}
\item[Operations:] An operation, $\mathsf A$, corresponds to a single use of an apparatus where (a) we identify inputs with some apertures, (b) we identify outputs with some apertures,  (c) we fix the setting (select particular knob settings) and restrict our attention to a specified set of outcomes,  (in the simplest case we restrict our attention to a set containing a single outcome).
\item[Inputs and outputs:] An input\index{inputs} corresponds to using an aperture to allow a system of a specified type to pass into the apparatus.  An output\index{outputs} corresponds to using an aperture to allow a system of a specified type to pass out of the apparatus.
\item[Types:] When\index{types} we specify an input or output we must also specify the type.  The type corresponds to the kind of systems (electrons, photons, small rocks, \dots) we use an aperture for.
\item[Settings:] The setting\index{settings} is part of the specification of the operation.  We denote it by $\mathsf{ s(A)}$.  If we have a different setting for the same apparatus then we have a different operation.
\item[Outcome set:] \index{outcome sets}Each operation has an outcome set, denoted by $\mathsf{ o(A)}$, associated with it.  If $\mathsf{x_{\mathcal A}\in o(A)}$ then we say operation $\mathsf{A}$ \lq\lq happened".  The outcome set is part of the specification of $\mathsf A$.
\item[Compatible operations:] \index{compatible operations}Operations are said to be compatible if they correspond to the same apparatus use and the same knob settings but have different outcome sets.  We will denote such operations by ${\mathsf A}[i]$ and the corresponding outcome sets by ${\mathsf o}_i({\mathsf A})$.
\item[A complete set of operations] \index{complete set! of operations}is a set of operations that are compatible, whose outcome sets are disjoint, and where the union of these outcome sets is equal to the set of all possible outcomes.
\end{description}
We use the following notation to represent operations
\begin{equation}\label{operationA}
\begin{Diagram}{0}{-0.80}
\Opbox{A}{3,3}
\inwire[-5]{A}{1}\Opsymbol{a}
\inwire{A}{2}\Opsymbol{b}
\inwire[5]{A}{3}\Opsymbol{b}
\outwire[-5]{A}{1.5}\Opsymbol{b}
\outwire[5]{A}{2.5}\Opsymbol{c}
\end{Diagram}
~~~~~~~~~~~~\Longleftrightarrow ~~~~~~~~~~~~~\mathsf{A_{a_1b_2b_3}^{b_4c_5}}
\end{equation}
On the left hand side we have diagrammatic notation, on the right hand side we have symbolic notation.  In the diagrammatic notation, inputs enter at the bottom of the box and outputs leave at the top of the box.  In the symbolic notation inputs appear as subscripts and outputs appear as superscripts.  When we use symbolic notation it is necessary to label the inputs and outputs with integers so we can identify which outputs are connected to which inputs (see below). In the diagrammatic notation these integers are not necessary since we can see, by looking at the diagram, where the wires go.  Note that the knob setting and outcome set are taken to be absorbed into the specification of the operation, $\mathsf A$, so we do not represent them explicitly in this notation.

\subsection{Wires}

\index{wires}Experiments are performed by placing apparatuses next to each other.  In practise we can place apparatuses next to each other in any way allowed by their physical geometry.  For example, all the apparatuses can be piled into a box kept in a dark dusty corner of the laboratory.  We do not, generally, expect useful physics to come out of such haphazard arrangements.  For this reason we have introduced the notion of operations which, in conjunction with the following wiring rules, prescribe the use of apparatuses when building circuits.
\begin{description}
\item[Wires:] A wire corresponds to placing two apertures next to each other.  For any collection of operations connected by wires we demand
\begin{description}
\item[\emph{Directed:}] A wire connects an output to an input.
\item[\emph{One wire:}] At most one wire can be connected to any given input or output.
\item[\emph{Type matching:}] \index{type matching}Wires can only connect outputs to inputs of the same type.  The wire therefore has an associated type, denoted $\mathsf a$, $\mathsf b$, \dots
\item[\emph{No closed loops:}] Wires are directed (they go from output to input). We demand that if we trace forward along wires through the operations
then we cannot get back to the same operation. Since operations correspond to single uses of apparatuses, this corresponds to ruling out closed time-like loops.
\end{description}
\end{description}
We use the following notation (diagrammatic on the left, symbolic on the right) to represent two operations being joined by a wire:
\begin{equation}\label{ABwired}
\begin{Diagram}{0}{-1.2}
\Opbox{A}{7,0}
\Opbox[2]{B}{3,7}
\inwire[-5]{A}{1}\Opsymbol{a}
\inwire{A}{2}\Opsymbol{a}
\inwire[5]{A}{3}\Opsymbol{b}
\wire{A}{B}{1.5}{2}\opsymbol{b}
\outwire[5]{A}{2.5}\Opsymbol{c}
\inwire[-5]{B}{1}\Opsymbol{a}
\outwire[-5]{B}{1}\Opsymbol{d}
\outwire[5]{B}{2}\Opsymbol{c}
\end{Diagram}
~~~~~~~~ \Longleftrightarrow ~~~~~~~~~ \mathsf{A_{a_1a_2b_3}^{b_4c_5}}\mathsf{B_{a_6b_4}^{d_7c_8}}
\end{equation}
In the symbolic notation, the wire is represented by a repeated index. This is the reason we have to label the inputs and outputs with integers.  These integers are just labels and have no meaning beyond the fact that they tell us which inputs and outputs are joined.  We could permute the integers in any way without changing the physical meaning.

The physical meaning of the wires is that they tell us which input and output apertures are placed immediately next to one another.  This is a little like the diagram that often accompanies a self-assembly piece of furniture.  This diagram shows an exploded view of the piece of furniture with lines drawn from one piece to another showing how it is to be assembled.  The main difference here that an experiment may have moving parts and so our diagrams constitute something that happens in spacetime, whereas the piece of furniture is a static object.   Note, in particular, that the wires in the diagram do not correspond to actual wires. If we have actual wires in an experiment (such as optical fibers) then we must treat them as operations (they have an input, and output, and are connected up to other operations).  Only in the idealized case where physical wires correspond to the identity transformation could we consider treating them as wires as considered here.

Type matching ensures that we use the apparatuses in the way they are intended to be used.  For example, it prohibits us from matching an output for small rocks with an input for photons.  In practice, there is nothing to prevent us from mismatching types in this way.  However, it would probably lead to the malfunctioning of the apparatuses.  In such a circumstance, any operational laws of physics we have will not enable us to make predictions.  This is a genuine limitation of the operational approach to physics.  Physics should tell us what will happen in any circumstance.

\subsection{Fragments}\label{fragments}

\index{fragments}The object in (\ref{ABwired}) above is an example of a fragment.
\begin{description}
\item[Fragments:] A fragment is formed by wiring together a bunch of operations.  We will denote fragments by uppercase sans serif $\mathsf A$, $\mathsf B$, etc. (just as for operations which are, in fact, special cases of fragments). Note that a fragment can consist of disjoint parts (not connected by wires).
\item[Example:]  For example, let the fragment $\mathsf E$ be given by
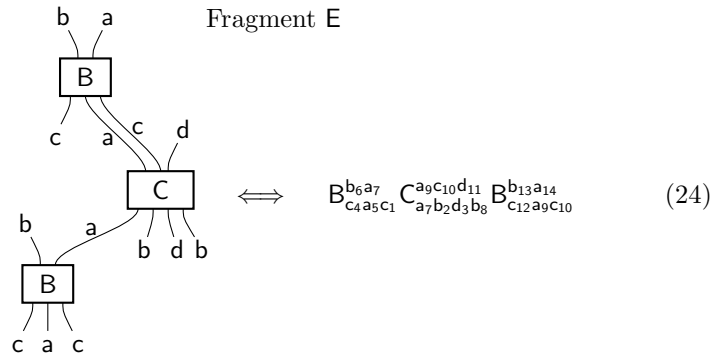
\begin{equation}\label{fragmentE}
\begin{Diagram}{-4}{-1}
%\Opbox{A}{4.1,-6}
\Opbox{B}{0,-1}\Opbox[4]{C}{6,4}\opbox{F}{2,10}\opsymbol{B}
\inwire[-5]{B}{1}\Opsymbol{c}\inwire{B}{2}\Opsymbol{a} \inwire[5]{B}{3}\Opsymbol{c}
%\wire{A}{B}{1}{3} \opsymbol{c}\wire{A}{C}{2}{2}\opsymbol{b}\wire{A}{C}{3}{3}\otherside\opsymbol{d}
\outwire[-5]{B}{1.5}\Opsymbol{b}\wire{B}{C}{2.5}{1}\opsymbol{a}
%\inwire[2]{H}{2}\Opsymbol{b}\inwire[4]{H}{3}\Opsymbol{d}
\inwire[-3]{C}{2} \Opsymbol{b} \inwire[3]{C}{3}\Opsymbol{d}\inwire[6]{C}{4}\Opsymbol{b}
\wire{C}{F}{1.5}{2}\opsymbol{a}\wire{C}{F}{2.5}{3}\otherside\opsymbol{c}\outwire[5]{C}{3}\Opsymbol{d}
\inwire[-5]{F}{1}\Opsymbol{c}
\outwire[-5]{F}{1.5}\Opsymbol{b}\outwire[5]{F}{2.5}\Opsymbol{a}
\placelatex{12,13}{\text{Fragment} \ensuremath{\mathsf E}}
\end{Diagram} \hspace{-1.6cm}
\Longleftrightarrow ~~~~ \mathsf{B_{c_4a_5c_1}^{b_6a_7} C_{a_7b_2d_3b_8}^{a_9c_{10}d_{11}} B_{c_{12} a_9 c_{10}}^{b_{13}a_{14}}}
\end{equation}
Notice that the operation $\mathsf B$ is used twice here (this corresponds to two separate uses of the same type of apparatus).
\item[Features:] A fragment will, in general, have some open inputs and outputs left over.  In particular, it may have outputs which could, in principle, be wired into inputs on the same fragment without violating the no closed loops assumption. It can be a part of a much bigger fragment.  {\it A fragment is therefore the circuit language equivalent of an arbitrary region of space-time.}  We allow a fragment to consist of disjoint parts (the circuit equivalent of an arbitrary region of space-time consisting of disjoint parts).  The outcome $\mathsf{ x_\mathcal{E}}$ is given by specifying the outcome at each operation making up the fragment.  In specifying the fragment, $\mathsf E$, we give an outcome set $\mathsf{ o(E)}$ (this is the cartesian product of the outcome sets for each operation), the settings $\mathsf {s(E)}$ (specifying this means specifying a setting at each operation), and the wiring $\mathsf{w(E)}$.  We will usually denote the settings and wiring by $\mathsf{sw(E)}$ for brevity.   The statement that a given fragment \lq\lq has happened" means that the set up with all the corresponding apparatuses placed together in accordance with the given settings and wiring, $\mathsf{sw(E)}$, was implemented and the outcome, $\mathsf{x_\mathcal{E}}$, was in the given outcome set, $\mathsf{o(E)}$.
\item[Setups:]  \index{setups}Each fragment naturally belongs to a class of fragments corresponding to the same apparatus uses with the same knob settings and the same wiring but having different outcome sets.  We will say that these fragments correspond to \lq\lq the same setup".  We will denote the members of such a set of fragments as $\mathsf{E}[i]$, the outcome sets by $\mathsf{ o}_i(\mathsf{ E})$, and the settings and wirings by $\mathsf{ sw}_i(\mathsf{E})$.
\item[Deterministic fragments:] \index{deterministic fragments}A deterministic fragment is one for which the set of outcomes is equal to the the set of all possible outcomes.  The set of all outcomes is the cartesian product of the sets of all outcomes at each operation constituting the fragment.  Since the outcome must be in the set of all possible outcomes, deterministic fragments always happen when the corresponding setup is put in place (we borrow this terminology from \cite{chiribella2010probabilistic}).
\end{description}

\subsection{Circuits}

\index{circuits}We now introduce an important notion.
\begin{description}
\item[A circuit] is formed when we wire together a bunch of operations and {\it have no open inputs or outputs left over}. Circuits are special cases of fragments. We will denote them by uppercase sans serif font, $\mathsf A$, $\mathsf B$, \dots.  For example, let the circuit $\mathsf H$ be
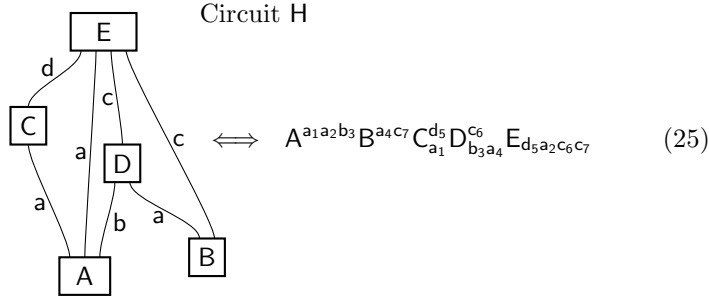
\begin{equation}
\begin{Diagram}{0}{-1.4}
%\constructiongrid{-2,-4}{19,15}
\Opbox{A}{8,-2}\Opbox[2]{B}{14.5,-1}\Opbox[2]{C}{5,6}\Opbox[2]{D}{10,4}\Opbox[4]{E}{9,11}
\wire{A}{C}{1}{1.5}\opsymbol{a}\wire{A}{E}{2}{2}\opsymbol{a}\wire{A}{D}{3}{1}\otherside\opsymbol[5,0]{b}
\wire{B}{D}{1}{2}\opsymbol{a}\wire{B}{E}{2}{4}\otherside\opsymbol{c}\wire{D}{E}{1.5}{3}\opsymbol{c}\wire{C}{E}{1.5}{1}\opsymbol[0,8]{d}
\placelatex{17,12}{\text{Circuit} \ensuremath{\mathsf H} }
\end{Diagram}
\hspace{-1.5cm}
\Longleftrightarrow ~~  \mathsf{   A^{a_1a_2b_3} B^{a_4c_7} C_{a_1}^{d_5} D_{b_3a_4}^{c_6} E_{d_5a_2c_6c_7}  }
\end{equation}
\item A circuit can consist of disjoint parts.
\item[The outcome set] for a circuit is given by the cartesian product of the outcome sets for each of the operations making up the circuit.  We say that the circuit has \lq\lq happened" if the outcome seen at each apparatus use is in the outcome set of the corresponding operation.
\end{description}

\subsection{Proto-systems}

We will later give a definition of what constitutes a \emph{system} using the notion of \emph{filters}. The idea is that a system is what we have after a filter. One example of a filter is the \lq\lq do nothing" filter (i.e.\ just the identity).   This allows us to define a special case of system which we will call \emph{ proto-systems}.  A proto-system is what we have after a \lq\lq do nothing" filter.  All proto-systems are systems.

To define proto-systems first we think about how we can break up circuits into fragments.  In fact we can do this in any arbitrary way.  One particular way we can do it is with synchronous sets of wires.
\begin{quote}
{\bf A synchronous set of wires:} \index{synchronous set of wires}is a set of wires having the property that it is impossible to trace forward from any wire in the set to any other wire in the set.  By tracing forward we mean tracing along wires from output to input through the operations.
\end{quote}
For example, the three wires picked out below
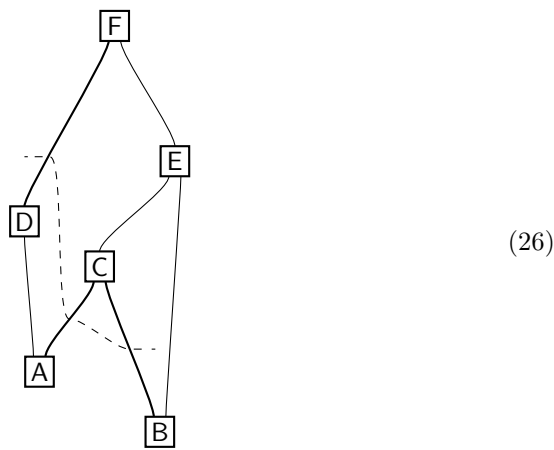
\begin{equation}
\begin{Diagram}[0.8]{0}{-1.8}
\Opbox[2]{A}{0,0}
\Opbox[2]{B}{8,-4}
\Opbox[2]{D}{-1,10}
\Opbox[2]{C}{4,7}
\Opbox[2]{E}{9,14}
\Opbox[2]{F}{5,23}
\wire{A}{D}{1}{1.5} \thickwire{A}{C}{2}{1}  \thickwire{B}{C}{1}{2}  \wire{B}{E}{2}{2}
\wire{C}{E}{1.5}{1}\thickwire{D}{F}{1.5}{1}  \wire{E}{F}{1.5}{2}
\begin{foliation}{-1}{7.7}
\Startfoliate{D}{F}{1.5}{1}{-0.2}\continuefoliate{A}{C}{2}{1}\finishfoliate{B}{C}{1}{2}
\end{foliation}
\end{Diagram}
\end{equation}
constitute a synchronous set.
A complete set of synchronous wires, or a hypersurface, is a synchronous set of wires which partitions the circuit into two.  We can foliate a circuit with hypersurfaces.  For example, a complete foliation (one which includes every wire at least once) is given below.
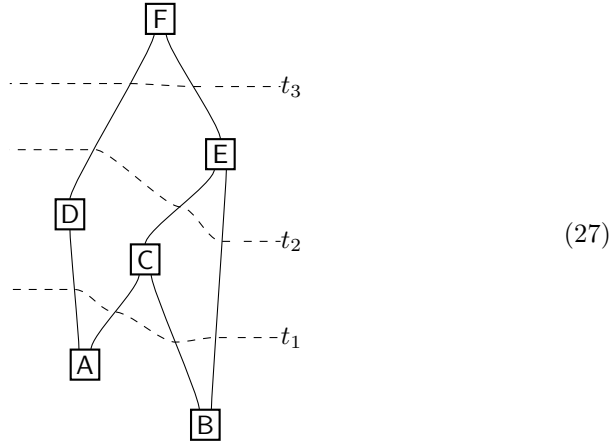
\begin{equation}
\begin{Diagram}[0.8]{0}{-1.8}
\Opbox[2]{A}{0,0}
\Opbox[2]{B}{8,-4}
\Opbox[2]{D}{-1,10}
\Opbox[2]{C}{4,7}
\Opbox[2]{E}{9,14}
\Opbox[2]{F}{5,23}
\wire{A}{D}{1}{1.5} \wire{A}{C}{2}{1}  \wire{B}{C}{1}{2}  \wire{B}{E}{2}{2}
\wire{C}{E}{1.5}{1}\wire{D}{F}{1.5}{1}  \wire{E}{F}{1.5}{2}
\begin{foliation}{-5}{13}
\startfoliate{A}{D}{1}{1.5}\continuefoliate{A}{C}{2}{1}\continuefoliate{B}{C}{1}{2}\Finishfoliate{B}{E}{2}{2}{-0.2} \otherside\putlatex{\ensuremath{t_1}}
\Startfoliate{D}{F}{1.5}{1}{-0.2}\continuefoliate{C}{E}{1.5}{1}\Finishfoliate{B}{E}{2}{2}{0.2}
\otherside\putlatex{\ensuremath{t_2}}
\Startfoliate{D}{F}{1.5}{1}{0.2}\finishfoliate{E}{F}{1.5}{2}
\otherside\putlatex{\ensuremath{t_3}}
\end{foliation}
\end{Diagram}
\end{equation}
It is always possible to provide a complete foliation for a circuit \cite{hardy2009foliable}.

We now define
\begin{quote}
{\bf A proto-system} \index{proto-system}is associated with a wire or synchronous set of wires.  The system type is determined by the wire types.
\end{quote}
For example, a proto-system of type $\mathsf{ aabc}$ is associated with a set of wires of type $\mathsf a$, $\mathsf a$, $\mathsf b$, and $\mathsf c$.  The proto-system $\mathsf{aabc}$ can, for example, be regarded as a composite with components of types $\mathsf a$, $\mathsf{ab}$, and $\mathsf c$.  As a matter of notation, we will sometimes refer to a proto-system that may be composite with a single letter.  For example, the system $\mathsf{aabc}$ could be denoted by the symbol $\mathsf d$.

\subsection{Preparations, transformations, and results}\label{preparationstransformationseffects}

We can partition a circuit into parts using systems.  The resulting fragments are
\begin{description}
\item[Preparations.] \index{preparations}Any fragment having open output but no open inputs is a preparation.  Here are some examples:
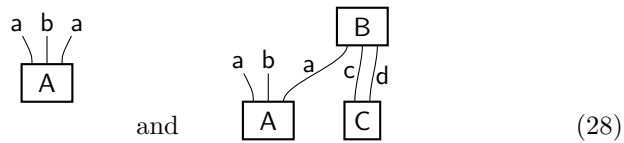
\begin{equation}
\begin{Diagram}{0}{0.5}
\Opbox{A}{0,0}
\outwire[-5]{A}{1} \Opsymbol{a} \outwire{A}{2} \Opsymbol{b} \outwire[5]{A}{3} \Opsymbol{a}
\end{Diagram}
~~~~~\text{and}~~~~~
\begin{Diagram}{0}{0}
\Opbox{A}{0,0} \Opbox{B}{5,5} \Opbox[2]{C}{5,0}
\wire{A}{B}{3}{1} \opsymbol{a}   \wire{C}{B}{1}{2} \opsymbol{c} \wire{C}{B}{2}{3}\otherside\opsymbol{d}
\outwire[-5]{A}{1} \Opsymbol{a} \outwire{A}{2} \Opsymbol{b}
\end{Diagram}
\end{equation}
The outputs necessarily constitute a synchronous set of wires (when a preparation is wired up to another fragment).  Hence, preparations can be thought of as preparing a proto-system in some given state.
\item[Transformations.] \index{transformations}A transformation is a fragment having open inputs and outputs that is used in \emph{transformation mode} (we will explain what this means below).  Here are some examples:
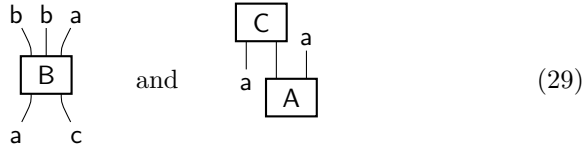
\begin{equation}
\begin{Diagram}{0}{0}
\Opbox{B}{0,0}
\inwire[-5]{B}{1}\Opsymbol{a}\inwire[5]{B}{3} \Opsymbol{c}
\outwire[-5]{B}{1}\Opsymbol{b}\outwire{B}{2}\Opsymbol{b}\outwire[5]{B}{3}\Opsymbol{a}
\end{Diagram}
~~~~~\text{and}~~~~~~
\begin{Diagram}{0}{-0.3}
\Opbox{A}{0,0} \Opbox{C}{-1.6,4} \wire{A}{C}{1}{3}
\inwire{C}{1} \Opsymbol{a} \outwire{A}{3} \Opsymbol{a}
\end{Diagram}
\end{equation}
The fragment on the right could have the output fed either directly into the input or indirectly (via some other operations).  If this is the case then the fragment is not being used in transformation mode.  For us to say that a fragment is being used in transformation mode it must be the case that the outputs of that fragment lie on a later hypersurface than the inputs.  This means that we should not be able to trace forward from an output along wires to an input on the fragment.  Any fragment having some open inputs and outputs can be put in transformation mode.
\item[Results.] \index{results}These are fragments having open inputs and no open outputs.  Here are some examples
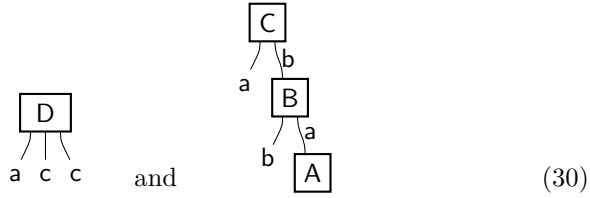
\begin{equation}
\begin{Diagram}{0}{0.8}
\Opbox{D}{0,0}
\inwire[-5]{A}{1} \Opsymbol{a} \inwire{A}{2} \Opsymbol{c} \inwire[5]{A}{3} \Opsymbol{c}
\end{Diagram}
~~~~~\text{and}~~~~~~
\begin{Diagram}{0}{0}
\Opbox[2]{A}{0,0} \Opbox[2]{B}{-1.2,4} \Opbox[2]{C}{-2.4,8}
\wire{A}{B}{1}{2}\otherside \opsymbol{a} \wire{B}{C}{1}{2} \otherside\opsymbol{b}
\inwire[-5]{B}{1} \Opsymbol{b} \inwire[-5]{C}{1} \Opsymbol{a}
\end{Diagram}
\end{equation}
The inputs necessarily constitute a synchronous set of wires (when wired up to another fragment).  Results can be thought of as corresponding to measurement outcomes.
\end{description}
We note that there are fragments which cannot be understood as being equal to one of these special types (namely those fragments not bounded by input and output synchronous sets of wires).  Nevertheless, any circuit can be broken up into preparations, transformations, and results.  This is clearly true because, at the most fine grained level, operations always correspond to a preparation, transformation, or result.   Having identified a given preparation, transformation, or result, we can use it to build different circuits.

Deterministic preparations, transformations, and results are ones for which the set of outcomes is equal to the set of all possible outcomes.

A measurement is made up of compatible results.  We define:
\begin{quote}
{\bf A measurement}\index{measurements}, $\{ \mathsf{B}_\mathsf{a_1}[l]: l=1, 2, \dots\}$,  is a collection of results corresponding to the same setup having outcomes sets, $\mathsf o_l(B)$, which are disjoint and whose union is the set of all outcomes. We can simply say that the outcome of the measurement is $l$ (corresponding to the label of the outcome set).
\end{quote}

\section{Probabilities in the circuit framework}

Up to now we have simply discussed the operational description of experiments.  Physical theories go beyond mere description.  They make predictions. In the context of the the circuit framework predictions concern probabilities.  In this section we will introduce probabilities.   We need this so that we can introduce the notion of state and other related concepts in the section.

Modern physics makes extensive use of probability.  However, there is much debate about what the correct interpretation of probability is.  There are various options.  It could be interpreted as a relative frequency, as a propensity, as an objective degree of belief, as a subjective degree of belief, or some hybrid of these \cite{gillies2002philosophical}.  There are serious problems with all these approaches.  Fuchs, who adopts point of view that probabilities are subjective degrees of belief, believes that the quantum formalism may follow from a proper understanding of how to interpret probability \cite{fuchs2010qbism}. This is a significantly deeper point of view than adopted in the present work.  We will simply define probabilities to have certain mathematical properties (the usual mathematical properties) without attempting to provide interpretation of probability or justification for these properties.  Ultimately, however, the goal of understanding the nature of probability in the context of physics is likely to play an important role in resolving foundational issues and pushing physics forward beyond quantum theory.

\subsection{Assigning probabilities}

Whichever interpretation of probability we adopt, at some point we will wish to \emph{assign} a probability for something to happen.  In our context this means assigning a probability for a fragment to happen.  If we do alow ourselves to assign a probability with a fragment then we denote this by
\begin{equation}
\text{Prob}(\mathsf{A}) :=\text{Prob}(\mathsf{x}_\mathcal{A}\in\mathsf{o(A)}|\mathsf{sw(A)})
\end{equation}
Here $\mathsf{sw(A)}$ stands for the settings and wiring of the fragment $\mathsf{A}$.
Importantly,
\begin{quote}
\emph{~~~~ We do not assume we can always assign probabilities. }
\end{quote}
In particular, we may have a fragment with open inputs and outputs.  The probability of the fragment happening may depend, for example, on what we send into the open inputs.  That is, the probability may depend on conditions that are not given.  In these circumstances the actions of some adversary (who has control over what is sent into the inputs for example) may influence how likely the fragment is to happen and so we should not assign a probability.  In general we will allow ourselves to assign a probability with a fragment when we know it is independent of the actions of any adversary who has control over parts of the world not associated with the given fragment (e.g.\ the choice of settings, wiring, and outcome sets for operations that are not part of the given fragment).

We will consider fragments resulting from wiring together two fragments.  Thus, if we have fragments $\mathsf{A}$ and $\mathsf{B}$ then we will represent a fragment resulting from wiring them together as $\mathsf{AB}$.   We have suppressed the input and output labels here because we are dealing with the general case.  There may be more than one way of wiring together two fragments to obtain a bigger fragment. One special case that is always possible is where we do not place any wires between them.   For example, putting back in the subscripts and superscript, we could have fragments $\mathsf{A_{a_1b_2}^{c_3}}$ and $\mathsf{B_{c_4}^{a_5c_6}}$.  In this case, $\mathsf{AB}$ could represent $\mathsf{A_{a_1b_2}^{c_3}} \mathsf{B_{c_3}^{a_5c_6}}$, $\mathsf{A_{a_1b_2}^{c_3}} \mathsf{B_{c_3}^{a_1c_6}}$, or $\mathsf{A_{a_1b_2}^{c_3}} \mathsf{B_{c_4}^{a_5c_6}}$ to list just three possibilities.  We will consider joint probabilities for such fragments.  We will denote the joint probability by
\begin{equation}
\text{Prob}(\mathsf{AB}) :=\text{Prob}(\mathsf{x_\mathcal{AB}\in o(AB)|sw(AB)})
\end{equation}
We will also consider assigning conditional probabilities.  We denote conditional probabilities by
\begin{equation}
\text{Prob}(\mathsf{A|B}) :=\text{Prob}(\mathsf{x_\mathcal{A}\in o(A)|x_\mathcal{B}\in o(B), sw(AB)})
\end{equation}
We have suppressed the subscripts and superscripts in the notation $\mathsf{AB}$ because we are dealing with the general case.

The idea of \emph{assigning a probability} is essentially a primitive here.  We will allow ourselves to associate a probability (or, more generally, a conditional probability) in certain yet to be specified circumstances (when the actions of an adversary would make no difference).  Exactly what it means to assign a probability will depend on which interpretation of probability one adopts.

The idea that one should not always allow oneself to assign a probability does not seem to be much discussed in the literature.  However, it is a fact that the physical theories we have only allow us to calculate probabilities for very special situations (see examples in Sec.\ \ref{wellconditionedprobabilities}).  These situations are, generally, defined by the causal structure assumed to be operating in the background. For example, we would not make a probabilistic prediction for the 20th time step given \emph{only} information about the 5th time step (and no information about what happened in between).  But we may be able to make a probabilistic prediction for the 20th time step given information about the 19th time step (though even here we have to be careful - see the second example in Sec.\ \ref{wellconditionedprobabilities}).   The fact that physical theories only allow us to calculate probabilities for rather special situations is not normally made explicit.   In this paper we wish to take a more general point of view and so we do need to be explicit about this.

\subsection{Properties of probabilities}\label{propertiesofprobabilities}

\index{probabilities, properties of}We wish to demand the following  properties of probabilities.
\begin{description}
\item[Non-negative.]  If we can assign a probability $\text{Prob}(\mathsf{A})$  for some fragment $\mathsf{A}$ then
    \begin{equation}\label{probpropgeqzero}
    0\leq \text{Prob}(\mathsf{A})
    \end{equation}
\item[Deterministic fragments.]  For any deterministic fragment, $\mathsf{A}[\Omega]$ (where $\Omega$ is understood to be the set of all outcomes), we can assign a probability and this probability is equal to one, i.e.\
    \begin{equation}\label{probonefordetfrag}
    \text{Prob}(\mathsf{A}[\Omega]) =1
    \end{equation}
\item[Additivity.]  Let $\mathsf{A}[\mu]$, $\mathsf{A}[\nu]$, and $\mathsf{A}[\mu\cup\nu]$ be three fragments associated with the same setup having outcome sets equal to $\mu$, $\nu$, and $\mu\cup\nu$ respectively. Let $\mu$ and $\nu$ be disjoint.   If we can assign probabilities to any two of $\text{Prob}(\mathsf{A}[\mu])$, $\text{Prob}(\mathsf{A}[\nu])$, and  $\text{Prob}(\mathsf{A}[\mu\cup \nu])$, then we can assign a probability to the third and, further,
    \begin{equation}\label{additivityofprobs}
    \text{Prob}(\mathsf{A}[\mu\cup\nu ])
    = \text{Prob}(\mathsf{A}[\mu]) + \text{Prob}(\mathsf{A}[\nu])
    \end{equation}
\item[Joint probabilities] Let $\mathsf{A}[\Omega]$ be a deterministic fragment and $\mathsf B$ be another fragment. Let $\mathsf{A}[\Omega]\mathsf{B}$ be a fragment resulting from wiring together these two fragments.  Then if we can assign a probability to either of $\text{Prob}(\mathsf{A}[\Omega]\mathsf{B})$ and $\text{Prob}(\mathsf{B})$ then we can assign a probability to the other and, further,
    \begin{equation}\label{jointprobproperty}
    \text{Prob}(\mathsf{A}[\Omega]\mathsf{B}) = \text{Prob}(\mathsf{B})
    \end{equation}
\item[Conditional probabilities.]  Let $\mathsf{A}$, $\mathsf{B}$, be two fragments and let $\mathsf{AB}$ be a fragment that results from wiring them together.  If we can assign probabilities $\text{Prob}(\mathsf{AB})$ and $\text{Prob}(\mathsf{B})$ and if, further, $\text{Prob}(\mathsf{B})\not=0$, then we can assign a conditional probability $\text{Prob}(\mathsf{A|B})$ where
    \begin{equation}\label{condprobgeneral}
    \text{Prob}(\mathsf{A|B})=\frac{\text{Prob}(\mathsf{AB})}{\text{Prob}(\mathsf{B})}
    \end{equation}
    (in some treatments this is regarded as a definition of conditional probability).
\end{description}

These are the standard properties required of probabilities with the important additions that we have take some care to deal with the fact that we cannot always assign probabilities and we have stated them for fragments (comprised out of operations and wires).  Note that we only consider finite $|\Omega|$ so we do not concern ourselves with measure theoretic concerns that might result from having an infinite number of outcomes.  This is consistent with the operational approach taken in this paper - the data taken in any real experiment can only have finite $|\Omega|$.  We note the following
\begin{enumerate}
\item It follows from (\ref{probpropgeqzero}), (\ref{probonefordetfrag}) and (\ref{additivityofprobs}) that probabilities, must be less than or equal to one.
%\item From these properties it is fairly straightforward to prove that conditional probabilities, such as $\text{Prob}(\mathsf{A}[l]|\mathsf{B})$, satisfy the same properties as probabilities such as $\text{Prob}(\mathsf{A})$ given under the \lq\lq disjoint outcome sets" heading above (as long as $\text{Prob}(\mathsf{A}[\Omega]\mathsf{B})\not=0$).  % should I prove this somewhere (an appendix?).
\item We can write
\begin{equation}\label{condproducteqn}
\text{Prob}(\mathsf{AB})={\text{Prob}(\mathsf{A|B})}{\text{Prob}(\mathsf{A}[\Omega]\mathsf{B})}
\end{equation}
but only with a caveat.  In the case where $\text{Prob}(\mathsf{B})\not=0$ equation (\ref{condproducteqn}) follows from (\ref{condprobgeneral}).  The caveat concerns the case when $\text{Prob}(\mathsf{B})=0$.  It follows from the properties of probabilities (in particular, using (\ref{jointprobproperty}) and (\ref{additivityofprobs})) that we must also have $\text{Prob}(\mathsf{A}\mathsf{B})=0$.   In this case we cannot associate a conditional probability $\text{Prob}(\mathsf{A|B})$.  However, no matter what value we put in the place of $\text{Prob}(\mathsf{A|B})$ (as long as it in the range 0 to 1)  equation (\ref{condproducteqn}) will be satisfied (with both sides being equal to zero).  We will understand this equation in this way. Note, by way of example, that in the proof given in Sec.\ \ref{probsfactorsection} below works well with this understanding of (\ref{condproducteqn}).
\end{enumerate}

\subsection{Assumption 1 for the circuit framework}\label{Assumponesec}

To get ourselves going we need some situation in which we can assign probabilities.  A circuit does not have open inputs or outputs so it is reasonable to assume that it is not subject to outside influences.  We make the following assumption.
\begin{quote}
{\bf Assump 1.}  \index{assumptions!Assump 1@\textbf{Assump 1}}We can assign a probability with any given circuit (the probability that the circuit \lq\lq happens"), and this probability depends only on the specification of the given circuit (the knob settings and outcome sets at the operations, and the wiring).
\end{quote}
Hence we can meaningfully speak of $\text{Prob}(\mathsf{A})$ for any circuit, $\mathsf{A}$.

It follows from {\bf Assump 1} that we can assign a conditional probability $\text{Prob}(\mathsf{A|B})$ when $\mathsf{AB}$ is a circuit.  Let $\mathsf{A}[\Omega]$ be the deterministic fragment corresponding to same setup as $\mathsf A$.  Then, for the special case where $\mathsf{AB}$ is a circuit, (\ref{jointprobproperty}) and(\ref{condproducteqn}) give
\begin{equation}\label{condforcircuit}
\text{Prob}(\mathsf{A|B})=\frac{\text{Prob}(\mathsf{AB})}{\text{Prob}(\mathsf{A}[\Omega]\mathsf{B})}
\end{equation}
(as long as $\text{Prob}(\mathsf{A}[\Omega]\mathsf{B})$ is non-zero).  This is because we can, by {\bf Assump 1}, assign probabilities for both the numerator and the denominator (as they correspond to circuits).

\subsection{Probabilities factor for composite circuits}\label{probsfactorsection}

\index{probabilities for composite circuits}It follows from {\bf Assump 1} that if $\mathsf A$ and $\mathsf B$ are both circuits then, for the composite circuit $\mathsf{ AB}$ (consisting of the disconnected parts $\mathsf A$ and $\mathsf B$), we have
\begin{equation}\label{probfatorises}
\text{Prob}(\mathsf{AB})=\text{Prob}(\mathsf{A})\text{Prob}(\mathsf{ B})
\end{equation}
because
\begin{flalign*}
\text{Prob}&(\mathsf{AB}) = \text{Prob}\big(\mathsf{x_\mathcal{A}\in o(A), x_\mathcal{B}\in o(B)|\,sw(A), sw(B)}\big)  \\
 =& \text{Prob}\big(\mathsf{x_\mathcal{A}\in o(A)| x_\mathcal{B}\in o(B), sw(A), sw(B)}\big) \text{Prob}\big(\mathsf{x_\mathcal{B}\in o(B)|\, sw(A), sw(B)}\big)  \\
 =& \text{Prob}\big(\mathsf{x_\mathcal{A}\in o(A)|\, sw(A)}\big) \text{Prob}\big(\mathsf{ x_\mathcal{B}\in o(B)|\, sw(B)}\big)
\end{flalign*}
where we use (\ref{condproducteqn}) in the second line and {\bf Assump 1} in the third line.  We can write this proof more succinctly as
\begin{equation} \label{probfactorisessuccinct}
\text{Prob}(\mathsf{AB})=\text{Prob}(\mathsf{A|B}) \text{Prob}(\mathsf{B}) = \text{Prob}(\mathsf{A})\text{Prob}(\mathsf{ B})
\end{equation}
Note that if $\text{Prob}(\mathsf{B})=0$ then we have to be careful using the conditional probability since then it is not clear we can assign it using (\ref{jointprobproperty}).  However, we still have $\text{Prob}(\mathsf{AB})=\text{Prob}(\mathsf{A})\text{Prob}(\mathsf{ B})$ since it follows from (\ref{additivityofprobs}) and (\ref{jointprobproperty}) that $\text{Prob}(\mathsf{B})=0$ implies $\text{Prob}(\mathsf{AB})=0$.

Chiribella, D'Ariano, and Perinotti  take the factorization of probabilities for disjoint circuits as a starting point in their circuit model \cite{chiribella2010probabilistic, chiribella2010informational}.

\subsection{Well conditioned probabilities}\label{wellconditionedprobabilities}

The probability $\text{Prob}(\mathsf{A|B})$ makes sense if $\mathsf{AB}$ is a circuit but what if it is a fragment?  We cannot assume that we can assign probabilities with arbitrary fragments in general as they may have open inputs and outputs. However, there may be particular situations in which we can meaningfully speak of probabilities for fragments even when do they have open inputs and outputs.  For this end, the following definition is useful.
\begin{quote}
{\bf Well conditioned probabilities:} \index{well conditioned probabilities}If $\mathsf A$ and $\mathsf B$ are fragments, we will say that we have a \emph{well conditioned probability}, $\text{Prob}\mathsf{(A|B)}$, if
\begin{equation}
\text{Prob}\mathsf{(A|BC)} = \text{Prob}\mathsf{(A|BD)}
\end{equation}
for all fragments $\mathsf C$ and $\mathsf D$ such that $\mathsf{ABC}$ and $\mathsf{ABD}$ are circuits.
\end{quote}
If we have a well conditioned probability, $\text{Prob}\mathsf{(A|B)}$, it is is fully determined by $\mathsf{A}$ and $\mathsf{B}$ as long as the fragment $\mathsf{AB}$ is part of a bigger circuit.  In such circumstances we can meaningfully assign a probability with $\text{Prob}\mathsf{(A|B)}$.  Note, however, the important caveat that this is true as long as $\mathsf{AB}$ is part of a bigger circuit.   If we have a setup in which we do not close all open inputs and outputs (so the fragment is not part of a bigger circuit) then it is not clear whether we can expect any theory to be useful.     If we had open inputs not connected to some type-matched outputs then, in principle, anything might enter the apparatus through them.  There is no guarantee that such things would not damage the experimental equipment (in which case we could not expect our theory to make useful predictions at all).  For example, we may leave open inputs that are meant for photons but may happen to have small rocks impinging on them.  Given such considerations, it is good experimental practice to close all open inputs.  In general, in operational theories, we can only expect reasonable predictions when all inputs are type-matched.  What about if we have no open inputs but still have open outputs? In causal theories there are no influences from the future (see {\bf T\ref{deteffect}} in Sec.\ \ref{uniqueimplications}).  In such theories we can ignore the future.  The action of ignoring the future implements a deterministic result and, under this, the outputs are effectively closed.  However, (i) the theory under consideration may not be causal and, (ii) we may want to allow the situation in which an adversary can condition on something that happens in the future (so then we cannot ignore what actually happens in the future - see the second example below).  For these reasons it is good experimental practise to close all outputs as well as inputs.  We will assume that any fragment we consider is part of an experiment corresponding to a bigger circuit.

For the special case where $\mathsf{AB}$ is a circuit we saw in Sec.\ \ref{Assumponesec} that the conditional probability $\text{Prob}(\mathsf{A|B})$ is given by the standard equation
\begin{equation}
\text{Prob}(\mathsf{A|B})=\frac{\text{Prob}(\mathsf{AB})}{\text{Prob}(\mathsf{A}[\Omega]\mathsf{B})}
\end{equation}
since, by {\bf Assump 1}, we can assign the probabilities given in the numerator and denominator.  What happens when $\mathsf{AB}$ is a fragment?  If we have a well conditioned probability $\text{Prob}(\mathsf{A|B})$ then, for any $\mathsf C$ such that $\mathsf{ABC}$ is a circuit,
\begin{equation}
\text{Prob}(\mathsf{A|B})= \text{Prob}(\mathsf{A|BC}) =\frac{\text{Prob}(\mathsf{ABC})}{\text{Prob}(\mathsf{A}[\Omega]\mathsf{BC})}
\end{equation}
Hence,
\begin{equation}
\text{Prob}(\mathsf{A|B})=\frac{\text{Prob}(\mathsf{AB})}{\text{Prob}(\mathsf{A}[\Omega]\mathsf{B})}
\end{equation}
if we have well conditioned probabilities $\text{Prob}(\mathsf{AB})$ and $\text{Prob}(\mathsf{A}[I]\mathsf{B})$ (and the latter probability is non-zero).   In other words, we can use the standard equation for calculating conditional probabilities as long as we can assign probabilities as required.   This is consistent with the properties of probabilities given in Sec.\ \ref{propertiesofprobabilities}.

For generic fragments $\mathsf{AB}$ we should not expect to have a well conditioned probability $\text{Prob}\mathsf{(A|B)}$.  To illustrate this, consider an experiment from quantum physics.  Imagine we we have a device $\mathsf{ A^{a_1}}$ which prepares a spin half particle (which we take to be of type $\mathsf a$) in the up state followed, in sequence, by two spin measurements $\mathsf{ B_{a_1}^{a_2}}$ and  $\mathsf{ C_{a_2}^{a_3}}$, along some directions, and then followed by an operation $\mathsf{ D_{a_3}^{b_4}}$ which may be a spin measurement or something else (these spin measurements are \lq\lq non-demolition" measurements that allow the system to emerge out the other end - i.e. they are transformations in the terminology of this paper).
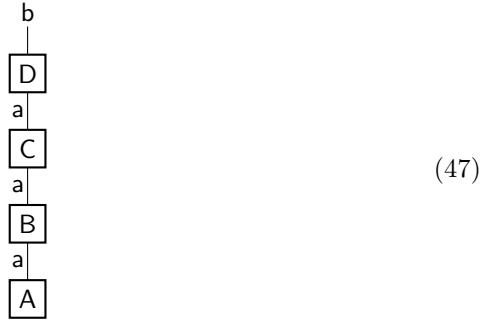
\vspace{-10pt}
\begin{equation}
\begin{Diagram}{0}{-1.8}
\Opbox[2]{A}{0,0}
\Opbox[2]{B}{0,4}
\Opbox[2]{C}{0,8}
\Opbox[2]{D}{0,12}
\wire{A}{B}{1.5}{1.5}\opsymbol{a}
\wire{B}{C}{1.5}{1.5}\opsymbol{a}
\wire{C}{D}{1.5}{1.5}\opsymbol{a}
\outwire{D}{1.5}\Opsymbol{b}
\end{Diagram}
\end{equation}
The following three examples illustrate the notion of a well conditioned probability.
\begin{enumerate}
\item There is not a well conditioned probability $\text{Prob}\mathsf{(C_{a_2}^{a_3}|A^{a_1})}$ since what happens depends on which direction the spin is measured along at $\mathsf B$, i.e.\
\begin{equation}
\text{Prob}\mathsf{(C_{a_2}^{a_3}|A^{a_1}B_{a_1}^{a_2})} \not=\text{Prob}\mathsf{(C_{a_2}^{a_3}|A^{a_1}\tilde{B}_{a_1}^{a_2})}
\end{equation}
where $\mathsf{\tilde{B}_{a_1}^{a_2}}$ is a spin measurement along a different direction from $\mathsf{B_{a_1}^{a_2}}$.
\item Perhaps a little more surprisingly, there is not a well conditioned probability $\text{Prob}\mathsf{(B_{a_1}^{a_2}|A^{a_1})}$ because
\begin{equation}
\text{Prob}\mathsf{(B_{a_1}^{a_2}|A^{a_1}C_{a_2}^{a_3})} \not=\text{Prob}\mathsf{(B_{a_1}^{a_2}|A^{a_1}\tilde{C}_{a_2}^{a_3})}
\end{equation}
(where $\mathsf{\tilde{C}_{a_2}^{a_3}}$ is a spin measurement along a different direction to $\mathsf{C_{a_2}^{a_3}}$) because postselection effects the probability (as a simple calculation will show).
\item The pre- and post-selected probability $\text{Prob}\mathsf{(B_{a_1}^{a_2}|A^{a_1}C_{a_2}^{a_3})}$ is well conditioned since
\begin{equation}
\text{Prob}\mathsf{(B_{a_1}^{a_2}|A^{a_1}C_{a_2}^{a_3}D_{a_3}^{a_4})}=\text{Prob}\mathsf{(B_{a_1}^{a_2}|A^{a_1}C_{a_2}^{a_3}\tilde{D}_{a_3}^{a^4})} \end{equation}
for any $\mathsf{D_{a_3}^{b_4}}$ and $\mathsf{\tilde{D}_{a_3}^{b_4}}$.  This is true because $\mathsf C$ is a complete spin measurement (corresponding to a non-degenerate observable) and so subsequent post-selection does not effect this probability.
\end{enumerate}

\section{Systems, states, effects, and transformations}

Having incorporated the notion of probability into the circuit framework we are now in a position to define a number of important notions.  In particular, we can associate states with preparations, transformation matrices with transformations, and effects with results.  In this section we will see how to do this and define a number of related concepts.

\subsection{Equivalence of fragments}\label{equivalenceoffragments}

\index{equivalence!of fragments}We can define a rather useful notion of equivalence of fragments.  Any circuit can be broken up into fragments.  If we have a particular fragment, $\mathsf{A}$, then it can be completed into a circuit with another fragment, $\mathsf{C}$ (note we are suppressing type labels since we are discussing general fragments).  In general, there will be many fragments that complete a given fragment into a circuit.  We will say that fragments $\mathsf A$ and $\mathsf B$ are equivalent if either one can replace the other in any circuit and the probability of the circuit remains unchanged when we perform such a replacement.  That is
\begin{quote}
$\mathsf A \equiv \mathsf B$ iff (i) for every circuit $\mathsf{AC}$ there exists a circuit $\mathsf{BC}$ and vice versa and (ii)   $\text{Prob}(\mathsf{AC})=\text{Prob}(\mathsf{BC})$.
\end{quote}
Two fragments can only be equivalent if they have the same types of inputs and outputs with comparable causal structure (so we can plug one fragment in the place of the other).

We will also define a restricted notion of equivalence for transformations.  Any fragment having some open inputs and outputs can be put in transformation mode (see Sec.\ \ref{preparationstransformationseffects}).  It is useful to define a notion of equivalence for the case that we only consider transformation mode.  For this we will use the symbol $\cong$.  We will say $\mathsf A\cong \mathsf B$ iff, for every circuit $\mathsf{AC}$ in which $\mathsf A$ is in transformation mode there exists a circuit $\mathsf{BC}$ where $\mathsf B$ is in transformation mode such that $\text{Prob}(\mathsf{AC})=\text{Prob}(\mathsf{BC})$.   We can extend this to include preparations and results.  For them, restricted equivalence is the same as equivalence (since there is no issue of having to put them into transformation mode).  In Part \ref{theduotensorframework} we will introduce the assumption of \emph{full decomposability} to set up the mathematical framework within which we give mathematical axioms for quantum theory.   In Part \ref{thereconstruction} we give operational postulates for quantum theory. The third of these is is the assumption of \emph{tomographic locality}.  In fact the assumption of full decomposability and tomographic locality will be shown to be equivalent.  We will show that it follows from tomographic locality that $\mathsf{A\cong B}$ if and only if $\mathsf{A\equiv B}$.

\subsection{Maximal sets of distinguishable preparations}\label{maximalsetsofdistinguishablepreps}

\index{distinguishable preparations}Consider a set of preparations, $\mathsf{A}^\mathsf{a_1}[n]$ where $n=1, 2, \dots$.  These preparations are said to form a distinguishable set if there exists a measurement $\{ \mathsf{B}_\mathsf{a}[n]: n=1, 2, \dots\}$ such that
\begin{equation}
\text{Prob}(\mathsf{A}^\mathsf{a_1}[n]\mathsf B_\mathsf{a_1}[n'] )=\delta_{nn'}.
\end{equation}
If there exists no distinguishable set of preparations having more elements, then this set is said to be maximal\index{maximal sets of distinguishable preparations}.  We let the number of elements in a maximal set of distinguishable preparations is $N_\mathsf{a}$ \index{N@$N_\mathsf{a}$} (for a proto-system of type $\mathsf a$).  The measurement that distinguishes them is called a \emph{maximal measurement}\index{maximal measurement}.  The elements of a maximal measurement are called \emph{maximal results}\index{maximal results}.  We will reserve the letters $\mathsf{U}$, $\mathsf{V}$, and $\mathsf{W}$ to denote maximal distinguishable sets and the corresponding results.  Thus, we have
\begin{equation}\label{uaua}
\text{Prob}(\mathsf{U}^\mathsf{a_1}[n]\mathsf U_\mathsf{a_1}[n'])=\delta_{nn'}
\end{equation}
where $\{ \mathsf{U}^\mathsf{a}[n]: n=1~\text{to}~ N\}$  are is the maximal distinguishable set of preparations, and $\{ \mathsf{U}[n]_\mathsf{a}: n=1~\text{to}~ N\}$ is the corresponding maximal measurement. Note there is no ambiguity using the same symbol for preparations and the corresponding results since the position of the label $\mathsf{a}$ tells us whether we have a preparation or an result.  We have similar properties for $\mathsf V$ and $\mathsf W$ (in some proofs we will need to refer to more than one maximal distinguishable set).

The maximum amount of classical information we can send with a proto-system is given by $N_\mathsf{a}$.  Measured in bits, it is equal to $\text{log}_2 N_\mathsf{a}$.  We will call $\text{log}_2 N_\mathsf{a}$ the \emph{information carrying capacity}\index{information carrying capacity}.

We can use the notion of a maximal measurement to define certain restricted classes of preparations which we will call \emph{informational subsets}.
\begin{quote}
{\bf Informational subsets.} \index{informational subset!of preparations} Let $\{ \mathsf{U}_\mathsf{a}[n]: n=1, 2, \dots\}$ be a maximal measurement. An informational subset, $S$, is associated with a subset of the outcomes $O(S) \subseteq (1, 2, \dots, N_\mathsf{a})$.   We define the informational subset $S$ as consisting of all preparations, $\mathsf A^{a_1}$ that, when this maximal measurement is performed, only give rise to outcomes in the associated subset, $O(S)$ .
\end{quote}
This means, in particular, that
\begin{equation}
\text{if}~~ \mathsf{A^{a_1}}\in S ~~\text{then}~~ \text{Prob}(\mathsf{A^{a_1}} \mathsf{U}_\mathsf{a}[n])=0 ~~ \text{for all}~~ n\in \overline O(S),
\end{equation}
where $\overline O(S)$ is the set of outcomes not in $O(S)$.  For each informational subset, $S$, we can define the complement subset $\overline S$ which is associated with the complement set of outcomes $\overline O(S)$.

We will say that system types having $N_\mathsf{a}=1$ are \emph{trivial}\index{trivial system types}.

\subsection{Systems and filters}\label{systemsandfilterssection}

We defined proto-systems as being identified with synchronous sets of wires.  We will now define what we mean by a system. To do this we need to define a filter. A filter is defined with repect to a maximal set of distinguishable preparations for a proto-system.
\begin{quote}
{\bf A filter}\index{filters} is associated with an informational subset of preparations, $S$.  It is defined to be a transformation, $\mathsf{F_{a_1}^{a_2}}$, that inputs and outputs a proto-system, $\mathsf a$ (which could be composite), which leaves preparations in the informational subset $S$ uneffected and blocks preparations in the complement informational subset $\overline S$. In symbolic form,
\begin{equation} \label{filtera}
\text{if}~~\mathsf{A^{a_1}}\in S ~~\text{then}~~ \text{Prob}(\mathsf{A^{a_1} F_{a_1}^{a_2} B_{a_2}})= \text{Prob}(\mathsf{A^{a_1}  B_{a_1}}) ~~ \text{for all} ~~ \mathsf{B_a},
\end{equation}
and
\begin{equation}\label{filterb}
\text{if}~~\mathsf{A^{a_1}}\in \overline S ~~\text{then}~~  \text{Prob}(\mathsf{A^{a_1} F_{a_1}^{a_2}B_{a_2}})=0 ~~ \text{for all}~~ \mathsf{B_a}.
\end{equation}
The \emph{capacity} \index{capacity of filter} of the filter is defined to be equal to $|O(S)|$.
\end{quote}
[An example of a filter from quantum theory would be a transformation that projects onto a particular subspace of the Hilbert space.  All states belonging to this subspace pass through unchanged.  States orthogonal to the subspace states are blocked.]

We now define
\begin{quote}
\index{systems}{\bf A system} is associated with a wire, or synchronous set of wires, after a filter.
The system type is determined by the filter (which in turn is determined by the associated informational subspace and maximal measurement) and the wire types the filter acts on.
\end{quote}
We will denote this system type by letter different from that of the (unfiltered) proto-system, for example $\mathsf b$, to denote that it has been filtered.
This definition of a system is consistent with actual practice. Typically, an experimentalist will make sure he gains some control over what passes through the apertures by filtering.   We can absorb the filters into the definition of the operations (on the outputs, and, if we want, the inputs as well since this makes no difference).  Hence, except when we are proving results which depend on using filters, we need not explicit about whether we have filters in place or not. It is worth noting a few points.
\begin{enumerate}
\item To indicate that we have a new system, $\mathsf b$, after a filter acting on a proto-system $\mathsf a$, we can write the filter as $\mathsf{ F_{a_1}^{b_2}}$ though often we will write $\mathsf{ F_{a_1}^{a_2}}$.
\item A filter, $\mathsf{ F_{a_1}^{a_2}}$, followed by another similar filter, $\mathsf{ F_{a_2}^{a_3}}$, is clearly also a filter which passes and blocks the same sets of preparations.
\item The \lq\lq do nothing" transformation (the identity filter) is clearly a filter.  Therefore proto-systems are systems.
\item A composite of two proto-systems is a system itself since the identity filter on each component clearly acts as a filter (the identity filter) on the composite.  However, it is not clear that a composite of two systems will, in general, also be a system. In Part \ref{thereconstruction} we prove that it follows from the operational postulates given there that filters acting on the components of a proto-system always comprise a filter on the composite proto-system. Hence, if the postulates are true, a composite of two systems is a system itself.
\item We will show that it follows from the postulates that we can build filters corresponding to filtering to any informational subsets, $S$ (having any number of elements in $O(S)$).  We will also show that, for these filters, $|O(S)|$ is the maximum number of preparations that can be perfectly distinguished after a filter.    This means we can construct a system, $\mathsf b$, having any given maximun number, $N_\mathsf{b}$, of distinguishable preparations.
\item We will say that one filter is \emph{smaller} than another if it has smaller $|O(S)|$.
\end{enumerate}

Filters are defined to be transformations that act on proto-system.  Having defined systems as what we have after a filter (on a proto-system), we can define a notion of a \emph{filter on a system} \index{filters!on a system} where the system, $\mathsf b$, in question may itself have been previously obtained by filtering a proto-system, $\mathsf a$.  A filter on a system, $\mathsf b$, is a filter on the underlying proto-system, $\mathsf a$,  defined with respect to a set of outcomes $O(S') \subseteq O(S)$ of some maximal measurement that can be used define a filter from $\mathsf a$ to $\mathsf b$ ($O(S)$ being the set of outcomes for this original filter).

\subsection{States and effects}\label{associatingstateswithpreparations}

We define
\begin{quote}
{\bf The state}\index{states} associated with a preparation for a given type of system is that thing given by any mathematical object which can be used to calculate the probability for any circuit built from this preparation followed by any result for this type of system.
\end{quote}
Note, since we are calculating the probability for the circuit,  we are using the state to calculate the \emph{joint probability} of the outcome belonging both to the outcome set for for the given preparation and for the given result (rather than the conditional probability of seeing an outcome in the result outcome set given an outcome in the preparation outcome set). Also note that the system may be filtered.

This definition ensures that the state is in one to one correspondence with equivalence classes of preparations. I.e.\ all preparations having the same state are equivalent to one another.

One mathematical object that would serve as a state for some preparation, $\mathsf{A_{a_1}}$, for a system of type $\mathsf a$ would simply be a list of all the probabilities, $\text{Prob}(\mathsf{A^{a_1}Z_{a_1}})$, for all circuits that can be build from $\mathsf{ A^{a_1}}$ followed by a result $\mathsf{Z_{a_1}}$ where $\mathsf{Z_{a_1}}$ runs over all possible results on a system of type $\mathsf a$.  In general this will be a very long list of probabilities.  However, in general, in a physical theory we expect there to be some relationships between the probabilities such that we can calculate all the probabilities from just a subset corresponding to a subset of the results (which we will call \emph{fiducial results})\index{fiducial results}.  We will, further, consider only linear relationships\index{linearity} (we will discuss this restriction below and in Appendix \ref{linearityfrommixing}).  Thus, we can consider a list of probabilities
\begin{equation}
A^{a_1}= \text{Prob}(\mathsf{A}^\mathsf{a_1}\mathsf{X}_\mathsf{a_1}^{a_1}) ~~~~\text{for}~~ a_1=1 ~~\text{to} ~~K_{\mathsf a}
\end{equation}
where $\mathsf{X}_\mathsf{a_1}^{a_1}$ are the fiducial results and, for a general result, $\mathsf{B_{a_1}}$, the probability is given by the linear relationship
\begin{equation}\label{probAB}
\text{Prob}(\mathsf{A^{a_1}B_{a_1}})= A^{a_1}B_{a_1}
\end{equation}
where summation over the index $a_1$ is implied and $B_{a_1}$ is a list of coefficients associated with the result $\mathsf{B_{a_1}}$.

Note that use sans serif font to represent fragments such as preparations (e.g.\ $\mathsf{A^{a_1}}$) and results (e.g.\ $\mathsf{B_{a_1}}$).  We also use sans serif for the type labels, $\mathsf{a}$, $\mathsf{b}$, \dots.  However, we use normal maths font to represent states (e.g.\ $A^{a_1}$) and the coefficients associated with results.  We also use normal maths font for the index $a_1$ over which we sum.

The list of coefficients, $B_{a_1}$, associated with the result $\mathsf{B_{a_1}}$ will play an important role.  We call this list of coefficients the \emph{effect} associated with the given result.

It is worth noting a few points at this stage.
\begin{enumerate}
\item The choice of fiducial results, $\mathsf{X}_\mathsf{{a_1}}^{a_1}$, will not in general be unique.  However, we always choose minimal such sets (so that there exists no other set of fiducial effects with fewer elements).
\item $A^{a_1}$ is a list of $K_\mathsf{a}$ \index{K@$K_\mathsf{a}$} probabilities giving the state associated with the preparation $\mathsf{A^{a_1}}$. Two preparations having the same state are equivalent.
\item $B_{a_1}$ is a list of $K_\mathsf{a}$ coefficients giving the effect \index{effects} associated with the result $\mathsf{B_{a_1}}$.  These coefficients can, in principle, be negative [and in quantum theory they will sometimes be negative].  Two results having the same effect are equivalent.
\item The integer $K_\mathsf{a}$ plays an important role.  It is is the dimension of the state (and effect) space for a system of type $\mathsf a$.
\item If $K_\mathsf{a}$ is finite \index{finite system types} then the state can be determined from finitely many probabilities. We call such system types \lq\lq finite"
\item It is clear that $K_\mathsf{a}\geq N_\mathsf{a}$ since we need $N_\mathsf{a}$ fiducial results just to deal with the distinguishable states (see (\ref{uaua})). [In quantum theory $K_\mathsf{a}=N_\mathsf{a}^2$.]
\end{enumerate}
In setting up these ideas we considered only linear relationships.  It is possible that nonlinear relationships will lead to a smaller number of fiducial effects.  However, in probability theories, it is very natural to allow arbitrary mixtures of preparations (where we toss a $\lambda$ weighted coin and prepare one preparation if we get heads, and another preparation if we get tails).  It is shown in Appendix \ref{linearityfrommixing} that, if we allow such mixtures, then we cannot do better than use linear relationships and, indeed, we will necessarily have linear relationships for calculating probabilities in any optimal representation of the state \cite{hardy2009foliable}.  Even if we do not allow arbitrary mixtures, we can still choose to use a linear representation as here.  The only caveat is that this may not be optimal [indeed, if we restrict ourselves to pure states in quantum theory a more optimal representation exists in terms of amplitudes rather than probabilities].

\subsection{Assumption 2 for the circuit framework}

We make the following assumption
\begin{quote}
{\bf Assump 2} \index{assumptions!Assump 2@\textbf{Assump 2}}There exists at least one finite and nontrivial system type.
\end{quote}
I.e.\ there exists at least one system type having $K_\mathsf{a}$ finite and $N_\mathsf{a}>1$.  This assumption concerns two issues: (1) existence of non-trivial systems and (2) finiteness.

We will see in Part \ref{thereconstruction} that it is sufficient to assume the existence of one finite nontrivial type to prove from the postulates that there exist types for every finite $N_\mathsf{a}$ and that these will be finite.

This is a much weaker assumption than we might have required.  We might have had to assume that there exists systems having $N=1, 2, 3, \dots$ and assume that, $K_\mathsf{a}$ is finite for finite $N_\mathsf{a}$.

It is impossible that we could prove that $K_\mathsf{a}=\infty$ experimentally. Thus, it might be said that a finite value follows from the operational approach taken in this paper.  However, the issue is a little more subtle than that.  If $K_\mathsf{a}$ is finite then, once we surpass a certain level of experimental sophistication, the measured value of $K_\mathsf{a}$ will stabilize at that finite value.  However, if $K_\mathsf{a}$ is infinite, then as we increase the sophistication of the experiment, the measured value of $K_\mathsf{a}$ will continue to increase.  We could assume that, for some fundamental reasons, there is a limit to how sophisticated the experiment can be.  In this case the maximum measured value of $K_\mathsf{a}$ would have to be finite.

In \cite{hardy2001quantum} finiteness of the state space dimension follows from the simplicity axiom.  It is a background assumption in \cite{chiribella2010informational} and \cite{dakic2009quantum}.  Masanes and M\"uller \cite{masanes2010derivation} actually include an explicit axiom in their list of axioms, namely \lq\lq In systems that carry one bit of information, each state is characterized by a finite set of outcome probabilities."

\subsection{Types of state}

There are many ways of classifying states.  We will collect a few useful classifications together here.

The \emph{null state}\index{states!null state}\index{null states}, $0^{a_1}$, has all entries equal to zero. One way (though not necessarily the only way) to prepare such a state is to add an outcome to our specification of a preparation that cannot happen and let the outcome set associated with the null preparation have this outcome as its only member.  Then the joint probability for a circuit consisting of this preparation followed by any effect must be zero (since the outcome on the preparation can never happen).

\index{states!mixed}\emph{Mixed states} are states which can be simulated by a probabilistic mixture of some distinct states. Thus, a mixed state, $A^{a_1}$, can be written
\begin{equation}
A^{a_1} = \lambda B^{a_1} + (1-\lambda) C^{a_1}
\end{equation}
where $B^{a_1}$ and $C^{a_1}$ are distinct and $0<\lambda<1$.  We include the null state, $0^{a_1}$, as a possible state in defining mixed states.

\index{states!pure}\index{pure states}\emph{Pure states} are states which are not mixed.

\index{states!parallel}Two states, $A^{a_1}$ and $B^{a_1}$ are said to be \emph{parallel} if $A^{a_1}=\mu B^{a_1}$ for some positive number $\mu$. It is always possible to write the shorter of the parallel states as a convex sum of the other and the null state.  Thus, if $\mu<0$ we can write $A^{a_1}=\mu B^{a_1} +(1-\mu)0^{a_1}$.

\index{states!pure parallel}Any state that is parallel to a pure state will be called a \emph{pure-parallel} state.  Pure-parallel states are, strictly speaking, mixed.  However, they can only be written as a convex combination of the null state and the pure state to which the state is parallel.  If pure-parallel states are normalized then they become pure.

\subsection{Types of effect}

There are many types of effect we may consider.  We will make particular use of the following types.

A \emph{maximal effect}\index{maximal effect} is one associated with a maximal result (corresponding to one outcome of a maximal measurement).

A \emph{deterministic effect}\index{deterministic effect} is one associated with a deterministic result (having outcome set equal to the set of all outcomes).  We will see that, in causal theories, the deterministic effect is unique for any given system type (see {\bf T\ref{deteffect}} in Sec.\ \ref{uniqueimplications} and \cite{chiribella2010probabilistic}).

\subsection{Sets of states}

Associated with any preparation is a state.  Hence, associated with any set of preparations is a set of states.  In Sec.\ \ref{maximalsetsofdistinguishablepreps} we defined two types of sets of preparations, namely maximal distinguishable sets of preparations and informational subsets of preparations.

Associated with a maximal distinguishable set of preparations is a \emph{maximal distinguishable set of states}\index{maximal sets of distinguishable states}. By convention, we always exclude the null state from such sets (if we did not do this then the maximum number of distinguishable states would be greater by one). This is reasonable since it gives rise to zero probability for any circuit it is included in.

Associated with an informational subset of preparations is an \emph{informational subset of states}.\index{informational subset!of states}  These are the states that only give rise to some given subset of outcomes of some given maximal measurement (and have probability zero for the complement set of outcomes).

We now define an important notion.  \index{non-flat set of states}.
\begin{quote}
{\bf Non-flat set of states.} A set of states is non-flat if it is a spanning subset of some informational subset of states.
\end{quote}
By a \emph{spanning} we simply mean that any state in the informational subset set can be written as a linear combination (possibly with negative coefficients) of the states in the non-flat set.  A set of states is said to be \emph{flat} \index{flat set of states} if it is not a spanning subset of any informational subset of states. The name flat is then justified since the set is missing at least one dimension.

If we are given a set of states and have to test to see whether it is non-flat then we need to search for some informational subset of states for which it is a spanning subset.  If there exists no such subset then the set of states is flat.

Once we have the five postulates ({\bf P1} to {\bf P5}) in place we will see that the notion of a non-flat sets of states turn out to be a kind of generalization of the notion of a pure-parallel state.   In particular, it follows from {\bf P1} that the state in a single member non-flat set must be pure-parallel (see {\bf T\ref{singlemembersets}}).  Further, non-flat sets share many properties with pure-parallel states: (i) A reversible transformation preserves both the pure-parallel property and the non-flatness property; (ii) The state of a composite system formed from components prepared in pure-parallel states is also pure-parallel (this is implied by {\bf T\ref{comppurestates}}, and set of states for a composite system formed from the product states taken from non-flat sets for the components is non-flat (this follows from the postulates, {\bf P3} in particular); (iii) It follows from the postulates ({\bf P5} in particular) that filters send pure-parallel states to pure-parallel states(see {\bf T\ref{filtersnonmixing}}) and, more generally, they send non-flat sets to non-flat sets.

\subsection{Transformations}   \label{transformationssection}

\index{transformations}In general, a transformation, $\mathsf{ B_{a_1}^{b_2}}$, inputs a system of some type $\mathsf a$ and outputs a system of some type $\mathsf b$.   If it acts on a preparation, $\mathsf{A^{a_1}}$, then $\mathsf{A^{a_1}B_{a_1}^{b_2}}$ constitutes a new preparation.  The fiducial probabilities for this new preparation are
\begin{equation}
\text{Prob}(\mathsf{X}_\mathsf{b_2}^{b_2}\mathsf{B_{a_1}^{b_2} A^{a_1}}) := B_{a_1}^{b_2} A^{a_1}
\end{equation}
where we must have a linear expression on the left hand side since we can regard $\mathsf{X}_\mathsf{b_2}^{b_2}\mathsf{B_{a_1}^{b_2}}$ as a result.  Hence, the transformation matrix $B_{a_1}^{b_2}$ is associated with the transformation $\mathsf{ B_{a_1}^{b_2}}$.

\index{transformations!on a system}We will say a transformation acts \emph{on a system} if it inputs and outputs the same type of system.  In particular, if the system has been filtered, this means that it will remain in the appropriate informational subset.

\index{transformations!identity}An \emph{identity transformation} is a transformation on a system that leaves things unchanged. I.e.\ if it is inserted on any system in any circuit the probability for that circuit remains the same.

A transformation, $\mathsf{B_{a_1}^{b_2}}$, is said to be \emph{reversible}\index{transformations!reversible}\index{reversible transformations} if there exists another transformation, $\mathsf{\tilde B^{a_3}_{b_2}}$, such that the transformation $\mathsf{B_{a_1}^{b_2}}\mathsf{\tilde B^{a_3}_{b_2}}$ is equivalent to the identity transformation.  Note that $\mathsf{\tilde B^{a_3}_{b_2}}$ is also, clearly, a reversible transformation.

If a reversible transformation, $\mathsf{B_{a_1}^{a_2}}$,  acts on a system in a pure state then the state afterwards must also be pure.  To see this assume the contrary. Assume $A^{a_1}$ is pure.
\begin{equation}
A^{a_1} \longrightarrow B_{a_1}^{a_2} A^{a_1} = \lambda C^{a_2} + (1-\lambda) D^{a_2}
\end{equation}
where $C^{a_1}$ and $D^{a_1}$ are distinct and $0<\lambda<1$.
Therefore
\begin{equation}
 A^{a_1} = \lambda\tilde B^{a_1}_{a_2} C^{a_2} + (1-\lambda)\tilde B^{a_1}_{a_2} D^{a_2}
\end{equation}
The states $ \tilde B^{a_1}_{a_2} C^{a_2}$ and $\tilde B^{a_1}_{a_2} D^{a_2}$ must be distinct since $\tilde B^{a_1}_{a_2}$ is reversible.  Hence $A^{a_1}$ is mixed which contradicts our starting point.

A \emph{non-mixing}\index{transformations!non-mixing}\index{non-mixing transformations} transformation is one that transforms pure states into pure-parallel states (i.e.\ states that are pure up to normalization).  Strictly speaking pure-parallel states are mixed (they are a mixture of the given pure state and the null state).  Consequently, calling such transformations non-mixing is potentially misleading. However, this name conveys the idea better than the alternatives.  [In quantum theory, filtering (i.e.\ projecting into a given subspace) is a non-mixing transformation since the projected state is pure (up to normalization) if the original state is.]

A transformation is \emph{non-flattening}\index{transformations!non-flattening}\index{non-flattening transformations} if, for any non-flat set of state we send in, the set of states coming out is also non-flat.  Note that the dimension of the space spanned by the output set may be different from the space spanned by the input set as the associated filter may be different.
In the prelude we illustrated how filters themselves effect non-flattening transformations in quantum theory.  In Appendix \ref{appendixflattening} we provide a proof that filters (and, indeed, all non-mixing transformations) are non-flattening in quantum theory.

A \emph{compound transformation}\index{transformations!compound}\index{compound transformations} on a system, $\mathsf{B_{a_1}^{a_2}}$,  is a transformation on a system which can be formed from two sequential transformations on a system neither of which is equal to the identity.  I.e.\ $\mathsf{B_{a_1}^{a_2}}$ is compound if we can write $\mathsf{B_{a_1}^{a_2}}=\mathsf{C_{a_1}^{a_3}} \mathsf{D_{a_3}^{a_2}}$.

\subsection{Assumption 3 for the circuit framework}\label{assumptionthreesec}

Fragments have a certain input-output structure\index{input-output structure}. This captures both the types of the inputs and outputs left open and also the causal structure between them.  Fragments with the same input-output structure can be plugged in place of each other in a bigger circuit.  Consider the set, $S$, of fragments having some given input-output structure.
A general fragment, $\mathsf{A} \in S$, is characterized by its probabilistic properties
\begin{equation}
\{\text{Prob}(\mathsf{AC}): \text{all}~\mathsf{C}~\text{such that}~ \mathsf{AC}~\text{is a circuit}\}
\end{equation}
(we are suppressing the type labels since we are dealing with the general case).
We can consider a hypothetical fragment\index{fragments!hypothetical}, $\mathsf{Q}$, with the same input-output structure as the fragments in $S$. By \lq\lq hypothetical" we mean that we are considering the possibility that a fragment with a given input-output structure having certain probabilistic properties may exist.

We say that $\mathsf{A}$ is \emph{operationally indiscernible}\index{operational indiscernibility} from $\mathsf{B}$ to an accuracy of $\delta$ if
\begin{equation}
|\text{Prob}(\mathsf{AC}) - \text{Prob}(\mathsf{BC})| \leq \delta
\end{equation}
for all fragments $\mathsf C$ such that $\mathsf{AC}$ and $\mathsf{BC}$ are circuits.

We make the following assumption in the circuit framework.
\begin{quote}
{\bf Assump 3}\index{assumptions!Assump 3@\textbf{Assump 3}} If, for any accuracy $\delta>0$, there exists a fragment $A[\delta]$ that is operationally indiscernible from a given hypothetical fragment, $\mathsf Q$, then there actually exists a fragment with the probabilistic properties of $\mathsf Q$.
\end{quote}
If $\mathsf{Q}$ is operationally indiscernible from some fragment in $S$ for any accuracy $\delta>0$ then, whatever level of accuracy we work to, there always exists some fragment that behaves in the same way as the given hypothetical fragment.  Since measurements cannot be arbitrarily accurate, there is no way of telling apart, by operational means, the case where we make {\bf Assump 3} from the case where we do not.  In this sense this assumption is one of mathematical convenience.   Chiribella, D'Ariano, and Perinotti have a equivalent background assumption in \cite{chiribella2010informational} (and they give a similar motivation).

In Appendix \ref{compactness} \index{compactness} we consider the vectors formed from fiducial probabilities that characterize fragments having the same input-output structure.  We show that, for the case where only a finite number of fiducial probabilities is required, it follows from {\bf Assump 3} that the space of such vectors is compact (i.e. they are bounded and closed).   This means, in particular, that the space of states, effects, and transformations are compact (so long as they are characterized by a finite number of fiducial probabilities).  More generally, it also implies that the sets of allowed duotensors considered in Part \ref{theduotensorframework} are compact for this finite case.

\newpage

\part{Mathematical reformulation of quantum theory}\label{theduotensorframework}

In this Part we will provide a reformulation of quantum theory mediated by the duotensor framework put forward in \cite{hardy2010formalism}.   In Sec.\ \ref{operationsandduotensors} we will review the framework showing how duotensors (tensor-like objects but with a bit more structure) can be associated with operations.  In Sec.\ \ref{operatorsandduotensors} we will show how a rather analogous framework can be set up for associating operators with duotensors.  In Sec.\ \ref{operationsandoperators} we show how to use duotensors to form a bridge between operations and operators and we use this to reformulate quantum theory in terms of two mathematical axioms.  The reformulation we obtain is very similar to that provided by Chiribella, D'Ariano, and Perinotti (CDP) in \cite{chiribella2009theoretical}.  The latter does not use duotensors as intermediate objects nor does it use the more succinct notation we adopt. However, the basic formulae of the two approaches are related by appropriate insertion of partial transposes (as discussed in Sec.\ \ref{combs}).  The reformulation provided in this paper and that of CDP have much in common with the quantum realization of the causaloid formulation given in \cite{hardy2005probability, hardy2007towards}.  All three consider associating mathematical objects with general circuit fragments.

This paper is written such that it is possible to follow the reconstruction of quantum theory from postulates in Part \ref{thereconstruction} up to  and including Sec.\ref{gebitssection} where we obtain the qubit without reading  Part \ref{theduotensorframework}.  The last part of the reconstruction, in which we obtain quantum theory for arbitrary Hilbert space dimension, employs ideas from Part \ref{theduotensorframework}.

\section{Operations and duotensors}\label{operationsandduotensors}

The duotensor framework was put forward in \cite{hardy2010formalism}.  It pertains to circuits and fragments built out of operations and wires.  It is built on two assumptions.  The first is the same as {\bf Assump 1} - that we can associate a probability with a circuit and this probability depends only on the specification of that circuit.  The second is that operations are {\it fully decomposable} (we will explain this below).  We will show that this second assumption is equivalent to the assumption of tomographic locality, {\bf P3}.  It has been developed for the case of finite dimensional state spaces (i.e.\ finite $K_\mathsf{a}$).  The duotensor framework is, therefore, applicable to the reconstruction from operational postulates to be given in Part \ref{thereconstruction}.   The notation employed so far in this paper is, for the most part,  the same as in \cite{hardy2010formalism}.

\subsection{Motivation}

The motivation for constructing the duotensor framework is to provide a formulation for physical theories having the following property
\begin{quote}
{\bf Formalism locality:}\index{formalism locality} A formalism for a physical theory is said to have the property of \lq\lq formalism locality" if we can do calculations pertaining to any region of spacetime employing only mathematical objects associated with that region.
\end{quote}
Note that this is a property of the way a theory is formulated rather than being an intrinsic property of the physics itself.  In the circuit framework an arbitrary region of spacetime refers to an arbitrary fragment.  In the duotensor framework we are able to associate a mathematical object, the duotensor, with any arbitrary fragment.  Further, by having a duotensor for different fragments pertaining to the same setup (i.e.\ the same apparatus uses but with different outcome sets - see Sec.\ \ref{fragments}) we can deduce whether there are any probabilities for this setup which happen to be independent of what is happening elsewhere and what these probabilities are equal to.  Hence, we have a formulation having the property of formalism locality.

In standard formulations of quantum theory we evolve a state in time with respect to some foliation of the spacetime.  Circuits can be foliated using complete synchronous sets of wires (a set of wires is complete if it partitions the circuit) \cite{hardy2010formalism}.  If we foliate a circuit in this way we are effectively partitioning it so that a preparation is followed by a bunch of sequential transformations followed, finally, by an effect.  However, preparations, transformations, and effects are special cases of fragments.  If our formalism requires that we use a foliation in this way then we do not have the formalism locality property.

Histories formulations have a similar problem - they generally pertain to the entire history of the system and do not allow us to consider different arbitrary space time regions by themselves.

In these formulations we effectively have equations that apply only to specially shaped spacetime regions for which the probabilities are necessarily well conditioned.  For example, in theories which evolve a state, these specially shaped regions must have an initial and a final space-like hypersurface for any interval over which the state is evolved.  To make statements about arbitrary spacetime regions in such theories, we need to first apply the theory to the specially shaped regions.  This is the reason that these formulations do not have the formalism locality property.

It would be impossible to identify special shaped regions if we had indefinite causal structure  as we expect in a theory of quantum gravity.  Given this, it is natural that the theory to be formulated in a way that it has formalism locality property.   In \cite{hardy2005probability, hardy2007towards, markes2009entropy, hardy2009quantum3, hardy2010bformalism}, the causaloid framework is developed. It provides a formalism local framework pertaining to situations where we do not have definite causal structure.  Dealing with indefinite causal structure is not an immediate issue for the present paper since we are considering circuits whose wires define a causal structure. However, quantum theory with well defined causal structure may ultimately be best understood as a limiting case of a more general theory of quantum gravity having indefinite causal structure.  In this case, it is likely that the formulation of quantum theory we would most naturally arrive at in such a limiting procedure would be one having the formalism locality property.  Consequently, a formalism local formulation of quantum theory, as provided in this work, is more likely to provide insight into the problem of quantum gravity.

\subsection{Equivalence}\label{equivalenceforduotensors}

In Sec.\ \ref{equivalenceoffragments} we defined two fragments to be equivalent if the probability for any circuit containing one of these fragments is unchanged when the other fragment is substituted in its place.   The duotensor framework uses a more general notion of equivalence that applies to linear sums of fragments.  This more general notion reduces to the notion of Sec.\ \ref{equivalenceoffragments} in the appropriate special cases.

First we define the function $p(\cdot)$ \index{pfunction@$p(\cdot)$ function} as follows
\begin{equation}
p(\alpha \mathsf{A} + \beta \mathsf{B} + \dots) := \alpha\text{Prob}(\mathsf{A}) + \beta\text{Prob}(\mathsf{B}) +\dots
\nonumber\end{equation}
for circuits $\mathsf A$, $\mathsf B$, \dots. and real numbers $\alpha$, $\beta$, \dots (these can be negative).   Note that the $p(\cdot)$ function is only defined for linear sums of circuits.  We cannot define something like this for general linear sums of fragments because fragments do not, in genaral, have a probability associated with them.

We will consider expressions like
\[
\text{expression}_1= \alpha+ \beta\mathsf{C}+\gamma\mathsf{D}+\dots
\]
where $\alpha$, $\beta$, \dots are real numbers and $\mathsf{C}$, $\mathsf{D}$, \dots are fragments (they may be circuits).   Equivalence is defined in the following way
\begin{quote}
{\bf Equivalence:} \index{equivalence!of expressions} We write
\[
\text{expression}_1 \equiv \text{expression}_2
\]
if
\[
p(\text{expression}_1 \mathsf{E}) \equiv p(\text{expression}_2 \mathsf{E})
\]
for any fragment $\mathsf E$ that makes the contents of the argument on both sides of this equation into a linear sum of circuits (note that we are suppressing the subscripts and superscripts on the symbol $\mathsf E$ since we are talking about a general fragment).
\end{quote}
One example illustrating this is the following.  We have
\[
\mathsf{ \alpha A^{a_1} + \beta B^{a_1}  \equiv \gamma C^{a_1} +\delta D^{a_1} }
\]
if
\[
p(\mathsf{ [\alpha A^{a_1} + \beta B^{a_1}] E_{a_1} })  = p(\mathsf{[\gamma C^{a_1} +\delta D^{a_1}] E_{a_1} })  ~~~~\text{for all} ~~\mathsf{ E_{a_1}}
\]
Here is another (important) example.  In general, we have
\begin{equation}\label{probAequivA}
\mathsf{ A} \equiv \text{Prob}(\mathsf{A}) ~~~ \text{for any circuit} ~~ \mathsf{A}
\end{equation}
The proof of this is simple.  The most general fragment, $\mathsf E$, that completes a circuit into a circuit is another circuit.  For any circuit $\mathsf E$
\[ p(\mathsf{ A E} ) = p(\mathsf{A})p(\mathsf{E}) = p(\text{Prob}(\mathsf{A}) \mathsf{E}) \]
This example emphasizes that equivalence is a weaker notion than equality.  Clearly a circuit is not, itself, equal to a number.

In general, there are only two types of equivalence:
\begin{enumerate}
\item Each expression is a real number plus a linear combination of circuits:
\[
\alpha + \beta\mathsf{A} + \gamma \mathsf{B} + \dots \equiv \delta + \epsilon \mathsf{ C}+ \zeta \mathsf{D} + \dots
\]
where $\mathsf{A}$, $\mathsf B$, \dots, $\mathsf C$, $\mathsf D$, \dots, are all circuits.
\item Each expression is a linear combination of fragments
\[
\alpha\mathsf{A} + \beta \mathsf{B} + \dots \equiv \gamma\mathsf{ C}+ \delta \mathsf{D} + \dots
\]
where $\mathsf{A}$, $\mathsf B$, \dots, $\mathsf C$, $\mathsf D$, \dots, are all fragments having the same causal structure (so that any one can be substituted for any other of these fragments in any circuit).
\end{enumerate}

\subsection{Fiducial results}

We introduced the \index{fiducial results} fiducial set of results $\{ \mathsf{X}_\mathsf{a_1}^{a_1} : a_1=1 ~\text{to}~ K_\mathsf{a}\}$ in Sec.\ \ref{associatingstateswithpreparations}. We can write
\begin{equation}\label{BfidXB}
\mathsf{B}_\mathsf{a_1}\equiv B_{a_1} \mathsf{X}_\mathsf{a_1}^{a_1}
\end{equation}
Since we are employing summation convention with respect to $a_1$ we have here a linear sum of fiducial results weighted by the coefficients in $B_{a_1}$.
To see that (\ref{BfidXB}) is correct note that
\begin{equation}\label{pABAB}
p(\mathsf{A^{a_1}B_{a_1}}) = p(\mathsf{A^{a_1}}\mathsf{X}_\mathsf{a_1}^{a_1} B_{a_1}) = p(\mathsf{A^{a_1}}\mathsf{X}_\mathsf{a_1}^{a_1}) B_{a_1}
= A^{a_1}B_{a_2}
\end{equation}
which is the equation we obtained in Sec.\ \ref{associatingstateswithpreparations}.

It will be useful to associate fiducial results with graphical elements
\begin{equation}
\begin{Diagram}{0}{0.1}
\fideffect{X}{0,0}   \outblack{X}{1}\Duosymbol{a} \inwire{X}{1}\Opsymbol{a}
\end{Diagram}
~~~\Longleftrightarrow ~~\mathsf{X}_\mathsf{a_1}^{a_1} ~~ \text{where}~~a_1=1~~\text{to}~~K_\mathsf{a}
\end{equation}
When represented graphically, fiducial elements have a black dot.  The reason for this will become clear.
We represent $B_{a_1}$ with a graphical element:
\begin{equation}
B_{a_1} ~~~~\Longleftrightarrow ~~
\begin{Diagram}{0}{-0.1}
\Duobox[2]{B}{0,0}
\inwhite{B}{1.5}\Duosymbol{a}
\end{Diagram}
\end{equation}
This has a white dot on it.
Equation (\ref{BfidXB}) can be written in both symbolic and diagrammatic form
\begin{equation}\label{BfidXBwithdiagram}
\mathsf{B}_\mathsf{a_1}\equiv B_{a_1} \mathsf{X}_\mathsf{a_1}^{a_1}~~~~~\Longleftrightarrow  ~~~~~
\begin{Diagram}{0}{0}
\Opbox[2]{B}{0,0}\inwire{B}{1.5}\Opsymbol{a}
\end{Diagram}
~~\equiv~~
\begin{Diagram}{0}{0.07}
\fideffect{X}{0,-7pt}\Duobox[2]{B}{5,0}
\inwire{X}{1}\Opsymbol{a} \linkbw{X}{B}{1}{1.5} \duosymbol{a}
\end{Diagram}
\end{equation}
When we link up two elements we require that black dots are placed next to white dots.  The horizontal link in (\ref{BfidXBwithdiagram}), interupted by a black and a white dot, corresponds to summing over the index $a_1$.  We define
\begin{equation}
\begin{Diagram}{0}{0.07}
\fideffect{X}{0,-7pt}\Duobox[2]{B}{5,0}
\inwire{X}{1}\Opsymbol{a} \linkbw{X}{B}{1}{1.5} \duosymbol{a}
\end{Diagram}
~~ := ~~
\begin{Diagram}{0}{0.07}
\fideffect{X}{0,-7pt}\Duobox[2]{B}{4,0}
\inwire{X}{1}\Opsymbol{a} \link{X}{B}{1}{1.5} \duosymbol{a}
\end{Diagram}
\end{equation}
This is unambiguous at this stage since the fiducial result must have a black dot so the $B$ box must have a white dot.  This is a hybrid diagram.  Hybrid diagrams have wires running up for operational description and links running to the left for the mathematics. Horizontal links between boxes represent the summation over the corresponding index.

\subsection{Fiducial preparations}

We can introduce a \index{fiducial preparations} fiducial set of preparations, $\{{}_{a_1}\!\mathsf{X}^\mathsf{a_1} : a_1=1 ~\text{to}~ K_\mathsf{a}\}$.  These correspond to a linearly independent spanning set of states.  We define ${}^{a_1}\! A$ by
\begin{equation}\label{AfidXA}
\mathsf{A^{a_1}} \equiv {}^{a_1}\!\!A \,\,\,{}_{a_1}\!\mathsf{X}^\mathsf{a_1}
\end{equation}
This can be related to $A^{a_1}$.  Using (\ref{BfidXB}) and (\ref{AfidXA}) we obtain
\begin{equation}\label{pABgAB}
p(\mathsf{A^{a_1}B_{a_1}}) =  {}^{a'_1}\!\!A  \,\, p({}_{a'_1}\!\mathsf{X}^\mathsf{a_1}\mathsf{X}_\mathsf{a_1}^{a_1}) \, B_{a_1}
= {}_{a'_1}\! g^{a_1} \,\,\, {}^{a'_1}\!\!A B_{a_1}
\end{equation}
where we define
\begin{equation}
{}_{a'_1}\! g^{a_1} := p({}_{a'_1}\!\mathsf{X}^\mathsf{a_1}\mathsf{X}_\mathsf{a_1}^{a_1})
\end{equation}
We call this object the \emph{hopping metric}\index{hopping metric}.  Since both the fiducial effects and the fiducial preparations correspond to linearly independent sets of vectors, the hopping metric must be non-singular.  Comparing (\ref{pABAB}) and (\ref{pABgAB}) we must have
\begin{equation}
A^{a_1} = {}_{a'_1}\! g^{a_1} \,\,  {}^{a'_1}\! A
\end{equation}
Since the hopping metric is non-singular this shows that ${}^{a_1}\! A$ is an alternative way of representing the state.
The effect of the hopping metric is to cause the indices to hop across the central symbol.   With this in mind, we define
\begin{equation}
{}_{a_1} B := {}_{a_1}\! g^{a'_1}\,\, B_{a'_1}
\end{equation}
This is an alternative way of representing the effect.   We can write
\begin{equation}\label{probABmultiversions}
\text{Prob}(\mathsf{A^{a_1}B_{a_2}}) = A^{a_1}B_{a_1} = {}^{a_1}\! A \,\,\, {}_{a_1}\! B = {}_{a'_1}\! g^{a_1} \,\, {}^{a'_1}\!\!A B_{a_1}
=  {}^{a'_1}\! g_{a_1} \,\, A^{a'_1} \,\,{}_{a_1}\! B
\end{equation}
where
\begin{equation}
{}^{a'_1}\! g_{a_1} := \left( {}_{a'_1}\! g^{a_1} \right)^{-1}
\end{equation}
is the inverse of the hopping metric.  The hopping metric is guaranteed to have only positive entries (less than  or equal to one).  Its inverse, however, can have negative entries.  [In quantum theory the inverse of the hopping metric does have negative entries.  Indeed, this can be regarded as the origin of the negative numbers appearing in convex probabilities frameworks.  These negative numbers have conceptual implications \cite{spekkens2008negativity, ferrie2009framed}.]

\subsection{Diagrammatic representation of the simple circuit}\label{diagrammaticrepresentationsimplecircuit}

We can represent (\ref{AfidXA}) as
\begin{equation}\label{ABcircuit}
\begin{Diagram}{0}{-0.2}
\Opbox[2]{A}{0,-7pt} \outwire{A}{1.5} \Opsymbol{a}
\end{Diagram}
~~ \equiv ~~
\begin{Diagram}{0}{-0.2}
\Duobox[2]{A}{0,-7pt} \fidprep{X}{5,0} \linkwb{A}{X}{1.5}{1}\duosymbol{a} \outwire{X}{1}\Opsymbol{a}
\end{Diagram}
~~~\Longleftrightarrow ~~~ \mathsf{A^{a_1}} \equiv {}^{a_1}\!\!A \,\,\,{}_{a_1}\!\mathsf{X}^\mathsf{a_1}
\end{equation}
The elements of this equation are
\begin{equation}
\begin{Diagram}{0}{-0.1}
\fidprep{X}{0,0} \inblack{X}{1} \Duosymbol{a} \outwire{X}{1} \opsymbol{a}
\end{Diagram}
~~\Longleftrightarrow ~~ {}_{a_1}\!\mathsf{X}^\mathsf{a_1} ~~~~~~~~~~~~~~~~
\begin{Diagram}{0}{-0.05}
\Duobox[2]{A}{0,0} \outwhite{A}{1.5}\duosymbol{a}
\end{Diagram}
\Longleftrightarrow ~~  {}^{a_1}\!\!A
\end{equation}
Note that, again, we have a black dot on the fiducial element and we match black and white dots.  We define
\begin{equation}
\begin{Diagram}{0}{0}
\Duobox[2]{A}{0,-7pt} \fidprep{X}{5,0} \linkwb{A}{X}{1.5}{1}\duosymbol{a} \outwire{X}{1}\Opsymbol{a}
\end{Diagram}
~~ := ~~
\begin{Diagram}{0}{0}
\Duobox[2]{A}{0,-7pt} \fidprep{X}{4,0} \link{A}{X}{1.5}{1}\duosymbol{a} \outwire{X}{1}\Opsymbol{a}
\end{Diagram}
\end{equation}
Once again, this is not ambiguous at this stage since the fiducial preparation must have a black dot and so the $A$ box must have a white dot.

The hopping metric is represented by $\bbdots$
\begin{equation}\label{defbb}
\begin{Diagram}{0}{-0.1}
\bbmetric{g}{0,0}\duosymbol{a}
\end{Diagram}
~:=~
p\left(\!\!\!\!
\begin{Diagram}{0}{-0.6}
\fidprep{X2}{4,0}
\fideffect{X}{4,4}
\wire{X2}{X}{1}{1}\opsymbol{a}
\inblack{X2}{1} \Duosymbol{a}
\outblack{X}{1}\Duosymbol{a}
\end{Diagram}
\!\!\!\!\right)
~~~~~\Leftrightarrow ~~~~~
{}_{a'_1}\! g^{a_1} := p({}_{a'_1}\!\mathsf{X}^\mathsf{a_1}\mathsf{X}_\mathsf{a_1}^{a_1})
\end{equation}
and its inverse is represented by $\wwdots$.
\begin{equation}
\wwdots ~ \Leftrightarrow ~ {}^{a'_1}\! g_{a_1}
\end{equation}
With these diagrammatic conventions we can redo some of the manipulations above
\begin{equation}
\text{Prob}\left(
\begin{Diagram}{0}{-0.6}
\Opbox[2]{A}{0,0}\Opbox[2]{B}{0,4} \wire{A}{B}{1.5}{1.5}\opsymbol{a}
\end{Diagram}
\,\right)
=
\text{Prob}\left(
\begin{Diagram}{0}{-0.6}
\Duobox[2]{A}{0.5,-7.5pt} \fidprep{X2}{5,0} \linkwb{A}{X2}{1.5}{1}\duosymbol{a}
\fideffect{X}{5,4}\Duobox[2]{B}{9.5,4cm+7pt}
\linkbw{X}{B}{1}{1.5} \duosymbol{a}
\wire{X2}{X}{1}{1}\opsymbol{a}
\end{Diagram}
\right)
=
\begin{Diagram}{0}{0}
\Duobox[2]{A}{0,0} \bbinsert[0.8]{g}{4,0} \duosymbol{a} \Duobox[2]{B}{8,0}
\link{A}{g}{1.5}{1}\link{g}{B}{1}{1.5}
\end{Diagram}
\end{equation}
Further, we put
\begin{equation}
\begin{Diagram}{0}{-0.1}
\Duobox[2]{A}{0,0} \outblack{A}{1.5}
\end{Diagram}
\!\!\!\!:=~
\begin{Diagram}{0}{-0.1}
\Duobox[2]{A}{0,0} \outwhite{A}{1.5} \bbmetric[0.7]{g}{3.7,0}
\end{Diagram}
~~~~~~~~~~
\begin{Diagram}{0}{-0.1}
\Duobox[2]{B}{0,0} \inblack{A}{1.5}
\end{Diagram}
~:=
\begin{Diagram}{0}{-0.1}
\Duobox[2]{B}{0,0} \inwhite{A}{1.5} \bbmetric[0.7]{g}{-3.7,0}
\end{Diagram}
\end{equation}
With these definitions we can redo the RHS of (\ref{probABmultiversions}) - the probability, $\text{Prob}(\mathsf{A^{a_1}B_{a_1}})$, is equal to
\begin{equation}
%\text{Prob}\left(
%\begin{Diagram}{0}{-1}
%\Opbox[2]{A}{0,0} \Opbox[2]{B}{0,4} \wire{A}{B}{1.5}{1.5} \opsymbol{a}
%\end{Diagram}
%\right)
%~=~
\begin{Diagram}{0}{-0.1}
\Duobox[2]{A}{0,0} \Duobox[2]{B}{5,0}
\linkbw{A}{B}{1.5}{1.5}\duosymbol{a}
\end{Diagram}
~=~
\begin{Diagram}{0}{-0.1}
\Duobox[2]{A}{0,0} \Duobox[2]{B}{5,0}
\linkwb{A}{B}{1.5}{1.5}\duosymbol{a}
\end{Diagram}
~=~
\begin{Diagram}{0}{-0.1}
\Duobox[2]{A}{0,0} \bbinsert[0.8]{g}{4,0} \duosymbol{a} \Duobox[2]{B}{8,0}
\link{A}{g}{1.5}{1}\link{g}{B}{1}{1.5}
\end{Diagram}
~=~
\begin{Diagram}{0}{-0.1}
\Duobox[2]{A}{0,0} \wwinsert[0.8]{g}{4,0} \duosymbol{a} \Duobox[2]{B}{8,0}
\link{A}{g}{1.5}{1}\link{g}{B}{1}{1.5}
\end{Diagram}
\end{equation}
We can simply represent this as
\begin{equation}
\begin{Diagram}{0}{-0.1}
\Duobox[2]{A}{0,0} \Duobox[2]{B}{5,0}
\link{A}{B}{1.5}{1.5}\duosymbol{a}
\end{Diagram}
\end{equation}
By virtue of the notational conventions we have adopted, we can introduce pairs of black and white dots, or delete them (in pairs) as we wish:
\begin{equation}
\nwdots\bndots = \nndotslong = \nbdots\wndots
\end{equation}
This implies that both $\bwdots$ and $\wbdots$ are equal to the identity since
\begin{equation}
\bbdots\wwdots = \bwdots ~~~~~~~\text{and}~~~~~~~ \wwdots\bbdots = \wbdots
\end{equation}
as $\wwdots$ is the inverse of $\bbdots$.

Recall that, by virtue of the way we have defined equivalence, $\mathsf{A}\equiv \text{Prob}(\mathsf {A})$ for any circuit $\mathsf A$ (this is equation (\ref{probAequivA}) above).  This means that
\begin{equation}\label{hoppingequivdiagram}
\begin{Diagram}{0}{-0.6}
\fidprep{X2}{4,0}
\fideffect{X}{4,4}
\wire{X2}{X}{1}{1}\opsymbol{a}
\inblack{X2}{1} \Duosymbol{a}
\outblack{X}{1}\Duosymbol{a}
\end{Diagram}
~~\equiv~~
\begin{Diagram}{0}{-0.1}
\bbmetric{g}{0,0}\duosymbol{a}
\end{Diagram}
\end{equation}
by virtue of the way that $\bbdots$ is defined in (\ref{defbb}).  Hence, we can write
\begin{equation}
\begin{Diagram}{0}{-0.6}
\Opbox[2]{A}{0,0}\Opbox[2]{B}{0,4} \wire{A}{B}{1.5}{1.5}\opsymbol{a}
\end{Diagram}
~~\equiv~~
\begin{Diagram}{0}{-0.6}
\Duobox[2]{A}{0.5,-7.5pt} \fidprep{X2}{5,0} \linkwb{A}{X2}{1.5}{1}\duosymbol{a}
\fideffect{X}{5,4}\Duobox[2]{B}{9.5,4cm+7pt}
\linkbw{X}{B}{1}{1.5} \duosymbol{a}
\wire{X2}{X}{1}{1}\opsymbol{a}
\end{Diagram}
~~\equiv~~
\begin{Diagram}{0}{-0.1}
\Duobox[2]{A}{0,0} \bbinsert[0.8]{g}{4,0} \duosymbol{a} \Duobox[2]{B}{8,0}
\link{A}{g}{1.5}{1}\link{g}{B}{1}{1.5}
\end{Diagram}
~~=~~
\begin{Diagram}{0}{-0.1}
\Duobox[2]{A}{0,0} \Duobox[2]{B}{5,0}
\link{A}{B}{1.5}{1.5}\duosymbol{a}
\end{Diagram}
\end{equation}
We will give a sequence of equivalent diagrams like this when we consider a more complicated circuit.

\subsection{Full decomposability}\label{fulldecomposabilitysubsection}

The second assumption used to set up the duotensor framework is the following
\begin{quote} {\bf Full decomposability}. \index{full decomposability! of operations} We assume that any operation is equivalent to a linear combination of operations each of which consists of an effect for each input and a preparation for each output.  We do not lose any generality by choosing these to be fiducial sets  since any other set could be written as a linear combination of the fiducial set. Hence, this assumption is equivalent to the statement that any operation, $\mathsf{A_{a_1b_2\dots c_3}^{d_4e_5\dots f_6}}$, can be written as
\begin{equation}\label{decomposibility}
\mathsf{A_{a_1b_2\dots c_3}^{d_4e_5\dots f_6}} \,\,\equiv {}^{d_4e_5\dots f_6}\!A_{a_1b_2\dots c_3}\,\, \mathsf{X}_\mathsf{a_1}^{a_1} \mathsf{X}_\mathsf{b_2}^{b_2} \cdots \mathsf{X}_\mathsf{c_3}^{c_3} \,\,{}_{d_4}\!\mathsf{X}^\mathsf{d_4}{}_{e_5}\!\mathsf{X}^\mathsf{e_5}\cdots {}_{f_6}\!\mathsf{X}^\mathsf{f_6}
\end{equation}
in symbolic notation, or
\begin{equation}\label{decomposibilitydiagram}
\begin{Diagram}[1.4]{0}{0}
\Opbox[5]{A}{0,0}
\inwire{A}{1}\Opsymbol{a} \inwire{A}{2.2}\Opsymbol{b} \putlatex[30,15]{\ensuremath{\dots}} \inwire{A}{5}\Opsymbol{c}
\outwire{A}{1}\Opsymbol{d}\outwire{A}{2.2}\Opsymbol{e}\putlatex[30,-50]{\ensuremath{\dots}}   \outwire{A}{5}\Opsymbol{f}
\end{Diagram}
~~~\equiv ~~~
\begin{Diagram}[1.2]{0}{0}
\Duobox[5]{A}{0,0}
\putlatex[-44,-11]{\ensuremath{\vdots}} \putlatex[-135,-20]{\ensuremath{\ddots}}
\putlatex[45,-11]{\ensuremath{\vdots}} \putlatex[170,-5]{\ensuremath{\ddots}}
\linkedeffect[0.8]{A}{5}{c}{-3.5}{0}\duosymbol[-13,-4]{c} \thispoint{cbase}{-3.5,-6}\Opsymbol{c} \wire{cbase}{c}{1}{1}
\linkedeffect[0.8]{A}{2.2}{b}{-6.1}{0}\duosymbol[22,-3]{b}  \thispoint{bbase}{-6.1,-6}\Opsymbol{b} \wire{bbase}{b}{1}{1}
\linkedeffect[0.8]{A}{1}{a}{-7.5}{0}\duosymbol[41,-1]{a}  \thispoint{abase}{-7.5,-6}\Opsymbol{a} \wire{abase}{a}{1}{1}
\placelatex[-20, 16]{-4,-6}{\ensuremath{\dots}}
\linkedprep[0.8]{A}{1}{d}{3.5}{0}\duosymbol[4,1]{d}  \thispoint{dbase}{3.5,6}\Opsymbol[0,38]{d} \wire{d}{dbase}{1}{1}
\linkedprep[0.8]{A}{2.2}{e}{4.9}{0}\duosymbol[-17,-5]{e} \thispoint{ebase}{4.9,6}\Opsymbol[0,38]{e} \wire{e}{ebase}{1}{1}
\linkedprep[0.8]{A}{5}{f}{7.5}{0}\otherside\duosymbol[-57,-4]{f}  \thispoint{fbase}{7.5,6}\Opsymbol[0,38]{f} \wire{f}{fbase}{1}{1}
\placelatex[9, -16]{6,6}{\ensuremath{\dots}}
\end{Diagram}
\end{equation}
in diagrammatic notation.
\end{quote}
We will prove in Sec.\ \ref{SecTomographiclocality} that this assumption is equivalent to the assumption of tomographic locality (i.e.\ {\bf P3}).

\subsection{What are duotensors?}\label{whatareduotensors}

The object, ${}^{d_4e_5\dots f_6}\!A_{a_1b_2\dots c_3}$, in (\ref{decomposibility}) above is an example of a duotensor\index{duotensors}.  Diagrammatically , it corresponds to a box with all white dots on it
\begin{equation}
\begin{Diagram}{0}{0}
\Duobox[5]{A}{0,0}
\putlatex[-44,-9]{\ensuremath{\vdots}} \putlatex[44,-9]{\ensuremath{\vdots}}
\inwhite{A}{1}\Duosymbol{a}\inwhite{A}{2.2}\Duosymbol{b} \inwhite{A}{5}\Duosymbol{c}
\outwhite{A}{1}\Duosymbol{d}\outwhite{A}{2.2}\Duosymbol{e} \outwhite{A}{5}\Duosymbol{f}
\end{Diagram}
~~~\Leftrightarrow~~~ {}^{d_4e_5\dots f_6}\!A_{a_1b_2\dots c_3}
\end{equation}
since we want to have black dots next to the fiducial elements in (\ref{decomposibilitydiagram}).  We can put a white dot on a fiducial element
\begin{equation}\label{whitedotfids}
\begin{Diagram}{0}{0.1}
\fideffect{X}{0,0} \outblack{X}{1}\Duosymbol[65,0]{a} \inwire{X}{1}\Opsymbol{a}
\wwmetric[0.7]{g}{3.1,7pt}
\end{Diagram}
~=~
\begin{Diagram}{0}{0.1}
\fideffect{X}{0,0} \outwhite{X}{1}\Duosymbol{a} \inwire{X}{1}\Opsymbol{a}
\end{Diagram}
~~~~~~~~~~
\begin{Diagram}{0}{-0.1}
\fidprep{X}{0,0} \inblack{X}{1} \Duosymbol[-65,0]{a} \outwire{X}{1} \opsymbol{a}
\wwmetric[0.7]{g}{-3.1,-7pt}
\end{Diagram}
~=~
\begin{Diagram}{0}{-0.1}
\fidprep{X}{0,0} \inwhite{X}{1} \Duosymbol{a} \outwire{X}{1} \opsymbol{a}
\end{Diagram}
\end{equation}
A white dot on a fiducial element therefore corresponds to a sum over fiducial elements weighted by the relevent entries in the inverse of the hopping metric.  With this understanding, we can place black and white dots on the links in (\ref{decomposibilitydiagram}).  In this way we can extract a box with inputs and outputs having black and white dots.  For example
\begin{equation}\label{duotensorexample}
\begin{Diagram}{0}{0}
\Duobox[4]{A}{0,0}
\outwhite[-4]{A}{1.5}\Duosymbol{a} \outwhite{A}{2.5}\Duosymbol{b} \outblack[4]{A}{3.5}\Duosymbol{d}
\inblack[-6]{A}{1}\Duosymbol{b}    \inblack[-2]{A}{2}\Duosymbol{c}   \inwhite[2]{A}{3}\Duosymbol{b} \inwhite[6]{A}{4}\Duosymbol{c}
\end{Diagram}
~~~~~\Leftrightarrow ~~~~~   {}^{a_1b_2}_{b_3c_4} \! A^{d_5}_{b_6c_7}
\end{equation}
The map between black and white dots and the placement of indices is given by
\begin{equation}
{}_\bullet^\circ A^\bullet_\circ
\end{equation}
Subscripts and pre-subscripts correspond to the inputs on the left of the box.  Superscripts and pre-superscripts correspond to outputs on the right of the box.
The object in (\ref{duotensorexample}) is tensor-like with a bit more structure, indices can appear on the left as well as the right.  The reason for this is that there are two independently chosen basis sets associated with every index - a fiducial set of effects and a fiducial set of preparations.  (For tensors we only have one choice of basis set associated with each index.)  Given this, we will call this mathematical object a {\it duotensor}.  We can put an index on the right or hop it over to the left (using the hopping tensor), or vice versa.  For example,
\begin{equation}
\begin{Diagram}{0}{-0.1}
\Duobox[4]{A}{0,0}
\outwhite[-4]{A}{1.5}\Duosymbol{a} \outblack{A}{2.5}\Duosymbol{b} \outblack[4]{A}{3.5}\Duosymbol{d}
\inblack[-6]{A}{1}\Duosymbol{b}    \inwhite[-2]{A}{2}\Duosymbol{c}   \inblack[2]{A}{3}\Duosymbol{b} \inwhite[6]{A}{4}\Duosymbol{c}
\end{Diagram}
\!\! ~=~ \,
\begin{Diagram}{0}{-0.1}
\Duobox[4]{A}{0,0}
\outwhite[-4]{A}{1.5}\Duosymbol{a} \outwhite{A}{2.5}\Duosymbol[70,0]{b} \outblack[4]{A}{3.5}\Duosymbol{d}
\inblack[-6]{A}{1}\Duosymbol{b}    \inblack[-2]{A}{2}\Duosymbol[-70,0]{c}   \inwhite[2]{A}{3}\Duosymbol[-70,0]{b} \inwhite[6]{A}{4}\Duosymbol{c}
\wwmetric[0.8]{g1}{-3.8,0.53} \bbmetric[0.8]{g2}{-3.8,-0.53} \bbmetric[0.8]{g3}{3.8,0}
\end{Diagram}
\end{equation}
or, in symbolic form,
\begin{equation}
{}^{a_1}_{b_3b_6} \! A^{b_2d_5}_{c_4c_7}   = {}_{b'_2} g^{b_2}\,{}_{b_6} g^{b'_6}\, {}^{c'_4}\! g_{c_4}\,\, {}^{a_1b'_2}_{b_3c'_4} \! A^{d_5}_{b'_6c_7}
\end{equation}
Subscripts always correspond to inputs and superscripts always correspond to outputs.  For the diagrammatic representation, subscripts go on left and superscripts on the right of the boxes.

If we change to a new set of fiducial effects then we perform a transformation that effects the subscripts and superscripts.  If we change to a new set of fiducial preparations then we perform a transformation that effects the pre-subscripts and pre-superscripts.  It is shown in Appendix \ref{transformingduotensors} that these transformations have the properties one would expect of a tensor-like object.

We can use the hopping metric to put the indices on the left or the right (or correspondingly, change their colour in the diagrammatic representation).  It is interesting to consider a few possibilities
\begin{enumerate}
\item \emph{All white dots} \index{duotensors!all white dots}corresponds to the weighting in the sum over fiducial elements. For example,
\begin{equation}\label{allwhitedots}
\begin{Diagram}{0}{0}
\Opbox[2]{A}{0,0}
\inwire[-4]{A}{1}\inwire[4]{A}{2} \outwire[-4]{A}{1}\outwire[4]{A}{2}
\end{Diagram}
~~\equiv~~
\begin{Diagram}{0}{0}
\Duobox[2]{A}{0,0}
\fideffect[0.6]{1A}{-5.4,0.3}\inwire{1A}{1}
\fideffect[0.6]{2A}{-4,-0.6}\inwire{2A}{1}
\fidprep[0.6]{A1}{4,0.6}\outwire{A1}{1}
\fidprep[0.6]{A2}{5.4,-0.3}\outwire{A2}{1}
\linkbw{1A}{A}{1}{1} \linkbw{2A}{A}{1}{2}
\linkwb{A}{A1}{1}{1} \linkwb{A}{A2}{2}{1}
\end{Diagram}
\end{equation}
\item \emph{All black dots} \index{duotensors!all black dots}corresponds to the fiducial probabilities (when we place fiducial preparations on each input and fiducial effects on each output). For example
\begin{equation}\label{allblackduotensor}
\begin{Diagram}{0}{0}
\Duobox[2]{A}{0,0}
\inblack{A}{1}\Duosymbol{a}\inblack{A}{2}\Duosymbol{b}
\outblack{A}{1}\Duosymbol{c}\outblack{A}{2}\Duosymbol{d}
\end{Diagram}
~=~
\text{Prob}\left(
\begin{Diagram}{0}{0}
\Opbox[2]{A}{0,0}
\fidprep[0.65]{1X}{-0.4,-3}\wire{1X}{A}{1}{1}\opsymbol{a}\inblack{1X}{1}\Duosymbol{a}
\fidprep[0.65]{2X}{0.4,-4.5}\wire{2X}{A}{1}{2}\otherside\opsymbol{b}\inblack{2X}{1}\Duosymbol{b}
\fideffect[0.65]{X1}{-0.4,4.5}\wire{A}{X1}{1}{1}\opsymbol{c}\outblack{X1}{1}\Duosymbol{c}
\fideffect[0.65]{X2}{0.4,3}\wire{A}{X2}{2}{1}\otherside\opsymbol{d}\outblack{X2}{1}\Duosymbol{d}
\end{Diagram}
\right)
\nonumber\end{equation}
This follows by capping the inputs outputs in (\ref{allwhitedots}) with fiducial elements.
\item \emph{Standard form} \index{duotensors!standard form}is when we have all the indices on the right hand side so we have only superscripts and subscripts.  In standard form we are only invoking the use of fiducial effects (and not fiducial preparations).  Diagrammatically this corresponds to having all white dots on the left and all black dots on the right
\begin{equation}
\begin{Diagram}{0}{0}
\Duobox[2]{A}{0,0}\inwhite{A}{1}\inwhite{A}{2}\outblack{A}{1}\outblack{A}{2}
\end{Diagram}
\end{equation}
In earlier sections of this work we were, effectively, only using the standard form. We can clearly connect duotensors up in standard form without using the hopping metric since black dots will always be put next to white dots.
\end{enumerate}

\subsection{The identity transformation and the hopping metric}\label{identitytranssec}

The hopping metric is, itself, a duotensor.  We can put this duotensor in standard form.  Then we have $\wbdots$.  This is equal to the identity (as we noted in
Sec.\ \ref{diagrammaticrepresentationsimplecircuit}).  Hence, the operation corresponding to this duotensor must effect the identity map.  This operation is simply given by tagging on the fiducial elements.  Hence,
\begin{equation}\label{identityfidsequation}
\begin{Diagram}{0}{-0.1}
\fideffect[0.7]{X}{0,-0.15} \fidprep[0.7]{Y}{7,0.2}
\linkbwbw{X}{Y}{1}{1}
\inwire{X}{1} \outwire{Y}{1}
\end{Diagram}
~=~
\begin{Diagram}{0}{-0.1}
\fideffect[0.7]{X}{0,-0.15} \fidprep[0.7]{Y}{2,0.2}
\link{X}{Y}{1}{1}
\inwire{X}{1} \outwire{Y}{1}
\end{Diagram}
~\equiv~
\begin{Diagram}{0}{0}
\thispoint{A}{-1,-3}\thispoint{B}{+1,+3}
\wire{A}{B}{1}{1}
\end{Diagram}
\end{equation}
We could have used $\bbdots$, $\wwdots$, or $\bwdots$ on the LHS with matching colours on the dots on the fiducial elements.  Interestingly, (\ref{identityfidsequation}) is equivalent to the assumption of full decomposability (and therefore equivalent to the assumption of tomographic locality in {\bf P2}). This is clear since, by (\ref{identityfidsequation}), we can insert the LHS of (\ref{identityfidsequation}) on the end of every wire coming into or going out of an operation.  Then we can apply the $p(\cdot)$ function to obtain full decomposability.   We have already shown that (\ref{identityfidsequation}) follows from full decomposability so we have established equivalence.

\subsection{General circuits}

We can apply full decomposability to any circuit to convert the circuit into an equivalent duotensor calculation. We will show how to do this by example.  Consider the circuit
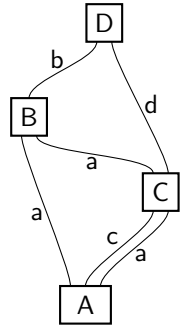
\begin{equation}
\begin{Diagram}{0}{-2}
\Opbox{A}{0,0} \Opbox[2]{C}{4,6} \Opbox[2]{B}{-3,10} \Opbox[2]{D}{1,15}
\wire{A}{B}{1}{1}\opsymbol{a} \wire{A}{C}{2}{1}\opsymbol[0,6]{c} \wire{A}{C}{3}{2}\otherside\opsymbol{a} \wire{C}{B}{1}{2}\opsymbol{a}
\wire{C}{D}{2}{2}\otherside\opsymbol[4,0]{d} \wire{B}{D}{1.5}{1}\opsymbol[0,6]{b}
\end{Diagram}
\end{equation}
Applying full decomposability we obtain the equivalent diagram
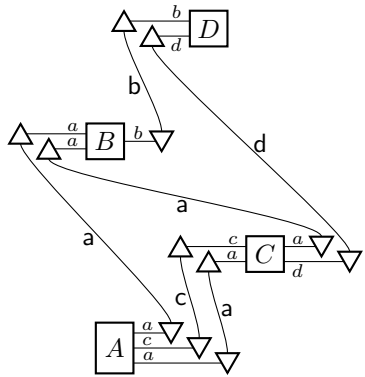
\begin{equation}
\begin{Diagram}{0}{-3.3}
\begin{move}{-2,4}
\Duobox{A}{0,0}
\linkedprep[0.7]{A}{1}{A1}{3}{0}\duosymbol[-3,-6]{\scriptstyle a}
\linkedprep[0.7]{A}{2}{A2}{4.5}{0}\duosymbol[-25,-6]{\scriptstyle c}
\linkedprep[0.7]{A}{3}{A3}{6}{0}\duosymbol[-44,-6]{\scriptstyle a}
\end{move}
\begin{move}{-2.5,15}
\Duobox[2]{B}{0,0}
\linkedeffect[0.7]{B}{1}{1B}{-4.5}{0}\duosymbol[25,-6]{\scriptstyle a}
\linkedeffect[0.7]{B}{2}{2B}{-3}{0}\duosymbol[3,-6]{\scriptstyle a}
\linkedprep[0.7]{B}{1.5}{B15}{3}{0} \duosymbol[-3,-3]{\scriptstyle b}
\end{move}
\begin{move}{6,9}
\Duobox[2]{C}{0,0}
\linkedeffect[0.7]{C}{1}{1C}{-4.5}{0}\duosymbol[25,-6]{\scriptstyle c}
\linkedeffect[0.7]{C}{2}{2C}{-3}{0}\duosymbol[3,-6]{\scriptstyle a}
\linkedprep[0.7]{C}{1}{C1}{3}{0} \duosymbol[-3,-6]{\scriptstyle a}
\linkedprep[0.7]{C}{2}{C2}{4.5}{0} \otherside\duosymbol[-25,3]{\scriptstyle d}
\end{move}
\begin{move}{3,21}
\Duobox[2]{D}{0,0}
\linkedeffect[0.7]{D}{1}{1D}{-4.5}{0}\duosymbol[25,-2]{\scriptstyle b}
\linkedeffect[0.7]{D}{2}{2D}{-3}{0}\otherside\duosymbol[3,3]{\scriptstyle d}
\end{move}
\wire{A1}{1B}{1}{1}\opsymbol{a} \wire{A2}{1C}{1}{1}\opsymbol{c} \wire{A3}{2C}{1}{1}\otherside\opsymbol{a}
\wire{C1}{2B}{1}{1}\opsymbol{a} \wire{C2}{2D}{1}{1}\otherside\opsymbol[4,0]{d} \wire{B15}{1D}{1}{1}\opsymbol{b}
\end{Diagram}
\end{equation}
which is the same as
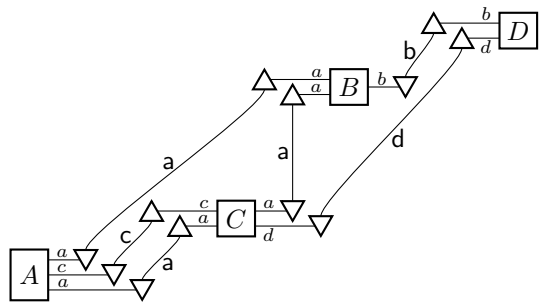
\begin{equation}
\begin{Diagram}{0}{-3}
\begin{move}{-7,5}
\Duobox{A}{0,0}
\linkedprep[0.7]{A}{1}{A1}{3}{0}\duosymbol[-3,-6]{\scriptstyle a}
\linkedprep[0.7]{A}{2}{A2}{4.5}{0}\duosymbol[-25,-6]{\scriptstyle c}
\linkedprep[0.7]{A}{3}{A3}{6}{0}\duosymbol[-44,-6]{\scriptstyle a}
\end{move}
\begin{move}{10,15}
\Duobox[2]{B}{0,0}
\linkedeffect[0.7]{B}{1}{1B}{-4.5}{0}\duosymbol[25,-6]{\scriptstyle a}
\linkedeffect[0.7]{B}{2}{2B}{-3}{0}\duosymbol[3,-6]{\scriptstyle a}
\linkedprep[0.7]{B}{1.5}{B15}{3}{0} \duosymbol[-3,-3]{\scriptstyle b}
\end{move}
\begin{move}{4,8}
\Duobox[2]{C}{0,0}
\linkedeffect[0.7]{C}{1}{1C}{-4.5}{0}\duosymbol[25,-6]{\scriptstyle c}
\linkedeffect[0.7]{C}{2}{2C}{-3}{0}\duosymbol[3,-6]{\scriptstyle a}
\linkedprep[0.7]{C}{1}{C1}{3}{0} \duosymbol[-3,-6]{\scriptstyle a}
\linkedprep[0.7]{C}{2}{C2}{4.5}{0} \otherside\duosymbol[-25,3]{\scriptstyle d}
\end{move}
\begin{move}{19,18}
\Duobox[2]{D}{0,0}
\linkedeffect[0.7]{D}{1}{1D}{-4.5}{0}\duosymbol[25,-2]{\scriptstyle b}
\linkedeffect[0.7]{D}{2}{2D}{-3}{0}\otherside\duosymbol[3,3]{\scriptstyle d}
\end{move}
\wire{A1}{1B}{1}{1}\opsymbol{a} \wire[0.4]{A2}{1C}{1}{1}\opsymbol{c} \wire[0.4]{A3}{2C}{1}{1}\otherside\opsymbol{a}
\wire{C1}{2B}{1}{1}\opsymbol{a} \wire{C2}{2D}{1}{1}\otherside\opsymbol{d} \wire[0.4]{B15}{1D}{1}{1}\opsymbol[-3,0]{b}
\end{Diagram}
\end{equation}
We can insert black and white dots in each link (with a black dot next to the fiducial element) then, using (\ref{hoppingequivdiagram}), insert the hopping metric to obtain the equivalent diagram
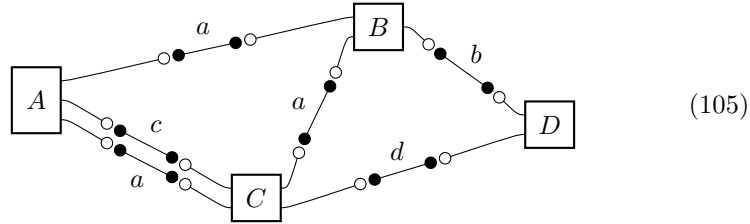
\begin{equation}
\begin{Diagram}[1.3]{0}{0}
\Duobox{A}{-3,0} \Duobox[2]{B}{11,3} \Duobox[2]{C}{6,-4}  \Duobox[2]{D}{18,-1}
\linkwbbw[0.6]{A}{B}{1}{1}\duosymbol{a} \linkwbbw[0.6]{A}{C}{2}{1}\duosymbol{c} \linkwbbw[0.6]{A}{C}{3}{2}\otherside\duosymbol{a}
\linkwbbw[0.6]{C}{B}{1}{2}\duosymbol{a} \linkwbbw[0.6]{C}{D}{2}{2}\duosymbol{d} \linkwbbw[0.6]{B}{D}{1.5}{1}\duosymbol{b}
\end{Diagram}
\end{equation}
Note that we are implicitly using the fact that the $p(\cdot)$ function factorizes over disjoint circuits.   Hence we obtain
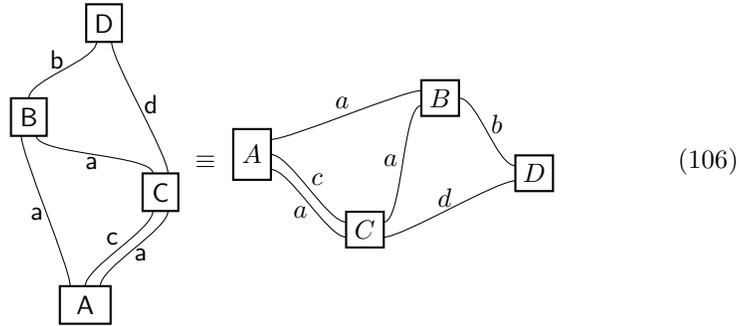
\begin{equation}
\begin{Diagram}{0}{-2}
\Opbox{A}{0,0} \Opbox[2]{C}{4,6} \Opbox[2]{B}{-3,10} \Opbox[2]{D}{1,15}
\wire{A}{B}{1}{1}\opsymbol{a} \wire{A}{C}{2}{1}\opsymbol[0,6]{c} \wire{A}{C}{3}{2}\otherside\opsymbol{a} \wire{C}{B}{1}{2}\opsymbol{a}
\wire{C}{D}{2}{2}\otherside\opsymbol[4,0]{d} \wire{B}{D}{1.5}{1}\opsymbol[0,6]{b}
\end{Diagram}
~ \equiv ~
\begin{Diagram}{0}{0}
\Duobox{A}{0,0} \Duobox[2]{B}{10,3} \Duobox[2]{C}{6,-4}  \Duobox[2]{D}{15,-1}
\link{A}{B}{1}{1}\duosymbol{a} \link{A}{C}{2}{1}\duosymbol{c} \link{A}{C}{3}{2}\otherside\duosymbol{a}
\link{C}{B}{1}{2}\duosymbol{a} \link{C}{D}{2}{2}\duosymbol{d} \link{B}{D}{1.5}{1}\duosymbol{b}
\end{Diagram}
\end{equation}
where we have canceled over pairs of black and white dots. Using (\ref{probAequivA}), we obtain
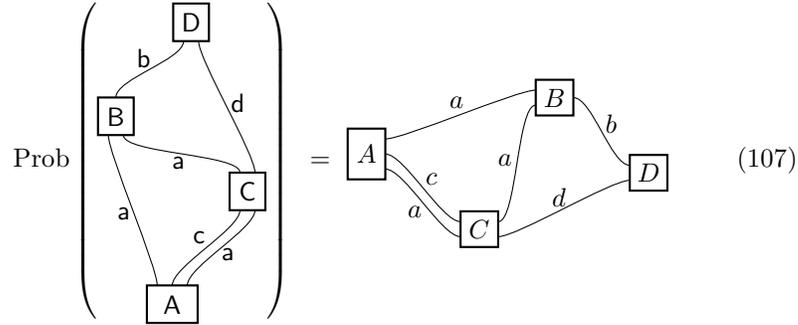
\begin{equation}
\text{Prob}\left(
\begin{Diagram}{0}{-2}
\Opbox{A}{0,0} \Opbox[2]{C}{4,6} \Opbox[2]{B}{-3,10} \Opbox[2]{D}{1,15}
\wire{A}{B}{1}{1}\opsymbol{a} \wire{A}{C}{2}{1}\opsymbol[0,6]{c} \wire{A}{C}{3}{2}\otherside\opsymbol{a} \wire{C}{B}{1}{2}\opsymbol{a}
\wire{C}{D}{2}{2}\otherside\opsymbol[4,0]{d} \wire{B}{D}{1.5}{1}\opsymbol[0,6]{b}
\end{Diagram}
\right)
~ = ~
\begin{Diagram}{0}{0}
\Duobox{A}{0,0} \Duobox[2]{B}{10,3} \Duobox[2]{C}{6,-4}  \Duobox[2]{D}{15,-1}
\link{A}{B}{1}{1}\duosymbol{a} \link{A}{C}{2}{1}\duosymbol{c} \link{A}{C}{3}{2}\otherside\duosymbol{a}
\link{C}{B}{1}{2}\duosymbol{a} \link{C}{D}{2}{2}\duosymbol{d} \link{B}{D}{1.5}{1}\duosymbol{b}
\end{Diagram}
\end{equation}
It is striking that the probability for a circuit is given by a duotensor calculation that looks the same as the circuit itself.  A similar thing will be true if we use symbolic notation (putting all the duotensors in standard form).  This will clearly be true for any circuit.  In the diagrammatic case we need only rotate the diagram through $90^\circ$ and change the font from sans serif to normal maths font.  In the symbolic case we need only change the font.

\subsection{Formalism locality}\label{formalismlocality}

\index{formalism locality}In this subsection we will show how we can formulate physical theories that can be put into the duotensor framework in a formalism local fashion.  So far we have only shown how to calculate probabilities for circuits.   In this section we will be interested in calculating probabilities for fragments.  It is reasonable to assume that circuits have probabilities associated with them that are independent of what is happening elsewhere (this is {\bf Assump 1}).  However, it is not reasonable to assume the same thing of fragments in general since the probability associated with a fragment may depend on what is outside the fragment (for example, a fragment has open inputs into which some system may be sent).  However, in special circumstances, a fragment may have a probability that is independent (or approximately independent) of what is outside the fragment.

In Sec.\ \ref{wellconditionedprobabilities} we introduced the notion of well conditioned probabilities.  We can meaningfully speak of a probability $\text{Prob}(\mathsf{A|B})$ only if it is well conditioned (so that $\text{Prob}(\mathsf{A|BC})$ is independent of $\mathsf C$ where $\mathsf C$ completes $\mathsf{AB}$ into a circuit).  We now introduce a further notion.
\begin{quote}
{\bf Well conditioned probability ratio:} \index{probability ratio}We will say the probability ratio
\begin{equation}\label{probratio}
\frac{\text{Prob}(\mathsf{A}[i])}{\text{Prob}(\mathsf{A}[j])}
\end{equation}
is well conditioned if
\[ \frac{\text{Prob}\mathsf{(A[\mathnormal{i}]C)}}{\text{Prob}\mathsf{(A[\mathnormal{j}]C)}} \]
is independent of $\mathsf C$ where $\mathsf C$ is any fragment which completes $\mathsf A[i]$ (and $\mathsf A[j]$) into a circuit.   Here $\mathsf A[i]$ and $\mathsf A[j]$ correspond to different outcome sets for the same setup.
\end{quote}
According to {\bf Assump 1} we can associate probabilities with both the numerator and denominator of this expression (as they correspond to circuits) so we this is a test we can run.
Note that, by (\ref{condproducteqn}), this equivalent to demanding that
\[ \frac{\text{Prob}(\mathsf{A}[i]\,|\,\mathsf{C})}{\text{Prob}(\mathsf{A}[j]\,|\,\mathsf{C})}  \]
is independent of $\mathsf C$ (the conditional probabilities in this expression are well conditioned since $\mathsf{A}[i]\mathsf{C}$ and $\mathsf{A}[j]\mathsf{C}$ are circuits).

If $\text{Prob}(\mathsf{A}[i])$ and $\text{Prob}(\mathsf{A}[j])$ are each well conditioned then it follows that their ratio is.  However, it is possible that, taken separately, they are not well conditioned but that the ratio is.  If we define probabilities as long-run relative frequencies then the probability ratio is equal to the number of times $\mathsf{A}[i]$ happens divided by the number of times $\mathsf{A}[j]$ happens in the long run.

A special case is where the outcome set $\mathsf{o}_j$ for the fragment $\mathsf{A}[j]$ is the set of all possible outcomes. We denote this $\mathsf{A}[I]$. Since some outcome must happen we know that $\text{Prob}\mathsf{(A[\mathnormal{I}]\,|\, B)}=1$ for any $\mathsf B$.  We can, then, regard the probability $\text{Prob}(\mathsf{ A[\mathnormal{i}]\,|\,B })$ as a probability ratio (using (\ref{condprobgeneral}))
\begin{equation}\label{ratiotocondition}
\text{Prob}\mathsf{(A[\mathnormal{i}]\,|\,B)}=\frac{\text{Prob}\mathsf{(A[\mathnormal{i}]B)}}{\text{Prob}\mathsf{(A[\mathnormal{I}]B)}}
\end{equation}
Hence, the idea of a probability ratio is more general than that of a conditional probability.

A probability ratio that is not well conditioned is not well defined.  Whatever value we write down for it could be made to be wrong by an adversary who has control over other conditions that would effect the outcome we are looking at.  Therefore, we cannot expect a physical theory to predict the values of probability ratios that are not well conditioned.  However, it is reasonable to expect our physical theory to tell us whether a probability ratio is well conditioned.

\emph{Our objective} is to construct a mathematical framework for theories which
\begin{enumerate}
\item Associates mathematical objects with all fragments.
\item Provides a mathematical condition for saying whether the probability ratio
\[\frac{\text{Prob}\mathsf{(A[i])}}{\text{Prob}\mathsf{(A[j])}}   \]
for any $\mathsf{A[i]}$ and $\mathsf{A[j]}$ (for the same setup) is well conditioned employing only mathematical objects associated with these fragments.
\item In the case that the probability ratio is well conditioned, provides an expression saying what it is equal to employing only mathematical objects associated with the fragments $\mathsf{A[i]}$ and $\mathsf{A[j]}$.
\end{enumerate}
If we can do this we will have the formalism locality property.

We will now see how to achieve this objective in the duotensor framework. First we state the result.
\begin{quote} {\bf The probability ratio}
\begin{equation}\label{wellconditionedconditioncandidate}
\frac{\text{Prob}(\mathsf{E[i]})}{\text{Prob}(\mathsf{E[j]})}
\end{equation}
where $\mathsf{E[i]}$ and $\mathsf{E[j]}$ are two fragments corresponding to different outcome sets for the same setup
is
\begin{description}
\item[well conditioned] if and only if the corresponding duotensors, $E[i]$ and $E[j]$, are proportional, and
\item[equal to] the constant of proportionality $k$ in $E[i]=k E[j]$ (if well conditioned).
\end{description}
\end{quote}
To prove this we note that, for the probability ratio (\ref{wellconditionedconditioncandidate}) to be well conditioned, we require that
\begin{equation}\label{wellconditionedcondition}
\frac{\text{Prob}(\mathsf{E[i]F})}{\text{Prob}(\mathsf{E[j]F})}
\end{equation}
be independent of $\mathsf F$ for any choice of fragment $\mathsf F$ that completes the circuit.  One set of fragments for $\mathsf F$ we can consider are the fragments that consist of simply putting fiducial preparations on each input of $\mathsf{ E[i]}$, and $\mathsf{ E[j]}$, and fiducial effects on each output.  This gives us the fiducial probabilities for these two fragments.   Let us consider
(\ref{wellconditionedcondition}) with respect to these choices for $\mathsf F$.  We saw in Sec.\ \ref{whatareduotensors} that the entries of the duotensor with all black dots is equal to the fiducial probabilities.  Hence, for (\ref{wellconditionedcondition}) to hold, we require that the elements of $ E[i]$ are proportional to the corresponding elements of $ E[j]$ with the same constant of proportionality.  Hence the two duotensors must be proportional.  (This is clearly necessary when these two duotensors are in \lq\lq all black dots form" but it must also be true when they are both in any other given form since multiplication by hopping tensors, which are non-singular, will not effect such a proportionality relationship).   This is a necessary condition but it is also a sufficient condition since, clearly, (\ref{wellconditionedcondition}) is independent of the choice of $\mathsf F$ (where this completes the circuit) when $E[i]$ and $E[j]$ are proportional.   Further, when the two duotensors are parallel then the probability ratio is simply given by the proportionality constant.

This result is in accordance with our objectives as stated above.  In particular, note that we have the property of formalism locality since we employ only the duotensors associated with the fragments $\mathsf{E}[i]$ and $\mathsf{E}[j]$ corresponding to the set up we are interested in (which might be part of a much bigger set up).

\section{Operators and duotensors}\label{operatorsandduotensors}

The duotensor framework was originally built to be applied to circuits and fragments built out of operations.  However, as we will see, it can also be applied in a fairly analogous way to objects built out of operators acting on complex Hilbert spaces.  In this section we present this as a purely mathematical structure.  Of course, our interest in this structure is that quantum theory fits into it very comfortably.  The duotensor framework for operators is based on two facts (since we are in a mathematical rather than physical setting we have facts rather than assumptions).  The first fact is that a circuit built from operators is equal to a real number which depends only on the details of this circuit (this is analogous to {\bf Assump 1} that we can associate a probability with a circuit that depends only on the details of that circuit).  The second is that operators are fully decomposable (which is, of course, analogous to the assumption that operations are fully decomposable).  We will explain these facts more fully below.

\subsection{Operators}

\index{operators}We start by introducing types, $\mathsf a$, $\mathsf b$, \dots.  We have composite types such as $\mathsf{aab}$, etc. We may sometimes represent a composite type by a single letter.  Next we introduce complex Hilbert spaces, ${\cal H}_\mathsf{a_1}$, ${\cal H}_\mathsf{b_2}$, \dots having dimensions $N_\mathsf{a}$, $N_\mathsf{b}$, \dots (the type determines the Hilbert space dimension).  We define the space of Hermitian operators acting on these Hilbert spaces as ${\cal V}_\mathsf{a_1}$, ${\cal V}_\mathsf{b_2}$, \dots.  These spaces have dimension $N_\mathsf{a}^2$, $N_\mathsf{b}^2$, \dots (we can deduce this simply by counting the number of real parameters in a Hermitian matrix).  We represent an operator in ${\cal V}_\mathsf{a_1}$ by $\hat A_\mathsf{a_1}$, $\hat B_\mathsf{a_1}$ \dots We will call this a {\it result operator}\index{result operators}.  We will introduce complex Hilbert spaces, ${\cal H}^\mathsf{a_1}$, ${\cal H}^\mathsf{b_2}$, \dots having dimensions $N_\mathsf{a}$, $N_\mathsf{b}$, \dots. These are isomorphic to the earlier introduced Hilbert spaces (as they have the same dimensions) but we introduce them also with a superscript label to enable us to define certain structures later.  The space of Hermitian operators on these Hilbert spaces are ${\cal V}^\mathsf{a_1}$, ${\cal V}^\mathsf{b_2}$, \dots.  These spaces have dimension $N_\mathsf{a}^2$, $N_\mathsf{b}^2$, \dots.  Operators in ${\cal V}^\mathsf{a_1}$ will be written $\hat A^\mathsf{a_1}$, $\hat B^\mathsf{a_1}$ \dots We will call this a {\it preparation operator}\index{preparation operators}.

We define the Hilbert space ${\cal H}_\mathsf{a_1b_2} := {\cal H}_\mathsf{a_1}\otimes{\cal H}_\mathsf{b_2}$ for composite type $\mathsf{ ab}$.  This has dimension $N_\mathsf{ab}=N_\mathsf{a}N_\mathsf{b}$.  We can also define the Hilbert space ${\cal H}_\mathsf{a_1}^\mathsf{b_2}= {\cal H}_\mathsf{a_2}\otimes{\cal H}^\mathsf{b_2}$. This also has dimension $N_\mathsf{ab}=N_\mathsf{a}N_\mathsf{b}$.  In general we have
\begin{equation}
{\cal H}_\mathsf{a_1b_2\dots c_3}^\mathsf{d_4e_5\dots f_6} := {\cal H}_\mathsf{a_1} \otimes {\cal H}_\mathsf{b_2} \otimes \dots \otimes {\cal H}_\mathsf{c_3}
\otimes {\cal H}^\mathsf{d_4} \otimes {\cal H}^\mathsf{e_5} \otimes \dots \otimes {\cal H}^\mathsf{f_6}
\end{equation}
The dimension of this space is $N_\mathsf{ab\dots cde\dots f}= N_\mathsf{a}N_\mathsf{b}\dots N_\mathsf{c}N_\mathsf{d}N_\mathsf{e}\dots N_\mathsf{f}$.

We define ${\cal V}_\mathsf{a_1b_2}$ as the space of Hermitian operators acting on ${\cal H}_\mathsf{a_1b_2}$.  In fact we have ${\cal V}_\mathsf{a_1b_2}={\cal V}_\mathsf{a_1}\otimes{\cal V}_{b_2}$ (this is true for complex Hilbert spaces but not for real Hilbert spaces).  The dimension of   ${\cal V}_\mathsf{a_1b_2}$ is $N_\mathsf{ab}^2 = N^2_\mathsf{a} N^2_\mathsf{b}$.   We define ${\cal V}_\mathsf{a_1b_2\dots c_3}^\mathsf{d_4e_5\dots f_6}$ as the space of Hermitian operators acting on ${\cal H}_\mathsf{a_1b_2\dots c_3}^\mathsf{d_4e_5\dots f_6}$.  We have
\begin{equation}\label{VequalsVVV}
{\cal V}_\mathsf{a_1b_2\dots c_3}^\mathsf{d_4e_5\dots f_6} := {\cal V}_\mathsf{a_1} \otimes {\cal V}_\mathsf{b_2} \otimes \dots \otimes {\cal V}_\mathsf{c_3}
\otimes {\cal V}^\mathsf{d_4} \otimes {\cal V}^\mathsf{e_5} \otimes \dots \otimes {\cal V}^\mathsf{f_6}
\end{equation}
Note, again, that this is true when the underling spaces are complex (rather than real) Hilbert spaces. We write
\begin{equation}
\hat A_\mathsf{a_1b_2\dots c_3}^\mathsf{d_4e_5\dots f_6}
~~~~~\Leftrightarrow~~~~~
\begin{Diagram}[1.4]{0}{0}
\Dopbox[5]{A}{0,0}
\inwire{A}{1}\Opsymbol{a} \inwire{A}{2.2}\Opsymbol{b} \putlatex[30,15]{\ensuremath{\dots}} \inwire{A}{5}\Opsymbol{c}
\outwire{A}{1}\Opsymbol{d}\outwire{A}{2.2}\Opsymbol{e}\putlatex[30,-50]{\ensuremath{\dots}}   \outwire{A}{5}\Opsymbol{f}
\end{Diagram}
\end{equation}
for an operator in ${\cal V}_\mathsf{a_1b_2\dots c_3}^\mathsf{d_4e_5\dots f_6}$.   We have given both the symbolic notation (on the left) and diagrammatic notation (on the right).  We will use the terminology of inputs and outputs corresponding to the subscripts and superscripts respectively.

We introduce a fiducial operator \index{fiducial operators} basis for ${\cal V}_\mathsf{a_1}$
\begin{equation}
\hat{X}_\mathsf{a_1}^{a_1}
~~\Longleftrightarrow ~~
\begin{Diagram}{0}{0.1}
\dfideffect{X}{0,0}  \outblack{X}{1}\Duosymbol{a} \inwire{X}{1}\Opsymbol{a}
\end{Diagram}
~~~~~ \text{where}~~a_1=1~~\text{to}~~K_\mathsf{a}
\end{equation}
These are a spanning set of operators for the space ${\cal V}_\mathsf{a_1}$.  This means that any $\hat A_\mathsf{a_1}\in{\cal V}_\mathsf{a_1}$ can be written
\begin{equation}
\hat{B}_\mathsf{a_1}= B_{a_1} \hat{X}_\mathsf{a_1}^{a_1}~~~~~\Longleftrightarrow  ~~~~~
\begin{Diagram}{0}{0}
\Dopbox[2]{B}{0,0} \inwire{B}{1.5}\Opsymbol{a}
\end{Diagram}
~~=~~
\begin{Diagram}{0}{0.07}
\dfideffect{X}{0,-7pt}  \Duobox[2]{B}{5,0}
\inwire{X}{1}\Opsymbol{a} \linkbw{X}{B}{1}{1.5} \duosymbol{a}
\end{Diagram}
\end{equation}
We define
\begin{equation}
\begin{Diagram}{0}{0.07}
\dfideffect{X}{0,-7pt} \Duobox[2]{B}{5,0}
\inwire{X}{1}\Opsymbol{a} \linkbw{X}{B}{1}{1.5} \duosymbol{a}
\end{Diagram}
~~ := ~~
\begin{Diagram}{0}{0.07}
\dfideffect{X}{0,-7pt} \Duobox[2]{B}{4,0}
\inwire{X}{1}\Opsymbol{a} \link{X}{B}{1}{1.5} \duosymbol{a}
\end{Diagram}
\end{equation}
(i.e. we can cancel over black and white dots).  As before, the horizontal line indicates that we are summing over the associated index ($a_1$ in this case).

Similarly, we introduce a fiducial set of operators for the space ${\cal V}^\mathsf{a_1}$
\begin{equation}
{}_{a_1}\!\hat{X}^\mathsf{a_1}~~\Longleftrightarrow ~~
\begin{Diagram}{0}{-0.1}
\dfidprep{X}{0,0} \inblack{X}{1} \Duosymbol{a} \outwire{X}{1} \opsymbol{a}
\end{Diagram}
~~ \text{where}~~a_1=1~~\text{to}~~K_\mathsf{a}
\end{equation}
We can write any operator, $\hat {A^{a_1}}\in {\cal V}^\mathsf{a_1}$, as
\vspace{-20pt}
\begin{equation}
\hat A^{a_1} = {}^{a_1}\!\!A \,\,\,{}_{a_1}\!\hat{X}^\mathsf{a_1}
~~~\Longleftrightarrow ~~~
\begin{Diagram}{0}{-0.2}
\Dopbox[2]{A}{0,-7pt}  \outwire{A}{1.5} \Opsymbol{a}
\end{Diagram}
~~ = ~~
\begin{Diagram}{0}{-0.2}
\Duobox[2]{A}{0,-7pt} \dfidprep{X}{5,0}  \linkwb{A}{X}{1.5}{1}\duosymbol{a} \outwire{X}{1}\Opsymbol{a}
\end{Diagram}
\end{equation}
We define
\begin{equation}
\begin{Diagram}{0}{0}
\Duobox[2]{A}{0,-7pt} \dfidprep{X}{5,0}  \linkwb{A}{X}{1.5}{1}\duosymbol{a} \outwire{X}{1}\Opsymbol{a}
\end{Diagram}
~~ := ~~
\begin{Diagram}{0}{0}
\Duobox[2]{A}{0,-7pt} \dfidprep{X}{4,0}  \link{A}{X}{1.5}{1}\duosymbol{a} \outwire{X}{1}\Opsymbol{a}
\end{Diagram}
\end{equation}
so we can cancel over black and white dots.

\subsection{Tensor product notation}\label{notation}

\index{tensor product notation}The standard tensor product symbol, $\otimes$, is both redundant and potentially obstructive.  We will not generally use it.  If we have two operators, $\hat A^\mathsf{a_1}$ and $\hat B^\mathsf{b_2}$ for example, we could write $\hat A^\mathsf{a_1} \otimes \hat B^\mathsf{b_2}$ since this is an operator in ${\cal V}^\mathsf{a_1}\otimes {\cal V}^\mathsf{b_2}$.  However, we will simply write $\hat A^\mathsf{a_1} \hat B^\mathsf{b_2}$.  The fact that we have a tensor product is clear because we have the labels $\mathsf{a_1}$ and $\mathsf{b_2}$ with different integers.  We will adopt this notation in general.  For example, $\hat A_\mathsf{a_1}\otimes \hat C^\mathsf{b_2}$ will be written $\hat A_\mathsf{a_1} \hat C^\mathsf{b_2}$.  The order is not important, so we could also write this as $\hat C^\mathsf{b_2}\hat A_\mathsf{a_1}$.  We could have two instances of the same type, for example, $\hat A^\mathsf{a_1}\otimes \hat B^\mathsf{a_2}$ is written $\hat A^\mathsf{a_1}\hat B^\mathsf{a_2}$. We know the latter is a tensor product because we have different integer labels as subscripts to the types.

Note, in particular, $\hat A^\mathsf{a_1} \hat B_\mathsf{a_2}$ is a tensor product (it has different integer labels) whereas $\hat A^\mathsf{a_1} \hat B_\mathsf{a_1}$ is not.  Expressions like the latter appear later when we put wires between operators (we will discuss this in more detail in Sec.\ \ref{operatorwiresfragscircuits}).  The interpretation of $\hat A^\mathsf{a_1} \hat B_\mathsf{a_1}$ is that it corresponds to taking the trace after the direct multiplication of the operators (if the operators were represented by matrices then we would multiply the matrices together).  In more conventional notation we would write $\hat A^\mathsf{a_1} \hat B_\mathsf{a_1}$ as $\text{Trace}(\hat A^\mathsf{a_1} \hat B_\mathsf{a_1})$ making the trace explicit.   We impose type matching when we put wires between operators (we are then guaranteed that spaces, ${\cal V}^\mathsf{a_1}$ and ${\cal V}_\mathsf{a_1}$, are of the same dimension so such multiplication is possible).  Note that, although these operators do not commute under multiplication, the trace is the same which ever way we multiply them together.  Hence, we can safely write $\hat A^\mathsf{a_1} \hat B_\mathsf{a_1}= \hat B_\mathsf{a_1}\hat A^\mathsf{a_1}$.  More generally we may have something like
$\hat A^\mathsf{a_1} \hat B_\mathsf{b_2} \hat C_\mathsf{a_1} \hat D^\mathsf{b_3}$.  In more conventional notation this is equal to
$\text{Trace}(\hat A^\mathsf{a_1} C_\mathsf{a_1}) \hat B_\mathsf{b_2}\otimes \hat D^\mathsf{b_3}$.

Even more generally we can have expressions such as $\hat{A}^\mathsf{a_1b_2}\hat{B}_\mathsf{b_2}^\mathsf{c_3a_4}\hat{C}_\mathsf{a_5c_3a_4}^\mathsf{b_6}$ where we have more than one wire.   In this case the wire (or repeated label) indicates that we take the partial trace over the corresponding spaces.   This means that, for example,
\begin{equation}
\hat{A}^\mathsf{a_1b_2}\hat{B}_\mathsf{b_2}^\mathsf{c_3a_4}\hat{C}_\mathsf{a_5c_3a_4}^\mathsf{b_6} \in {\cal V}_\mathsf{a_5}^\mathsf{a_1b_6}
\end{equation}
This is similar to Einstein's summation convention.  Here the partial trace is implicit where ever we have a repeated index (or wire).  We will explain below how such expressions can be calculated using full decomposability. We will refer to taking the partial trace over all the wires as the \emph{circuit trace}\index{circuit trace}.

There is, in general, no ambiguity in dropping the tensor product symbol, $\otimes$, as the integer labels on the type symbols carry all the necessary information.  Keeping the tensor symbol, on the other hand, would require that we took care to keep the symbols in the right order and will often require padding expressions with the identity.  These requirements would add unnecessary complications in what follows and would go very much against the spirit of the circuit framework (where circuits are interpreted graphically).  In particular, the usual notation with the tensor product symbol requires that we foliate the circuit first.  A foliation is additional structure put in by hand.

\subsection{Operators are fully decomposable}

One of the beautiful features of complex Hilbert spaces is that operators acting on them (i.e. in the space ${\cal V}$) are fully decomposable. This fact follows from (\ref{VequalsVVV}). This is not true of operators acting on real Hilbert spaces.
\begin{quote} {\bf Full decomposability of operators}.  \index{full decomposability! of operators}It is a fact that any operator is equal to a linear combination of operations each of which consists of an result operator for each input and a preparation operator for each output.  We do not lose any generality by choosing these to be fiducial sets  since any other set could be written as a linear combination of the fiducial set. Hence, this assumption is equivalent to the statement that any operator, $\hat A_\mathsf{a_1b_2\dots c_3}^\mathsf{d_4e_5\dots f_6}$, can be written as
\begin{equation}\label{decomposibilityhat}
\hat A_\mathsf{a_1b_2\dots c_3}^\mathsf{d_4e_5\dots f_6} \,\,= {}^{d_4e_5\dots f_6}\!A_{a_1b_2\dots c_3}\,\, \hat {X}_\mathsf{a_1}^{a_1} \hat{X}_\mathsf{b_2}^{b_2} \cdots \hat{X}_\mathsf{c_3}^{c_3} \,\,{}_{d_4}\!\hat{X}^\mathsf{d_4}{}_{e_5}\!\hat{X}^\mathsf{e_5}\cdots {}_{f_6}\!\hat{X}^\mathsf{f_6}
\end{equation}
in symbolic notation, or
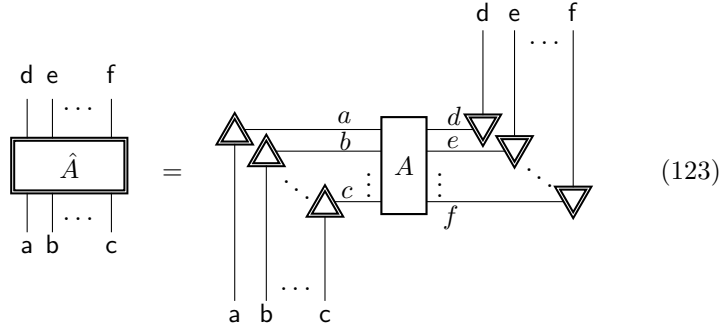
\begin{equation}\label{decomposibilitydiagramhat}
\begin{Diagram}[1.4]{0}{0}
\Dopbox[5]{A}{0,0}
\inwire{A}{1}\Opsymbol{a} \inwire{A}{2.2}\Opsymbol{b} \putlatex[30,15]{\ensuremath{\dots}} \inwire{A}{5}\Opsymbol{c}
\outwire{A}{1}\Opsymbol{d}\outwire{A}{2.2}\Opsymbol{e}\putlatex[30,-50]{\ensuremath{\dots}}   \outwire{A}{5}\Opsymbol{f}
\end{Diagram}
~~~= ~~~
\begin{Diagram}[1.2]{0}{0}
\Duobox[5]{A}{0,0}
\putlatex[-44,-14]{\ensuremath{\vdots}} \putlatex[-135,-20]{\ensuremath{\ddots}}
\putlatex[45,-14]{\ensuremath{\vdots}} \putlatex[170,-5]{\ensuremath{\ddots}}
\dlinkedeffect[0.8]{A}{5}{c}{-3.5}{0}  \duosymbol[-13,-4]{c} \thispoint{cbase}{-3.5,-6}\Opsymbol{c} \wire{cbase}{c}{1}{1}
\dlinkedeffect[0.8]{A}{2.2}{b}{-6.1}{0}\duosymbol[22,-3]{b}
\thispoint{bbase}{-6.1,-6}\Opsymbol{b} \wire{bbase}{b}{1}{1}
\dlinkedeffect[0.8]{A}{1}{a}{-7.5}{0}\duosymbol[41,-1]{a} \thispoint{abase}{-7.5,-6}\Opsymbol{a} \wire{abase}{a}{1}{1}
\placelatex[-20, 16]{-4,-6}{\ensuremath{\dots}}
\dlinkedprep[0.8]{A}{1}{d}{3.5}{0}\duosymbol[4,1]{d}  \thispoint{dbase}{3.5,6}\Opsymbol[0,38]{d} \wire{d}{dbase}{1}{1}
\dlinkedprep[0.8]{A}{2.2}{e}{4.9}{0}\duosymbol[-17,-5]{e}  \thispoint{ebase}{4.9,6}\Opsymbol[0,38]{e} \wire{e}{ebase}{1}{1}
\dlinkedprep[0.8]{A}{5}{f}{7.5}{0}\otherside\duosymbol[-57,-4]{f} \thispoint{fbase}{7.5,6}\Opsymbol[0,38]{f} \wire{f}{fbase}{1}{1}
\placelatex[9, -16]{6,6}{\ensuremath{\dots}}
\end{Diagram}
\end{equation}
in diagrammatic notation.
\end{quote}
Note that we have equality here since the space under consideration, ${\cal V}_\mathsf{a_1b_2\dots c_3}^\mathsf{d_4e_5\dots f_6}$, is a linear space.  In the case of operations in the previous section we had equivalence rather than equality for the assumption of full decomposability.  We will develop a notion of equivalence for operators in the duotensor framework.  Things look a little bit different than for operations.  However, this makes no difference in the end. The analogy between the duotensor framework for operations and the duotensor framework for operators is good but not perfect.

\subsection{Wires, fragments, and circuits}\label{operatorwiresfragscircuits}

We have added a certain structure to the operators we have considered. Namely, that they have inputs and outputs labeled by types.  We will now see how this structure enables us to wire these operators together to build operators with even more structure.  We can use a wire to join an output to an input of the same type. Consider, first, a simple example
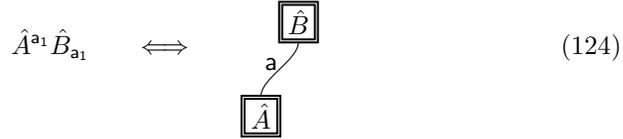
\begin{equation}
\hat A^\mathsf{a_1} \hat B_\mathsf{a_1}~~~~~ \Longleftrightarrow ~~~~~
\begin{Diagram}{0}{-1}
\Dopbox[2]{A}{0,0} \Dopbox[2]{B}{2,5} \wire{A}{B}{1.5}{1.5} \opsymbol[-1,0]{a}
\end{Diagram}
\end{equation}
The wire here means that we take the trace of the product of $\hat A^\mathsf{a_1}\in {\cal V}^\mathsf{a_1}$ and $\hat B_\mathsf{a_1}\in{\cal V}_\mathsf{a_1}$.  I.e.\ this is equal to $\text{Trace}(\hat A^\mathsf{a_1} \hat B_\mathsf{a_1})$ in more standard notation.  Note that the trace of the product of two Hermitean operators is always real.

If we have a more complicated situation then the best way to understand what it means to place a wire is to use the full decomposability of the operators.  Thus, we can write
\begin{equation}\label{operatorwiresymbolic}
\hat A_\mathsf{a_1b_2}^\mathsf{b_3c_4}\hat B_\mathsf{a_5b_3}^\mathsf{d_6c_7} =
A_{a_1b_2}^{b_3c_4}\, \hat X^{a_1}_\mathsf{a_1} \hat X^{b_2}_\mathsf{b_2}  \, {}_{b_3}\! \hat X^\mathsf{b_3} \, {}_{c_4} \! \hat X^\mathsf{c_4} \,
B_{a_5b_{3'}}^{d_6c_7}\, \hat X^{a_5}_\mathsf{a_5} \hat X^{b_{3'}}_\mathsf{b_3}  \, {}_{d_6}\! \hat X^\mathsf{d_6} \, {}_{c_7} \! \hat X^\mathsf{c_7}
\end{equation}
or, in diagrammatic notation,
\begin{equation}\label{operatorwirediagramatic}
\begin{Diagram}{0}{-1.2}
\Dopbox[2]{A}{7,0}
\Dopbox[2]{B}{3,7}
\inwire[-5]{A}{1}\Opsymbol{a}
%\inwire{A}{2}\Opsymbol{a}
\inwire[5]{A}{2}\Opsymbol{b}
\wire{A}{B}{1}{2}\opsymbol{b}
\outwire[5]{A}{2}\Opsymbol{c}
\inwire[-5]{B}{1}\Opsymbol{a}
\outwire[-5]{B}{1}\Opsymbol{d}
\outwire[5]{B}{2}\Opsymbol{c}
\end{Diagram}
~~~~~~ = ~~~~~~
\begin{Diagram}{0}{-1.5}
\Duobox{A}{0,0}
\dlinkedeffect{A}{1}{1A}{-5}{0} \duosymbol{a}  \inwire{1A}{1} \Opsymbol{a}
\dlinkedeffect{A}{3}{2A}{-3}{0} \duosymbol[0,1]{b}  \inwire{2A}{1} \Opsymbol{b}
\dlinkedprep{A}{1}{A1}{3}{0} \otherside \duosymbol{b}
\dlinkedprep{A}{3}{A2}{5}{0}   \otherside \duosymbol{c}  \outwire{A2}{1} \Opsymbol{c}
\Duobox{B}{-3,10}
\dlinkedeffect{B}{1}{1B}{-8}{0}  \duosymbol{a}  \inwire{1B}{1} \Opsymbol{a}
\dlinkedeffect{B}{3}{2B}{-6}{0} \duosymbol[0,1]{b}
\dlinkedprep{B}{1}{B1}{0}{0}    \otherside \duosymbol{d} \outwire{B1}{1} \Opsymbol{d}
\dlinkedprep{B}{3}{B2}{2}{0}     \otherside\duosymbol{c} \outwire{B2}{1} \Opsymbol{c}
\wire{A1}{2B}{1}{1} \opsymbol{b}
\end{Diagram}
\end{equation}
This operator is in the space
\begin{equation}
{\cal V}_\mathsf{a_1b_2a_5b_3}^\mathsf{b_3c_4d_6c_7} = {\cal V}_\mathsf{a_1}\otimes{\cal V}_\mathsf{b_2} \otimes {\cal V}_\mathsf{a_5} \otimes
{\cal V}^\mathsf{c_4} \otimes {\cal V}^\mathsf{d_6} \otimes {\cal V}^\mathsf{c_7}
\end{equation}
Note that we have taken the trace over the $\mathsf{b_3}$ space.

An \emph{operator fragment} \index{operator fragments} is the object resulting from wiring a bunch of operators together. This will be a Hermitean operator.  We may have disjoint parts. We may have some open inputs and outputs.

An \emph{operator circuit} \index{operator circuits} is the object resulting from wiring a bunch of operators together such that we have no open inputs or outputs left over.  This will be equal to a real number.  Since there are no open inputs or outputs it is not possible to join an operator circuit to an operator fragment with wires.

\subsection{Evolution}\label{evolutionsection}

If we have appropriate causal structure in the wiring - namely no closed loops - then we can impose an evolution picture on this operation structure at the level of equivalent operators.

The box
\begin{equation}
\hat B_\mathsf{a_1}^\mathsf{b_2}
~~~~~~~ \Leftrightarrow ~~~~~~~
\begin{Diagram}{0}{0}
\Dopbox{B}{0,0}
\inwire{B}{2} \Opsymbol{a} \outwire{B}{2}\Opsymbol{b}
\end{Diagram}
\end{equation}
corresponds to an operator on ${\cal V}_\mathsf{a_1}\otimes{\cal V}^\mathsf{b_2}$.  However, we can use it to obtain a map acting on operators in ${\cal V}^\mathsf{a}$ and producing operators in ${\cal V}^\mathsf{b}$.  Consider
\begin{equation}
\hat A^\mathsf{a_1} \hat B_\mathsf{a_1}^\mathsf{b_2}
~~~~\Leftrightarrow~~~~
\begin{Diagram}{0}{-0.8}
\Dopbox{A}{0,0}
\Dopbox{B}{0,4}
\wire{A}{B}{2}{2} \opsymbol{a}
\outwire{B}{2} \Opsymbol{b}
\end{Diagram}
\end{equation}
This is an operator in ${\cal V}^\mathsf{b_2}$.  We define the map $\$_B[\cdot]$ by
\begin{equation}\label{evolutionequation}
\$_B[\hat A^\mathsf{a_1}]:= \hat A^\mathsf{a_1} \hat B_\mathsf{a_1}^\mathsf{b_2}
\end{equation}
This is a linear map from ${\cal V}^\mathsf{a_1}$ to ${\cal V}^\mathsf{b_2}$.  We call this a \emph{superoperator}\index{superoperators}.

If we wish to think in the standard way of having a state that evolves in time we can.   We break the operator circuit up into fragments along some foliation lines and think of an initial state, represented by an operator, as evolving through the circuit.  If we evolve using superoperators then we will get an operator at each stage that is equal to the operator corresponding to the accumulated preparation fragment up to the given foliation line.  It is important to note that we do not need to think in terms of a time evolving state.  In particular, if we wish to calculate what some given operator circuit is equal to we can break it up into fragments in any way we wish. If we know the operator fragments then we can combine them (using the implicit circuit trace) and obtain the value of the operator circuit.

We can, by this method, also define an evolution that evolves an operator \lq\lq backwards" from output to input. Thus, define the map $\tilde \$_B[\cdot]$ by
\begin{equation}
\tilde \$_B[\hat C_\mathsf{b_2}] := \hat C_\mathsf{b_2} \hat B_\mathsf{a_1}^\mathsf{b_2}
\end{equation}
This is a linear map from ${\cal V}_\mathsf{b_2}$ to ${\cal V}_\mathsf{a_1}$.

The existence of the maps $\$_B$ and $\tilde \$_B$ does not imply that these maps are invertible.

\subsection{The operator hoping metric}

If we obtain duotensors from operator structures then the hopping metric\index{hopping metric! for operator structures} is given by
\begin{equation}\label{defbboperator}
\begin{Diagram}{0}{-0.1}
\bbmetric{g}{0,0}\duosymbol{a}
\end{Diagram}
~:=~
\begin{Diagram}{0}{-0.6}
\dfidprep{X2}{4,0}
\dfideffect{X}{4,4}
\wire{X2}{X}{1}{1}\opsymbol{a}
\inblack{X2}{1} \Duosymbol{a}
\outblack{X}{1}\Duosymbol{a}
\end{Diagram}
~~~~~\Leftrightarrow ~~~~~
{}_{a'_1}\! g^{a_1} := {}_{a'_1}\!\hat{X}^\mathsf{a_1}\hat{X}_\mathsf{a_1}^{a_1}
\end{equation}
and its inverse is represented by $\wwdots$.   The entries in the hopping metric must be real because we are taking the trace of the product of Hermitian operators.  The inverse will, therefore, also have real entries.

We can simplify any expression for an operator fragment by replacing matched fiducial pairs (i.e.\ those joined by a wire) by the hopping metric.  We can then cancel over black and white dots as we did in the previous section. Consider the example in (\ref{operatorwirediagramatic}) above. We obtain
\begin{equation}\label{operatorwirediagramatictwo}
\begin{Diagram}{0}{-1.2}
\Dopbox[2]{A}{7,0}
\Dopbox[2]{B}{3,7}
\inwire[-5]{A}{1}\Opsymbol{a}
%\inwire{A}{2}\Opsymbol{a}
\inwire[5]{A}{2}\Opsymbol{b}
\wire{A}{B}{1}{2}\opsymbol{b}
\outwire[5]{A}{2}\Opsymbol{c}
\inwire[-5]{B}{1}\Opsymbol{a}
\outwire[-5]{B}{1}\Opsymbol{d}
\outwire[5]{B}{2}\Opsymbol{c}
\end{Diagram}
=
\begin{Diagram}{0}{-1.5}
\Duobox{A}{0,0}
\dlinkedeffect{A}{1}{1A}{-5}{0} \duosymbol{a}  \inwire{1A}{1} \Opsymbol{a}
\dlinkedeffect{A}{3}{2A}{-3}{0} \duosymbol{b}  \inwire{2A}{1} \Opsymbol{b}
\dlinkedprep{A}{1}{A1}{3}{0} \otherside \duosymbol{b}
\dlinkedprep{A}{3}{A2}{5}{0}   \otherside \duosymbol{c}  \outwire{A2}{1} \Opsymbol{c}
\Duobox{B}{-3,10}
\dlinkedeffect{B}{1}{1B}{-8}{0}  \duosymbol{a}  \inwire{1B}{1} \Opsymbol{a}
\dlinkedeffect{B}{3}{2B}{-6}{0} \duosymbol{b}
\dlinkedprep{B}{1}{B1}{0}{0}    \otherside \duosymbol{d} \outwire{B1}{1} \Opsymbol{d}
\dlinkedprep{B}{3}{B2}{2}{0}     \otherside\duosymbol{c} \outwire{B2}{1} \Opsymbol{c}
\wire{A1}{2B}{1}{1} \opsymbol{b}
\end{Diagram}
=
\begin{Diagram}{0}{-1.5}
\Duobox{A}{0,0}
\dlinkedeffect{A}{1}{1A}{-5}{0} \duosymbol{a}  \inwire{1A}{1} \Opsymbol{a}
\dlinkedeffect{A}{3}{2A}{-3}{0} \duosymbol{b}  \inwire{2A}{1} \Opsymbol{b}
%\dlinkedprep{A}{1}{A1}{3}{0} \otherside \duosymbol{b}
\dlinkedprep{A}{3}{A2}{5}{0}   \otherside \duosymbol{c}  \outwire{A2}{1} \Opsymbol{c}
\Duobox{B}{-3,10}
\dlinkedeffect{B}{1}{1B}{-8}{0}  \duosymbol{a}  \inwire{1B}{1} \Opsymbol{a}
%\dlinkedeffect{B}{3}{2B}{-6}{0} \duosymbol{b}
\dlinkedprep{B}{1}{B1}{0}{0}    \otherside \duosymbol{d} \outwire{B1}{1} \Opsymbol{d}
\dlinkedprep{B}{3}{B2}{2}{0}     \otherside\duosymbol{c} \outwire{B2}{1} \Opsymbol{c}
\link[3]{A}{B}{1}{3} \duosymbol{b}
\end{Diagram}
\end{equation}
Associated with the fragment in (\ref{operatorwirediagramatictwo}) is the duotensor
\begin{equation}
\begin{Diagram}{0}{0}
\Duobox{A}{0,0} \Duobox{B}{-3,10} \link[3]{A}{B}{1}{3} \duosymbol{b}
\inwhite{A}{1} \Duosymbol{a} \inwhite{A}{3} \Duosymbol{b}
\outwhite{A}{3} \Duosymbol{c}
\inwhite{B}{1}  \Duosymbol{a}
\outwhite{B}{1} \Duosymbol{d} \outwhite{B}{3} \Duosymbol{c}
\end{Diagram}
\end{equation}
This duotensor provides the coefficients for the sum over fiducials.  Two operator fragments that have the same duotensors after replacing all matched fiducial pairs with the hopping metric must be equal.   Note that, to determine whether two fragments are equal, we must put the corresponding duotensors in the same form (for example, all white dots as shown here).

In the above example there is only one wire.  In a more complicated example we would have many. However many matched fiduial pairs we replace by the hopping metric, we will continue to have equality.  This is one way to explicitly calculate the circuit trace implicit in the wiring for a general operator fragment.

When we explicitly calculate the circuit trace \index{circuit trace} implicit in the wiring of an operator circuit we will get a real number.  For example, the operator circuit
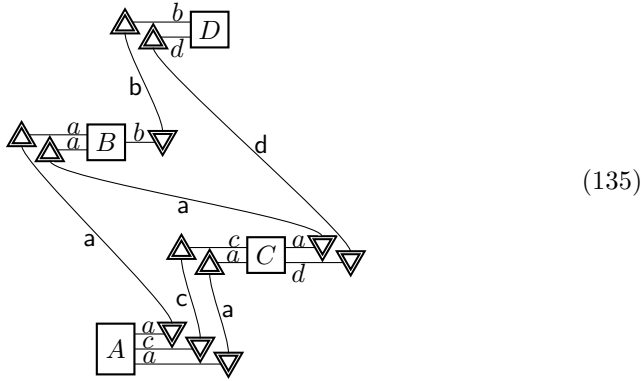
\begin{equation}
\begin{Diagram}{0}{-3.3}
\begin{move}{-2,4}
\Duobox{A}{0,0}
\dlinkedprep[0.7]{A}{1}{A1}{3}{0}\duosymbol[-3,-6]{  a}
\dlinkedprep[0.7]{A}{2}{A2}{4.5}{0}\duosymbol[-25,-6]{  c}
\dlinkedprep[0.7]{A}{3}{A3}{6}{0}\duosymbol[-44,-6]{  a}
\end{move}
\begin{move}{-2.5,15}
\Duobox[2]{B}{0,0}
\dlinkedeffect[0.7]{B}{1}{1B}{-4.5}{0}\duosymbol[25,-6]{  a}
\dlinkedeffect[0.7]{B}{2}{2B}{-3}{0}\duosymbol[3,-6]{  a}
\dlinkedprep[0.7]{B}{1.5}{B15}{3}{0} \duosymbol[-3,0]{  b}
\end{move}
\begin{move}{6,9}
\Duobox[2]{C}{0,0}
\dlinkedeffect[0.7]{C}{1}{1C}{-4.5}{0}\duosymbol[25,-6]{  c}
\dlinkedeffect[0.7]{C}{2}{2C}{-3}{0}\duosymbol[3,-6]{  a}
\dlinkedprep[0.7]{C}{1}{C1}{3}{0} \duosymbol[-3,-6]{  a}
\dlinkedprep[0.7]{C}{2}{C2}{4.5}{0} \otherside\duosymbol[-25,0]{  d}
\end{move}
\begin{move}{3,21}
\Duobox[2]{D}{0,0}
\dlinkedeffect[0.7]{D}{1}{1D}{-4.5}{0}\duosymbol[25,0]{  b}
\dlinkedeffect[0.7]{D}{2}{2D}{-3}{0}\otherside\duosymbol[3,0]{  d}
\end{move}
\wire{A1}{1B}{1}{1}\opsymbol{a} \wire{A2}{1C}{1}{1}\opsymbol{c} \wire{A3}{2C}{1}{1}\otherside\opsymbol{a}
\wire{C1}{2B}{1}{1}\opsymbol{a} \wire{C2}{2D}{1}{1}\otherside\opsymbol[4,0]{d} \wire{B15}{1D}{1}{1}\opsymbol{b}
\end{Diagram}
\end{equation}
involves taking the trace over the entire space.   Note that when we have more than one wire we are implicitly using the mathematical fact that, in conventional notation, $\text{Trace}(\hat A\otimes \hat B) = \text{Trace}(\hat A) \text{Trace}(\hat B)$.

\section{Operations and operators}\label{operationsandoperators}

In this section we will discuss four related topics.  First we will discuss how to set up a correspondence between operations and operators such that the probability for a circuits is given by the corresponding operator circuit. Second, we will provide some mathematical definitions and theorems for the case of operators in the duotensor framework that are motivated by this correspondence.  Third, we will show how to formulate quantum theory as a theory relating operations and operators within the duotensor framework.  Finally we will discuss how to formulate quantum theory in a formalism local fashion.

\subsection{Correspondence}

In this section we will see how to use duotensors to link operations and operators.
\begin{quote}
{\bf Operation-operator correspondence.} \index{correspondence} We will say that operations correspond to operators if there is a mapping from operations, $\mathsf{A}_\mathsf{a_1b_2\cdots c_3}^\mathsf{d_4e_5\cdots f_6}$, to operators, $\hat{A}_\mathsf{a_1b_2\cdots c_3}^\mathsf{d_4e_5\cdots f_6}$, such that the probability for \emph{any} circuit comprised of operations is equal to the operator circuit obtained under this mapping.
\end{quote}
It is important that, under this correspondence mapping, we have the same input-output structure.   If operations correspond to operators then, for example,
\begin{equation}
\text{Prob}(\mathsf{A}^\mathsf{a_1b_2}\mathsf{B}_\mathsf{b_2}^\mathsf{c_3a_4}\mathsf{C}_\mathsf{a_1c_3a_4})
=\hat{A}^\mathsf{a_1b_2}\hat{B}_\mathsf{b_2}^\mathsf{c_3a_4}\hat{C}_\mathsf{a_1c_3a_4}
\end{equation}
This same example in diagrammatic form
\begin{equation}\label{probtraceexample}
\text{Prob}\left(~
\begin{Diagram}{0}{-1.4}
\Opbox[2]{A}{0,0} \Opbox[2]{B}{2,4} \Opbox{C}{-1,10}
\wire{A}{C}{1}{1} \opsymbol{a} \wire{A}{B}{2}{1.5}\opsymbol[-3,1]{b} \wire{B}{C}{1}{2} \opsymbol{c} \wire{B}{C}{2}{3}\otherside\opsymbol{a}
\end{Diagram}
~\right)
~=~
\begin{Diagram}{0}{-1.4}
\Dopbox[2]{A}{0,0} \Dopbox[2]{B}{2,4} \Dopbox{C}{-1,10}
\wire{A}{C}{1}{1} \opsymbol{a} \wire{A}{B}{2}{1.5}\opsymbol[-3,1]{b} \wire{B}{C}{1}{2} \opsymbol{c} \wire{B}{C}{2}{3}\otherside\opsymbol{a}
\end{Diagram}
\end{equation}
We will now prove
\begin{T}\label{correspondencetheorem}
If we can associate fiducial result and preparation operators with fiducial result and preparation operations,
\[
\hat X_\mathsf{a_1}^{a_1} \leftrightarrow \mathsf{X}_\mathsf{a_1}^{a_1} ~~\text{and}~~
{}_{a_1}\! \hat X^\mathsf{a_1} \leftrightarrow {}_{a_1}\! \mathsf{X}^\mathsf{a_1}
~~~ \text{for} ~ a_1=1 ~\text{to}~K_\mathsf{a}, ~~\text{for all types}~\mathsf{a}
\]
such that
\begin{equation}\label{equalhoppingmetrics}
{}_{a_1}\! \hat X^\mathsf{a_1}\hat X_\mathsf{a_1}^{a'_1}=\text{Prob}({}_{a_1}\! \mathsf{X}^\mathsf{a_1}\mathsf{X}_\mathsf{a_1}^{a'_1})
\end{equation}
then we can set up a correspondence from operations to operators such that the operation
\begin{equation}\label{operationcorrespondsto}
\mathsf{A_{a_1b_2\dots c_3}^{d_4e_5\dots f_6}} \,\,\equiv {}^{d_4e_5\dots f_6}\!A_{a_1b_2\dots c_3}\,\, \mathsf{X}_\mathsf{a_1}^{a_1} \mathsf{X}_\mathsf{b_2}^{b_2} \cdots \mathsf{X}_\mathsf{c_3}^{c_3} \,\,{}_{d_4}\!\mathsf{X}^\mathsf{d_4}{}_{e_5}\!\mathsf{X}^\mathsf{e_5}\cdots {}_{f_6}\!\mathsf{X}^\mathsf{f_6}
\end{equation}
corresponds to the operator
\begin{equation}\label{operatorcorrespondsto}
\hat A_\mathsf{a_1b_2\dots c_3}^\mathsf{d_4e_5\dots f_6} \,\,= {}^{d_4e_5\dots f_6}\!A_{a_1b_2\dots c_3}\,\, \hat {X}_\mathsf{a_1}^{a_1} \hat{X}_\mathsf{b_2}^{b_2} \cdots \hat{X}_\mathsf{c_3}^{c_3} \,\,{}_{d_4}\!\hat{X}^\mathsf{d_4}{}_{e_5}\!\hat{X}^\mathsf{e_5}\cdots {}_{f_6}\!\hat{X}^\mathsf{f_6}
\end{equation}
This defines a correspondence map from operations to operators.
\end{T}
Note that (\ref{equalhoppingmetrics}) is saying that the hopping metric for operations is the same as the hopping metric for operators.  If we have equal hopping metrics then, for example, both sides of (\ref{probtraceexample}) are equal to
\begin{equation}
\begin{Diagram}{0}{0}
\Duobox[2]{A}{0,0} \Duobox[2]{B}{8,-3} \Duobox{C}{20,1.5}
\linkwbbw{A}{B}{2}{1.5} \duosymbol{b}
\linkwbbw{A}{C}{1}{1}   \duosymbol{a}
\linkwbbw{B}{C}{1}{2}   \duosymbol{c}
\linkwbbw{B}{C}{2}{3}  \otherside \duosymbol{a}
\end{Diagram}
\end{equation}
and therefore equal to each other.  A similar result clearly follows for any circuit and hence the map in {\bf T\ref{correspondencetheorem}} does induce a correspondence from operations to operators.

A second useful result follows from this
\begin{T}\label{anyfidswork}
If we have correspondence between operators and operations with respect to one association of fiducial operations with fiducial operators then we will have correspondence with respect to any other association of fiducial operations with \emph{corresponding} fiducial operators.
\end{T}
Since we have said that the second association of fiducial operations is with corresponding fiducial operators (i.e.\ as prescribed by (\ref{operationcorrespondsto}) and (\ref{operatorcorrespondsto})) we must have equal hopping metrics. Hence, the result follows from {\bf T\ref{correspondencetheorem}}.

%We have defined one type of equivalence for operations, using the $p(\cdot)$ function and another for operators using the $t(\cdot)$ function.  We can employ both types of equivalence at the same time if we wish so long as (i) we have equal hopping metrics as in (\ref{equalhoppingmetrics}) and (ii) operations are only wired together with other operations and operators are only wired together with other operators.  With regard to the second point note that operations and operators can be linked via duotensors.

\subsection{Physical operators}\label{physicaloperatorssection}

In this subsection we provide some mathematical definitions and prove a few simple mathematical theorems for the operator-duotensor framework that are motivated by the physical considerations of the previous section.  We will apply these mathematical definitions to give a succinct statement of quantum theory in the next subsection.

A correspondence from operations to operators gives rise a subset of operators, ${\cal O}_\mathsf{a}^\mathsf{b}$, for each system type pair, $(\mathsf{a}, \mathsf{b})$. This subset consists of all the operators for which there is a corresponding operation for this given type pair.  Note that $\mathsf{a}$ and $\mathsf{b}$ may be composite.  We will call the collection of these subsets, $\{ {\cal O}_\mathsf{a}^\mathsf{b}: \text{all types}~\mathsf{a, b} \}$, an \emph{operator superset}\index{operator supersets}.

We define
\begin{quote}
{\bf An operator superset is physical}\index{operator supersets!physical} if
\begin{enumerate}
\item The value of any operator circuit formed from operators in the operator superset is between 0 and 1.
\item The operator superset contains preparation and result operators equal to all rank one projectors for every type.
\item The operator superset contains result operators corresponding to the identity operator, $\hat I_\mathsf{a_1}$, for every type.
\end{enumerate}
\end{quote}
This is a purely mathematical definition.  It is, however, motivated by physical considerations.  The motivation for the first property is so that the value of the operator circuit can be equal to a probability. The motivation for the second property comes from quantum theory.  In quantum theory, pure states and maximal effects are represented by rank one projection operators. The motivation for the third property also comes from quantum theory.  The identity operator, $\hat I_\mathsf{a_1}$ corresponds to the deterministic result $\mathsf T_\mathsf{a_1}$ (where the outcome set consists of all outcomes on the associated measurement).
\begin{quote}\index{physical operators}
{\bf Physical operators.} An operator,
\begin{equation}
\hat B_\mathsf{a_1b_2\dots c_3}^\mathsf{d_4e_5\dots f_6}
\end{equation}
is said to be \emph{physical} if
\begin{equation}\label{positiveoperatorcondition}
0\leq \hat A^\mathsf{a_1b_2\dots c_3g_7} \hat B_\mathsf{a_1b_2\dots c_3}^\mathsf{d_4e_5\dots f_6} \hat C_\mathsf{d_4e_5\dots f_6g_7}
\end{equation}
and
\begin{equation}\label{traceoperatorcondition}
\hat A^\mathsf{a_1b_2\dots c_3g_7} \hat B_\mathsf{a_1b_2\dots c_3}^\mathsf{d_4e_5\dots f_6}\hat I_\mathsf{d_4e_5\dots f_6g_7} \leq 1
\end{equation}
for all rank one projection operators $\hat A^\mathsf{a_1b_2\dots c_3g_7}$ and $C_\mathsf{d_4e_5\dots f_6g_7}$ and for all types $\mathsf g$.
\end{quote}
For emphasis, we have written our general operator, $\hat B_\mathsf{a_1b_2\dots c_3}^\mathsf{d_4e_5\dots f_6}$, with the possible composite nature of the input ($\mathsf{ab\dots c}$) and the output ($\mathsf{de\dots f}$) shown explicitly.  We could equally write a general operator as $\hat B_\mathsf{a_1}^\mathsf{b_2}$ where it is understood that $\mathsf a$ and $\mathsf b$ may be composite types.  Note, incidently, that the type $\mathsf{g}$ in the above definition also may be composite.

We prove the following theorem
\begin{T}\label{physicalimpliespositive}
Physical preparation operators, $\hat B^\mathsf{a_1}$, are positive and have trace less than or equal to one.  Physical result operators, $\hat B_\mathsf{a_1}$ are positive and less than or equal to the identity, $\hat I_\mathsf{a_1}$.
\end{T}
An arbitrary physical preparation operator, $\hat B^\mathsf{a_1}$, must satisfy
\begin{equation}
0\leq \hat A^\mathsf{g_2}\hat B^\mathsf{a_1}\hat C_\mathsf{a_1g_2}
\end{equation}
for all rank one projectors $\hat A^\mathsf{g_2}$ and $\hat C_\mathsf{a_1g_2}$.  In particular, we can choose $\hat C_\mathsf{a_1g_2}=\hat D_\mathsf{a_1}\hat E_\mathsf{g_2}$ where $\hat A^\mathsf{g_2}\hat E_\mathsf{g_2}=1$ and $\hat D_\mathsf{a_1}$ is an arbitrary rank one projector.  Therefore,
\begin{equation}
0\leq \hat B^\mathsf{a_1}\hat D_\mathsf{a_1}
\end{equation}
for all rank one projectors, $\hat D_\mathsf{a_1}$.  Hence, $\hat B^\mathsf{a_1}$ must be positive.  That it must have trace less than or equal to one follows from (\ref{traceoperatorcondition}) and the fact that $\hat I_\mathsf{a_1g_2}=\hat I_\mathsf{a_1}\hat I_\mathsf{g_2}$.  This gives
\begin{equation}
\hat A^\mathsf{g_2}\hat B^\mathsf{a_1}\hat I_\mathsf{a_1}\hat I_\mathsf{g_2}\leq 1
\end{equation}
which gives
\begin{equation}
\hat B^\mathsf{a_1} \hat I_\mathsf{a_1}\leq 1
\end{equation}
Hence, $\hat B^\mathsf{a_1}$ has trace less than or equal to one.  Now consider an arbitrary physical result operator, $\hat B_\mathsf{a_1}$.  By definition, we have
\begin{equation}
0\leq \hat A^\mathsf{a_1g_1} \hat B_\mathsf{a_1} \hat C_\mathsf{g_2}
\end{equation}
for all rank one projectors, $\hat A^\mathsf{a_1g_2}$ and $\hat C_\mathsf{g_2}$.  We can choose $\hat A^\mathsf{a_1g_2} = \hat D^\mathsf{a_1}\hat E^\mathsf{g_2}$ where $\text{Trace}(\hat D^\mathsf{g_2}\hat C_\mathsf{g_2})=1$.  Hence,
\begin{equation}
0\leq \hat B_\mathsf{a_1} \hat D^\mathsf{a_1}
\end{equation}
Hence, $\hat B_\mathsf{a_1}$ must be positive.  From (\ref{traceoperatorcondition}) we have
\begin{equation}
\hat A^\mathsf{a_1g_2} \hat B_\mathsf{a_1} \hat I_\mathsf{g_2}\leq 1
\end{equation}
for all $\hat A^\mathsf{a_1}$ equal to rank one projectors.  In particular, we can choose $\hat A^\mathsf{a_1g_2}= \hat F^\mathsf{a_1}\hat G^\mathsf{g_2}$ where both $\hat F^\mathsf{a_1}$ and $\hat G^\mathsf{g_2}$ are rank one projectors.  It then follows that
\begin{equation}
\hat F^\mathsf{a_1} \hat B_\mathsf{a_1}\leq 1
\end{equation}
for all rank one projectors, $\hat F^\mathsf{a_1}$.  This means that $\hat I_\mathsf{a_1}-\hat B_\mathsf{a_1}$ is positive or, equivalently, that $\hat B_\mathsf{a_1}\leq \hat I_\mathsf{a_1}$.  This proves {\bf T\ref{physicalimpliespositive}}.

For operators with an input and an output we can prove the following
\begin{T}\label{CPmaps}
The superoperator\index{superoperators} $\$_B(\cdot)$, given by $\$_B(\hat E^\mathsf{a_1})=\hat E^\mathsf{a_1}\hat B_\mathsf{a_1}^\mathsf{b_2}$, is a completely positive trace non-increasing map if and only if the operator $\hat B_\mathsf{a_1}^\mathsf{b_2}$ is physical.
\end{T}
First we note that
\begin{equation}
\hat A^\mathsf{a_1g_3} \hat B_\mathsf{a_1}^\mathsf{b_2}=\$_B\otimes I(\hat A^{a_1g_3})
\end{equation}
where $\$_B\otimes I$ acts as $\$_B$ on ${\cal V}_\mathsf{a_1}$ and as the identity on ${\cal V}_\mathsf{g_3}$.
We note that
\begin{equation}\label{Btodollar}
\hat A^\mathsf{a_1g_3}\hat B_\mathsf{a_1}^\mathsf{b_2}\hat C_\mathsf{b_2g_3} =\text{Trace}( \$_B\otimes I(\hat A^{a_1g_3}) \hat C_\mathsf{b_2g_3})
\end{equation}
(in an obvious but slightly ad hoc notation).  Assume that $\hat B_\mathsf{a_1}^\mathsf{b_2}$ is physical.  Hence, it follows from (\ref{positiveoperatorcondition}) that
\begin{equation}
0\leq  \text{Trace}(\$_B\otimes I(\hat A^\mathsf{a_1g_3})\hat C_\mathsf{b_2g_3})
\end{equation}
for all rank one projectors $\hat A^\mathsf{a_1g_3}$ and $\hat C_\mathsf{a_1g_3}$.  Hence, $\$_B(\cdot)$ is completely positive.  It follows from (\ref{traceoperatorcondition}) that
\begin{equation}
\text{Trace}(\$_B\otimes I(\hat A^\mathsf{a_1g_3})\hat I_\mathsf{b_2g_3}) \leq 1
\end{equation}
for all rank one projectors, $\hat A^\mathsf{a_1g_3}$.  Consider $\hat A^\mathsf{a_1g_3}=\hat E^\mathsf{a_1}\hat F^\mathsf{g_3}$ where $\hat E^\mathsf{a_1}$ and $\hat F^\mathsf{g_3}$ are rank any one projectors.  Since we have $\hat I_\mathsf{b_2g_3}=\hat I_\mathsf{b_2}\hat I_\mathsf{g_3}$ we have
\begin{equation}
\text{Trace}(\$_B(\hat E^\mathsf{a_1})\hat I_\mathsf{b_2}) \leq 1
\end{equation}
Rank one projectors have trace equal to one and they span the space of operators in ${\cal V}^\mathsf{a_1}$.  Hence $\$_B(\cdot)$ is a completely positive trace non-increasing function if $\hat B_\mathsf{a_1}^\mathsf{b_2}$ is physical.   Using (\ref{Btodollar}) in a similar way we easily obtain the result that $\hat B_\mathsf{a_1}^\mathsf{b_2}$ is physical if $\$_B(\cdot)$ is a completely positive trace non-increasing function.
This proves {\bf T\ref{CPmaps}}.

We can now prove the following theorem.
\begin{T}\label{supersettheorem}
An operator superset is physical if and only if every operator in it is physical (and it contains preparation and result operators equal to all rank one projectors and result operators equal to the identity for every type).
\end{T}
The parenthetical remark is necessary because of the way physical operator supersets are defined.   The \lq\lq only if" part follows immediately because the conditions (\ref{positiveoperatorcondition}, \ref{traceoperatorcondition}) are imposed if the operator superset is physical.  To prove that if an operator superset has has only physical operators it must be a physical operator superset we need to prove that any operator circuit built out of physical operators will be equal to a number between zero and one.  To this end we note that any operator circuit containing $\hat B_\mathsf{a_1b_2\dots c_3}^\mathsf{d_4e_5\dots f_6}$ can be written in the form
\begin{equation}\label{anycircuitoperator}
\hat D^\mathsf{a_1b_2\dots c_3g_7} \hat B_\mathsf{a_1b_2\dots c_3}^\mathsf{d_4e_5\dots f_6}\hat E_\mathsf{d_4e_5\dots f_6g_7}
\end{equation}
for some ancillary system $\mathsf{g_7}$ which may be composite.  The operator $\hat D^\mathsf{a_1b_2\dots c_3g_7}$ must be positive and have trace less than or equal to one (by {\bf T\ref{physicalimpliespositive}}).   Hence it can be written as a convex sum of rank one projectors.   By {\bf T\ref{physicalimpliespositive}}, the operator $\hat E^\mathsf{d_4e_5\dots f_6g_7}$ must be positive.  Hence it can be written as a sum of rank one projectors weighted by positive numbers.    Therefore, it follows from the fact that $\hat B_\mathsf{a_1b_2\dots c_3}^\mathsf{d_4e_5\dots f_6}$ is physical that (\ref{anycircuitoperator}) has trace between zero and one.  This proves {\bf T\ref{supersettheorem}}.   This theorem is useful in that it allows us to characterize physical operator supersets by a condition on each of the elements.

Recall from Sec.\ \ref{operations} that a complete set of operations is a set of operations associated with the same apparatus use having a given apparatus setting having disjoint outcome sets whose union is the set of all outcomes.  Motivated by this we define
\begin{quote}
{\bf A complete set of physical operators}\index{complete set! of physical operators}, $\{\hat B_\mathsf{a_1b_2\dots c_3}^\mathsf{d_4e_5\dots f_6}[l]: l=1~\text{to}~L\}$, is set for which each operator is physical and, further,
\begin{equation}\label{completesetsum}
\sum_{l=1}^L \hat B_\mathsf{a_1b_2\dots c_3}^\mathsf{d_4e_5\dots f_6}[l] \hat I_\mathsf{d_4e_5\dots f_6} = \hat I_\mathsf{a_1b_2\dots c_3}
\end{equation}
\end{quote}
The condition (\ref{completesetsum}) is equivalent to the requirement that
\begin{equation}\label{completesetofoperators}
\sum_{l=1}^L \hat A^\mathsf{a_1b_2\dots c_3} \hat B_\mathsf{a_1b_2\dots c_3}^\mathsf{d_4e_5\dots f_6}[l] \hat I_\mathsf{d_4e_5\dots f_6}=1
\end{equation}
for all rank one projectors, $\hat A^\mathsf{a_1b_2\dots c_3}$.  This condition is motivated by the physical constraint that the sum of probabilities over all outcomes should add to one.  We note that all physical operators belong to at least one complete set of physical operators.

Sometimes we will be interested in using completely positive maps instead of the operators described here.  For this purpose we note the following
\begin{T}\label{completephysicalcompletecpmaps}
The operators $\{\hat B_\mathsf{a_1}^\mathsf{b_2}[l]: l=1~\text{to}~L\}$ are a complete set of physical operators if and only if the superoperators in the set $\{ \$^l_B(\cdot): l=1~\text{to} ~ L\}$, given by $\$^l_B(\hat E^\mathsf{a_1})=\hat E^\mathsf{a_1}\hat B_\mathsf{a_1}^\mathsf{b_2}[l]$, are each completely positive and their sum, $\sum_l \$_B^l(\cdot) $, is trace preserving.
\end{T}
This follows immediately from {\bf T\ref{CPmaps}} and the fact that the condition (\ref{completesetofoperators}) is equivalent to the requirement that
$\sum_l \$_B^l(\cdot) $ be trace preserving.

\subsection{Positivity of operators under input transpose}\label{positivityunderinputtranspose}

The physical motivation for considering physical operators (and complete sets of these) is clear.  However, it is not so clear how to recognize whether a particular operator is physical (and whether a set of these is complete) without exhaustively checking the conditions for all rank one projectors and all auxiliary systems $\mathsf g$.  In fact they have a very simple characterization as we will now see. This characterization is inspired by the use of the Choi-Jamio\l kowski operator by Chiribella, D'Ariano, and Perinotti in their \lq\lq quantum combs" approach.  We will review their approach in Sec.\ \ref{combs}.  Rather than using the Choi-Jamio\l kowski operator, we will stick with $\hat B_\mathsf{a_1}^\mathsf{b_2}$.  We show that the partial transpose over the input space must be positive by employing a theorem ({\bf T\ref{ACprojectortheorem}} below) adapted from a similar theorem in the work of Aharonov, Popescu, Tollaksen, and Vaidman \cite{aharonov2009multiple} (these authors considered vectors in the Hilbert space rather than ${\cal V}$).

We will be interested in the partial transpose \index{partial transpose}of operators.  To define a transpose we need to work in a given basis for the underlying Hilbert space (to see that the transpose depends on the basis note that, if the basis is such that the matrix is diagonalized, taking the transpose leaves the matrix unaffected). Thus, when we take the transpose over a space, we will work in some standard basis.  Fortunately, the results we obtain will not depend on which basis we choose. We fix a basis for each system type.  We fix the same basis in ${\cal H}_\mathsf{a}$ and ${\cal H}^\mathsf{a}$.  We define the \emph{input transpose} \index{input transpose} of an operator $\hat B_\mathsf{a_1}^\mathsf{b_2}\in {\cal V}_\mathsf{a_1}\otimes{\cal V}^\mathsf{b_2}$ to be the partial transpose of $\hat B_\mathsf{a_1}^\mathsf{b_2}$ over the input space ${\cal V}_\mathsf{a_1}$ in the standard basis.  We will denote the input transpose of $\hat B_\mathsf{a_1}^\mathsf{b_2}$ by $\hat B_\mathsf{a_1^T}^\mathsf{b_2}$. The \emph{output transpose} \index{output transpose} is defined to be the partial transpose over the output space ${\cal V}^\mathsf{b_2}$ in the standard basis.  We denote the output transpose of $\hat B_\mathsf{a_1}^\mathsf{b_2}$ by $\hat B_\mathsf{a_1}^\mathsf{b_2^T}$.  The input transpose of $\hat C_\mathsf{a_1b_2c_3}^\mathsf{d_4}$ can be written as $\hat C_\mathsf{a_1^Tb_2^Tc_3^T}^\mathsf{d_4}$ or $\hat C_\mathsf{[a_1b_2c_3]^T}^\mathsf{d_4}$.  We can also define objects such as $\hat D_\mathsf{a_1b_2^Tc_3}^\mathsf{d_4e_5^T}$ where we take the partial transpose of some of the input spaces and some of the output spaces.  We note that   $[\hat D_\mathsf{a_1b_2^Tc_3}^\mathsf{d_4e_5^T}]^T =
\hat D_\mathsf{a^T_1b_2c^T_3}^\mathsf{d^T_4e_5}$ and that positivity of $\hat D_\mathsf{a_1b_2^Tc_3}^\mathsf{d_4e_5^T}$ is equivalent to positivity of $\hat D_\mathsf{a^T_1b_2c^T_3}^\mathsf{d^T_4e_5}$.
We will prove
\begin{T}\label{partialtransposeinsert}
An operator fragment is unchanged if we take the partial transpose over spaces corresponding to any subset of the matched wires.  For example,
\begin{equation}
\hat A_\mathsf{a_1b_2}^\mathsf{c_3d_4c_7} B_\mathsf{a_5c_3c_7}^\mathsf{a_1d_6} = \hat A_\mathsf{a^T_1b_2}^\mathsf{c^T_3d_4c_7} B_\mathsf{a_5c^T_3c_7}^\mathsf{a^T_1d_6}
\end{equation}
\end{T}
To prove this we can expand out each operator into its fully decomposed form.  We can then consider matched fiducial pairs such as ${}_{a_1}\! \hat X^\mathsf{a_1}$ with $\hat X_\mathsf{a_1}^{a'_1}$.  Since we are taking the circuit trace, we are taking the trace over such matched pairs. Hence, we have
\begin{equation}
{}_{a_1}\! \hat X^\mathsf{a_1}\hat X_\mathsf{a_1}^{a'_1}
\end{equation}
entering into the expression.  This is just the hopping metric.  It is easy to show that
\begin{equation}
{}_{a_1}\! \hat X^\mathsf{a_1}\hat X_\mathsf{a_1}^{a'_1} =
   =  {}_{a_1}\! \hat X^\mathsf{a^T_1}\hat X_\mathsf{a^T_1}^{a'_1}
\end{equation}
using well known matrix properties (recall that it is implicit in the notation that we are taking the trace).  This means the hopping metric is invariant under taking the transpose over the space associated with the given system.  Hence, {\bf T\ref{partialtransposeinsert}} follows.

% !!! in proof below I may want to switch to the output trace in main proof.

Consider the operator fragment
\begin{equation}
\hat A^\mathsf{a_1g_3} \hat C_\mathsf{b_2g_3}
~~~~~\leftrightarrow~~~~~
\begin{Diagram}{0}{-1.2}
\Dopbox{A}{0,0}  \Dopbox{C}{0,9}
\wire{A}{C}{3}{3} \otherside\opsymbol{g}
\outwire{A}{1}\Opsymbol{a} \inwire{C}{1}\Opsymbol{b}
\end{Diagram}
\end{equation}
where both $\hat A^\mathsf{a_1g_g}$ and $\hat C_\mathsf{b_2g_3}$ are rank one projectors.   We have
\begin{equation}
\hat A^\mathsf{a_1g_3} \hat C_\mathsf{b_2g_3}  \in {\cal V}^\mathsf{a_1}\otimes{\cal V}_\mathsf{b_2}
\end{equation}
We will prove
\begin{T}\label{ACprojectortheorem}
There exist rank one projectors $\hat A^\mathsf{a_1g_g}$ and $\hat C_\mathsf{b_2g_3}$ such that the input transpose of $\hat A^\mathsf{a_1g_3} \hat C_\mathsf{b_2g_3}$ is equal to any rank one projector in ${\cal V}^\mathsf{a_1}\otimes{\cal V}_\mathsf{b_2}$.  A similar result is true for the output transpose.  These results are true for any choice of standard basis for taking the transpose.
\end{T}
First we need to develop a little notation.  We use the notation developed in \cite{aharonov2009multiple}.  Any rank one preparation  projector can be written
\begin{equation}
\hat D^\mathsf{a_1} = |D^\mathsf{a_1}\rangle \langle D^\mathsf{a_1}|  \in \overleftarrow{\cal H}^\mathsf{a_1}\otimes\overrightarrow{\cal H}^\mathsf{a_1}
\end{equation}
where
\begin{equation}
|D^\mathsf{a_1}\rangle \in \overleftarrow{\cal H}^\mathsf{a_1} ~~~~\text{and}~~~~ \langle D^\mathsf{a_1} | \in \overrightarrow{\cal H}^\mathsf{a_1}
\end{equation}
These Hilbert spaces have dimension $N_\mathsf{a}$.  We have
\begin{equation}\label{ACCA}
\hat A^\mathsf{a_1g_3} \hat C_\mathsf{b_2g_3}  = \langle A^\mathsf{a_1g_3}|C_\mathsf{b_2g_3}\rangle \langle C_\mathsf{b_2g_3} | A^\mathsf{a_1g_3}\rangle
\end{equation}
To understand the RHS note that
\begin{equation}
 \langle C_\mathsf{b_2g_3} | A^\mathsf{a_1g_3}\rangle \in \overleftarrow{\cal H}^\mathsf{a_1}\otimes \overrightarrow{\cal H}_\mathsf{b_2} ~~~~\text{and} ~~~~ \langle A^\mathsf{a_1g_3}|C_\mathsf{b_2g_3}\rangle \in \overleftarrow{\cal H}_\mathsf{b_2}\otimes \overrightarrow{\cal H}^\mathsf{a_1}
\end{equation}
because, in each case, we take the inner product over the space corresponding to $\mathsf{g_3}$.  Hence, the RHS of (\ref{ACCA}) is proportional to a rank one projector.  We will now see that, by appropriate choice of $|A^\mathsf{a_1g_3}\rangle$ and $\langle C_\mathsf{b_2g_3} |$, $ \langle C_\mathsf{b_2g_3} | A^\mathsf{a_1g_3}\rangle$ can be proportional to any vector in $\overleftarrow{\cal H}^\mathsf{a_1}\otimes \overrightarrow{\cal H}_\mathsf{b_2}$.  A general vector in $\overleftarrow{\cal H}^\mathsf{a_1}\otimes \overrightarrow{\cal H}_\mathsf{b_2}$ can be written
\begin{equation}\label{HLHRprojector}
\sum_n |E^\mathsf{a_1}[n]\rangle \otimes \langle V_\mathsf{b_2}[n] |
\end{equation}
where $|E^\mathsf{a_1}[n]\rangle$ is some set of vectors (not necessarily normalized or orthogonal) in $\overleftarrow{\cal H}^{a_1}$ and $\langle V_\mathsf{b_2}[n]|$ is the orthonormal basis in $\overrightarrow{\cal H}_\mathsf{b_2}$ with respect to which we will take the transpose.  We choose
\begin{equation}
|A^\mathsf{a_1g_3}\rangle = \sum_n |E^\mathsf{a_1}[n]\rangle|W^\mathsf{g_3}[n]\rangle ~~~~\text{and}~~~~
\langle C_\mathsf{b_2g_3}| = \sum_n  \langle V_\mathsf{b_2}[n]|\langle W^\mathsf{g_3}[n]|
\end{equation}
where $|W^\mathsf{g_3}[n]\rangle$ is an orthonormal basis in $\overleftarrow{\cal H}^\mathsf{g_3}$ and we choose $\mathsf g$ such that $N_\mathsf{g} \geq N_\mathsf{a}$.  This choice immediately gives (\ref{HLHRprojector}).  Hence, we can have
\begin{equation}
\hat A^\mathsf{a_1g_3} \hat C_\mathsf{b_2g_3} = \left( \sum_n |E^\mathsf{a_1}[n]\rangle \otimes \langle V_\mathsf{b_2}[n] | \right)\otimes \left(\sum_m  \langle E^\mathsf{a_1}[m]| \otimes | V_\mathsf{b_2}[m] \rangle
\right)
\end{equation}
the input transpose of this (in the $|V_\mathsf{b_2}[n]\rangle$ basis) is
\begin{equation}\label{aftertranspose}
\hat A^\mathsf{a_1g_3} \hat C_\mathsf{b^T_2g_3} = \left( \sum_n |E^\mathsf{a_1}[n]\rangle \otimes | V_\mathsf{b_2}[n] \rangle \right)\otimes \left(\sum_m  \langle E^\mathsf{a_1}[m]| \otimes \langle V_\mathsf{b_2}[m] |\right)
\end{equation}
As long as
\begin{equation}
\sum_n \langle E^\mathsf{a_1}[n]|E^\mathsf{a_1}[n] \rangle =1
\end{equation}
the operator in (\ref{aftertranspose}) is a rank one projector in ${\cal V}^\mathsf{a_1}\otimes{\cal V}_\mathsf{b_2}$.  Further, by appropriate choice of the $|E^\mathsf{a_1}[n] \rangle$'s, we can obtain any rank one projector in this way.  The transpose of a projection operator is also a projection operator.  Hence
\begin{equation}
\hat A^\mathsf{a^T_1g_3} \hat C_\mathsf{b_2g_3}
\end{equation}
is also a projection operator. This proves {\bf T\ref{ACprojectortheorem}}.

We can now prove the following important theorem
\begin{T}\label{positivesameasphysical} \index{physicality}
{\bf Physicality.}  An operator, $\hat B_\mathsf{a_1}^\mathsf{b_2}$, is physical if and only if (a) its input transpose is positive ($0\leq B_\mathsf{a_1^T}^\mathsf{b_2}$) and (b) it satisfies
\begin{equation}\label{ctraceBIcondition}
\hat B_\mathsf{a_1}^\mathsf{b_2}\hat I_\mathsf{b_2} \leq \hat I_\mathsf{a_1}
\end{equation}
\end{T}
The first point follows from {\bf T\ref{partialtransposeinsert}} and {\bf T\ref{ACprojectortheorem}}.   By {\bf T\ref{partialtransposeinsert}} we note that
\begin{equation}\label{insertBintester}
\hat A^\mathsf{a_1g_3} \hat B_\mathsf{a_1}^\mathsf{b_2} \hat C_\mathsf{b_2g_3}
= \hat B_\mathsf{a^T_1}^\mathsf{b_2} (\hat A^\mathsf{a^T_1g_3} \hat C_\mathsf{b_2g_3})
\end{equation}
For $\hat B_\mathsf{a_1}^\mathsf{b_2}$ to be physical we require $0\leq \hat A^\mathsf{a_1g_3} \hat B_\mathsf{a_1}^\mathsf{b_2} \hat C_\mathsf{b_2g_3}$.  By (\ref{insertBintester}) and {\bf T\ref{ACprojectortheorem}} we see that this is equivalent to requiring that $\hat B_\mathsf{a^T_1}^\mathsf{b_2}$ is positive.  To prove the second point, first note that $\hat I_\mathsf{b_2g_3}=\hat I_\mathsf{b_2}\hat I_\mathsf{g_3}$.  Hence
\begin{equation}\label{traceoutg}
\hat A^\mathsf{a_1g_3}\hat  B_\mathsf{a_1}^\mathsf{b_2} \hat I_\mathsf{b_2g_3}  = \hat E^\mathsf{a} (\hat B_\mathsf{a}^\mathsf{b_2} \hat I_\mathsf{b_2})
\end{equation}
where $\hat E^\mathsf{a_1}= \hat A^\mathsf{a_1g_3}\hat I_\mathsf{g_3}$.  Since we can write $\hat A^\mathsf{a_1g_3}$ as a product of rank one projectors, $\hat E^\mathsf{a_1}$ can be equal to any rank one projector.   It follows that either side of (\ref{traceoutg}) is less than or equal to one if and only if (\ref{ctraceBIcondition}) is satisfied.  This proves {\bf T\ref{positivesameasphysical}}.

Theorem {\bf T\ref{positivesameasphysical}} is important because it means that can we define physical operators in the following way (this being equivalent to the previous definition).
\begin{quote}\index{physical operators}
{\bf Physical operators.}  An operator, $\hat B_\mathsf{a_1}^\mathsf{b_2}$,  is said to be physical if its input transpose, $\hat B_\mathsf{a^T_1}^\mathsf{b_2}$, is positive and it satisfies
\begin{equation}\label{asymmetry}
B_\mathsf{a_1}^\mathsf{b_2}\hat I_\mathsf{b_2} \leq \hat I_\mathsf{a_1}
\end{equation}
\end{quote}
While the physical motivation for the former definition was clearer, it is easier to check whether a particular operator is physical with this definition.  Note \begin{itemize}
\item We could equivalently have stated that the output transpose must be positive.
\item We can use any basis for taking the partial transpose.  Given that this condition is both necessary and sufficient when used with respect to any given standard basis, it follows that positivity of the input transpose is independent of which standard basis we adopt.
\item There are two distinct interesting things happening here that single out time from space.  Time is represented here by the fact that we have an input-output structure. Space is represented by the fact that we can have composite systems.
    \begin{itemize}
    \item Consider a general operator such as $\hat D_\mathsf{a_1b_2c_3}^\mathsf{e_4f_5g_6}$.   The positivity of $\hat D_\mathsf{a^T_1b^T_2c^T_3}^\mathsf{e_4f_5g_6}$ (or equivalently of $\hat D_\mathsf{a_1b_2c_3}^\mathsf{e^T_4f^T_5g^T_6}$) implies that time is different from space. This is similar to the fact that we have a different sign associated with the time coordinate in the Minkowski metric.  This analogy is quite strong because, apart from this difference (the partial transpose, or the minus sign) we otherwise treat space and time on the same footing.  While this fact shows that time is distinct from space, it does not impose any time asymmetry.
    \item The second requirement, given in (\ref{asymmetry}), is time asymmetric.  It indicates that the future does not influence the past.  We do not, however, wish to assume that the past does not influence the future.  However, in a fully time symmetric formulation of quantum theory, we would wish to treat the past and the future on the same footing.
    \end{itemize}
    At this stage, it should be pointed out, we are just exhibiting some interesting mathematics.  These remarks concerning time will become relevant to the physical situation in Sec.\ \ref{QTaxioms} when we show how to use these mathematical results to reformulate quantum theory.
\end{itemize}

We can give the following definition for a complete set of physical operators which is equivalent to the (slightly different) definition we gave earlier.
\begin{quote}\index{complete set!of physical operators}
{\bf A complete set of physical operators}, $\{ \hat B_\mathsf{a_1}^\mathsf{b_2}[l]: l=1~\text{to}~L\}$, is a set for which the input transpose of every operator is positive and
\begin{equation}\label{completepositiveI}
\sum_{l=1}^L \hat B_\mathsf{a_1}^\mathsf{b_2}[l]\hat I_\mathsf{b_2} = \hat I_\mathsf{a_1}
\end{equation}
\end{quote}
Note that it follows from the fact that $\hat B_\mathsf{a_1}^\mathsf{b_2}[l]$ has positive input transpose that $\hat B_\mathsf{a_1}^\mathsf{b_2}[l]\hat I_\mathsf{b_2}$ is positive (using {\bf T\ref{partialtransposeinsert}} and the fact that $I_\mathsf{b^T_2}=I_\mathsf{b_2}$).  It follows that each operator in a complete set of physical operators (by this new definition) satisfies $\hat B_\mathsf{a_1}^\mathsf{b_2}[l]\hat I_\mathsf{b_2} \leq \hat I_\mathsf{a_1}$.  Hence, the operators in a complete set of physical operators are physical.

It is clear that operator fragments must also have positive input transpose and satisfy (\ref{ctraceBIcondition}) as they can be put in transformation mode so that they are, effectively, operators.   However, there will, in general, be further constraints on general operator fragments coming from the shape of the circuit.  The analogous issue in the quantum combs framework has been considered in \cite{chiribella2009theoretical}.

\subsection{Two mathematical axioms for quantum theory}\label{QTaxioms}

Using the definitions and theorems of the previous two subsections, we can give the following statement of quantum theory.
\begin{quote}\index{axioms}
{\bf QUANTUM THEORY.}  The following two mathematical axioms specify quantum theory:
\begin{description}
\item[Axiom 1] \index{axioms!\textbf{Axiom 1}}Operations correspond to operators.
\item[Axiom 2] \index{axioms!\textbf{Axiom 2}}Every complete set of physical operators corresponds to a complete set of operations.
\end{description}
The operators here are understood to act on a complex Hilbert space.
\end{quote}
We could replace Axiom 1 by the requirement \emph{for every system type we can associate a fiducial set of operators with a fiducial set of operations such that we get equal hopping metrics}.  Axiom 1 as given then follows as a consequence of {\bf T\ref{correspondencetheorem}}.  This restatement is mathematically simpler but verbally more demanding.  We could combine the two axioms into the more pithy statement:
\begin{quote}
{\bf QUANTUM THEORY:} Every complete set of positive operators corresponds to a complete set of operations and vice versa.
\end{quote}
The \lq\lq vice versa" part of this statement here is a little stronger than Axiom 1 above. We will stick with the two axioms version in this paper. This provides a rather succinct statement of quantum theory.  We will unpack it into a more familiar form.  We note that these axioms imply the following statements
\begin{enumerate}
\item \emph{The trace formula follows.}  The definition of the word \lq\lq corresponds" implies that the probability for a circuit is equal to the operator circuit (in which the trace is implicit).
\item \emph{Operations correspond to operators with positive input transpose having} $\hat B_\mathsf{a_1}^\mathsf{b_2}\hat I_\mathsf{b_2} \leq \hat I_\mathsf{a_1}$. If we had a single operation for which this were not true then it would follow from Axiom 2 and {\bf T\ref{ACprojectortheorem}} that we could form corresponding operator circuits having values less than zero or greater than one.  Since the probability of a circuit is given by this the operator circuit we demand that such a trace must be between zero and one.  Hence all operations must correspond to physical operators.
\item \emph{All operators having positive input transpose and} $\hat B_\mathsf{a_1}^\mathsf{b_2}\hat I_\mathsf{b_2} \leq \hat I_\mathsf{a_1}$ \emph{correspond to operations.}  This follows immediately from Axiom 2 since all such operators belong to at least one complete set of positive operators.
\item \emph{Preparations correspond to positive operators having trace less than or equal to one.}  To see this we must regard the input as the trivial type (having $N_\mathsf{a}=1$).  The identity is then just equal to one.  This property then follows from $\hat B^\mathsf{b_2}\hat I_\mathsf{b_2} \leq \hat I_\mathsf{a_1}$
\item \emph{All results correspond to positive operators that are less than or equal to $\hat I_\mathsf{a_1}$.}  To see this we regard the output as the trivial type.  The result operator must clearly be positive (since the output is trivial).  This property then follows from having $\hat B_\mathsf{a_1}^\mathsf{b_2}\hat I_\mathsf{b_2} \leq \hat I_\mathsf{a_1}$.
\item \emph{The transformation associated with any operation is a completely positive trace non-increasing map.}  Since all operators correspond to physical operators this result follows from {\bf T\ref{CPmaps}}.
\end{enumerate}

We have given mathematical axioms for quantum theory above. The real objective of this paper is to provide operational natural postulates for quantum theory.  We will provide such a set of postulates in the next Part.   We will reconstruct quantum theory in three parts.  The first part will set up the basics concerning filters and systems.  In the second part we will obtain the qubit.  We will not need the machinery of the duotensor formalism for these first two parts.  In the third part we will use the duotensor formalism to obtain quantum theory as characterized by the above two mathematical axioms.   At this stage it is worth recalling the results that have to be proven in order to obtain these axioms.  These are
\begin{enumerate}
\item That we can find an association of fiducial operators with fiducial operations such that we have equal hopping metrics for each type.
\item That we have preparations and results corresponding to  all rank one projectors for every type.
\item That we have a result corresponding to the identity operator for each type.
\item That we can realize a complete set of operations corresponding to every complete set of physical operators.
\end{enumerate}
The first property guarantees, by {\bf T\ref{correspondencetheorem}}, that we have a correspondence between operations and operators (and hence we have the trace rule for calculating probabilities).  The second and third properties actually follow from the fourth property but it is useful to list them separately since they motivate the requirement that supersets of operators should be physical.  Further, we will derive these properties from the postulates before we prove the fourth property.  The fourth property is simply axiom 2.

\subsection{Choi-Jamio\l kowski isomorphism and quantum combs}\label{combs}

In this subsection we note the strong similarity between the duotensor mediated operator formulation of quantum theory given above and the \lq\lq quantum combs" framework due to Chiribella, D'Ariano, and Perinotti (CDP) \cite{chiribella2009theoretical}.  In particular, the Choi-Jamio\l kowski operator used by CDP is the input transpose of the operator used here and the \emph{link product} used by CDP is equivalent to the circuit trace formula used here after some notational translation and the appropriate insertion of partial transposes.  A precursor to both the link product of CDP and the \lq\lq circuit trace product" used here is the \emph{causaloid product} as applied to quantum theory in \cite{hardy2005probability, hardy2007towards}.  All three provide a way of finding, in general, the mathematical object associated with a fragment from the mathematical objects associated with smaller fragments that comprise this bigger fragment.  These approaches pertain to the general mixed state case (with general transformations and general measurements).  The multi-time approach of Aharonov, Popescu, Tollaksen, and Vaidman \cite{aharonov2009multiple} and the general boundary approach of Oeckl \cite{oeckl2003general} can be understood to be doing something similar but are restricted to the pure state case.  In particular, the notation of \cite{aharonov2009multiple} was employed in proving {\bf T\ref{ACprojectortheorem}}.

First we will review the basic notation and equations of the quantum combs \index{quantum combs}approach.   CDP take ${\cal L}({\cal H})$ to be the set of linear operators on the finite dimensional Hilbert space ${\cal H}$.  The set of linear maps from ${\cal L}({\cal H}_0)$ to ${\cal L}({\cal H}_1)$ is denoted ${\cal L}({\cal L}({\cal H}_0), {\cal L}({\cal H}_1))$.   A linear map in ${\cal L}({\cal L}({\cal H}_0), {\cal L}({\cal H}_1))$ is denoted ${\mathscr M}$.  A superoperator is an example of such a linear map.   There is a one to one correspondence from linear maps, ${\mathscr M}$, in ${\cal L}({\cal L}({\cal H}_0), {\cal L}({\cal H}_1))$ to linear operators, $M$, on ${\cal L}({\cal H}_1 \otimes {\cal H}_0)$ given by
\begin{equation}\label{CJmap}
M = {\mathscr M} \otimes {\mathscr I}_{{\cal L}({\cal H}_0)} |I_{{\cal H}_0}\rangle\rangle \langle\langle I_{{\cal H}_0}|
\end{equation}
where ${\mathscr I}_{{\cal L}({\cal H})}$ is the identity map on ${\cal L}({\cal H})$ and
\begin{equation}
|I_{\cal H}\rangle\rangle = \sum_n |n\rangle|n\rangle
\end{equation}
Here $\{ |n\rangle \}$ is a fixed orthonormal basis for the complex Hilbert space ${\cal H}$.  The map between ${\mathscr M}$ and $M$ in (\ref{CJmap}) is the Choi-Jamio\l owski isomorphism.   One can prove
\begin{enumerate}
\item a linear map ${\mathscr M}$ is trace-preserving if and only if its Choi-Jamio\l owski operator  satisfies the property
\begin{equation}
\text{Tr}_{{\cal H}_1}[M] = I_{{\cal H}_0}
\end{equation}
where $\text{Tr}_{\cal H}$ denotes the partial trace over ${\cal H}$ and $I_{\cal H}$ is the identity operator in ${\cal H}$.  More generally, a trace non-increasing map has Choi-Jamio\l owski operator satisfying
\begin{equation}\label{conditiononCJ}
\text{Tr}_{{\cal H}_1}[M] \leq  I_{{\cal H}_0}
\end{equation}
\item A linear map ${\mathscr M}$ is Hermitian preserving if and only if its Choi-Jamio\l owski operator, $M$, is Hermitian.
\item A linear map ${\mathscr M}$ is completely positive if and only if its Choi-Jamio\l owski operator, $M$, is positive.
\end{enumerate}
The link product for finding the Choi-Jami\l owski operator associated with the composition ${\mathscr N}\circ{\mathscr M}$ (i.e.\ two sequential operations) is given by
\begin{equation}\label{linkproduct}
N * M := \text{Tr}_{{\cal H}_1} [( I_{{\cal H}_2} \otimes M^{T_1} )(N\otimes I_{{\cal H}_0}) ]
\end{equation}
Here $M\in {\cal L}({\cal H}_1 \otimes {\cal H}_0)$ and $N\in {\cal L}({\cal H}_2 \otimes {\cal H}_1)$.  The notation $M^{T_i}$ means we take the partial transpose of $M$ in the space ${\cal H}_i$.  The \emph{general link product} applies to the more general case in which we have maps with input and output spaces that are tensor products of Hilbert spaces and where these maps are only composed through some of these spaces (in the language of this paper this corresponds to placing some wires between two general fragments).  The general link product is given by
\begin{equation}\label{generallinkproduct}
N * M :=  \text{Tr}_{{\cal M}\cap {\cal N}}
[( I_{{\cal M} \backslash {\cal N}} \otimes M^{T_{{\cal M}\cap {\cal N}}} )(N\otimes I_{{\cal M}\backslash{\cal N}}) ]
\end{equation}
Here $M\in {\cal L}(\bigotimes_{m\in{\cal M}} {\cal H}_m)$ and $N\in {\cal L}(\bigotimes_{n\in{\cal N}} {\cal H}_n)$.   The set-subscript ${\cal X}$ refers to the Hilbert space $\bigotimes_{i\in {\cal X}} {\cal H}_i$.

The tensor product notation is rather cumbersome (as pointed out in Sec.\ \ref{notation}).   An important part of the reformulation in this paper is to provide notation that is sympathetic to the graphical structure of operator fragments.  In particular, with the notation introduced in Sec.\ \ref{notation}, it is not necessary to pad equations with identity operators.   We will now translate the equations of CDP into the notation of the present paper.
Associated with any operator, $\hat B_\mathsf{a_1}^\mathsf{b_2}$, is the Choi-Jamio\l kowski operator defined as
\begin{equation}
(\$_B\otimes I) \hat J^\mathsf{a_1a_3} = \hat B_\mathsf{a_1}^\mathsf{b_2} \hat J^\mathsf{a_1a_3}
\end{equation}
where $\$_B $ acts on ${\cal V}^\mathsf{a_1}$, $I$ acts as the identity on ${\cal V}^\mathsf{a_3}$, and we define
\begin{equation}
\hat J^\mathsf{a_1a_3} := |J^\mathsf{a_1a_3}\rangle\langle J^\mathsf{a_1a_3}| ~~~~~\text{with}~~~~~
|J^\mathsf{a_1a_3}\rangle = \sum_{n=1}^\mathsf{N_\mathsf{a}} |U^\mathsf{a_1}[n]\rangle|U^\mathsf{a_3}[n]\rangle
\end{equation}
It is an easy calculation to show that
\begin{equation}
\hat B_\mathsf{a_1}^\mathsf{b_2} \hat J^\mathsf{a_1a_3} = \hat B_\mathsf{a^T_3}^\mathsf{b_2}
\end{equation}
(where the equality is understood to be numerical - strictly speaking this equation is illegal since the subscripts and superscripts do not match).  Here the standard basis with respect to which the input trace is taken is $\{|U^\mathsf{a_1}[n]\rangle\}$ as used in the definition of $J^\mathsf{a_1a_3}$.   Hence the Choi-Jamio\l owski operator is the equal to $\hat B_\mathsf{a^T_1}^\mathsf{b_2}$, which we know to be positive for physical operators.  Physicality also imposes that $\hat B_\mathsf{a_1}^\mathsf{b_2} \hat I_\mathsf{b_2} \leq \hat I_\mathsf{a_1}$ (where $\hat I_\mathsf{a_1}$ is the identity operator on ${\cal H}_\mathsf{a_1}$).  Using {\bf T\ref{partialtransposeinsert}} and the fact that $\hat I_\mathsf{a^T_1}=\hat I_\mathsf{a_1}$, we obtain
\begin{equation}
\hat B_\mathsf{a^T_1}^\mathsf{b_2} \hat I_\mathsf{b_2} \leq  \hat I_\mathsf{a_1}
\end{equation}
which is the same as (\ref{conditiononCJ}) in the present notation.   If we have two operators $\hat A_\mathsf{a_1}^\mathsf{b_2}$ and $\hat B_\mathsf{b_2}^\mathsf{c_3}$ which are wired together to form
\begin{equation}
\hat A_\mathsf{a_1}^\mathsf{b_2}\hat B_\mathsf{b_2}^\mathsf{c_3}
\end{equation}
The input transpose of this can, using {\bf T\ref{partialtransposeinsert}}, be written as
\begin{equation}
\hat A_\mathsf{a^T_1}^\mathsf{b^T_2}\hat B_\mathsf{b^T_2}^\mathsf{c_3}
\end{equation}
which is the same as the link product (\ref{linkproduct}) in the present notation.  More generally, if we can have two operator fragments such as $\hat A_\mathsf{a_1b_2}^\mathsf{c_3d_4}$ and $\hat B_\mathsf{d_4e_5}^\mathsf{b_2f_6}$ comprising the bigger operator fragment
\begin{equation}
\hat A_\mathsf{a_1b_2}^\mathsf{c_3d_4}\hat B_\mathsf{d_4e_5}^\mathsf{b_2f_6}
\end{equation}
We can think of this as the \lq\lq circuit trace product" (as it is implicit in the notation that we are taking the circuit trace).  Using {\bf T\ref{partialtransposeinsert}}, the input transpose of this can be written as
\begin{equation}
\hat A_\mathsf{a^T_1b_2}^\mathsf{c_3d^T_4}\hat B_\mathsf{d^T_4e^T_5}^\mathsf{b_2f_6}
\end{equation}
which is the same as the general link product defined in (\ref{generallinkproduct}) in the present notation.

The main difference between the approach here and that of CDP is that CDP work with an operator that is positive.  To compensate for this they have to introduce partial transposes into the equation for the link product.  Here, instead, the basic object is not positive (though its input transpose is).  The pay-off for working with a non-positive object is that the equation for putting operators together (with implicit circuit trace) does not involve taking any partial transposes.
This does raise the question of which object is more natural, $\hat B_\mathsf{a_1}^\mathsf{b_2}$, or its input transpose $\hat B_\mathsf{a^T_1}^\mathsf{b_2}$.  The argument for $\hat B_\mathsf{a^T_1}^\mathsf{b_2}$ being more natural is that it is positive. This is a mathematical argument. The argument for $\hat B_\mathsf{a_1}^\mathsf{b_2}$ being more natural is that it arises naturally as a sum over fiducial operators (via its fully decomposed form).  This is a physical argument since these fiducial operators correspond to operations. Added mathematical support for adopting $\hat B_\mathsf{a_1}^\mathsf{b_2}$ as the more natural object comes from the fact that we get simpler equations for combining operators.  This is particularly the case when we have more than two operators to combine.

There are a few more subtle differences between the quantum combs and the duotensor mediated approaches that are worth mentioning here.  First, the subscripts and superscripts play a deep and essential role in the reformulation presented in this paper.  It is the added structure associated with these subscripts and superscripts that distinguish plain old operators from the rather more useful operator structures used here and set up the connection with circuits comprised of operations.   Further, this subscript/superscript notation allows us to make taking the partial trace implicit in the notation in a natural way.  Such a calculus of subscripts and superscripts does not appear in the quantum combs approach. It is this calculus (with the accompanying abandonment of the $\otimes$ symbol) which accounts for the fact that the \lq\lq circuit trace product" formula appears much simpler than the link product formula.   Second, we work with spaces, ${\cal V}_\mathsf{a_1}$, of Hermitian operators rather than with the general spaces, ${\cal L}(\cal H)$, of linear maps as in the Choi-Jamio\l owski approach.  Such a restriction is motivated from the beginning by the physics since we wish to use full decomposability of operators in analogy to the full decomposability of operations.

The duotensor framework is a consequence of pursuing the reasoning of the causaloid framework \cite{hardy2005probability} (itself motivated by quantum gravity) in the context of the circuit model.  The quantum combs approach of CDP  was motivated by thinking about quantum information processing.  It is interesting that these two routes should lead to similar formulations of quantum theory.  CDP have proven numerous results in the quantum combs framework.  It should be possible to take these over into the duotensor operator framework.

\subsection{Formalism locality for quantum theory}

\index{formalism locality}We note that we can write $E_{a_1b_2}^{c_3}[i]= r E_{a_1b_2}^{c_3}[j]$ (for duotensors) if and only if $\hat E_\mathsf{a_1b_2}^\mathsf{c_3}[i]=r\hat E_\mathsf{a_1b_2}^\mathsf{c_3}[j]$.  Hence, it follows from the results of Sec.\ \ref{formalismlocality} that the probability ratio
\begin{equation}
\frac{\text{Prob}(\mathsf{E[i]})}{\text{Prob}(\mathsf{E[j]})}
\end{equation}
is well conditioned if and only if $\hat E_\mathsf{a_1b_2}^\mathsf{c_3}[i]=r\hat E_\mathsf{a_1b_2}^\mathsf{c_3}[j]$ (i.e.\ these two operators are proportional to each other).  If these operators are proportional then the probability ratio is given by
\begin{equation}
\frac{\text{Prob}(\mathsf{E[i]})}{\text{Prob}(\mathsf{E[j]})} = r
\end{equation}
Clearly this works for operator fragments in general and so we have a formalism local formulation of quantum theory.

We can consider \emph{operator ratios} such as
\begin{equation}\label{operatorratio}
\frac{\hat E_\mathsf{a_1b_2}^\mathsf{c_3}[i]}{\hat E_\mathsf{a_1b_2}^\mathsf{c_3}[j]}
\end{equation}
We will say that two such operator ratios are equal if they are equal after canceling over all scalar factors.  Further, we will say that two operator ratios are equal if one can be made equal to the other by multiplying one of the objects by an object of the form $\hat A_\mathsf{a_1}^\mathsf{b_2} / \hat A_\mathsf{a_1}^\mathsf{b_2}$ (this means we can cancel over an operator factor that appears in both the numerator and the denominator).
In the case that the operator ratio is equal to a number we have a well conditioned probability ratio that is equal to this number.    In the case that the operator ratio is not equivalent to a number then the probability ratio is not well conditioned.  However, we can think of the operator ratio as providing a measure of \lq\lq how well conditioned" the probability is.   If the numerator and denominator are almost proportional then the probability ratio cannot vary too much.   Hence, the operator ratio is still quantifying something physically meaningful.  This provides an answer to the the question posed by Arkani-Hamed in the introduction to \cite{arkanihamed2011quantum}. Operator ratios provide a way to do quantitative physics even when we do not have well conditioned probabilities.

%!!!! work out what the range of possible prob ratios in indefinite case.  Think about whether circuit operator is a kind of spacetime wave-function - can we trace over rest of world to get appropriate object for any arbitrary region?

\subsection{Disanalogies  between operations and operators}

The operator framework was set up by analogy with the operation framework.   However, there are a few points of disanalogy between operations and operators when we come to put them into the duotensor framework. First, as we noted already,  operations are equivalent to a fully decomposed form whereas operators are equal to a fully decomposed form.  In general, the notion of equivalence plays an important role in the operation framework but plays no role in the operator framework. Second, we note that if we are given  $\mathsf{A^{a_1}B^{b_2}}$ for example, it contains full information as to what $\mathsf{A^{a_1}}$ and $\mathsf{B^{b_2}}$ are. This expression can be interpreted simply as a \emph{list} of operations.   However, if we are provided with $\hat{A}^\mathsf{a_1}\hat B^\mathsf{b_2}$ (which is the tensor product of the two operators) then we cannot quite get our hands on $\hat{A}^\mathsf{a_1}$ and $\hat B^\mathsf{b_2}$ because it is not clear which term any overall factors belong to.  The disanalogy is stronger when we consider circuits because then we actually multiply operators together and then take the trace.  Given only the trace of the product of two operators, such as $\hat A^\mathsf{a_1}\hat C_\mathsf{a_1}$, we cannot return the original operators.  These disanalogies do not matter for formulating quantum theory because, in the end, we are always interested in the case where we use operator circuits to calculate probabilities.  However, for future applications, it may be useful to have a way formulating operators within the duotensor framework that is fully analogous to the way operations are formulated.  One way to do this is to reinterpret an expression like $\hat{A}^\mathsf{a_1}\hat B^\mathsf{b_2} \hat C_\mathsf{a_1}$ as a list. We could expand this out as $(\hat{A}^\mathsf{a_1},\hat B^\mathsf{b_2}, \hat C_\mathsf{a_1})$.  In this list wires are indicated by repeated integer labels on the type symbols.  We could use then proceed in exact analogy with the operation case but defining a $t(\cdot)$ as a linear extension of $\text{Trace}(\cdot)$ in the same way that the the $p(\cdot)$ function is defined as a linear extension of $\text{Prob}(\cdot)$.

\newpage

\part{Operational postulates for quantum theory}\label{thereconstruction}

In this part of the paper we will present a set of five operational postulates, {\bf P1, P2, P3, P4}$'$, and {\bf P5}.  We then show how to reconstruct classical probability theory and quantum theory from them.  If {\bf P4}$'$ is replaced by the stronger postulate, {\bf P4}, then we will see that classical probability theory is ruled out.  Hence, quantum theory follows from {\bf P1-5}.

In Sec.\ \ref{thepostulates} we give the postulates and discuss each of them in turn.  In Sec.\ \ref{filtersandsystemssection}-\ref{partIII} we give the reconstruction.  Sec.\ \ref{filtersandsystemssection} we extract a number of general properties of theories satisfying {\bf P1, P2, P3} and {\bf P4}$'$.  In Sec.\ \ref{gebitssection} we use {\bf P5} in addition to the other postulates to obtain the qubit of quantum theory in the non-classical case.   In Sec.\ \ref{partIII} we use the formalism developed in Part \ref{theduotensorframework} of this paper to obtain quantum theory for the general case.

\section{The postulates}\label{thepostulates}

\index{postulates}We will prove that, within the circuit framework presented in Part \ref{thecircuitframework}, the following postulates are consistent with classical probability theory and quantum theory only.
\begin{description}
\item[P1] \emph{Sharpness.} \index{sharpness}\index{postulates!\textbf{P1} \emph{sharpness}} Associated with any given pure state is a unique maximal effect giving probability equal to one.  This maximal effect does not give probability equal to one for any other pure state.
\item[P2] \emph{Information locality.}\index{information locality}\index{postulates!\textbf{P2} \emph{information locality}} A maximal measurement on a composite system is effected if we perform maximal measurements on each of the components.
\item[P3] \emph{Tomographic locality.}\index{tomographic locality}\index{postulates!\textbf{P3} \emph{tomographic locality}} The state of a composite system can be determined from the statistics collected by making measurements on the components.
\item[P4$'$] \emph{Permutability.} \index{permutability}\index{postulates!\textbf{P4}$'$ \emph{permutability}}There exists a reversible transformation on any system effecting any given permutation of any given maximal set of distinguishable states for that system.
\item[P5] \emph{Sturdiness.} \index{sturdiness}\index{postulates!\textbf{P5} \emph{sturdiness}} Filters are non-flattening.
\end{description}
To single out quantum theory it suffices to add anything that is consistent with quantum theory and inconsistent with classical probability theory. One way to do this is to add the word \lq\lq compound" to postulate {\bf P4$'$}:
\begin{description}
\item[P4] \emph{Compound permutability.} \index{compound permutability}\index{postulates!P6@\textbf{P4} \emph{compound permutability}} There exists a compound reversible transformation on any system effecting any given permutation of any given maximal set of distinguishable states for that system.
\end{description}
Recall that a compound transformation on as system is one which can be formed from two sequential transformations on the system neither of which is equal to the identity transformation.  Postulates {\bf P1-5} give rise to quantum theory.

We will discuss the meaning and motivation of each postulate along with alternative statements in some cases.

\subsection{{\bf P1}: Sharpness}

\index{sharpness}\index{postulates!\textbf{P1} \emph{sharpness}}Postulate {\bf P1} says that there is a one to one correspondence (a bijection) between the set of pure states and the set of maximal effects such that when we send a pure state onto a measurement having the associated maximal effect (under this map) as one of its effects we will certainly get the outcome corresponding to this maximal effect.  In the case that we follow a pure state by some other maximal effect (than the one given under this correspondence) we do not get probability equal to one.

This is true in classical probability theory.  Consider a system consisting of a ball that can be in one of $N$ boxes.  The pure states correspond to the ball being in a particular box with probability one.  The maximal effects consist looking into a particular box.  Clearly {\bf P1} is true here.

In quantum theory the pure states correspond to rank one projectors.  The maximal effects also correspond to rank one projectors.  The probability is given by taking the trace of the product of the state and effect projectors.  Postulate {\bf P1} is clearly satisfied.

Pure states are, in some sense, the most refined states.  We can think of them as corresponding to the most basic statements we can make about the world.  Maximal measurements are, in some sense, the most refined measurements.  We can think of maximal effects as corresponding to the most basic propositions about the world we can have.  It makes sense then, that there should be a unique correspondence between the most basic statements about the world and the most basic propositions as given by {\bf P1}.

Note that there is a certain time asymmetry in this postulate. Purity is a different concept from maximality.  This postulate associates purity coming from a preparation (the past) with maximality coming from a result (the future).  Of course we could consider a postulate assuming that pure effects are associated with states in a maximal set of distinguishable states but we do not need this for the purposes of the reconstruction.

A number of results follow from {\bf P1}.  We will provide these in Sec.\ \ref{uniqueimplications} below.  In particular, we will see that {\bf P1} implies that all pure states belong to some maximal set of distinguishable states.  The number of states in a maximal distinguishable set of states, $N_\mathsf{a}$, is a constant associated with the system type.  This imposes the property that distinguishable sets of states that consist only of pure states have the same number of elements.   This rules out odd shaped convex sets of states which do not have this property.  We will also see that causality follows from {\bf P1} (that the future does not influence the past).  This is an important property that is often taken as a background assumption in this kind of work.

The name, \lq\lq sharpness" is taken from Wilce's work \cite{wilce2009four}.  Wilce has a similar (though not exactly equivalent) postulate.

\subsection{{\bf P2}: Information locality}

% could also include here an alternative P3c: A maximal effect on a composite is effected if we have maximal effects on the compoenents.

\index{information locality}\index{postulates!\textbf{P2} \emph{information locality}} An alternative statement of this postulate is
\begin{description}
\item[P2a] For a composite system of type $\mathsf{ab}$ composed of systems of types $\mathsf{a}$ and $\mathsf{b}$ we have $N_\mathsf{ab}=N_\mathsf{a}N_\mathsf{b}$.
\end{description}
To see this is equivalent we first note that {\bf P2} clearly implies {\bf P2a} by counting.  To see that {\bf P2a} implies {\bf P2} we note that we can perform maximal measurements on each of the components.  Such a composite measurement will distinguish at least one set of $N_\mathsf{a}N_\mathsf{b}$ states (corresponding to preparing distinguishable states in the original sets for each of the components).  But, by {\bf P2a}, this must constitute a maximal set.  Hence the measurement on the composite is a maximal measurement as stated in {\bf P2}.

\index{information carrying capacity}Since information carrying capacity is defined as $\text{log}_2 N_\mathsf{a}$ for a system of type $\mathsf a$, we have another alternative statement
\begin{description}
\item[P2b] Information carrying capacity is additive for systems made up of components.
\end{description}
This corresponds very well with our usual intuition.  If we have a memory stick that carries 2 gigabytes and another that carries 8 gigabytes then, combined, they can carry 10 gigabytes of memory.

We call this property \emph{information locality} since the amount of global information a system can carry is simply given by adding together the local amounts.

While this postulate is rather innocent looking, it is very powerful when used in conjunction with {\bf P4} as we will see.  Further, it is possible to imagine situations in which {\bf P2} is not true \cite{hardy2010limited}.  For example, imagine systems of type $\mathsf a$ consist of a die which has a locked door on one side, and systems of type $\mathsf b$ consist of a key having a head on one side and a tails on the other (like a coin).  The die has $N_\mathsf{a}=6$ and the key has $N_\mathsf{b}=2$.  But if we suppose that the key unlocks the door on the die (when the two are proximate), and further, inside the die is another key then, $N_\mathsf{ab}=24$ rather than $12$.

\subsection{{\bf P3}: Tomographic Locality}\label{SecTomographiclocality}

\index{tomographic locality}\index{postulates!\textbf{P3} \emph{tomographic locality}}Postulate {\bf P3} is often referred to as \emph{local tomography} and has been much discussed in the literature \cite{araki1980characterization, bergia1980actual, wootters1990local, mermin1998quantum}. We call it \emph{tomographic locality} to contrast it with \emph{information locality}.  We can translate it into mathematical language.  Let $\mathsf{A^{a_1b_2}}$ be a preparation for a composite system of type $\mathsf{ab}$.  The postulate says that the probabilities
\begin{equation}\label{jointprobs}
\text{Prob}(\mathsf{X}_\mathsf{{a_1}}^{a_1} \mathsf{X}_\mathsf{{b_2}}^{b_2} \mathsf{A^{a_1b_2}}) := A^{a_1b_2}
\end{equation}
are sufficient to determine the state. Here $\mathsf{X}_\mathsf{{a_1}}^{a_1}$ and $\mathsf{X}_\mathsf{{b_2}}^{b_2}$ are fiducial results with $a_1=1$ to $K_\mathsf{a}$ and $b_2=1$ to $K_\mathsf{b}$ for systems of types $\mathsf a$ and $\mathsf b$ respectively.  Note
\begin{enumerate}
\item We need at least a full set of fiducial results at each end since our preparations could be the product preparations of the form $\mathsf{A^{a_1b_2}}= \mathsf{B^{a_1} C^{b_2}}$
and we could complete this preparation into a circuit with a product result, $\mathsf{D_{a_1} E_{b_2}}$.  For the resulting circuit
\begin{equation}
\text{Prob}(\mathsf{B^{a_1} C^{b_2}}\mathsf{D_{a_1} E_{b_2}}) = \text{Prob}(\mathsf{B^{a_1}}\mathsf{D_{a_1}})
\text{Prob}(\mathsf{ C^{b_2}}\mathsf{E_{b_2}}) = (B^{a_1}C^{b_2})( D_{a_1} E_{b_2})
\end{equation}
from the factorization property of Sec.\ \ref{probsfactorsection}.  There are $K_\mathsf{a}K_\mathsf{b}$ linearly independent joint states of the form $(B^{a_1}C^{b_2})$.  Therefore we need at least $K_\mathsf{a}K_\mathsf{b}$  results to determine the state. These could be the product of the fiducial results, $\mathsf{X}_\mathsf{a_1}^{a_1} \mathsf{X}_\mathsf{b_2}^{b_2}$.  Incidently, this is true independently of the assumption of local tomography. Hence, in general, we have $K_\mathsf{ab}\geq K_\mathsf{a}K_\mathsf{b}$.
\item We do not need more results at each end than the full set of fiducial results, $\mathsf{X}_\mathsf{{a_1}}^{a_1} \mathsf{X}_\mathsf{{b_2}}^{b_2}$ .  To see this assume the contrary.  Assume we also need, at least one more result, $\mathsf{Y_{a_1}} \mathsf{Y_{b_2}}$, consisting an result at each end (in accord with {\bf P3}).  Thus, we imagine we also need, at least, the probability
    \begin{equation}
    \text{Prob}(\mathsf{Y_{a_1}} \mathsf{Y_{b_2}} \mathsf{A^{a_1b_2}})
    \end{equation}
    But
    \begin{flalign}\label{bringoutreasoning}
    \text{Prob}(\mathsf{Y_{a_1}} \mathsf{Y_{b_2}} \mathsf{A^{a_1b_2}})
   &  = Y_{a_1} \text{Prob}(\mathsf{X_{a_1}}^{a_1} \mathsf{Y_{b_2}} \mathsf{A^{a_1b_2}}) \nonumber \\
    &  = Y_{a_1} Y_{b_2} \text{Prob}(\mathsf{X_{a_1}}^{a_1} \mathsf{X_{b_2}}^{b_2} \mathsf{A^{a_1b_2}}) = Y_{a_1} Y_{b_2} A^{a_1b_2}
    \end{flalign}
    where, to complete the first step we regard the circuit as the result $\mathsf{Y_{a_1}}$ acting on the preparation $\mathsf{Y_{b_2}} \mathsf{A^{a_1b_2}}$, and to complete the second step, we regard the circuit in each term as the result $\mathsf{Y_{b_2}}$ acting on the preparation
    $\mathsf{X_{a_1}}^{a_1} \mathsf{A^{a_1b_2}}$.   We see that this additional probability we conjectured needing can actually be calculated from the probabilities we already have.  Hence, we do not need any additional results at either end.
\end{enumerate}
The reasoning above leads to an alternative formulation of postulate {\bf P3}.
\begin{description}
\item[P3a] For a composite system of type $\mathsf{ab}$ composed of systems of types $\mathsf{a}$ and $\mathsf{b}$ we have $K_\mathsf{ab}=K_\mathsf{a}K_\mathsf{b}$.
\end{description}

Now, since the probabilities $A^{a_1b_2}$ determine the state, this object can be taken to represent the state.  For a result, $\mathsf{B_{a_1b_2}}$ acting on this preparation the probability is therefore given by
\begin{equation}
\text{Prob}( \mathsf{A^{a_1b_2}F_{a_1b_2}} ) = A^{a_1b_2}B_{a_1b_2}
\end{equation}
where $B_{a_1b_2}$ are the coefficients in the sum and represent the effect associated with the result $\mathsf{B_{a_1b_2}}$.  For transformations the same reasoning goes through as in Sec.\ \ref{transformationssection}.  Hence we have
\begin{equation}\label{compositewithtrans}
\text{Prob}( \mathsf{A^{a_1b_2}G_{a_1b_2}^{c_3d_4} C_{c_3 d_4}} ) = A^{a_1b_2}G_{a_1b_2}^{c_3d_4} F_{c_3 d_4}
\end{equation}
These results generalize to more than two systems in the obvious way.  We see that, in this case the probability for the circuit is given by an expression that results from changing the sans serif font (inside the argument on the LHS) to normal maths font (as seen on the RHS).  This works in general as we will now prove.  Readers who read part Part \ref{theduotensorframework} have seen this result already for the duotensor framework (where the duotensors are in standard form) using full decomposability rather than local tomography.
\begin{T}\label{sanseriftomathnormaltheorem}
The probability for a general circuit is given by changing the sans serif font in the description of the circuit to normal maths font.
\end{T}
To prove this consider a general operation $\mathsf{B_{a_1}^{c_3}}$ ($\mathsf{a}$ and $\mathsf{c}$ can be composite so this is a general operation).  The most general circuit containing this operation can be put in the form $\mathsf{A^{a_1b_2}B_{a_1}^{c_3}C_{c_3b_2}}$.   Consider performing a product result, $\mathsf{Y_{c_3}Z_{b_2}}$ on this preparation.   Using reasoning similar to that used in (\ref{bringoutreasoning}) we see that
\begin{flalign}
\text{Prob}(\mathsf{A^{a_1b_2}B_{a_1}^{c_3}} \mathsf{Y}_\mathsf{c_3} \mathsf{Z}_\mathsf{b_2})
 & =  Y_{c_3} B_{a_1}^{c_3} \text{Prob}(\mathsf{A^{a_1b_2}} \mathsf{X}_\mathsf{a_1}^{a_1} \mathsf{Z}_\mathsf{b_2}) \nonumber \\
& =  Y_{c_3} B_{a_1}^{c_3} Z_{b_2} \text{Prob}(\mathsf{A^{a_1b_2}} \mathsf{X}_\mathsf{a_1}^{a_1} \mathsf{X}_\mathsf{b_2}^{b_2})
= Y_{c_3} B_{a_1}^{c_3} Z_{b_2} A^{a_1b_2}
\end{flalign}
From tomographic locality, probabilities of this form (corresponding to product results) determine the state.  Hence, the state associated with the preparation $\mathsf{A^{a_1b_2}B_{a_1}^{c_3}}$ is $A^{a_1b_2}B_{a_1}^{c_3}$.    This means that the probability for the circuit $\mathsf{A^{a_1b_2}B_{a_1}^{c_3}C_{c_3b_2}}$ is given by $A^{a_1b_2}B_{a_1}^{c_3}C_{c_3b_2}$.  From this {\bf T\ref{sanseriftomathnormaltheorem}} follows because we can apply this iteratively to replace each operation by the corresponding transformation matrix.

There is another formulation of {\bf P3}.  This is the assumption of full decomposability discussed in Sec.\ \ref{fulldecomposabilitysubsection} (and introduced in \cite{hardy2010formalism}).  It is immediately clear that full decomposability implies tomographic locality since it could be applied to the special case where the operation is a preparation.  That  tomographic locality implies full decomposability follows from {\bf T\ref{sanseriftomathnormaltheorem}} and the fact, as seen in (\ref{compositewithtrans}) above that we can associate any operation $\mathsf{A_{a_1b_2\dots c_3}^{d_4e_5\dots f_6}}$ with a transformation matrix $A_{a_1b_2\dots c_3}^{d_4e_5\dots f_6}$.  It follows that the probability for a circuit is linear in this matrix.  Further, this matrix can be converted into
${}^{d_4e_5\dots f_6}\!A_{a_1b_2\dots c_3}$ by applying the hopping metric. This corresponds to the duotensor having all white dots. These are the coefficients for the expansion of an operation in terms of fiducials.  Hence full decomposability follows from {\bf P3}.  Yet another formulation of tomographic locality is afforded by equation (\ref{identityfidsequation}) as discussed in Sec.\ \ref{identitytranssec}.

In Sec.\ \ref{equivalenceoffragments} we defined two fragments to be equivalent ($A\equiv B$) if they gave the same probabilities when one is substituted for the other in any circuit.  We also defined a restricted notion for two fragments to be equivalent ($A\cong B$) if they given the same probabilities when one is substituted for the other in any circuit in which they are restricted to be in transformation mode.   We prove
\begin{T}\label{equivequalsrestrictedequiv}
It follows from tomographic locality that $A\equiv B$ if and only if $A\cong B$ for any two fragments $A$ and $B$.
\end{T}
This means equivalence under the restricted situation where the fragments are in transformation mode  implies equivalence in the general situation.
Consider fiducial preparations $\{ {}_{a_1}\! X^\mathsf{a_1}: a_1=1~\text{to}~K_\mathsf{a} \}$ (these are a set of preparations whose states constitute a spanning set for the given type).  Consider a general fragment $\mathsf{B_{a_1b_2\dots c_3}^{d_4e_5\dots f_6}}$.  It follows from {\bf T\ref{sanseriftomathnormaltheorem}} that there is a matrix, $B_{a_1b_2\dots c_3}^{d_4e_5\dots f_6}$ associated with this fragment given multiplying together the matrices corresponding to the operations that compose this fragment in accordance with the given wiring.  Consider the fiducial probabilities
\begin{equation}\label{fidprobsofBhere}
\text{Prob}( \mathsf{B_{a_1b_2\dots c_3}^{d_4e_5\dots f_6}} {}_{a_1}\! X^\mathsf{a_1} {}_{b_2}\! X^\mathsf{b_2} \dots {}_{c_3}\! X^\mathsf{c_3} \,\,
\mathsf{X}_\mathsf{d_4}^{d_4} \mathsf{X}_\mathsf{e_5}^{e_5} \dots \mathsf{X}_\mathsf{f_6}^{f_6})
\end{equation}
(in the duotensor framework these correspond to duotensors with all black dots).  The product preparations, ${}_{a_1}\! X^\mathsf{a_1} {}_{b_2}\! X^\mathsf{b_2} \dots {}_{c_3}\! X^\mathsf{c_3}$, corresponds to a spanning set of states for the input $\mathsf{a_1b_2\dots c_3}$. The product results,
$\mathsf{X}_\mathsf{d_4}^{d_4} \mathsf{X}_\mathsf{e_5}^{e_5} \dots \mathsf{X}_\mathsf{f_6}^{f_6}$, corresponds to a spanning set of effects for the output $\mathsf{d_4e_5\dots f_6}$.   Hence, these fiducial probabilities in (\ref{fidprobsofBhere}) determine the matrix $B_{a_1b_2\dots c_3}^{d_4e_5\dots f_6}$ associated with the fragment $\mathsf{B_{a_1b_2\dots c_3}^{d_4e_5\dots f_6}}$.  Therefore, two fragments are equivalent if and only if they have the same fiducial probabilities.   Now we note that the circuit in (\ref{fidprobsofBhere}) is in transformation mode so {\bf T\ref{equivequalsrestrictedequiv}} follows.

%!!! mention identity identity as another formulation of tomloc.

\subsection{{\bf P4}$'$: Permutability}

\index{permutability}\index{postulates!\textbf{P4}$'$ \emph{permutability}} Let $\{ \mathsf{U^{a_1}}[n]: n=1~ \text{to} ~ N_\mathsf{a} \}$ be a maximal distinguishable set of states with corresponding maximal measurement $\{\mathsf U_{a_1}[n]: n=1 ~\text{to} ~N_\mathsf{a}\}$.  Then we have
\begin{equation}
\text{Prob}(\mathsf{U}^\mathsf{a_1}[n]\mathsf U_\mathsf{a_1}[n'] )=\delta_{nn'}
\end{equation}
Postulate {\bf P4} says that, for any given permutation $\pi$ of the integers $n=1$ to $N_\mathsf{a}$, there exists a reversible transformation, $\mathsf{P_{a_1}^{a_2}}$ such that
\begin{equation}
\text{Prob}(\mathsf{U}^\mathsf{a_1}[n]\mathsf{P_{a_1}^{a_2}}\mathsf U_\mathsf{a_2}[n'] )=\delta_{\pi(n)n'}
\end{equation}
The permutation transformation can be thought of as permutating the states corresponding to the preparations.  Alternatively, it can be thought of as acting on the maximal measurement giving rise to a new maximal measurement with correspondingly permuted effects.

This is a natural requirement and, indeed, a very classical one since it applies to states in a maximal distinguishable set.  From the point of view of information theory we can think of the states in the maximally distinguishable set as letters in an alphabet. Then {\bf P4}$'$ says we can perform perform a lossless arbitrary translation of a message encoded with respect to one such alphabet to one encoded with respect to any permutation of this alphabet.

An alternative way of stating {\bf P4}$'$ is that there exists a reversible transformation permuting any pair of states in a maximal distinguishable set of states while leaving the other states in the set undisturbed.  We could then implement a general permutation by many pairwise permutations.

\index{compound permutability}\index{postulates!P6@\textbf{P4} \emph{compound permutability}}
To single out quantum theory we add the word \lq\lq compound" to {\bf P4}$'$ to get {\bf P4}.  For the case where $N_\mathsf{a}=2$ the only permutation is the one that swaps the states.  Hence a compound transformation must consist of two (reversible) transformations each of which do something other than effect the identity or swap the states.  This forces there to exist at least one more maximal set of distinguishable states so we cannot be in the classical situation.
That such transformations should be compound is well motivated. In general, reversible transformations are implemented by letting the system pass through some field or through some piece of matter (such as a piece of glass).  If, after passing through length $L$ of this field or matter, a reversible permutation is effected, then after passing through length $l<L$ some other reversible transformation must be effected.  Hence, the permutation transformation is compound. The transformation after $l$ and the remaining transformation after a further $L-l$ of the field or matter.  In fact, to a very good approximation, $l$ can be varied continuously and so we can argue that the transformation should be continuous.  However, this stronger requirement is not needed to reconstruct quantum theory within the present postulate set (though it is used in \cite{hardy2001quantum}).

\subsection{{\bf P5}: Sturdiness}

\index{sturdiness}\index{postulates!\textbf{P5} \emph{sturdiness}}
Postulate {\bf P5} states that filters are non-flattening.  A filtering transformation is a pretty dramatic transformation.  It completely kills that part of the state that is not in the support of the filter.  The name \lq\lq sturdiness" is apt since {\bf P5} asserts that the states are as sturdy as they can be in the circumstances (i.e.\ when subject to such a dramatic transformation).  An instructive metaphor is the following.  Consider a terraced row of houses numbered 1 through 5.  Imagine that we suddenly destroy houses 1, 4, and 5 with some large mechanical device that simply flattens them.   If houses 2 and 3 remained intact (even though they had been adjoined to now demolished houses) then we would rightly think that terraces of houses of this type were as sturdy as they could be in the circumstances.

Quantum theory satisfies {\bf P5} as discussed in the prelude and proven in Appendix \ref{appendixflattening}. While we may be accustomed to thinking of quantum states as being a little bit delicate, in fact they are pretty tough.

Classical probability theory also satisfies {\bf P5} (as is obvious after a little thought).

It is clearly possible for theories to be non-flattening as we have two examples.  Any theory in which filters did sometimes flatten would clearly be impoverished in a certain respect.  They would have the property that, in a certain sense, filters destroy more information than necessary.  

We conjecture in the postlude that {\bf P5}  can be replaced by the postulate that filters are non-mixing.  There is certainly a very close link between the non-mixing property and the non-flattening property.  With a little help from {\bf P1} we can prove (see {\bf T\ref{nonflatimpliesnonmix}} below) that non-flattening transformations are also non-mixing (by considering single member non-flat sets). It is proved in Appendix \ref{appendixflattening} that, in quantum theory, non-mixing transformations are also non-flattening.  Further, any transformation that is non-mixing must ensure that pure states do not end up inside the convex set of states.  Flattening transformations are likely to have the property that, up on flattening the set of states, they send some pure states to mixed states.  However, a proof of this for the case treated in this paper is missing.

\section{Filters and systems}\label{filtersandsystemssection}

In this section we extract some general properties of theories satisfying {\bf P1, P2, P3} and {\bf P4}$'$.

\subsection{Implications of sharpness}\label{uniqueimplications}

\index{sharpness!implications of}
We will say that an effect identifies a state if we get probability one for the circuit comprised of the corresponding result and preparation.
Postulate {\bf P1} says that each pure state is identified by one, and only one, maximal effect.  Further, it says that there is one, and only one, pure state that is identified by any given maximal effect.  This postulate allows us to prove a number of useful theorems.
\begin{T}\label{origP2}
Every pure state for a system is a member of at least one maximal set of distinguishable states for that system.
\end{T}
Any pure state is associated with some maximal effect (by {\bf P1}).  Every maximal effect must be associated with at least one (though possibly more than one) maximal measurement.  There must exist a maximal set of distinguishable states distinguished by any such maximal measurement.  We can substitute the particular element of the maximal set of distinguishable states by the pure state that is identified by the associated maximal effect.  This gives us (what might be) a new maximal set of distinguishable states that the given pure state is a member of.  This proves {\bf T\ref{origP2}}.

We also obtain
\begin{T}\label{distingpure}
Every state in a maximal distinguishable set is pure.
\end{T}
Assume a given state, $U^{a_1}[n]$, in a maximal distinguishable set is mixed.  Then we can write it as a mixture of distinct pure states. Each pure state in this mixture must be identified by the same maximal effect, $U^{a_1}[n]$.  But this contradicts {\bf P1}.  This proves {\bf T\ref{distingpure}}.

A simple but useful result is
\begin{T}\label{meffectidentpureonly}
The only state that is identified by a given maximal effect is the associated pure state.
\end{T}
We already know, from {\bf P1}, that this is true for pure states.  Assume that some (mixed) state other than the associated pure state is identified by the given maximal effect.  Then it would follow that each pure state in the decomposition of this state is also identified by this maximal effect.  However, this would imply that there is more than one pure state identified by the given maximal effect contradicting {\bf P1}.

Associated with any maximal measurement is a maximal distinguishable set of preparations.  In fact we can prove
\begin{T}\label{uniquemaximal}
All maximal measurements that distinguish the states in any given maximal set of distinguishable preparations are equivalent.
\end{T}
We know from {\bf T\ref{distingpure}} that every state in any given maximal set of distinguishable preparations is pure.  We know from {\bf P1} that there is a unique effect associated with any pure state.  Hence, there is a unique set of effects associated with the maximal measurement that distinguishes the given maximal set of distinguishable states.  This proves {\bf T\ref{uniquemaximal}}.

A very useful result follows from {\bf P1}.
\begin{T}\label{redundantoutcomes}
For any maximal measurement, an outcome can only fire if it can fire for some state in the associated maximal set of distinguishable states.
\end{T}
To see this consider a maximal measurement with outcome sets $\mathsf{o}[n]$ associated with the effects.  Assume that all the outcomes in $\mathsf{o}[n]$ have a nonzero probability of firing if the corresponding state from the maximal set of distinguishable states is sent in.  Let $\mathsf{o}[0]$ be all the outcomes on the measurement that are not in any of the sets $\mathsf{o}[n]$.   We can append this set to $\mathsf{o}[1]$ to get a new maximal effect associated with the first of the distinguishable states.  This state must be pure (by {\bf T\ref{distingpure}}).  Now, we know from {\bf P1} that there is one and only one maximal effect identifying this pure state.  It follows that we have the same maximal effect whether we append $\mathsf{o}[0]$ to $\mathsf{o}[1]$ or not.  The only way this can be true is if the effect associated with $\mathsf{o}[0]$ is the null effect - its outcomes never happen. This proves {\bf T\ref{redundantoutcomes}}

We can now prove the following.
\begin{T}\label{puredet}
All pure states can correspond to deterministic preparations.
\end{T}
Recall that a deterministic preparation is one for which the set of outcomes is equal to the set of all possible outcomes.   First we will show that we can construct a deterministic preparation in a maximal distinguishable set.  Let $\mathsf{U^{a_1}}[n]$ be a preparation in a maximal distinguishable set with outcome set $\mathsf{o_U}[n]$.  We have
\begin{equation}
\text{Prob}(\mathsf{U^{a_1}}[n]\mathsf{U_{a_1}}[m]) = \delta_{nm}
\end{equation}
Now let  $\mathsf{ o'_U}[n]$ be the full set of possible outcomes on the apparatus use corresponding to $\mathsf{U^{a_1}}[n]$.  Let $\mathsf{\bar U^{a_1}}[n]$ be the corresponding preparation with outcome set $\mathsf{o'_ U}[n]$ (rather than $\mathsf{o_U}[n]$).  Then we can show that we must have
\begin{equation} \label{deterministicbasis}
\text{Prob}(\mathsf{\bar U^{a_1}}[n]\mathsf{U_{a_1}}[m]) = \delta_{nm}
\end{equation}
This is true since appending the extra outcomes to our specification of the preparation cannot change the fact that we get probability 1 when we use result $\mathsf{U_{a_1}}[n]$. Hence, the other results in the maximal measurement must have probability zero (as probabilities are non-negative and add up to one over a mutually exclusive set of outcomes). Therefore (\ref{deterministicbasis}) follows.  Since every pure state belongs to some maximal set of distinguishable states (by {\bf T\ref{origP2}}) any pure state can correspond to a deterministic preparation. Hence we have proved {\bf T\ref{puredet}}.

We can now use {\bf T\ref{puredet}} to prove the following important result \index{causality}
\begin{T}\label{deteffect}
{\bf Causality.} All deterministic results on a given system type are equivalent (i.e.\ they have the same effect).
\end{T}
Recall that a deterministic result is one whose set of outcomes is equal to the full set of possible outcomes.  To prove {\bf T\ref{deteffect}} we note that any state, $A^{a_1}$, can be written as a convex sum over some set of pure states, $B^{a_1}[j]$ (with $j=1$ to $J$) which we can take to correspond to deterministic preparations by {\bf T\ref{puredet}} and the null state (corresponding to $j=0$).
\begin{equation}
A^{a_1} = \sum_{j=0}^J \lambda_j B^{a_1}[j] ~~~\text{where} \lambda_j \geq 0 ~~\text{and}~~ \sum_{j=0}^J \lambda_j = 1.
\end{equation}
Let $\mathsf{C_{a_1}}$ and $\mathsf{D_{a_2}}$ be two deterministic results.  Then
\begin{equation}
\text{Prob}(\mathsf{ A^{a_1} C_{a_1} }) = A^{a_1} C_{a_1} = \sum_{j=0}^J  \lambda_j B^{a_1}[j]C_{a_1} = \sum_{j=1}^J \lambda_j = 1-\lambda_0
\end{equation}
since $B^{a_1}[j]C_{a_1}=1$ as $\mathsf{B^{a_1}[j]C_{a_1} }$ is a deterministic circuit for $j=1$ to $J$ (this is where we are using {\bf T\ref{puredet}}).  For similar reasons,
\begin{equation}
\text{Prob}(\mathsf{ A^{a_1} D_{a_1} }) = 1-\lambda_0
\end{equation}
This is true for any preparation $\mathsf{A^{a_1}}$ and hence the two deterministic results are equivalent.  This proves {\bf T\ref{deteffect}}.  There is a connection between this proof and Lemma 6 of \cite{chiribella2010probabilistic} which states that a theory is causal if every state is proportional to a state obtained by a deterministic preparation.  In particular, this proof uses the fact that that every state is proportional to a state associated with a deterministic preparation if all pure states can be regarded as corresponding to deterministic preparations.

Consider a circuit.  We can partition it into two parts with a synchronous set of wires.  One way to implement a deterministic result after the synchronous set of wires is to ignore the outcomes on the apparatuses after these wires (this amounts to having an outcome set consisting of all outcomes for these apparatuses).  {\bf T\ref{deteffect}} says that there is a unique deterministic effect. Hence, had we a quite different set of apparatuses after the synchronous set of wires, or the same apparatuses but with different knob settings, we would have the same effect.  Hence, we get the same probability.  Thus, {\bf T\ref{deteffect}} implies that there is no backward in time influence.  The probability associated with the outcomes up to any synchronous set of wires that partitions the circuit is, according to {\bf T\ref{deteffect}}, independent of what choices we make after this set of wires.  This justifies calling such sets of wires \lq\lq synchronous".  In \cite{chiribella2010probabilistic, chiribella2010informational} (and also, effectively, in\cite{hardy2009foliable}) the equivalence of deterministic effects is taken as a basic assumption in the framework.

Another way of reading {\bf T\ref{deteffect}} is that it implies \emph{no-signalling}\index{no-signalling}.  To see this consider
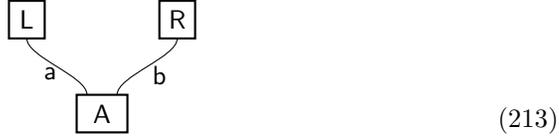
\begin{equation}
\begin{Diagram}{0}{0}
\Opbox{A}{0,0} \Opbox[2]{L}{-4,5} \Opbox[2]{R}{4, 5}
\wire{A}{L}{1}{1.5} \opsymbol{a} \wire{A}{R}{3}{1.5}\otherside\opsymbol[9,-2]{b}
\end{Diagram}
\end{equation}
The wire $\mathsf{b_2}$ on the left hand side, by itself, constitutes a synchronous set.  Hence, we can think of operation $\mathsf R$ as being after operations $\mathsf{A^{a_1b_2}L_{a_1}}$ (recall that such diagrams are interpreted graphically so there is no significance to the vertical position of the boxes on the page). If we ignore outcomes on the right hand side (this is the same thing as taking our set of outcomes to be the full set of outcomes), then $\mathsf R$ is a deterministic result.  In this case {\bf T\ref{deteffect}} implies the knob setting on $\mathsf R$ does not influence the probability of for the outcome on $\mathsf L$.   More precisely, {\bf T\ref{deteffect}} implies no-signalling in circumstances where there is no wire.  Conversely, it implies that if there is signalling from part of the setup to another there must be at least one wire going from the first part to the second part.

We now prove
\begin{T}\label{revtransdeterministic}
Any reversible transformation on a system is equivalent to some deterministic transformation.
\end{T}
What this means is that, if the outcome set associated with the reversible transformation does not include all outcomes, then the outcomes which are not included can never happen and, consequently, they can be included in the outcome set and give rise to an equivalent transformation.  Let $\mathsf{A^{a_1b_2}}$ be a deterministic preparation and let $\mathsf{T_{a_1b_2}}$ be a deterministic result.  Then $\text{Prob}(\mathsf{A^{a_1b_2}T_{a_1b_2}})=1$.  Let $\mathsf{B_{a_1}^{a_3}}$ be a reversible transformation on a system of type $\mathsf a$.  Let $\mathsf{\tilde B_{a_3}^{a_4}}$ be the inverse (so that
$\mathsf{B_{a_1}^{a_3} \tilde B_{a_3}^{a_4}}$ is the identity transformation).  Let $\mathsf{\bar{\tilde B}_{a_3}^{a_4}}$ be the transformation associated with the same setup as $\mathsf{\tilde B_{a_3}^{a_4}}$ but with an outcome set equal to the set of all outcomes associated with this setup. Finally, let $\mathsf{\bar{ B}_{a_1}^{a_3}}$ be the transformation associated with the same setup as $\mathsf{{ B}_{a_1}^{a_3}}$  but with an outcome set equal to the set of all outcomes associated with this setup.  Then we can show that
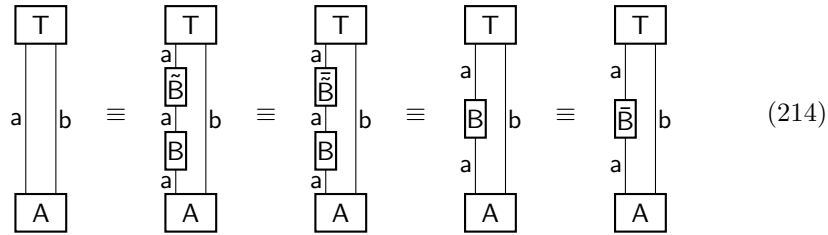
\begin{equation}
\begin{Diagram}{0}{-1.4}
\Opbox{A}{0,0} \Opbox{T}{0,10}
\wire{A}{T}{1}{1} \opsymbol{a}
\wire{A}{T}{3}{3} \otherside\opsymbol{b}
\end{Diagram}
~~ \equiv ~~
\begin{Diagram}{0}{-1.4}
\Opbox{A}{0,0} \Opbox{T}{0,10} \Opbox[1]{B}{-0.8,3.3} \opbox[1]{C}{-0.8,6.7} \opsymbol{\tilde B}
\wire{A}{B}{1}{1} \opsymbol{a} \wire{B}{C}{1}{1} \opsymbol{a} \wire{C}{T}{1}{1} \opsymbol{a}
\wire{A}{T}{3}{3} \otherside\opsymbol{b}
\end{Diagram}
~~\equiv ~~
\begin{Diagram}{0}{-1.4}
\Opbox{A}{0,0} \Opbox{T}{0,10} \Opbox[1]{B}{-0.8,3.3} \opbox[1]{C}{-0.8,6.7} \opsymbol{\bar{\tilde B}}
\wire{A}{B}{1}{1} \opsymbol{a} \wire{B}{C}{1}{1} \opsymbol{a} \wire{C}{T}{1}{1} \opsymbol{a}
\wire{A}{T}{3}{3} \otherside \opsymbol{b}
\end{Diagram}
~~\equiv~~
\begin{Diagram}{0}{-1.4}
\Opbox{A}{0,0} \Opbox{T}{0,10} \Opbox[1]{B}{-0.8,5}
\wire{A}{B}{1}{1} \opsymbol{a} \wire{B}{T}{1}{1} \opsymbol{a}
\wire{A}{T}{3}{3} \otherside \opsymbol{b}
\end{Diagram}
~~\equiv~~
\begin{Diagram}{0}{-1.4}
\Opbox{A}{0,0} \Opbox{T}{0,10} \opbox[1]{B}{-0.8,5} \opsymbol{\bar{B}}
\wire{A}{B}{1}{1} \opsymbol{a} \wire{B}{T}{1}{1} \opsymbol{a}
\wire{A}{T}{3}{3} \otherside \opsymbol{b}
\end{Diagram}
\end{equation}
The first step follows since $\mathsf{B_{a_1}^{a_3} \tilde B_{a_3}^{a_4}}$ is the identity transformation.  The second step follows since the probability for this circuit is already equal to one, so adding extra outcomes to $\mathsf{\tilde B_{a_3}^{a_4}}$ cannot change anything (these extra outcomes cannot happen).  The third step follows since $\mathsf{\tilde B_{a_3}^{a_4}}$ is deterministic and so $\mathsf{T_{a_4b_2}\tilde B_{a_3}^{a_4}}$ must be deterministic. But it follows from {\bf T\ref{deteffect}} that all deterministic results on a given type of system are equivalent.  The last step follow since the probability for the circuit is already equal to one so the extra outcomes associated with $\mathsf{\bar{ B}_{a_1}^{a_3}}$ over $\mathsf{{ B}_{a_1}^{a_3}}$ cannot happen.  Given these equivalences, it follows that the probability for the last circuit is also equal to one.  Since the afore mentioned extra outcomes on $\mathsf{\bar{B}_{a_1}^{a_3}}$ cannot happen, it follows that
\begin{equation}
\begin{Diagram}{0}{-1.4}
\Opbox{A}{0,0} \opbox{T}{0,10}  \opsymbol{C} \Opbox[1]{B}{-0.8,5}
\wire{A}{B}{1}{1} \opsymbol{a} \wire{B}{T}{1}{1} \opsymbol{a}
\wire{A}{T}{3}{3} \otherside \opsymbol{b}
\end{Diagram}
~~\equiv~~
\begin{Diagram}{0}{-1.4}
\Opbox{A}{0,0} \opbox{T}{0,10} \opsymbol{C} \opbox[1]{B}{-0.8,5} \opsymbol{\bar B}
\wire{A}{B}{1}{1} \opsymbol{a} \wire{B}{T}{1}{1} \opsymbol{a}
\wire{A}{T}{3}{3} \otherside \opsymbol{b}
\end{Diagram}
\end{equation}
for any result $\mathsf{C_{a_3b_2}}$.    If this equivalence is true for deterministic preparations, $\mathsf{A^{a_1b_2}}$, then it must be true for general preparations in place of $\mathsf{A^{a_1b_2}}$ also since the states associated with deterministic preparations span the full space of states (this follows from {\bf T\ref{puredet}}).  With a general transformation in place of $\mathsf{A^{a_1b_2}}$, it is true that any circuit containing the transformation $\mathsf{B_{a_1}^{a_3}}$ can be put into the form of the circuit on the on the left.  Hence {\bf T\ref{revtransdeterministic}} follows.

We now prove a few results concerning non-flat sets of states.   First we note that
\begin{T}\label{singlemembersets}
The state in any single member non-flat set is parallel to a pure state.
\end{T}
Consider a single member set of states comprised of a mixed state.  We can write the mixed state as a mixture of distinct pure states.   Since these pure states are distinct it follows from {\bf P1} that they cannot all be associated with the same maximal effect.  Hence, the mixed state must give rise to more than one outcome for any maximal measurement (and, in particular, they must be the outcomes associated with maximal effects since, according to {\bf T\ref{redundantoutcomes}} other outcomes cannot happen).  Hence, it can only belong to an informational subset having capacity greater than or equal to 2.  Since $K_\mathsf{a}\geq N_\mathsf{a}$, this means that this single member set of states is necessarily flat.  Hence, {\bf T\ref{singlemembersets}} follows.
From this it follows that
\begin{T}\label{nonflatimpliesnonmix}
All non-flattening transformations are non-mixing.
\end{T}
Consider a single member non-flat set of states.  The state in this must be parallel to a pure state.  If this non-flat set is sent through a non-flattening transformation then it must be non-flat afterwards also.  Hence, the state in it must continue to be parallel to a pure state.  Hence {\bf T\ref{nonflatimpliesnonmix}} follows.   This result is useful since it means that {\bf P5} implies that filters are non-mixing.

\subsection{Composite of two systems is a system}

\index{composite systems}
Systems are defined to be the thing we have after a filter.  A proto-system is the thing we have after a \lq\lq do-nothing" filter (i.e.\ the identity transformation).  If we have two proto-systems then they clearly constitute a system since the composition of two identity filters in parallel like this is a filter itself.  We will prove that this is generally true for filters in parallel.  We start with two proto-systems. We apply filters to each of them so that we now have a composite of two systems that are not necessarily proto-systems. We will show
\begin{T}\label{compfilt}
If we filter two proto-systems then we effect a filter on the composite.
\end{T}
Let $\mathsf{F_{a_1}^{a_3}}$ be a filter on a proto-system of type $\mathsf a$. Let the associated informational subset be $S_\mathsf{F}$ formed from the maximal measurement $\{ \mathsf{U_{a_1}}[m]: m=1~\text{to}~N_\mathsf{a} \}$.  Let  $\mathsf{G_{b_2}^{b_4}}$ be a filter on a proto-system of type $\mathsf{b}$.  Let the associated informational subset be $S_\mathsf{G}$ formed from the maximal measurement $\{ \mathsf{V_{b_1}}[n]: n=1~\text{to}~N_\mathsf{b} \}$.  It follows from ${\bf P2}$ that $\{ \mathsf{U_{a_1}}[m]\mathsf{V_{b_2}}[n]: mn= 11, 12, \dots, N_\mathsf{a}N_\mathsf{b} \}$ is a maximal measuement on the composite proto-system (of type $\mathsf{ab}$).   We define the informational subset $S_\mathsf{FG} $ to be associated with this maximal measurement on the composite and outcome set
\begin{equation}
O(S_\mathsf{FG}) = \{ mn: m\in O(S_\mathsf{F}), n\in O(S_\mathsf{G}) \}
\end{equation}
(this is just the cartesian product of $O(S_\mathsf{F})$ and $O(S_\mathsf{G})$).
We will now show that $\mathsf{F_{a_1}^{a_3}}\mathsf{G_{b_2}^{b_4}}$ is a filter on the composite proto-system and has $S_\mathsf{FG} $ as the associated informational subset.  We have
\begin{equation}\label{FGone}
\text{If} ~~ \mathsf{A^{a_1b_2}} \in S_\mathsf{FG} ~~ \text{then}~~
\text{Prob}\left(
\begin{Diagram}{0}{-0.6}
\Opbox{A}{0,0}
\opbox{Um}{-2,5}  \opsymbol{U\negs[\negs\mathnormal{m}\negs]}
\opbox{Vn}{2,5}   \opsymbol{V\negs[\mathnormal{n}]}
\wire{A}{Um}{1}{2}  \opsymbol{a}
\wire{A}{Vn}{3}{2}  \otherside\opsymbol{b}
\end{Diagram}
\right)>0
~~\Rightarrow ~~ mn\in O(S_\mathsf{FG})
\end{equation}
By causality ({\bf T\ref{deteffect}}), we must have $m\in S_\mathsf{F}$ (for the circuit to have non-zero probability) when we perform maximal measurement $\{ \mathsf{U_{a_1}}[m]: m=1~\text{to}~N_\mathsf{a} \}$ on the left whatever effect we have on the right.  And likewise, we must have $n\in S_\mathsf{G}$ when we perform maximal measurment $\{ \mathsf{V_{b_1}}[n]: n=1~\text{to}~N_\mathsf{b} \}$ on the right whatever effect we have on the left.
We can regard $\mathsf{A^{a_1b_2}V_{b_2}}[n]$ as a preparation of a system of type $\mathsf a$. Further, since $m\in S_\mathsf{F}$, this prepares a state in $S_\mathsf{F}$ which will pass through the filter unchanged.  Hence, for any effect $\mathsf{ C_a}$
\begin{equation}\label{FGtwo}
\text{If} ~~ \mathsf{A^{a_1b_2}} \in S_\mathsf{FG} ~~\text{then}~~~
\begin{Diagram}{0}{-1}
\Opbox{A}{0,0}
\opbox{F}{-2,5}  \opsymbol{F}
\opbox{C}{-2,10}  \opsymbol{C}
\wire{A}{F}{1}{2}  \opsymbol{a}
\wire{F}{C}{2}{2} \opsymbol{a}
\opbox{Vn}{2,5}   \opsymbol{V\negs[\mathnormal{n}]}
\wire{A}{Vn}{3}{2}  \otherside\opsymbol{b}
\end{Diagram}
\equiv
\begin{Diagram}{0}{-1}
\Opbox{A}{0,0}
\opbox{C}{-2,10}  \opsymbol{C}
\wire{A}{C}{1}{2}  \opsymbol{a}
\opbox{Vn}{2,5}   \opsymbol{V\negs[\mathnormal{n}]}
\wire{A}{Vn}{3}{2}  \otherside\opsymbol{b}
\end{Diagram}
~~ \text{and}~~n\in O(S_\mathsf{G})
\end{equation}
where the $\equiv$ sign means that the probability for the two circuits is the same.  It follows that both $\mathsf{A^{a_1b_2}F_{a_1}^{a_2}C_{a_2}}$ and $\mathsf{A^{a_1b_2}C_{a_1}}$ prepare a state in $S_\mathsf{G}$.  Therefore, for any effect $\mathsf{D_{b}}$,
\begin{equation}\label{FGthree}
\text{If} ~~ \mathsf{A^{a_1b_2}} \in S_\mathsf{FG}~~\text{then}~~
\begin{Diagram}{0}{-1}
\Opbox{A}{0,0}
\opbox{F}{-2,5}  \opsymbol{F}
\opbox{C}{-2,10}  \opsymbol{C}
\wire{A}{F}{1}{2}  \opsymbol{a}
\wire{F}{C}{2}{2} \opsymbol{a}
\opbox{G}{2,5}   \opsymbol{G}
\opbox{D}{2,10} \opsymbol{D}
\wire{A}{G}{3}{2}  \opsymbol{b}
\wire{G}{D}{2}{2} \opsymbol{b}
\end{Diagram}
\equiv
\begin{Diagram}{0}{-1}
\Opbox{A}{0,0}
\opbox{C}{-2,10}  \opsymbol{C}
\wire{A}{C}{1}{2}  \opsymbol{a}
\opbox{D}{2,10} \opsymbol{D}
\wire{A}{D}{3}{2}  \opsymbol{b}
\end{Diagram}
\end{equation}
Thus we see that for product effects $\mathsf{C_a D_b}$, the state is uneffected by the presence of the filters.  But, according to {\bf P3}, the state is fully characterized by product effects.  Hence, for any effect $\mathsf{B_{ab}}$
\begin{equation}\label{FGfour}
\text{If} ~~ \mathsf{A^{a_1b_2}} \in S_\mathsf{FG} ~~\text{then}~~
\begin{Diagram}{0}{-1}
\Opbox{A}{0,0} \Opbox{B}{0,10}
\opbox{F}{-2,5}  \opsymbol{F}
\wire{A}{F}{1}{2} \opsymbol{a}
\wire{F}{B}{2}{1} \opsymbol{a}
\opbox{G}{2,5}   \opsymbol{G}
\wire{A}{G}{3}{2}  \opsymbol{b}
\wire{G}{B}{2}{3} \opsymbol{b}
\end{Diagram}
=
\begin{Diagram}{0}{-1}
\Opbox{A}{0,0} \Opbox{B}{0,10}
\wire{A}{B}{1}{1}  \opsymbol{a}
\wire{A}{B}{3}{3}  \opsymbol{b}
\end{Diagram}
\end{equation}
This proves that $\mathsf{FG}$ passes states in $S_\mathsf{FG}$.  To complete the proof of {\bf T\ref{compfilt}} we need to show it blocks states in $\overline S_\mathsf{FG}$.
\begin{equation}\label{FGfive}
\text{If}~\mathsf{A^{a_1b_2}} \in \overline S_\mathsf{FG} ~~ \text{then}~~
\text{Prob}\left(
\begin{Diagram}{0}{-0.6}
\Opbox{A}{0,0}
\opbox{Um}{-2,5}  \opsymbol{U\negs[\negs\mathnormal{m}\negs]}
\opbox{Vn}{2,5}   \opsymbol{V\negs[\mathnormal{n}]}
\wire{A}{Um}{1}{2}  \opsymbol{a}
\wire{A}{Vn}{3}{2}  \opsymbol{b}
\end{Diagram}
\right)>0
~~\Rightarrow ~~ mn\in \overline O(S_\mathsf{FG})
\end{equation}
Hence,
\begin{equation}\label{FGsix}
\mathsf{A^{a_1b_2}} \in \overline S_\mathsf{FG}~ \text{and} ~ n\in O(S_\mathsf{G}) \Rightarrow
\begin{Diagram}{0}{-0.6}
\Opbox{A}{0,0}
\outwire[-4]{A}{1} \Opsymbol{a}
\opbox{Vn}{2,5}   \opsymbol{V\negs[\mathnormal{n}]}
\wire{A}{Vn}{3}{2}  \otherside\opsymbol{b}
\end{Diagram}
\!\!\!  \in \overline S_\mathsf{F}  ~~ \Rightarrow \text{Prob}\left(~
\begin{Diagram}{0}{-1.3}
%\boundingbox{-2,0}{2,10}
\Opbox{A}{0,0}
\opbox{F}{-2,5}  \opsymbol{F}
\opbox{C}{-2,10}  \opsymbol{C}
\wire{A}{F}{1}{2}  \opsymbol{a}
\wire{F}{C}{2}{2} \opsymbol{a}
\opbox{Vn}{2,5}   \opsymbol{V\negs[\mathnormal{n}]}
\wire{A}{Vn}{3}{2}  \opsymbol{b}
\end{Diagram}
\right) = 0
\end{equation}
Therefore,
\begin{equation}
\mathsf{A^{a_1b_2}} \in \overline S_\mathsf{FG}  ~\Rightarrow ~
\begin{Diagram}{0}{-1.3}
\Opbox{A}{0,0}
\opbox{F}{-2,5}  \opsymbol{F}
\opbox{C}{-2,10}  \opsymbol{C}
\wire{A}{F}{1}{2}  \opsymbol{a}
\wire{F}{C}{2}{2} \opsymbol{a}
\outwire[4]{A}{3} \opsymbol{b}
\end{Diagram}
\in \overline S_\mathsf{G}
\Rightarrow
\text{Prob}\left( ~
\begin{Diagram}{0}{-1.3}
\Opbox{A}{0,0}
\opbox{F}{-2,5}  \opsymbol{F}
\opbox{C}{-2,10}  \opsymbol{C}
\wire{A}{F}{1}{2}  \opsymbol{a}
\wire{F}{C}{2}{2} \opsymbol{a}
\opbox{G}{2,5}   \opsymbol{G}
\opbox{D}{2,10} \opsymbol{D}
\wire{A}{G}{3}{2}  \opsymbol{b}
\wire{G}{D}{2}{2} \opsymbol{b}
\end{Diagram}
\right)=0
\end{equation}
This is true for all effects $\mathsf{C_a}$ and $\mathsf{D_b}$.  Therefore, using {\bf P3},
\begin{equation}
\text{If} ~ \mathsf{A^{a_1b_2}} \in \overline S_\mathsf{FG} ~\text{then} ~
\text{Prob}\left(~
\begin{Diagram}{0}{-1.3}
\Opbox{A}{0,0}
\opbox{B}{0,10}  \opsymbol{B}
\opbox{F}{-2,5}  \opsymbol{F}
\wire{A}{F}{1}{2}  \opsymbol{a}
\wire{F}{B}{2}{1} \opsymbol{a}
\opbox{G}{2,5}   \opsymbol{G}
\wire{A}{G}{3}{2}  \opsymbol{b}
\wire{G}{B}{2}{3} \opsymbol{b}
\end{Diagram}
\right)
=0
\end{equation}
for all effects $\mathsf{B_{ab}}$ on the  composite.  This completes the proof that $\mathsf{FG}$ acts as a filter on the composite of the two proto-systems.

We obtain
\begin{T}\label{compsyst}
A composite of two systems is, itself, a system.
\end{T}
This follows from from {\bf T\ref{compfilt}} and the definition of a system.

This result is important since it means that, so far as composites are concerned, we can reason about systems that have been obtained by filtering in the same way as we reason about proto-systems.   In particular, we can redo the proof of {\bf T\ref{compfilt}} for filters on two general systems (not just proto-systems - see the definition of a filter on a system given at the end of Sec.\ \ref{systemsandfilterssection}).  This  gives
\begin{T}\label{compfiltgeneral}
If we filter two systems then we effect a filter on the composite.
\end{T}
This result follows immediately by redoing the proof of {\bf T\ref{compfilt}} in the light of {\bf T\ref{compsyst}}.  This would be relevant if we had a composite formed from two systems that had been obtained by filtering and then we consider further filtering on them.

\subsection{Some results for composite systems}

\index{composite systems}
Consider a composite system comprised of systems each prepared in a pure state. We will prove the following theorem.
\begin{T}\label{comppurestates}
A composite system where each component is prepared in a pure state is, itself, in a pure state
\end{T}
Consider a two component system with each component prepared in a pure state.  By {\bf T\ref{origP2}} we know that each of these pure states belong to a maximal distinguishable set of pure states for each system taken separately.  Consequently, it follows from {\bf P3} that the composite of these two pure states is a member a maximal distinguishable set of distinguishable states for the composite system. It then follows from {\bf T\ref{distingpure}} that this state is pure.  A multipartite system can be regarded successively as a set bipartite systems.  Thus, $\mathsf{abc}$ can be taken as a composite of $\mathsf{ab}$ and $\mathsf c$.  Then $\mathsf{ab}$ can be taken as a composite of $\mathsf a$ and $\mathsf b$.  If $\mathsf a$ and $\mathsf{b}$ are prepared in pure states then $\mathsf{ab}$ is in a pure state and therefore, if $\mathsf c$ is in a pure state, $\mathsf{abc}$ is in a pure state. Similar reasoning goes through for any number of systems. Hence {\bf T\ref{comppurestates}} follows.

We can use the deterministic effect to define a notion of marginals \cite{chiribella2010probabilistic}.  Thus, if we have a composite system prepared by $\mathsf{C^{a_1b_2}}$ and we wish to ignore system $\mathsf{b_2}$ so we just have a preparation for system $\mathsf{a_1}$ then we can assume we have performed the deterministic result, $\mathsf{T_{b_2}}$, on $\mathsf{b_2}$. This gives us the state $C^{a_1b_2}T_{b_2}$ for system $\mathsf{a_1}$.  We know from {\bf T\ref{deteffect}} that the deterministic effect is unique (it does not matter what measurement we have performed on system $\mathsf{b_2}$, if we ignore its outcomes, we have the same effect).  Hence, we can think of $C^{a_1b_2}T_{b_2}$ as being the state of system $\mathsf{a_1}$ as it does not depend on we do to the other particle. Using this concept of the state of a component of a composite system, we have the following theorem.
\begin{T}\label{puregivesproduct}
If one system of a bipartite system is in a pure state then the state of the bipartite system is a product state
\end{T}
It follows from tomographic locality ({\bf P3}) that we can write the bipartite state as $C^{a_1b_2}$ (as shown in Sec.\ \ref{SecTomographiclocality}).  Consider an arbitrary measurement $\{\mathsf{E_{b_2}}[l]\} $ that we can perform on system $\mathsf{b_2}$.  We have $T_{b_2}=\sum_l E_{b_2}[l]$ (since the deterministic effect corresponds an outcome set containing all outcomes). Hence,
\begin{equation}
A^{a_1}:= C^{a_1b_2} T_{b_2} = \sum_l C^{a_1b_2} E_{b_2}[l]
\end{equation}
If the state, $A^{a_1}$ of system $\mathsf a_1$ is pure then we know that each term in the sum must be proportional to $A^{a_1}$. This is because only a mixed state could be a sum of distinct states (the convex weighting can be thought of as being absorbed already into these terms).  Hence,
\begin{equation}
C^{a_1b_2} E_{b_2}[l] = \lambda_l A^{a_1}
\end{equation}
Let the state of system $\mathsf{b_2}$ be $B^{b_2}=T_{a_1}C^{a_1b_2}$.  We know that $A^{a_1}T_{a_1}=1$ because pure states are deterministic by {\bf T\ref{puredet}}. Hence $\lambda_l = B^{b_2}E_{b_2}[l]$.  Therefore, if in addition to having result $\mathsf{E_{b_2}}[l]$ on $\mathsf{b_2}$, we also have result $\mathsf{D_{a_1}}$ on system $\mathsf a_1$, then we get
\begin{equation}
\text{Prob}( \mathsf{C^{a_1b_2}}\mathsf{D_{a_1} E_{b_2}}[l] ) =  (A^{a_1} D_{a_1})(B^{b_2}E_{b_2}[l])
= \text{Prob}(\mathsf{A^{a_1} D_{a_1}}) \text{Prob}(\mathsf{B^{b_2} E_{b_2}}[l])
\end{equation}
We see that the probabilities factorize if one system is pure. This is true for any results $\mathsf{D_{a_1}}$ and $\mathsf{E_{b_2}}[l]$. In particular, it would be true for the set of fiducial results on each system.  Hence, the joint probabilities for fiducial probabilities must factorize.  As these fully characterize the state by {\bf P3} the state must factorize
\begin{equation}
C^{a_1b_2} = A^{a_1}B^{b_2}
\end{equation}
This proves {\bf T\ref{puregivesproduct}}.

We also have
\begin{T}\label{comppurecomppure}
If a bipartite system is both (i) in a pure state, and (ii) in a product state, then each of the components must be in a pure state
\end{T}
To prove this, assume the contary. Thus, let the state be $A^{a_1}B^{b_2}$.  Assume that $B^{b_2}=\lambda C^{b_2}+ (1-\lambda) D^{b_2}$ where $C^{b_2}$ and $D^{b_2}$ are distinct and $0<\lambda<1$.  Then the state of the bipartite system becomes
\begin{equation}
\lambda A^{a_1}C^{b_2}+ (1-\lambda) A^{a_1}D^{b_2}
\end{equation}
which is not pure.  This proves {\bf T\ref{comppurecomppure}}.

We can now prove the following
\begin{T}\label{RdeltamnT}
In the case where
\begin{equation}\label{Rdeltamn}
\text{Prob}\left(
\begin{Diagram}{0}{0}
\Opbox[4]{R}{0,0}
\opbox[3.4]{Um}{0,-4} \opsymbol{U[\mathnormal{m} ]}
\opbox[3.4]{V1}{-2,4} \opsymbol{V[1]}
\opbox[3.4]{Un}{2,4}  \opsymbol{W[\mathnormal{n}]}
\wire{Um}{R}{2.2}{2.5} \opsymbol{a}
\wire{R}{V1}{1.5}{2.2} \opsymbol{b}
\wire{R}{Un}{3.5}{2.2} \otherside\opsymbol{c}
\end{Diagram}
\right)
~~~= ~\delta_{mn},
\end{equation}
where $m=1$ to $N_\mathsf{a}$ and $n=1$ to $N_\mathsf{c}\geq N_\mathsf{a}$, we have
\begin{equation}\label{RVWresult}
\begin{Diagram}{0}{0}
\Opbox[4]{R}{0,0}
\opbox[3.4]{Um}{0,-4} \opsymbol{A}
\wire{Um}{R}{2.2}{2.5} \opsymbol{a}
\outwire{R}{1.5} \Opsymbol{b}
\outwire{R}{3.5} \Opsymbol{c}
\end{Diagram}
~~~  \equiv  ~~
\begin{Diagram}{0}{0}
\opbox[3.4]{V1}{0,0} \opsymbol{V[1]} \outwire{V1}{2.2} \Opsymbol{b}
\opbox[3.4]{B}{4,0} \opsymbol{B} \outwire{B}{2.2} \Opsymbol{c}
\end{Diagram}
\end{equation}
for any preparation $\mathsf A^{a_1}$.  Furthermore, if $\mathsf R$ is reversible, then:  (i) $A^{a_1}$ is pure if $B^{c_3}$ is and (ii) $B^{c_3}$ is pure if $A^{a_1}$ is.
\end{T}
First, we notice it follows from (\ref{Rdeltamn}) that $\{ \mathsf{R_{a_1}^{b_2c_3}V_{b_2}}[1]\mathsf{W_{c_3}}[n]: n=1~\text{to}~N_a \}$ is a maximal measurement since it distinguishes the states $U^a[m]$.  By {\bf T\ref{redundantoutcomes}}, this means that the results $\mathsf{R_{a_1}^{b_2c_3}V_{b_2}}[i]\mathsf{W_{c_3}}[n]$ with $i\not=1$ must be null results (as the outcomes associated with $i\not=1$ can never happen for any state coming into the $a_1$ input). We must always get $i=1$ at the $V_{b_2}[i]$ effect.  By {\bf T\ref{deteffect}}, this remains true even if we do not put  $W_{c_3}[n]$ on the $c_3$ output.   By {\bf T\ref{meffectidentpureonly}}, we know that only the state that is identified by $V_{b_2}[1]$ is $V^{b_2}[1]$.  So the $b_2$ system must be pure.  The result (\ref{RVWresult}) now follows immediately from {\bf T\ref{puregivesproduct}}. Finally, if $\mathsf R$ is reversible we note the following two points: (i) If $A^{a_1}$ is pure and, then it follows that the output state from $\mathsf R$ must be pure.  It then follows from {\bf T\ref{comppurecomppure}} that $B^{c_3}$ must be pure.  (ii) If $B^{c_3}$ is pure then, since $V^{b_2}$ is pure, it follows from {\bf T\ref{comppurestates}} that the output of $\mathsf R$ is pure.  Since $\mathsf R$ is reversible, the input must be pure (a mixed input into a reversible transformation cannot lead to a pure output).  This proves {\bf T\ref{RdeltamnT}}.

A slight elaboration on {\bf T\ref{RdeltamnT}} is of some use.
\begin{T}\label{Elaboration}
In the case where
\begin{equation}
\text{Prob}\left(
\begin{Diagram}{0}{0}
\Opbox[4]{R}{0,0}
\opbox[3.4]{Um}{-2,-4} \opsymbol{U[m]}
\opbox[3.4]{C}{2,-4} \opsymbol{C}
\opbox[3.4]{V1}{-2,4} \opsymbol{V[1]}
\opbox[3.4]{Un}{2,4}  \opsymbol{W[n]}
\wire{Um}{R}{2.2}{1.5} \opsymbol{a}
\wire{C}{R}{2.2}{3.5}  \otherside\opsymbol[8,0]{d}
\wire{R}{V1}{1.5}{2.2} \opsymbol{b}
\wire{R}{Un}{3.5}{2.2} \otherside\opsymbol{c}
\end{Diagram}
\right)
~~~= ~\delta_{mn}
\end{equation}
for some given preparation $\mathsf C^a$, we have
\begin{equation}
\begin{Diagram}{0}{0}
\Opbox[4]{R}{0,0}
\opbox[3.4]{Um}{-2,-4} \opsymbol{A}
\opbox[3.4]{C}{2,-4} \opsymbol{C}
\wire{Um}{R}{2.2}{1.5} \opsymbol{a}
\wire{C}{R}{2.2}{3.5}  \otherside\opsymbol[8,0]{d}
\outwire{R}{1.5} \Opsymbol{b}
\outwire{R}{3.5} \Opsymbol{c}
\end{Diagram}
~~~  \equiv  ~~
\begin{Diagram}{0}{0}
\opbox[3.4]{V1}{0,0} \opsymbol{V[1]} \outwire{V1}{2.2} \Opsymbol{b}
\opbox[3.4]{B}{4,0} \opsymbol{B} \outwire{B}{2.2} \Opsymbol{c}
\end{Diagram}
\end{equation}
for any $\mathsf A^a$.  Furthermore, if $\mathsf R$ is reversible, then $B^{c_3}$ is pure if $A^{a_1}$ and $C^{d_3}$ are.
\end{T}
We can regard the preparation $\mathsf{A^{a_1}C^{d_2}}$ as playing the same role as the preparation $\mathsf{A^{a_1}}$ in {\bf T\ref{RdeltamnT}}.  Hence {\bf T\ref{Elaboration}} follows immediately from {\bf T\ref{RdeltamnT}} and {\bf T\ref{comppurestates}}.

\subsection{Reversible transformations between states}

\index{reversible transformations}
Postulate {\bf P4}$'$ says that there exists a reversible transformation effecting any permutation of a given maximal set of distinguishable states.  We can use the postulates to prove a stronger result
\begin{T} \label{revperm}  There exists a reversible transformation on a system that takes the members of any given maximal set of distinguishable states and transforms them into the members of any permutation of any other given maximal set of distinguishable states for the system.
\end{T}
Let $\{\mathsf{U^{a_1}}[m]: m=1~~\text{to}~~N_\mathsf{a}\}$ and $\{\mathsf{V^{a_1}}[m]: m=1~~\text{to}~~N_\mathsf{a}\}$ be maximal sets of distinguishable preparations for a system of type $\mathsf{a}$.  Let $\{\mathsf{W^{b_2}}[m]: m=1~~\text{to}~~N_\mathsf{b}\}$ be a maximal set of preparations for a system of type $\mathsf b$. We will assume that $N_\mathsf{b}=N_\mathsf{a}$ (so $\mathsf{b}$ could be another instance of $\mathsf a$).
Let $\{\mathsf{U_{a_1}}[m]: m=1~~\text{to}~~N_\mathsf{a}\}$, $\{\mathsf{U_{a_1}}[m]: m=1~~\text{to}~~N_\mathsf{a}\}$, and
$\{\mathsf{W_{b_2}}[m]: m=1~~\text{to}~~N_\mathsf{b}\}$ be the corresponding maximal measurements.  It follows from {\bf P2} that
\begin{equation}
\{ \mathsf{U_{a_1}}[n]\mathsf{W_{b_2}}[m]: nm=11, 12, \dots, N_\mathsf{a}N_\mathsf{b} \}
\end{equation}
is a maximal measurement on the composite system of type $\mathsf{ab}$.   Let $\mathsf{P_{a_1b_2}^{a_3b_4}}$ be a permutation transformation on the composite system that effects the permutation
\begin{equation}
\pi_\mathsf{P} = (nm \leftrightarrow mn )
%\left( \begin{array}{l} n1 \leftrightarrow 1n  ~~\text{for}~~ n=1~~\text{to}~~ N_\mathsf{a} \\ \text{other}~~ nm ~~ \text{not permuted}  \end{array} \right)
\end{equation}
on the preparations $\mathsf{U^{a_1}}[n]\mathsf{W^{b_2}}[m]$. This is possible according to {\bf P4}$'$ (strictly {\bf P4}$'$ says it is the states that are permuted but this implies that the preparations are permuted to equivalent ones).   Similarly, let $\mathsf{Q_{a_1b_2}^{a_3b_4}}$ be a permutation transformation on the composite system that effects the permutation
\begin{equation}
\pi_\mathsf{Q} = ( mn \leftrightarrow nm )
%\left( \begin{array}{l} 1m \leftrightarrow m1  ~~\text{for}~~ m=1~~\text{to}~~ N_\mathsf{a} \\ \text{other}~~ mn ~~ \text{not permuted}  \end{array} \right)
\end{equation}
on the preparations $\mathsf{V^{a_1}}[m]\mathsf{W^{b_2}}[n]$.   Then the following circuit will implement the reversible transformation in {\bf T\ref{revperm}}.
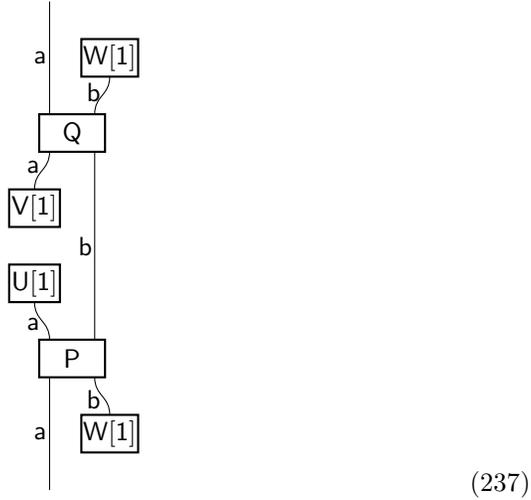
\begin{equation}\label{revtrans}
\begin{diagram}
\Opbox[4]{P}{0,0}
\opbox[4]{Q}{0,12} \opsymbol{Q}
\thispoint{IN}{-1.2, -7}   \wire{IN}{P}{1}{1} \opsymbol{a}
\opbox[3.4]{B1}{2,-4} \opsymbol{W[1]}
\wire{B1}{P}{2.2}{4} \opsymbol{b}
\opbox[3]{C1}{-2,4} \opsymbol{U[1]}
\wire{P}{C1}{1}{2} \opsymbol{a}
\wire{P}{Q}{4}{4} \opsymbol{b}
\opbox[3]{A1}{-2,8} \opsymbol{V[1]}
\wire{A1}{Q}{2}{1}  \opsymbol{a}
\opbox[3.4]{D1}{2,16} \opsymbol{W[1]}
\wire{Q}{D1}{4}{2.2} \opsymbol{b}
\thispoint{OUT}{-1.2, 19} \wire{Q}{OUT}{1}{1} \opsymbol{a}
\end{diagram}
\end{equation}
The permutation transformation, $\mathsf P$, swaps the state of the incoming system (of type $\mathsf a$) on to the intermediate system (of type $\mathsf b$) then the permutation transformation, $\mathsf Q$, swaps the state onto the outgoing system (of type $\mathsf a$) but with respect to a new set of distinguishable states.  By examining the effects of the permutation transformations, it is clear that this transformation will transform the state ${U^a}[n]$ into the state ${V^a}[n]$ (we can chose the labeling in the $\mathsf V$ set in any way so this corresponds to an arbitrary permutation).  Now we need to check that it is a reversible transformation on $\mathsf a$.  It is clearly a transformation on a system of type $\mathsf a$ since it has a system of type $\mathsf a$ going in and coming out.  To see that it reversible consider following it by the transformation
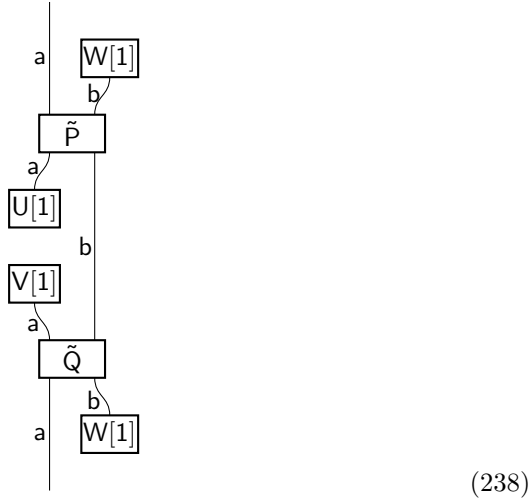
\begin{equation}\label{inverseofrevtrans}
\begin{diagram}
\opbox[4]{P}{0,0}  \opsymbol{\tilde Q}
\opbox[4]{Q}{0,12} \opsymbol{\tilde P}
\thispoint{IN}{-1.2, -7}   \wire{IN}{P}{1}{1} \opsymbol{a}
\opbox[3.4]{B1}{2,-4} \opsymbol{W[1]}
\wire{B1}{P}{2.2}{4} \opsymbol{b}
\opbox[3]{C1}{-2,4} \opsymbol{V[1]}
\wire{P}{C1}{1}{2} \opsymbol{a}
\wire{P}{Q}{4}{4} \opsymbol{b}
\opbox[3]{A1}{-2,8} \opsymbol{U[1]}
\wire{A1}{Q}{2}{1}  \opsymbol{a}
\opbox[3.4]{D1}{2,16} \opsymbol{W[1]}
\wire{Q}{D1}{4}{2.2} \opsymbol{b}
\thispoint{OUT}{-1.2, 19} \wire{Q}{OUT}{1}{1} \opsymbol{a}
\end{diagram}
\end{equation}
where $\mathsf{\tilde P}$ is the inverse transformation to $\mathsf P$ and $\mathsf{\tilde Q}$ is the inverse transformation to $\mathsf Q$.  Now consider sending in an arbitrary state, $A^{a_1}$, into the transformation in (\ref{revtrans}).   The state entering $\mathsf P$ is $A^{a_1}W^{b_2}[1]$.  The state leaving must, by {\bf T\ref{Elaboration}}, be of the form $\mathsf P$ be $U^{a_3}[1]B^{b_4}$.  The state entering $\mathsf Q$ is therefore $V^{a_5}[1]B^{b_4}$.  Using {\bf T\ref{Elaboration}} again, the state leaving $\mathsf Q$ must be of the form $D^{a_6}W^{b_7}[1]$. Therefore the state entering $\mathsf{\tilde Q}$ is $D^{a_6}W^{b_a}[1]$ (i.e.\ the same state as left $\mathsf Q$.  Since $\mathsf{\tilde Q}$ is the inverse transformation to $\mathsf Q$, the state leaving $\mathsf{\tilde Q}$ is the same as the state that entered $\mathsf Q$, i.e.\ $V^{a_8}[1]B^{b_4}$.  Hence, the state entering $\mathsf{\tilde P}$ is $U^{a_5}[1]B^{b_4}$.  This is the same as the state that left $\mathsf P$.   Since $\mathsf{\tilde P}$ is the inverse of $\mathsf P$ the state leaving $\mathsf{\tilde P}$ is the same as the state that entered $\mathsf P$, i.e.\ $A^{a_9}W^{b_{10}}[1]$. Hence the state leaving the apparatus finally is $A^{a_9}$ which is the same as the state that entered.  This works for all incoming states.  It also works if the incoming system is part of a composite system.  In this case the state for the composite is in the tensor product space (as follows from {\bf P3}) and so the identity acting on one component is equal to the identity acting on the whole.  Hence the transformation in (\ref{revtrans}) is reversible.

We immediately obtain\index{transitivity}
\begin{T}\label{revt}
{\bf Transitivity.}
There exists a reversible transformation between any pair of pure states.
\end{T}
This follows from {\bf T\ref{revperm}} and {\bf T\ref{origP2}} since any pure state is a member of some maximal distinguishable set of states.

We also obtain
\begin{T} \label{compositereversible}
If we perform reversible transformations on each component of a bipartite system then we effect a reversible transformation on the composite system.
\end{T}
Consider a composite transformaton $\mathsf{L_{a_1}^{c_3}R_{b_2}^{d_2}}$ where both $\mathsf{L_{a_1}^{c_3}}$ and $\mathsf{R_{b_2}^{d_2}}$ are reversible. If ${L_{a_1}^{c_3}}$ and ${R_{b_2}^{d_4}}$ are the transformation matrices for these transformations then there must exist a transformation with transformation matrices, ${\tilde L^{a_5}_{c_3}}$ and ${\tilde R^{b_6}_{d_4}}$ that are the inverse of ${L_{a_1}^{c_3}}$ and ${R_{b_2}^{d_4}}$.  Consider applying $\mathsf{L_{a_1}^{c_3}R_{b_2}^{d_4}}$ to a product state, $A^{a_1}B^{b_2}$.  The new state will be $A^{a_1}L_{a_1}^{c_3}B^{b_2} R_{b_2}^{d_2}$.  This evolution can be reversed by applying the inverse transformation on each component.  Hence, for product states, {\bf T\ref{compositereversible}} is true.  However, it follows from {\bf P3} that the product states form a spanning set for the full set of states (since the product states span a vector space of dimension $K_\mathsf{a}K_\mathsf{b}$).  Hence, ${L_{a_1}^{c_3}}{R_{b_2}^{d_4}}$ is non-singular having inverse equal to ${\tilde L^{a_5}_{c_3}}{\tilde R^{b_6}_{d_4}}$.  Since the latter transformation can be physically realized {\bf T\ref{compositereversible}} follows for all states.

\subsection{Constructing arbitrary filters}

We cannot assume that filters exist. Rather, we have to construct them from the postulates.

\index{filters}
We will now prove the following.
\begin{T} \label{arbfilt}
We can construct arbitrary filters.
\end{T}
This means that we can construct a filter associated with any informational subset $S$ defined with respect to any maximal measurement.   We will construct an arbitrary filter on a system of type $\mathsf a$. Consider two systems of types $\mathsf{a}$ and $\mathsf{b}$.  Together they form a composite of type $\mathsf{ab}$, which is also a system.  Let $\{\mathsf{U^{a_1}}[n]: n=1~ \text{to}~ N_\mathsf{a}\}$  be the maximal set of distinguishable preparations for $\mathsf{a}$ with respect to which define $S$ for the filter.   Let $\{\mathsf{U_{a_1}}[n]: n=1~ \text{to}~ N_\mathsf{a}\}$ be some maximal measurement that distinguishes this set.  Let $\{\mathsf{V^{b_1}}[m]: m=1~ \text{to}~ N_\mathsf{b}\}$ be any maximal set of distinguishable preparations for $\mathsf b$. Let $\{\mathsf{V_{b_1}}[m]: m=1~ \text{to}~ N_\mathsf{b}\}$ be the corresponding maximal measurement.  It follows from {\bf P2} that $\{\mathsf{U^{a_1}}[n]\mathsf{V^{b_1}}[m]: nm=11,12, \dots ,N_\mathsf{a}N_\mathsf{b}\}$ is a maximal set of distinguishable states for the composite system.  By {\bf P4} there exists a reversible permutation transformation, $\mathsf{P_{a_1b_2}^{a_3b_4}}$, on the composite system which effects the following permutation:
\begin{equation}
\pi = \left( \begin{array}{l} nm \leftrightarrow mn  ~~\text{if} ~~ n~\text{and}~m\in O(S) \\
                              nm \leftrightarrow nm ~~ \text{otherwise}                  \end{array} \right)
\end{equation}
We choose $\mathsf b$ such that $N_\mathsf{b}= N_\mathsf{a}$  allow such a permutation.
Since $\mathsf{P_{a_1b_2}^{a_3b_4}}$ is reversible, there exists another transformation, $\mathsf{\tilde P_{a_3b_4}^{a_5b_6}}$, such that
$\mathsf{P_{a_1b_2}^{a_3b_4}}\mathsf{\tilde P_{a_3b_4}^{a_5b_6}}$ is the identity transformation.  Note that although $\pi=\pi^{-1}$ for the above choice of $\pi$, it does not follow that $\mathsf{P_{a_1b_2}^{a_3b_4}}$ is equal to $\mathsf{\tilde P_{a_1b_2}^{a_3b_4}}$ because the transformation acts on $K_\mathsf{ab}$ (rather than just $N_\mathsf{ab}$) real parameters.   We choose some particular $n_1\in O(S)$. We will show that the set of results
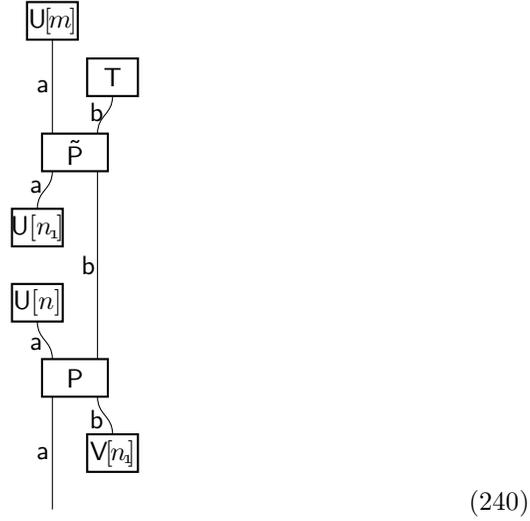
\begin{equation}\label{maximalmeas}
\begin{diagram}
\Opbox[4]{P}{0,0}
\opbox[4]{Q}{0,12} \opsymbol{\tilde P}
\thispoint{IN}{-1.2, -7}   \wire{IN}{P}{1}{1} \opsymbol{a}
\opbox[3]{B1}{2,-4} \opsymbol{V\negs[\negs \mathnormal{n_{\!1}\!}\negs]}
\wire{B1}{P}{2}{4} \opsymbol{b}
\opbox[3]{C1}{-2,4} \opsymbol{U[\mathnormal{n}]}
\wire{P}{C1}{1}{2} \opsymbol{a}
\wire{P}{Q}{4}{4} \opsymbol{b}
\opbox[3]{A1}{-2,8} \opsymbol{U[\mathnormal{n_{\!1}\!}]}
\wire{A1}{Q}{2}{1}  \opsymbol{a}
\opbox[3]{D1}{2,16} \opsymbol{T}
\wire{Q}{D1}{4}{2} \opsymbol{b}
\opbox{Um}{-1.2, 19} \opsymbol{U\negs[\negs \mathnormal{m}\negs]}
\wire{Q}{Um}{1}{2} \opsymbol{a}
\end{diagram}
\end{equation}
constitute a maximal measurement that distinguishes the maximal distinguishable set of preparations, $\{\mathsf{U^{a_1}}[n]: n=1~ \text{to}~ N_\mathsf{a}\}$, we started with.  Note that $\mathsf T$ is the deterministic result.  Here we have $N_a^2$ results since we have one for each value of $n$ and $m$.  Consider the input $U^{a_1}[n']$.  If $n'\in \overline O(S)$, then the output from $\mathsf P$ will be $U^{a_3}[n']V^{b_4}[n_1]$.  Hence only results having $n=n'$ will fire.  In fact, following the state $V[n']$ through $\mathsf{ \tilde P}$ we see that we must also have $m=n_1$ for the result to fire.  If $n'\in O(S)$ then, examining the effects of the permutation transformations, we see that only the effect that has $n=n_1$ and $m=n'$ will fire.  Hence we have a set of results that distinguishes the maximal set of distinguishable states.   Theoretically, we still have outcomes with $n=n_1$ and $m\in \overline O(S)$ which do not happen for any of the inputs $U^{a_1}[n]$.   We can bundle all such outcomes together and add them to the outcomes associated with, say, $U^{a_1}[n_1]$. We now have a maximal measurement that covers all outcomes which distinguishes the maximal set of preparations we started with.  By {\bf T\ref{uniquemaximal}}, this maximal measurement is equivalent to $\{\mathsf{U_{a_1}}[n]: n=1~ \text{to}~ N_\mathsf{a}\}$ with respect to which we have defined $S$.  We will now show that the transformation
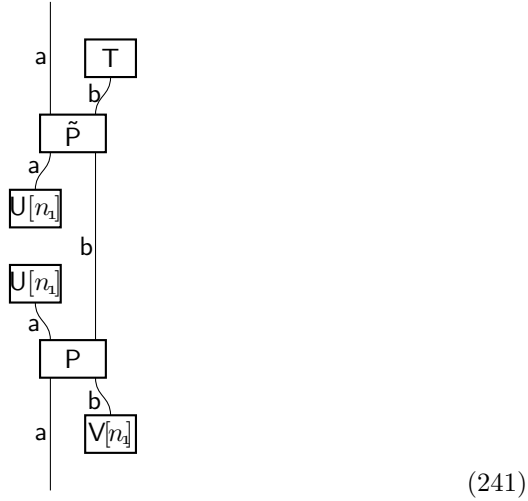
\begin{equation}\label{filter}
\begin{diagram}
\Opbox[4]{P}{0,0}
\opbox[4]{Q}{0,12} \opsymbol{\tilde P}
\thispoint{IN}{-1.2, -7}   \wire{IN}{P}{1}{1} \opsymbol{a}
\opbox[3]{B1}{2,-4} \opsymbol{V\negs[\negs \mathnormal{n_{\!1}\!}\negs]}
\wire{B1}{P}{2}{4} \opsymbol{b}
\opbox[3]{C1}{-2,4} \opsymbol{U[\mathnormal{n_{\!1}\!}]}
\wire{P}{C1}{1}{2} \opsymbol{a}
\wire{P}{Q}{4}{4} \opsymbol{b}
\opbox[3]{A1}{-2,8} \opsymbol{U[\mathnormal{n_{\!1}\!}]}
\wire{A1}{Q}{2}{1}  \opsymbol{a}
\opbox[3]{D1}{2,16} \opsymbol{T}
\wire{Q}{D1}{4}{2} \opsymbol{b}
\thispoint{OUT}{-1.2, 19} \wire{Q}{OUT}{1}{1} \opsymbol{a}
\end{diagram}
\end{equation}
is a filter with respect to  $S$. Assume we send in a system with preparation $\mathsf{ A^{a_1}}$.  If we do not see $n=n_1$ at the $\mathsf{U_{a_3}}[n_1]$ result after $\mathsf P$ then the transformation will have failed to happen.  We see that the transformation is certain to happen if $\mathsf{ A^{a_1}\in }S$  (so we will certainly get $n=n_1$ at the afore mentioned effect).  By {\bf T\ref{meffectidentpureonly}} we know that the only state which can be identified by the maximal result $\mathsf{U_{a_3}}[n_1]$ is the pure state $U^{a_3}[n_1]$. It follows from {\bf T\ref{puregivesproduct}} that the state after $\mathsf P$ is $U^{a_3}[n_1] B^{b_4}$ where $B^{b_4}$ is some state for system $\mathsf b$.  This is the same as the state going into $\mathsf{\tilde P}$.  Hence, since $\mathsf{\tilde P}$ is the inverse of $\mathsf P$, the state coming out of $\mathsf{\tilde P}$  must be the same as the state going into $\mathsf P$.  Hence, the the state that finally emerges is $A^{a_7}$.  This proves the first property required of a filter (that is passes states in $S$).  Since the set of effects in (\ref{maximalmeas}) constitute a maximal measurement it follows that $\mathsf{ A^{a_1}}\in \overline S$ then we will definitely not see $n=n_1$ at the effect after $\mathsf P$. By causality ({\bf T\ref{deteffect}}) this is true if if we send the system into the transformation in (\ref{filter}).  Hence, the incoming system will be blocked by this transformation.  This proves that this transformation is a filter with respect to $S$.  Since $S$ can be any informational subset (defined with respect to any maximal measurement) we have constructed an arbitrary filter.

It is worth noting that we have not proven that all filters for some given $S$ will process states that are not in either $S$ or $\overline S$ in the same way.  In fact, in both classical probability theory and quantum theory they do, and so it will follow from the postulates that all filters for a given $S$ are equivalent. This will only be apparent once the reconstruction is complete.  For the time being, however, we will see that a certain class of filters (special filters) have the property that they correspond to projective maps onto the subspace spanned by $S$ and so do process states not in $S$ or $\overline S$ in the same way.

\subsection{Special filters}\label{specialfiltersec}

We define\index{special filters}\index{filters!special}
\begin{quote}
{\bf A special filter}, $\mathsf F_{a_1}^{a_2}$, is a filter having the property that it belongs to a set of transformations, $\mathsf{F_{a_1}^{a_2}}[n]$, corresponding to the same setup having disjoint outcome sets labeled by $n$ such that
\begin{eqnarray}
\mathsf{F_{a_1}^{a_2}}[n_1]                       &\equiv& \mathsf{F_{a_1}^{a_2}}  ~~\text{for some particular} ~~n_1\in O(S) \\
\mathsf{F_{a_1}^{a_2}}[n] \mathsf{T_{a_2}}        &\equiv&  \mathsf{U_{a_1}}[n]  ~~\text{for} ~~ n\in\overline O(S)  \\
\mathsf{F_{a_1}^{a_2}}[n_1] \mathsf{U_{a_2}}[n]   &\equiv&  \mathsf{U_{a_1}}[n]  ~~\text{for} ~~ n\in O(S)
\end{eqnarray}
where $\{ \mathsf{U_{a_1}}[n]: n=1 ~~\text{to}~~ N_\mathsf{a} \}$ is the maximal measurement with respect to which $S$ for the filter is defined and $\mathsf{T_{a_2}}$ is the deterministic result.
\end{quote}
It follows from {\bf T\ref{deteffect}} and {\bf T\ref{redundantoutcomes}} that the deterministic result, $\mathsf{T_{a_2}}$, can be a maximal measurement where we course grain over outcomes to have only one outcome set.  This means that a special filter followed by the maximal measurement with respect to which the filter is defined is equivalent to the given maximal measurement (by {\bf T\ref{uniquemaximal}}).

We have the following theorem
\begin{T}\label{specialfiltersexist}
Arbitrary special filters exist.
\end{T}
This means that we can construct such a special filter for any informational subset defined with respect to any maximal measurement.
In fact we have already proven this. It is clear by inspection that the filter in (\ref{filter}) (see also (\ref{maximalmeas})) is an arbitrary special filter.

We now prove
\begin{T}\label{specialfilter}
Any preparation followed by a special filter, $\mathsf{ F_{a_1}^{a_2}}$, effects a preparation in the informational subset, $S$, associated with the filter. That is
\begin{equation}
\mathsf{A^{a_1} F_{a_1}^{a_2} }\in S
\end{equation}
for special filters.
\end{T}
Consider the set of results $\{\mathsf{F_{a_1}^{a_2}}[n] \mathsf{U_{a_2}}[m]\}$.  These results (with appropriate coursegraining as explained above) constitute a maximal measurement for the preparations $\{\mathsf{U^a}[n]: n=1 ~\text{to}~ N_\mathsf{a} \}$. By {\bf T\ref{uniquemaximal}} this maximal measurement is equivalent to any other for this set of preparations. The outcomes in $\{n_1m: m\in \overline O(S)\}$ on the results $\{\mathsf{F_{a_1}^{a_2}}[n_1] \mathsf{U_{a_2}}[m]\}$ do not happen for any of the preparations in $\{\mathsf{U^a}[n]: n=1 ~\text{to}~ N_\mathsf{a} \}$.  Hence, by {\bf T\ref{redundantoutcomes}}, they cannot happen for any incoming state.  This proves {\bf T\ref{specialfilter}}.

We can use {\bf T\ref{specialfilter}} to prove the following
\begin{T}\label{fidafterfilt}
Fiducial results for a system defined with respect to a special filter $\mathsf{ F_{a_1}^{a_2}}$ can all be of the form
\begin{equation}
\mathsf{ F_{a_2}^{a_3} B_{a_3}}
\end{equation}
\end{T}
This follows since the system, having emerged out of a special filter $\mathsf{ F_{a_1}^{a_2}}$ will, by {\bf T\ref{specialfilter}} and the definition of a filter, pass through $\mathsf{ F_{a_2}^{a_3}}$ unchanged. Hence, all fiducial results can be of the given form.

We will prove the following
\begin{T}\label{filtersproject}
Any special filter, $\mathsf{F}$, is represented by a projective map into the subspace spanned by states in the associated informational subset $S$.
\end{T}
Note that this theorem plays no role in the reconstruction though is mentioned in the postlude.
By {\bf T\ref{specialfilter}}, we have
\begin{equation}\label{fidfilters}
\begin{Diagram}{0}{-1.4}
\Opbox{A}{0,0}
\opbox{F1}{0,5} \opsymbol{F}
\opbox{X}{0,10} \opsymbol{X^\mathnormal{a_2}}
\wire{A}{F1}{2}{2} \opsymbol{a}
\wire{F1}{X}{2}{2} \opsymbol{a}
\end{Diagram}
~~~~\equiv~~~~
\begin{Diagram}{0}{-1.6}
\Opbox{A}{0,0}
\opbox{F1}{0,4} \opsymbol{F}
\opbox{F2}{0,8}  \opsymbol{F}
\opbox{X}{0,12} \opsymbol{X^\mathnormal{a_3}}
\wire{A}{F1}{2}{2} \opsymbol{a}
\wire{F1}{F2}{2}{2} \opsymbol{a}
\wire{F2}{X}{2}{2} \opsymbol{a}
\end{Diagram}
\end{equation}
Here $\{\mathsf{X}^{a_1}_\mathsf{a_1}: a_1=1 ~\text{to}~K_\mathsf{a}\}$ is a set of fiducial results for type $\mathsf{a}$.  Since this fiducial set is complete for the unfiltered set, it must also be complete (in fact over complete) for the filtered system (take this to be of type $\mathsf c$).  It follows from (\ref{fidfilters}) that $\{\mathsf{F_{a_2}^{a_3}}\mathsf{X}^{a_3}_\mathsf{a_3}: a_2=1 ~\text{to}~K_\mathsf{a}\}$ is an over complete set of fiducial results for $\mathsf c$. By choosing a subset of $K_\mathsf{c}$ of these that correspond to linearly independent effects, we can form a complete set of fiducial results, $\{\mathsf{X}_\mathsf{c_1}^{c_1}: c_1=1 ~\text{to}~ K_\mathsf{c}\}$, for $\mathsf c$.  The filter on the LHS of (\ref{fidfilters}) can be regarded as part of the fiducial effect so the LHS constitutes measuring the fiducial results before filtering.  The second filter on the RHS of (\ref{fidfilters}) can be regarded as part of the fiducial results so the RHS constitutes measuring the fiducial result after filtering.  Hence, it follows from (\ref{fidfilters}) that we get the same probabilities for these results whether we measure them before filtering or after filtering.  We also know by {\bf T\ref{specialfilter}} that, after filtering, the state is in the subspace associated with $\mathsf c$.   We can think of the initial state as having a component that is in the space spanned by the states of system $\mathsf c$ and an orthogonal component. It follows from the facts just established that the state after the filter is given by just the first component.  This proves {\bf T\ref{filtersproject}}.

We can regard the system after a filter, $\mathsf{F_{a_1}^{a_2}}$, as a new type of system, $\mathsf c$, say.  We can write the filter as $\mathsf{F_{a_1}^{c_2}}$, if we want to emphasize this fact.  Next we will prove
\begin{T}\label{NisO}
A system type, $\mathsf c$, created by applying a special filter has $N_\mathsf{c}$ equal to the capacity of the filter.
\end{T}
Recall that the capacity of a filter is $|O(S)|$ where $S$ is the informational subset associated with the filter.
Let $\{\mathsf{W_{c_2}[j]}: j=1 ~\to~ N_\mathsf{c}\}$ be a maximal measurement for $\mathsf c$.  Consider the set of effects
\begin{equation}
\mathsf{F_{a_1}^{c_2}}[n] \mathsf{W_{c_2}}[j]
\end{equation}
This set of results constitute a measurement which can distinguish $|\overline O(S)|+N_\mathsf{c}$ distinguishable states in the following way.  The state $U^{a_1}[n]$ for $n\in \overline O(S)$ gives outcome $n$ (which is not equal to $n_1$) at $\mathsf{F}[n]$.  Hence, by course graining over $\mathsf W$'s outcomes, we can distinguish these states. The states $W^{a_1}[j]$ are in $S$ and give rise to outcome $n_1$ at $\mathsf{F_{a_1}^{c_2}}[n]$ and outcome $j$ at $\mathsf{W_{c_2}}[j]$.  If $N_\mathsf{c}>|O(S)|$ then we have constructed a measurement which can distinguish more than $N_a$ states on systems of type $\mathsf a$ (which is impossible since this is the maximum number that can be distinguished).  We clearly can distinguish $|O(S)|$ states on $\mathsf c$ simply by using the measurement $\{\mathsf{U_{a_1}}[n]: n=1~ \text{to}~ N_\mathsf{a}\}$.  Hence, $N_\mathsf{c}=|O(S)|$.

We can use this to show the following important result\index{systems!with arbitrary $N_\mathsf{a}$}
\begin{T}\label{anyN}
We can create systems having arbitrary $N_\mathsf{a}$
\end{T}
Assume we want to create a system having a particular value of $N_\mathsf{a}$.  By {\bf Assump 2}, there exists at least one type of system, $\mathsf z$, having $1<N_\mathsf{z}<\infty$ (actually {\bf Assump 2} says that $K_\mathsf{z}<\infty$ but we must have $N_\mathsf{z}\leq K_\mathsf{z}$).  Hence we can create a composite system $\mathsf{zz}\dots\mathsf{z}$ having $N_\mathsf{zz\dots z}\geq N_\mathsf{a}$ (this uses {\bf P2} which implies $N_\mathsf{bc}=N_\mathsf{b}N_\mathsf{c}$).
By {\bf T\ref{arbfilt}} and {\bf T\ref{NisO}} we can filter down to a system having the required value of $N_\mathsf{a}$.

\subsection{Systems with same $N_\mathsf{a}$ are equivalent}

We can replace one system with another in a circuit by the following move
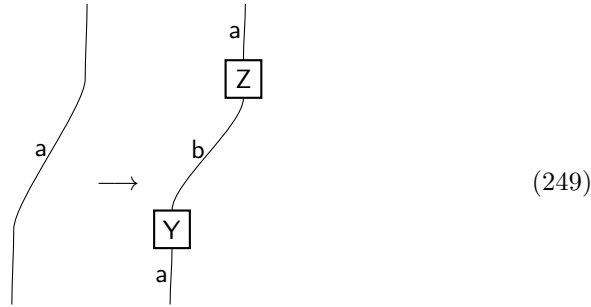
\begin{equation}
\begin{Diagram}{0}{-1.7}
\thispoint{IN}{0,0}
\thispoint{OUT}{4,16}
\thispoint{Y}{0.1,4}
\thispoint{Z}{3.9,12}
\wire[1.5]{Y}{Z}{1}{1}\opsymbol{a}
\wire{IN}{Y}{1}{1}
\wire{Z}{OUT}{1}{1}
\end{Diagram}
\longrightarrow
\begin{Diagram}{0}{-1.7}
\thispoint{IN}{0,0}
\thispoint{OUT}{4,16}
\Opbox[2]{Y}{0.1,4}
\Opbox[2]{Z}{3.9,12}
\wire[1.5]{Y}{Z}{1.5}{1.5}\opsymbol[-3,0]{b}
\wire{IN}{Y}{1}{1.5} \opsymbol{a}
\wire{Z}{OUT}{1.5}{1} \opsymbol{a}
\end{Diagram}
\end{equation}
The transformations $\mathsf Y$ and $\mathsf Z$ can be absorbed into the definitions of the operations from which the system of type $\mathsf a$ is outputted from and inputted into.  We will call this move a system substitution. We use this for the following definition.
\begin{quote}\index{equivalence!of system types}
{\bf Equivalence of system types}.  We say that two system types, $\mathsf a$ and $\mathsf b$, are equivalent if there exists a fixed system substitution by which any occurrence of a system of type $\mathsf a$ in any circuit can be replaced by a system of type $\mathsf b$ and another fixed system substitution by which any occurrence of a system of type $\mathsf b$ in a circuit can be replaced by a system of type $\mathsf a$ such the probability for the circuit is unchanged by these substitutions.
\end{quote}
We will now prove
\begin{T}\label{sameNequiv}
Systems $\mathsf a$ and $\mathsf b$ are equivalent if and only if $N_\mathsf{a}=N_\mathsf{b}$.
\end{T}
In other words, system types having the same information carrying capacity are equivalent (this was used as an axiom in \cite{hardy2001quantum}). Note this holds true whether the systems in question are proto-systems having the given information carrying capacity or have been obtained by filtering.  It is clear that $\mathsf a$ and $\mathsf b$ are not equivalent if $N_\mathsf{a}\not= N_\mathsf{b}$ since the system having smaller $N$ will not be able to support as many distinguishable states. Let $\{ \mathsf{U^{a_1}}[n]: n=1 ~\text{to}~N_\mathsf{a} \}$ be a maximal set of distinguishable states for a system of type $\mathsf a$ with associated maximal measurement Let $\{ \mathsf{U_{a_1}}[n]: n=1 ~\text{to}~N_\mathsf{a} \}$.  Let $\{ \mathsf{V^{b_2}}[m]: m=1 ~\text{to}~N_\mathsf{b} \}$ be a maximal set of distinguishable states for $\mathsf b$ with associated maximal measurement $\{ \mathsf{V_{b_2}}[m]: m=1 ~\text{to}~N_\mathsf{b} \}$.  We will now show that the following are system substitutions which prove equivalence when $N_\mathsf{a}=N_\mathsf{b}$.
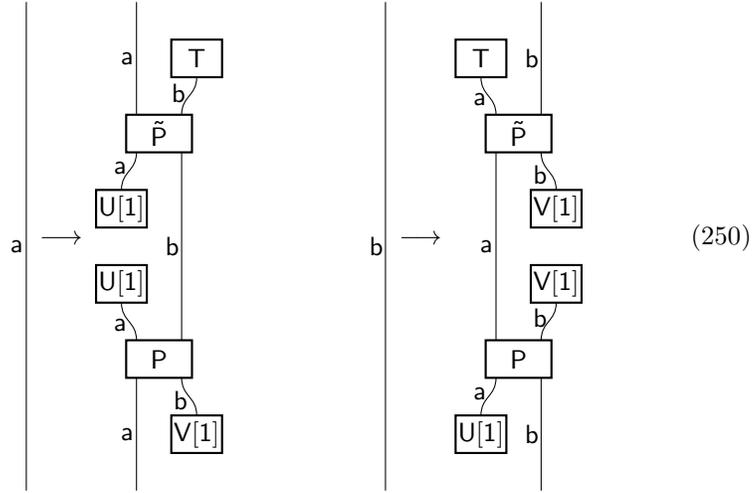
\begin{equation}\label{absubs}
\begin{Diagram}{0}{-1.7}
\thispoint{IN}{0,-7}
\thispoint{OUT}{0,19}
\wire{IN}{OUT}{1}{1} \opsymbol{a}
\end{Diagram}
\longrightarrow
\begin{Diagram}{0}{-1.7}
\Opbox[4]{P}{0,0}
\opbox[4]{Q}{0,12} \opsymbol{\tilde P}
\thispoint{IN}{-1.2, -7}   \wire{IN}{P}{1}{1} \opsymbol{a}
\opbox[3]{B1}{2,-4} \opsymbol{V[1]}
\wire{B1}{P}{2}{4} \opsymbol{b}
\opbox[3]{C1}{-2,4} \opsymbol{U[1]}
\wire{P}{C1}{1}{2} \opsymbol{a}
\wire{P}{Q}{4}{4} \opsymbol{b}
\opbox[3]{A1}{-2,8} \opsymbol{U[1]}
\wire{A1}{Q}{2}{1}  \opsymbol{a}
\opbox[3]{D1}{2,16} \opsymbol{T}
\wire{Q}{D1}{4}{2} \opsymbol[-3,0]{b}
\thispoint{OUT}{-1.2, 19} \wire{Q}{OUT}{1}{1} \opsymbol{a}
\end{Diagram}
~~~~~~~~~~~~~~~
\begin{Diagram}{0}{-1.7}
\thispoint{IN}{0,-7}
\thispoint{OUT}{0,19}
\wire{IN}{OUT}{1}{1} \opsymbol{b}
\end{Diagram}
\longrightarrow
\begin{Diagram}{0}{-1.7}
\opbox[4]{P}{0,0}  \opsymbol{P}
\opbox[4]{Q}{0,12} \opsymbol{\tilde P}
\thispoint{IN}{1.2, -7}
\wire{IN}{P}{1}{4} \opsymbol{b}
\opbox[3]{B1}{-2,-4} \opsymbol{U[1]}
\wire{B1}{P}{2}{1} \opsymbol{a}
\opbox[3]{C1}{2,4} \opsymbol{V[1]}
\wire{P}{C1}{4}{2} \opsymbol{b}
\wire{P}{Q}{1}{1} \opsymbol{a}
\opbox[3]{A1}{2,8} \opsymbol{V[1]}
\wire{A1}{Q}{2}{4}  \opsymbol{b}
\opbox[3]{D1}{-2,16} \opsymbol{T}
\wire{Q}{D1}{1}{2} \opsymbol{a}
\thispoint{OUT}{1.2, 19} \wire{Q}{OUT}{4}{1} \opsymbol{b}
\end{Diagram}
\end{equation}
where $\mathsf T$ is the deterministic result and where $\mathsf P$ is a reversible transformation effecting the permutation
\begin{equation}
\pi = ( nm \leftrightarrow mn  )
\end{equation}
of the preparations $\{ \mathsf{U_{a_1}}[n]\mathsf{V_{b_2}}[m]: nm=11, 12, \dots \}$ (by {\bf P2} such a set constitutes a maximal set of preparations for the composite and by {\bf P4} the reversible transformation, $\mathsf P$ exists). The inverse transformation to $\mathsf P$ is $\mathsf{\tilde P}$.
If the state going into the $\mathsf{a\rightarrow b}$ substitution circuit is $A^{a_1}$ then the state going into $\mathsf P$ is $A^{a_1}V^{b_2}[1]$.  Then it follows from {\bf T\ref{Elaboration}} that the state coming out of $\mathsf P$ is of the form $U^{a_1}[1]B^{b_2}$.  The state going into $\mathsf{\tilde P}$ will be of the same form.   Hence, the state coming out of $\mathsf{ \tilde P}$ will be of the form $A^{a_1}V^{b_2}[1]$ since $\mathsf{\tilde P}$ is the inverse of $\mathsf P$.  Hence, the state emerging from the substitution circuit is the same as the state that was sent in. This proves we can replace $\mathsf a$ with $\mathsf b$ for circuits that are partioned into two pieces by the system $\mathsf a$. However, in general, it could be the case that the circuit is of the form
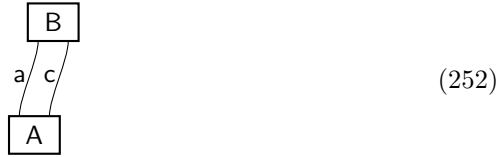
\begin{equation}\label{joinedform}
\begin{Diagram}{0}{-0.8}
\Opbox{A}{0,0}
\Opbox{B}{1,6}
\wire{A}{B}{1}{1} \opsymbol{a}
\wire{A}{B}{3}{3} \opsymbol{c}
\end{Diagram}
\end{equation}
(where $\mathsf c$ may denote a composite system and the preparation $\mathsf A$ and the result $\mathsf B$ may be comprised of many operations).
However, the statistics of such a circuit can, by {\bf P3}, be determined by the statistics of circuits of the form
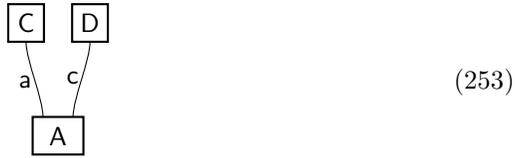
\begin{equation}\label{splitform}
\begin{Diagram}{0}{-0.8}
\Opbox{A}{0,0}
\Opbox[2]{C}{-1.7,6}
\Opbox[2]{D}{+1.7,6}
\wire{A}{C}{1}{1.5} \opsymbol{a}
\wire{A}{D}{3}{1.5} \opsymbol{c}
\end{Diagram}
\end{equation}
If circuits of the form (\ref{splitform}) have the same probabilities under the substitution, then it follows from {\bf P3} that circuits of the form
(\ref{joinedform}) will also.
Hence, it is sufficient to consider only circuits of this form.  In this circuit, we can regard $\mathsf{A^{a_1c_2} D_{c_2}}$ as a preparation of a system of type $\mathsf a$ up on which we perform effect $\mathsf{C_{a_1}}$.  For such circuits we have shown that we can substitute $\mathsf a$ by $\mathsf b$.  It is clear, by a similar argument, that we can substitute $\mathsf b$ by $\mathsf a$ using the substitution shown on the RHS in (\ref{absubs}).  This proves {\bf T\ref{sameNequiv}}.

From the above argumentation we also see that there exists a linear and invertible transformation between states of the $\mathsf a$ system and the corresponding states for the $\mathsf b$ system.  Hence, equivalence implies the following in general
\begin{T}\label{fidequiv}
We can find fiducial sets of results for equivalent systems with respect to which the set of allowed states, transformations, and effects on a system are the same.
\end{T}

\subsection{Two filters}

\index{filters!two filters}
We have already noted that if we have one filter, $\mathsf{F_{a_1}^{a_2}}$, followed by another filter, $\mathsf{F_{a_2}^{a_3}}$,  of the same type then the compound operation, $\mathsf{F_{a_1}^{a_2}F_{a_2}^{a_3}}$, is also a filter and filters with respect to the same informational subset.  What happens when we have two filters that are not of the same type?  We will treat a special case where both are special filters and filter with respect to the same maximal measurement.  We will prove
\begin{T}\label{overlappingfilters}
If $\mathsf{ F_{a_1}^{a_2}}$ is a special filter for informational subset $S_\mathsf{F}$ and $\mathsf{ G_{a_2}^{a_3}}$ is a special filter with informational subset $S_\mathsf{G}$ where both informational subsets are defined with respect to the same maximal measurement, then
\begin{equation}
\mathsf{ F_{a_1}^{a_2} G_{a_2}^{a_3}}
\end{equation}
is a filter with respect to the informational subset $S_\mathsf{F\cap G}$ defined with respect to the same maximal measurement where
\begin{equation}
O(S_\mathsf{F\cap G}) = O(S_\mathsf{F}) \cap O(S_\mathsf{G})
\end{equation}
\end{T}
Since $S_\mathsf{F\cap G}\subseteq S_\mathsf{F}, S_\mathsf{G}$, it is clear that if the input state is in $S_\mathsf{F\cap G}$ it will pass through both filters unchanged.  Now we need to prove that if the input state is in $\overline S_\mathsf{F\cap G}$ it will be blocked.  Consider the set of effects
\begin{equation}
\begin{Diagram}{0}{0}
\opbox{F}{0,0} \opsymbol{F[\mathnormal{p}]}
\opbox{G}{0,5} \opsymbol{G[\mathnormal{q}]}
\opbox{U}{0,10} \opsymbol{U[\mathnormal{r}]}
\inwire{F}{2}   \Opsymbol{a}
\wire{F}{G}{2}{2} \opsymbol{a}
\wire{G}{U}{2}{2}  \opsymbol{a}
\end{Diagram}
\end{equation}
where $\{ \mathsf{U_a}[r]: r=1~~\text{to}~~N_\mathsf{a}\}$ is the maximal measurement with respect to which $\mathsf F$ and $\mathsf G$ are defined.
We know that a special filter followed by a maximal measurement actually corresponds to an equivalent maximal measurement (with appropriate coursegraining).  Hence, $\mathsf{G_{a_2}^{a_3}}[q]\mathsf{U_{a_3}}[p]$, here corresponds to a maximal measurement.  And consequently, $\mathsf{F_{a_1}^{a_2}}[p]\mathsf{G_{a_2}^{a_3}}[q]\mathsf{U_{a_3}}[p]$ corresponds to a maximal measurement.   States in $\overline S_\mathsf{F\cap G}$ simply followed by the maximal measurement $\{ \mathsf{U_a}[n]: n=1~~\text{to}~~N_\mathsf{a}\}$ will give rise only to outcomes having $n\in \overline O(S_\mathsf{F\cap G})$.  Paying attention to the definition of special filters, this means that, for such states, we must have either $p\not=p_1$ or $q\not=q_1$ (or both) where $p_1$ and $q_1$ are the values of $p$ and $q$ for which the filtering is effected.    Consequently, such states must be blocked.  This proves {\bf T\ref{overlappingfilters}}.

\subsection{Relationship between $K_\mathsf{a}$ and $N_\mathsf{a}$}\label{woottershierachy}

\index{K@$K_\mathsf{a}$!relationship with $N_\mathsf{a}$}
We are now in a position to prove the following
\begin{T}\label{KNr}
The relationship between $K_\mathsf{a}$ and $N_\mathsf{a}$ for any system is of the form
\begin{equation}
K_\mathsf{a} = N_\mathsf{a}^r
\end{equation}
where $r=1, 2, \dots$ is a constant independent of the system type.
\end{T}
We will drop the subscript, $\mathsf a$, for the moment.  To prove {\bf T\ref{KNr}} we note that it follows from {\bf T\ref{fidequiv}} that $K$ is a function of $N$:
\begin{equation}\label{KofN}
K=K(N)
\end{equation}
We can filter any system having $N+1$ distinguishable states to have just $N$, or to have just $1$, such that the informational subsets associated with these two filtrations are nonoverlapping.  Hence,
\begin{equation}\label{Kincreases}
K(N+1) > K(N)
\end{equation}
From {\bf P2} we know that $N_\mathsf{ab} = N_\mathsf{a}N_\mathsf{b}$.  Hence {\bf P3} implies
\begin{equation}\label{multiplicative}
K(N_\mathsf{a}N_\mathsf{b})= K(N_\mathsf{a})K(N_\mathsf{b})
\end{equation}
In number theory we would say that $K(\cdot)$ is a completely multiplicative function.
Finally, we know that we have systems for which
\begin{equation}\label{allN}
N=1, 2, 3, \dots
\end{equation}
It is proven in Appendix 1 that {\bf T\ref{KNr}} follows from (\ref{KofN}-\ref{allN}).  This proof works by considering the prime factorisation of $N$.  Without the condition that $K$ is an increasing function of $N$ we could have completely multiplicative functions in which different prime factors are raised to different powers:
\begin{equation}
K(N)= \prod_{i} p_i^{r_im_i(N)}
\end{equation}
where $p_i$ is the $i$th prime number and $m_i(N)$ is the multiplicity of $p_i$ in the prime factorisation of $N$ (equal to 0 if this prime factor does not appear).

An alternative proof that $K_\mathsf{a}=N_\mathsf{a}^r$ is given in Sec.\ \ref{thesignature} below.

This relationship between $K$ and $N$ in {\bf T\ref{KNr}} was first suggested as a possible relationship by Wootters \cite{wootters1990local, wootters1986quantum}. It was first proved that it follows from the above conditions in \cite{hardy2001quantum}.  It suggests a hierarchy of classes of theories which we will call the \emph{Wootters hierarchy}.  We will see below that the first theory in the hiarachy, when $r=1$ is classical probability theory.  The next class of theories, when $r=2$, contains quantum theory (there are other toy theories having $r=2$ that are not consistent with the postulates in this paper \cite{kirkpatrick2003hardy, spekkens2007evidence}). We will see that there are no theories consistent with the postulates having $r>2$.  Kirkpatrick \cite{kirkpatrick2003hardy} has given a simple model for theories with a finite number of pure states having any value of $r$.  {\.Z}yczkowski \cite{karol2008quartic} has worked on constructing theories having $r=4$ which have a continuum of pure states.    Both Kirkpatrick's model and {\.Z}yczkowski's construction violate one or more of the postulates given here.

\subsection{Classical and non-classical cases}

\index{classical case}
We note the following theorem.
\begin{T}\label{classicalcase}
We have classical probability theory if and only if $K_\mathsf{a}=N_\mathsf{a}$.
\end{T}
If $K_\mathsf{a}=N_\mathsf{a}$ then one set of fiducial effects is simply the maximal effects, $\{\mathsf{U_a}[n]: n=1 ~\text{to}~ N_\mathsf{a}\}$, corresponding to a given maximal measurement.  Since these fiducial effects all belong to the same measurement, the probabilities, $A^{a_1}$, in a general state defined with respect to this fiducial set of effects must satisfy
\begin{equation}
\sum_{a=1}^{N_\mathsf{a}} A^{a_1} \leq 1
\end{equation}
The states in the maximal set of distinguishable states are represented by vectors, $U^{a_1}[n]$, in which one probability is equal to 1 and all the others are equal to zero.  These states are pure.  We have, then,
\begin{equation}
A^{a_1} = \sum_{n=1}^{N_\mathsf{a}} p^{a_1} U^{a_1}[n]
\end{equation}
where $p^{a_1}$ are the fiducial probabilities (numerically equal to $A^{a_1}$).   Hence a general state can be written as a convex combination of the pure states in this given maximal set of distinguishable states and the null state.  This means that these are the only pure states.  This is the defining characteristic of classical probability theory (that there is only one maximal set of distinguishable states, all these being pure).   It is a simple matter to show that we get the classical probability simplex, the correct rules for composite systems, for transformations, and so on (see \cite{hardy2009foliable} for example).  In the case that $K>N$ there must exist at least $N+1$ pure states.  This is inconsistent with classical probability theory.

\subsection{The signature}\label{thesignature}

The results in this subsection play no role in the reconstruction and can be skipped in a first reading (although we do give an alternative derivation of the relationship, $K_\mathsf{a}=N_\mathsf{a}^r$).

One particularly illuminating way of viewing the relationship between $K_\mathsf{a}$ and $N_\mathsf{a}$ is in terms of what we will call the \emph{signature}.  The signature tells us something about possible choices for the fiducial set of results.  We can construct a fiducial set of results by applying various (special) filters all defined with respect to a given maximal measurement $\{\mathsf{U}_\mathsf{a}[n]: n=1~~\text{to} ~~ N_\mathsf{a} \}$.   Let $\mathsf{F_{a_1}^{a_2}}\{n, n', n'',\dots \}[p]$ be the set of transformations associated with a special filter that filters with respect to informational subset $S_{nn'n''\dots}$ defined with respect to the given maximal measurement where $O(S_{nn'n''\dots })=\{ n, n', n'', \dots\}$.  The filter is effected for $p=p_1$.  We will first consider all such filters where $O$ has one element.  There are $N_\mathsf{a}$ such filters.  Then we will consider the cases where $O$ has two elements.  There are $\frac{N_\mathsf{a}(N_\mathsf{a}-1)}{2!}$ such filters.   And so on.  For each case we will consider the fiducial results that are formed by placing another special filter, $\mathsf{F_{a_2}^{a_3}}\{n, n', \dots \}[p]$, of the same type followed by some result (we know from {\bf T\ref{fidafterfilt}} that all fiducial results can be of this type).  At each stage we count only the additional fiducial results required beyond those that have been counted already.   We will, of course, make use of {\bf T\ref{overlappingfilters}} (the theorem concerning overlapping filters).  Now we will implement this procedure
\begin{description}
\item[Step 1] Consider special filters $\mathsf{F_{a_1}^{a_2}}\{n\}[p]$.  For systems passing through such a filter, we can form a fiducial set of results by letting them pass through another special filter, $\mathsf{F_{a_2}^{a_3}}\{n\}[p]$, of the same type and follow this by some result.  Let $x_1$ be the number of fiducial results required for such a system. Since there are $N_\mathsf{a}$ filters of this type, we have so far counted $x_1N_\mathsf{a}$ fiducial results.  In fact, we know that $x_1=1$ since it follows from {\bf T\ref{KNr}} that systems having $N=1$ have $K=1$.  Further, we can actually choose these fiducial results to be equal to the elements, $\mathsf{U_{a_2}}[n]$ of the given maximal measurement.
\item[Step 2] Consider special filters $\mathsf{F_{a_1}^{a_2}}\{n, n'\}[p]$.  For systems passing through such a filter, we can form a fiducial set of results by letting them pass through another special filter, $\mathsf{F_{a_1}^{a_2}}\{n, n'\}[p]$ of the same type and follow this by a some results.  We have already counted some contributions to this set of fiducial effects in Step 1.  Let $x_2$ be the number of \emph{additional} fiducial effects required for each such filtration.   We count $x_2\frac{N_\mathsf{a}(N_\mathsf{a}-1)}{2!}$ contributions in this step (since there are $\frac{N_\mathsf{a}(N_\mathsf{a}-1)}{2!}$ filters of this type).  These contributions are all independent by virtue of {\bf T\ref{overlappingfilters}} since $\mathsf{F_{a_1}^{a_2}}\{n,n'\}[p_1]\mathsf{F_{a_2}^{a_3}}\{n, n''\}[p_1]$ is equivalent to a filter with $O$ equal to the intersection of $\{n, n'\}$ and $\{ n, n'' \}$ and we have already counted such contributions in Step 1.
\item[Step 3]  And so on.
\end{description}
Adding up all these contributions to $K_\mathsf{a}$ we have
\begin{equation}\label{sigexpan}
K_\mathsf{a} = x_1 N_\mathsf{a} + x_2\frac{N_\mathsf{a}(N_\mathsf{a}-1)}{2!} + x_3\frac{N_\mathsf{a}(N_\mathsf{a}-1)(N_\mathsf{a}-2)}{3!} + \dots
\end{equation}
Since $K_\mathsf{a}$ is finite for finite $N_\mathsf{a}$ we must have $x_i=0$ for all $i>r$ where $r$ is a finite integer.  This means that $K_\mathsf{a}$ is a polynomial function of $N_\mathsf{a}$ of finite order:
\begin{equation}
K=\sum_n^r c_n N^n
\end{equation}
If we put this into $K(N_\mathsf{a}N_\mathsf{b})=K(N_\mathsf{a})K(N_\mathsf{b})$ and compare coefficients we see that we must get $K=N^r$.  This provides an alternative derivation of this relationship without using the number theoretic arguments of Appendix \ref{AppendixC}.

We will call the series of integers
\begin{equation}
\stackrel{\rightarrow}{x} = (x_1, x_2, x_3, \dots )
\end{equation}
the \emph{signature}\index{signature}.  Here are a few examples
\begin{eqnarray}
\stackrel{\rightarrow}{x}_{K=N} & = & (1, 0, 0, 0, 0, \dots) \\
\stackrel{\rightarrow}{x}_{K=N^2} & = & (1, 2, 0, 0, 0,\dots) \\
\stackrel{\rightarrow}{x}_{K=N^3} & = & (1, 6, 6, 0, 0, \dots) \\
\stackrel{\rightarrow}{x}_{K=N^4} & = & (1, 14, 36, 24, 0,  \dots)
\end{eqnarray}
We get these by putting $N_\mathsf{a}=1, 2, 3, 4$ in (\ref{sigexpan}).  This gives some insight into the Wootters hierarchy.  For the classical case there is nothing nontrivial happening beyond rank one filters.  For quantum theory ($K=N^2$) there is nothing nontrivial happening beyond rank two filters.  This appears to be related to the Sorkin hierarchy \cite{sorkin1994quantum} (see also \cite{ududec2010three, sinha2010ruling, niestegge2009sorkin}).   The Sorkin hierarchy concerns multi-slit interference experiments.  Classical interference has nothing non-trivial beyond one slit (two slit interference can be decomposed into one slit patterns).  Quantum interference has nothing non-trivial beyond two slits (three slit interference can be decomposed into two and one slit patterns).

\section{Gebits}\label{gebitssection}

In this section we will prove $N_\mathsf{a}=2$ case (the generalized bit or \emph{gebit}\index{gebit}) is in agreement with quantum theory (in particular that states belong to the Bloch sphere).  We do this in three basic steps. First we show that the pure states correspond to some subset of the points on a hypersphere. Second, we show that, in fact, every point on the hypersphere corresponds to a pure state. Third, we show that $K_\mathsf{a}=4$ when $N_\mathsf{a}=2$.  The second and third steps involve the use of {\bf P5} (for the first time in reconstruction).  In this third step we adapt an ingenious method developed by CDP involving teleportation.

\subsection{Gebits - basic properties}\label{gebitbasicproperties}

It follows from {\bf T\ref{redundantoutcomes}} that, for a gebit, we can write the deterministic effect, $T_a$, as
\begin{equation}
T_a= U_a[1] + U_a[\bar 1]
\end{equation}
for any maximal measurement $\{\mathsf{U_a}[n] : n=1, \bar 1 \}$ (in this subsection it is notationally convenient to use $1$ and $\bar 1$ as labels rather than $1$ and $2$).  We saw in {\bf T\ref{revtransdeterministic}} that any reversible transformation on a system is equivalent to a deterministic transformation.  It follows from this and the fact that the deterministic effect is unique ({\bf T\ref{deteffect}}) that the deterministic effect is is unchanged when preceded by any reversible (and therefore deterministic) transformation
\begin{equation}
\tilde R_{a_1}^{a_2} T_{a_1} = T_{a_2}
\end{equation}
(this is actually true for any deterministic transformation).   If a maximal effect, $U_a[n]$, is preceded by a reversible transformation we have another maximal effect
\begin{equation}
U'_{a_2}[n] = \tilde R_{a_1}^{a_2} U_{a_1}[n]
\end{equation}
since this set of effects distinguish the states $R_{a_1}^{a_2} U^{a_1}[n]$ where $\mathsf R$ is the inverse transformation to $\mathsf{\tilde R}$.

By {\bf T\ref{revt}}, we know that there exists a reversible transformation, $\mathsf{R_{a_1}^{a_2}}$, which takes any pure state to any other pure state.
If we apply two such transformations we get a third.  The reversible transformations must form a group, $\mathcal G$, whose elements can be represented by matrices.  It follows from {\bf Assump 3} that this matrix group is compact (see also Appendix \ref{compactness}).  Any compact matrix group admits an orthogonal representation \cite{boerner1970representations}.  If we go to the orthogonal representation then transformations will not change the length of the vectors (in this orthogonal representation) representing the state and hence all pure states must lie on (the surface of) a hypersphere. However, we do not know that all points on this hypersphere will have states at them.

In the orthogonal representation we will denote vectors by bold font lower case letters and indicate whether we have an effect or state by a lower or upper subscript for the system type. We will also include a constant factor, $\frac{1}{\sqrt{2}}$, for later convenience.  Thus, in orthogonal representation, the pure state $U^a[n]$ becomes ${\bf u}^\mathsf{a}[n]/\sqrt{2}$, the maximal effect $V_a[n]$ becomes ${\bf v}_\mathsf{a}[n]/\sqrt{2}$, and the identity effect $T_{a_1}$ becomes ${\bf t}_\mathsf{a}/\sqrt{2}$. We will keep upper case letters to represent transformations and drop the indices.   The reversible transformation $R_{a_1}^{a_2}$ becomes $R$ (we could write it as $R_\mathsf{a}^\mathsf{a}$ but this is unnecessary) and its inverse will be $R^T$ where $T$ denotes transpose (this is true because we are in an orthogonal representation so $R^T R = \ident$).  By {\bf T\ref{revt}}, a general pure state, ${\bf u}^\mathsf{a}$, is equal to $R{\bf u}^\mathsf{a}[1]$ for some particular pure state ${\bf u}^\mathsf{a}[1]$. This pure state is identified by the maximal effect ${\bf u}_\mathsf{a}=R{\bf u}_\mathsf{a}[1]$ since
\begin{eqnarray}
\text{Prob}(\mathsf{U_{a_1} U^{a_1}})& = & \frac{1}{2} {\bf u}_\mathsf{a}\cdot {\bf u}^\mathsf{a} \\
&=& \frac{1}{2}(R{\bf u}_\mathsf{a}[1])\cdot R{\bf u}^\mathsf{a}[1] \\
&=& \frac{1}{2}{\bf u}_\mathsf{a}[1]\cdot R^T R {\bf u}^\mathsf{a}[1] = 1
\end{eqnarray}
Note here that we need to include a factor of $\frac{1}{2}$ (this comes from the $\frac{1}{\sqrt{2}}$ factors introduced above) when calculating probabilities.
The maximal effect identifying a given pure state is, by {\bf P1}, unique.  Hence, all maximal effects can be written as ${\bf u}_\mathsf{a}=R{\bf u}_\mathsf{a}[1]$.
We know that the deterministic effect is unchanged by the action of $\mathcal{G}$.
\begin{equation}
R{\bf t}_\mathsf{a} = {\bf t}_\mathsf{a}:={\bf t}
\end{equation}
Hence we can expand a general maximal effect, ${\bf v}_\mathsf{a}$,
\begin{equation}\label{vsbarv}
{\bf v}_\mathsf{a} = v_0{\bf s} + {\bf \bar v}_\mathsf{a}
\end{equation}
and a general pure state, ${\bf u}^\mathsf{a}$, as
\begin{equation}\label{usbaru}
{\bf u}^\mathsf{a}= u_0 {\bf s} + {\bf \bar u}^\mathsf{a}
\end{equation}
where
\begin{equation}
{\bf s} = \frac{\bf t}{|{\bf t}|}
\end{equation}
and ${\bf \bar v}_\mathsf{a}$ and ${\bf \bar u}^\mathsf{a}$ are orthogonal to ${\bf s}$.   Since ${\bf s}$ is invarient under $\mathcal{G}$, $v_0$ and $u_0$ are constants. It follows from this fact and the fact that we have an orthogonal group that the lengths, $|{\bf \bar v}_\mathsf{a}|$ and $|{\bf \bar u}^\mathsf{a}$ are constant for these maximal effects and pure states. We choose $v_0=1$ for maximal effects.  We are free to make this choice since we can absorb any factor into $u_0$ (it is $v_0u_0$ that appears in the equation for the probability). This implies that ${\bf t}=2{\bf s}$ (so $|{\bf t}|=2$) because ${\bf t}={\bf v}_\mathsf{a}[1]+{\bf v}_\mathsf{a}[\bar 1]$.  The latter also implies
\begin{equation}
{\bf \bar v}_\mathsf{a}[1] + {\bf \bar v}_\mathsf{a}[\bar 1] =0
\end{equation}
Hence,
\begin{T}\label{maximaleffectsNtwo}
For a gebit, the maximal effects corresponding to a given maximal measurement are represented by antipodal points on the hypersphere.
\end{T}
We will prove a similar result for states in a maximal distinguishable set later (this is a more difficult thing to prove).

Now, $\text{Prob}(\mathsf{T_{a_1} U^{a_1}}=\frac{1}{2}{\bf t}_\mathsf a\cdot{\bf u}^\mathsf{a}=1$ and hence $u_0=1$.  This gives
\begin{equation}\label{probUV}
\text{Prob}(\mathsf{V_{a_1} U^{a_1}}) = \frac{1}{2}{\bf v}_\mathsf{a} \cdot {\bf u}^\mathsf{a} = \frac{1}{2}(1 + {\bf \bar v}_\mathsf{a}\cdot{\bf \bar u}^\mathsf{a})
\end{equation}
We can make ${\bf \bar v}_\mathsf{a}$ of unit length and then absorb the overall constant into ${\bf \bar u}^\mathsf{a}$.   Let $|{\bf \bar u}^\mathsf{a}|=u$.

Let us summarize these results. The maximal effects correspond to points on a unit $(K_\mathsf{a}-2)$-sphere (this is a sphere embedded in a $K_\mathsf{a}-1$ dimensional space).  The pure states correspond to points on a $(K_\mathsf{a}-2)$-sphere of radius $u$. We will see later that $u=1$. We do not know at this stage that all points on these hyperspheres correspond to allowed maximal effects and allowed pure states (we will prove this later).

The full set of states is in the convex hull of the pure states and the null state.  This means that they all live in (or on) a hyper-cone of length $1$ (since $u_0=1$) with the above hypersphere (of radius $u$) at the base. We do not know, at this stage, that all points on or in the cone actually correspond to states.  We can represent a general state corresponding to a preparation, $\mathsf{B^{a}}$, as
\begin{equation}
{\bf b}_\mathsf{a} = b_0 {\bf s} + {\bf \bar b}^\mathsf{a}
\end{equation}
so that $b_0=1$ for normalized states.

The information in ${\bf b}^\mathsf{a}$ is contained in the vector
\begin{equation}
( b_0,   b_1,  b_2, \dots, b_{K_\mathsf{a}-1})
\end{equation}
where $b_l$ are the components of ${\bf \bar b}^\mathsf{a}$ (numbered from $1$ to $K_\mathsf{a}-1$) in some basis.  Since this is a cone of length $1$ having a hypersphere of radius $u$ as base, we have
\begin{equation}\label{hyperconeconstraint}
\sum_{l=1}^{K_\mathsf{a}-1} b_l^2 \leq u^2 b_0
\end{equation}
for vectors on or in the cone.

The general effect $\mathsf{C_{a}}$ can be written
\begin{equation}
{\bf c}_\mathsf{a} = c_0 {\bf s} + {\bf \bar c}_\mathsf{a}
\end{equation}
The maximal effects have $c_0=1$.  We have
\begin{equation}
\text{Prob}(\mathsf{B^{a_1}C_{a_1}}) = \frac{1}{2}{\bf b}^\mathsf{a_1} \cdot {\bf c}_\mathsf{a_1}
\end{equation}
for a general effect and a general preparation.

\subsection{Going over to the non-classical case}\label{nonclassicalsection}

We saw in {\bf T\ref{KNr}} that $K_\mathsf{a}=N_\mathsf{a}^r$ with $r=1, 2, 3\dots $.  If $r=1$ then, for a gebit, we have $K_\mathsf{a}=2$.  Hence, the hypersphere is 0 dimensional (embedded in a one dimensional space). In other words, the pure states consist of two points, one pointing in the positive $b_1$ direction, and the other pointing in the negative $b_1$ direction.  The full set of states are convex combinations of these pure states and the null state (so the cone is a triangle).  This is the classical case where the gebit is simply a bit.  We have already shown that the $K_\mathsf{a}=N_\mathsf{a}$ case leads to classical probability theory.  Hence forth, we will assume that we are in the case $K_\mathsf{a}\not=N_\mathsf{a}$.   We can force this to be the case by using {\bf P4} (rather than {\bf P4}$'$).  For the classical gebit there is no compound permutation transformation.  Importantly, this is the only point at which we use {\bf P4} rather than {\bf P4}$'$ in the reconstruction.  We could force non-classicality with any such additional assumption that was inconsistent with classical probability theory.  Since we are intent on reconstructing quantum theory, any such additional assumption must, of course, be consistent with quantum theory.

\subsection{All points on hypersphere are populated}\label{hyperspherepopulated}

In this subsection we will prove that all points on the hypersphere correspond to (pure) states.  To do this we will use {\bf P5} (for the first time in the paper) and employ considerations involving a getrit (that is a system having $N_\mathsf{a}=3$)\index{getrit}.  The basic idea of the proof entails sending a non-flat set of states states associated with a gebit informational subset of the getrit through a filter associated with a different gebit informational subset. The net result is that the states move closer to a pole while remaining on the surface of the hypersphere (after being normalized).   If this is repeated we can get the non-flat set of states as close to a pole as we like. As they are non-flat, they span the surface of the hypersphere near this pole.  Given {\bf Assump 3} we know that the set must be closed and consequently there must exist an infinitesimal patch of pure states around the pole.  By transitivity {\bf T\ref{revt}} we can move this patch around to any other place there is a pure state.  Since there must be an infinitesimal patch around any such point the whole surface must be covered in pure states.  We will now fill out this argument in detail.

First we note
\begin{T}\label{filtersnonmixing}
Filters are non-mixing.
\end{T}
This is a trivial consequence of {\bf P5} and {\bf T\ref{nonflatimpliesnonmix}}.

%The following result will be useful.
%\begin{T}\label{gebitstates}
%For a gebit, there exists a set of $K_\mathsf{b}-1$ pure states whose projections into the subspace spanned by the hypersphere are linearly independent.
%\end{T}
%This has to be the case otherwise we could find a hypersphere of smaller dimension that included all the pure states.

Using special filters, we can prove
\begin{T}\label{getrit}
For a non-classical getrit (that is a system having $K_\mathsf{a}>N_\mathsf{a}$ and $N_\mathsf{a}=3$) we can have two distinct maximal measurements having one (and only one) element in common.
\end{T}
Let one maximal measurement be $\{\mathsf{U_{a_1}}[1], \mathsf{U_{a_1}}[2], \mathsf{U_{a_1}}[3]\}$.  We will show we can construct another, $\{\mathsf{U_{a_1}}[1], \mathsf{U_{a_1}}[2'], \mathsf{U_{a_1}}[3']\}$, which has one element in common with the first.  Note we are employing slightly different notation from before - the prime in $\mathsf{U_{a_1}}[2']$ indicates that this is a different effect from $\mathsf{U_{a_1}}[2]$ (previously we would have denoted this by a different letter, e.g. $\mathsf{V_{a_1}}[2]$).  Consider a special filter, $\mathsf{F_{a_1}^{a_2}}[n]$, associated with information subset $S_{23}$ having $O(S_{23})=\{2,3\}$.  On passing through this filter, the system becomes a gebit. If we follow this filter by a $\{\mathsf{U_{a_1}}[1], \mathsf{U_{a_1}}[2], \mathsf{U_{a_1}}[3]\}$ measurement then we effect the maximal measurement $\{\mathsf{U_{a_1}}[2], \mathsf{U_{a_1}}[3]\}$ on the gebit.  For this gebit, which is non-classical, $N_\mathsf{b}=2$ and $r\geq 2$ (in the nonclassical case) we have $K_\mathsf{b}\geq 4$ (using {\bf T\ref{KNr}}).  Hence there must be at least four pure states.  Since every pure state belongs so some maximal distinguishable set of states (by {\bf T\ref{origP2}} there must be at least two maximal sets of distinguishable states.  Consequently, there must be at least two distinct maximal measurements. Thus, in addition to $\{\mathsf{U_{a_1}}[2], \mathsf{U_{a_1}}[3]\}$, there must be at least one more.  Let this be $\{\mathsf{U_{a_1}}[2'], \mathsf{U_{a_1}}[3']\}$. Since maximal measurements correspond to antipodal points ({\bf T\ref{maximaleffectsNtwo}}) the four effects making up these two maximal measurements must be distinct.  If we follow the special filter defined above by  the maximal measurement, $\{\mathsf{U_{a_1}}[2'], \mathsf{U_{a_1}}[3']\}$, then, by the properties of the special filter, we effect the maximal measurement $\{\mathsf{U_{a_1}}[1], \mathsf{U_{a_1}}[2'], \mathsf{U_{a_1}}[3']\}$ on the original getrit. This has only one effect in common with the maximal measurement we started with.  This proves {\bf T\ref{getrit}}.

Next we will prove
\begin{T}\label{getrittwo}
If we have two maximal measurements $\{\mathsf{U_a}[1], \mathsf{U_a}[2], \mathsf{U_a}[3]\}$ and $\{\mathsf{U_a}[1], \mathsf{U_a}[2'], \mathsf{U_a}[3']\}$ for a getrit having one, and only one, effect in common then we have
\begin{equation}\label{probAU}
\text{Prob}(\mathsf{A^{a_1}U_{a_1}}[3]) =  \alpha \beta  ~~\text{for any} ~~ A^{a_1}\in S_{12'}
\end{equation}
where
\begin{equation}\label{defnalphabeta}
\alpha = 1-\text{Prob}(\mathsf{A^{a_1}U_{a_1}}[1])  ~~~ \text{and}~~~ \beta = 1-\text{Prob}(\mathsf{U^{a_1}}[2']\mathsf{U_{a_1}[2]})
\end{equation}
and $O(S_{12'})=\{1, 2'\}$
\end{T}
To prove this, consider a special filter, $\mathsf{G_{a_1}^{a_2}}[n]$, defined with respect to the maximal measurement $\{\mathsf{U_{a_1}}[1], \mathsf{U_{a_1}}[2], \mathsf{U_{a_1}}[3]\}$ having informational subset $S_{23}$ with $O(S_{23})=\{2,3\}$.  It follows from the properties of special filters that this is also a special filter with respect to the maximal measurement $\{\mathsf{U_{a_1}}[1], \mathsf{U_{a_1}}[2'], \mathsf{U_{a_1}}[3']\}$ with informational subset $S_{2'3'}$ where $O(S_{2'3'})=\{2',3'\}$.  Consider a pure state $A^{a_1}$ in $S_{12'}$.  This state has
\begin{equation}\label{Astateproperties}
\text{Prob}(\mathsf{A^{a_1}U_{a_1}}[1])= 1-\alpha  ~~\text{and} ~~ \text{Prob}(\mathsf{A^{a_1}U_{a_1}}[3']) =0
\end{equation}
The first property follows from the definition of $\alpha$ above.  The second by the fact that the state is in $S_{12'}$. From {\bf P1} and the first property above it follows that if $A^{a_1}\not= U^{a_1}[1]$ then $A^{a_1} G_{a_1}^{a_2}$ is not the null state null (it will not be blocked by the filter).  Since filters are non-mixing, by {\bf T\ref{filtersnonmixing}}, the state $A^{a_1} G_{a_1}^{a_2}$ must be proportional to a pure state.
\begin{equation}
A^{a_1} G_{a_1}^{a_2} = \alpha V^{a_2}
\end{equation}
That the constant of proportionality is equal to $\alpha$ follows from the fact that the probability of the state being absorbed by the filter is equal to $\text{Prob}(\mathsf{A^{a_1}U_{a_1}}[1])$.  Hence, the probability that it is not absorbed is $\alpha$ (and this must be equal to the normalization).
The state, $V^{a_2}$, must be in $S_{2'3'}$ using {\bf T\ref{specialfilter}} and the fact that $\mathsf{G_{a_1}^{a_2}}$ is associated with information subset $S_{2'3'}$ as shown above.  From the second property in (\ref{Astateproperties}) and the property of special filters that
\begin{equation}
\mathsf{U_{a_1}}[3'] \equiv \mathsf{G_{a_1}^{a_2} U_{a_2}}[3']
\end{equation}
we have
\begin{equation}
\text{Prob}(\mathsf{V ^{a_1}U_{a_1}}[3']) =0
\end{equation}
Hence,
\begin{equation}
\text{Prob}(\mathsf{V ^{a_1}U_{a_1}}[2']) =1
\end{equation}
as these two probabilities must add to one for a normalized gebit state in $S_{2'3'}$.   It follows from {\bf P1} that $V^{a_1}=U^{a_1}[2']$.  Hence
\begin{equation}\label{AGU}
A^{a_1} G_{a_1}^{a_2} = \alpha U^{a_2}[2']
\end{equation}
Now $U^{a_1}[2']U_{a_1}[2]=1-\beta $ and consequently $U^{a_1}[2']U_{a_1}[3] = \beta$ (as these two probabilities must add to 1). Hence,
\begin{equation}
\text{Prob}(\mathsf{A^{a_1}G_{a_1}^{a_2}U_{a_1}}[3]) = \alpha\beta
\end{equation}
Using the special filter property
\begin{equation}
\mathsf{U_{a_1}}[3] \equiv \mathsf{G_{a_1}^{a_2} U_{a_2}}[3]
\end{equation}
we have
\begin{equation}
\text{Prob}(\mathsf{A^{a_1}U_{a_1}}[3]) = \text{Prob}(\mathsf{A^{a_1}G_{a_1}^{a_2}U_{a_2}}[3]) =\alpha\beta
\end{equation}
This proves {\bf T\ref{getrittwo}}.

% any subset of K-1 states, under projection into hypersphere, will remain lin indep unless two of them correspond to opposite poles.  If U1 and U2 do correspond to opposite poles then we can discount U2.  If not then there cannot be a state opposite U1 because it would have prob greater than one.  Hence any subset of K-1 states excluding U2 (but including U1) are lin indep under projection.

We will now prove our main result of this subsection.
\begin{T}\label{hypersphere}\index{hypersphere}
{\bf The hypersphere.} The states of a gebit are given by the convex hull of the full set of points on a $(K_\mathsf{a}-2)$-sphere with the null state.
\end{T}
For the classical case this has already been shown in Sec.\ \ref{nonclassicalsection}.  Now consider the non-classical case.  Consider the gebit corresponding to the informational subset $S_{12'}$ where $O(S_{12'})=\{1, 2'\}$ (these are states which do not give rise to $\mathsf{U_{a_1}}[3']$).  There must be $K_\mathsf{b}$ linearly independent pure states, $\{A^{a_1}[k]: k=1~\text{to} K_\mathsf{b}\}$ in $S_{12'}$. These constitute a non-flat set of states. Let the first two of these states be $U^{a_1}[1]$ and $U^{a_1}[2']$ (these are the states which are identified by $\mathsf{U_{a_1}}[1]$ and $\mathsf{U_{a_1}}[2']$ respectively). We will call the other states the \lq\lq in-between" states. Now send this set of $K_\mathsf{b}$ states through a special filter $\mathsf{F_{a_1}^{a_2}}[n]$ with $O(S)=\{1,2\}$. We denote the new states by $B^{a_1}[k]$ (by {\bf T\ref{specialfilter}} these states belong to the informational subset $S_{12}$ where $O(S_{12})=\{1,2\}$).  By {\bf P5} and {\bf T\ref{filtersnonmixing}} this set of states remains non-flat and pure (up to normalization).  The $U^{a_1}[1]$ state will pass through the filter unchanged (as it is in $S_{12}$).  As we will see, the $B^{a_1}[2]$ state will be parallel to $U^{a_1}[2]$ while the in-between states will get closer to $U^{a_1}[1]$.  On passing through $\mathsf{F_{a_1}^{a_2}}$ we can measure $\{\mathsf{U_{a_1}}[1], \mathsf{U_{a_1}}[2]\}$.  By the properties of the special filter,
\begin{equation} \label{filterFU}
\mathsf{F_{a_1}^{a_2}}\mathsf{U_{a_2}}[1] \equiv\mathsf{U_{a_1}}[1]
\end{equation}
and
\begin{equation} \label{UequivFU}
\mathsf{U_{a_1}}[2] \equiv \mathsf{F_{a_1}^{a_2}}\mathsf{U_{a_2}}[2]  %\equiv \mathsf{F_{a_1}^{a_2}}[n_1](\mathsf{U_{a_1}}[2]+ \mathsf{U_{a_1}}[3])
\end{equation}
We have (using the same notation in the proof of {\bf T\ref{getrit}})
\begin{equation}\label{UUequivUU}
\mathsf{U_{a_1}}[2]+ \mathsf{U_{a_1}}[3] \equiv \mathsf{U_{a_1}}[2']+ \mathsf{U_{a_1}}[3']
\end{equation}
where the $+$ on the LHS indicates that we are course-graining over the outcomes of $\mathsf{U_{a_1}}[2]$ and $\mathsf{U_{a_1}}[3]$ to form a new result.
Equation (\ref{UUequivUU}) follows since if we add $\mathsf{U_{a_1}}[1]$ to both sides of this equation we get the deterministic effect which is unique (by {\bf T\ref{deteffect}}).
Hence,
\begin{equation}
\text{Prob}(\mathsf{A^{a_1}}[k]\mathsf{U_{a_1}}[2]) + \text{Prob}(\mathsf{A^{a_1}}[k]\mathsf{U_{a_1}}[3])
=\text{Prob}(\mathsf{A^{a_1}}[k]\mathsf{U_{a_1}}[2']) + \text{Prob}(\mathsf{A^{a_1}}[k]\mathsf{U_{a_1}}[3'])
\end{equation}
Using {\bf T\ref{getrittwo}} and the fact that $\text{Prob}(\mathsf{A^{a_1}}[k]\mathsf{U_{a_1}}[3'])=0$ (since $A^{a_1}[k]\in S_{12'}$) we obtain
\begin{equation}\label{AUltAU}
\text{Prob}(\mathsf{A^{a_1}}[k]\mathsf{U_{a_1}}[2]) +\alpha \beta= \text{Prob}(\mathsf{A^{a_1}}[k]\mathsf{U_{a_1}}[2'])
\end{equation}
Using $\mathsf{B^{a_2}}[k]= \mathsf{A^{a_1}}[k]\mathsf{F_{a_1}^{a_2}}$ and (\ref{UequivFU})
\begin{equation}\label{AUAFUBU}
\text{Prob}(\mathsf{A^{a_1}}[k]\mathsf{U_{a_1}}[2]) = \text{Prob}(\mathsf{A^{a_1}}[k]\mathsf{F_{a_1}^{a_2}}\mathsf{U_{a_2}}[2])
=  \text{Prob}(\mathsf{B^{a_1}}[k]\mathsf{U_{a_1}}[2])
\end{equation}
Putting all this together we obtain
\begin{flalign}
& \text{Prob}(\mathsf{B^{a_1}}[k]\mathsf{U_{a_1}}[1])  =  \text{Prob}(\mathsf{A^{a_1}}[k]\mathsf{U_{a_1}}[1]) \\
& \text{Prob}(\mathsf{B^{a_1}}[k]\mathsf{U_{a_1}}[2]) =   \text{Prob}(\mathsf{A^{a_1}}[k]\mathsf{U_{a_1}}[2'])- \alpha\beta
\end{flalign}
The first property follows from (\ref{filterFU}) and the second property follow from (\ref{AUltAU}) and (\ref{AUAFUBU}).
Recall that $A^{a_1}[k]\in S_{12'}$ while $B^{a_1}[k]\in S_{12}$ so it makes sense to compare these probabilities.  We see that while the probability for the $\mathsf{U_{a_1}}[1]$ outcome remains unchanged, the probability for the other outcome in the maximal measurement (associated with $\mathsf{U_{a_1}}[2]$ in $S_{12}$ and $\mathsf{U_{a_1}}[2']$ in $S_{12'}$) necessarily decreases. It follows that the states $B^{a_1}[k]$ are not normalized (except for the $k=1$ case).  However, by {\bf T\ref{filtersnonmixing}}, they are parallel to pure states (which are normalized).  Let these pure states be $C^{a_1}[k]$.   Following through the mathematics of normalization, we have $B^{a_1}[k]= (1-\alpha\beta) C^{a_1}[k]$.  Hence,
\begin{flalign}
&\text{Prob}(\mathsf{C^{a_1}}[k]\mathsf{U_{a_1}}[1]) = \frac{1}{1-\alpha\beta}  \text{Prob}(\mathsf{A^{a_1}}[k]\mathsf{U_{a_1}}[1]) \\
&\text{Prob}(\mathsf{C^{a_1}}[k]\mathsf{U_{a_1}}[2]) = \frac{1}{1-\alpha\beta}\big( \text{Prob}(\mathsf{A^{a_1}}[k]\mathsf{U_{a_1}}[2']) -\alpha\beta\big)
\end{flalign}
It follows from {\bf P1} that the states $C^{a_1}[k]$ are not equal to $U^{a_1}[1]$ (except for $k=1$).  Hence the smallest system that can support them is a gebit.  Since the set remains non-flat by {\bf P5}, the states $C^{a_1}[k]$ are linearly independent.
Since systems having the same $N_\mathsf{b}$ ($N_\mathsf{b}=2$ in this case) are equivalent (by {\bf T\ref{sameNequiv}}), there must exist a set of states $A^{a_1}[k,i=2]$ in $S_{12'}$ which bear the same relationship with $U^{a_1}[1]$ as the states $C^{a_1}[k]$ do with $U^{a_1}[1]$ in $S_{12}$ ($i$ is the iteration number).  We can iterate this process by sending the states $A^{a_1}[k,i=2]$ through the filter, normalizing to get a new set of states $C^{a_1}[k,i=3]$, then finding a set of states $A^{a_1}[k,i=3]$ in $S_{12'}$ which bear the same relationship with $U^{a_1}[1]$ as the states $C^{a_1}[k, i=3]$ do with $U^{a_1}[1]$ in $S_{12}$.  Hence,
\begin{flalign}
&\text{Prob}(\mathsf{A^{a_1}}[k, i+1]\mathsf{U_{a_1}}[1]) = \frac{1}{1-\alpha\beta}  \text{Prob}(\mathsf{A^{a_1}}[k, i]\mathsf{U_{a_1}}[1]) \\
&\text{Prob}(\mathsf{A^{a_1}}[k, i+1]\mathsf{U_{a_1}}[2']) = \frac{1}{1-\alpha\beta}\big( \text{Prob}(\mathsf{A^{a_1}}[k, i]\mathsf{U_{a_1}}[2']) -\alpha\beta\big)
\end{flalign}
Now $\alpha>0$ for any $A^{a_1}[k, i]$ that is distinct from $U^{a_1}[1]$ and we have $\beta>0$ by {\bf P1} as $U_{a_1}[2]$ does not identify $U^{a_1}[2']$ (see definitions of $\alpha$ and $\beta$ in (\ref{defnalphabeta})).  By application of {\bf P1} we see that, for $k=1,2$, the states remain unchanged on iteration.
For $k>2$, however, the states $A^{a_1}[k, i+1]$ are closer to $U^{a_1}[1]$ than the states $A^{a_1}[k, i]$ are by a finite amount since then
$\text{Prob}(\mathsf{A^{a_1}}[k, i+1]\mathsf{U_{a_1}}[1])$ is bigger than $\text{Prob}(\mathsf{A^{a_1}}[k, i]\mathsf{U_{a_1}}[1])$ by a finite amount and $\text{Prob}(\mathsf{A^{a_1}}[k, i+1]\mathsf{U_{a_1}}[2'])$ is smaller than $\text{Prob}(\mathsf{A^{a_1}}[k, i]\mathsf{U_{a_1}}[2'])$ by a finite amount.
Hence, in the limit, we get
\begin{eqnarray}
\text{Prob}(\mathsf{A^{a_1}}[k, i=\infty]\mathsf{U_{a_1}}[1])  =1 \\
\text{Prob}(\mathsf{A^{a_1}}[k, i=\infty]\mathsf{U_{a_1}}[2'])  =0
\end{eqnarray}
for $k>2$ where $i$ is the iteration number.   By {\bf P1} there is only one pure state having these properties, namely $U^{a_1}[1]$ itself.  Thus, in the limit, these states become equal to $U^{a_1}[1]$ (except for the $k=2$ case which remains unchanged).  However, after any finite number of steps we have a set of pure states which are linearly independent and (except for the $k=2$ case) as close to $U^{a_1}[1]$ as we wish.  We know by {\bf T\ref{revt}} that there exists a reversible transformation from $U^{a_1}[1]$ to each of these linearly independent states.  Since we can get these states as close to $U^{a_1}[1]$ as we wish, we represent these transformations as $I+\varepsilon_k X[k]$ (where $I$ is the identity) and regard $X[k]$ as generators.  We proved in Appendix \ref{compactness} that the space of states is compact (by {\bf Assump 3}).  This means it is closed.  Hence, we can generate all states in an infinitesimal patch around the pure state $U^{a_1}[1]$ on the surface of the hypersphere (the dimensionality of this patch being the same as that of the surface of the hypersphere as the states for $k>2$ are linearly independent and there are $K_\mathsf{b}-2$ or them). By {\bf T\ref{revt}}, this must be true for any pure state.  By moving this patch infinitesimally we can cover the surface of the hypersphere with pure states.  The full set of states is the convex sum of the pure states and the null state.  This proves {\bf T\ref{hypersphere}}.

We will now prove a few results to conclude this subsection on gebits.
\begin{T} \label{equalvectors}
There exists bases choices in which the vector representing any pure state for a gebit is equal to the vector representing the maximal effect which identifies it.
\end{T}
Consider a maximal effect represented by a point, ${\bf \bar v}_\mathsf{a}$ on a hypersphere of radius 1 (here we are using the notation of Sec.\ \ref{gebitbasicproperties}).  There must exist a pure state on the state hypersphere with vector ${\bf \bar v}^\mathsf{a}$ pointing in the same radial direction.  Any other pure state will, by equation (\ref{probUV}), have smaller probability.  Hence, this must be the state that is identified by the given effect.  It follows from equation (\ref{probUV}) that ${\bf \bar v}_\mathsf{a}\cdot{\bf \bar v}^\mathsf{a}=1$.  It follows that $u=1$, where $u$ is the radius of the hypersphere of pure states, and that ${\bf \bar v}_\mathsf{a}={\bf \bar v}^\mathsf{a}$.  Further, from (\ref{vsbarv}, \ref{usbaru}) and the fact that $u_0=v_0=1$ we have ${\bf  v}_\mathsf{a}\cdot{\bf  v}^\mathsf{a}=1$.  Every other pure state must be identified by some maximal effect which must, therefore, be represented by a vector that is equal to the vector representing the pure state in this basis.  This proves {\bf T\ref{equalvectors}}).

We can now prove a useful result concerning gebit effects.
\begin{T} \label{gebiteffects}
For gebits, effects that are proportional to any given maximal effect can only be written as a sum of effects that are also proportional to this same maximal effect.
\end{T}
By {\bf T\ref{equalvectors}} and {\bf T\ref{hypersphere}}, the maximal effects are cover the unit hypersphere.  The surface of the cone subtended by this hypersphere has effects which are proportional to maximal effects.  Consider one such effect, ${\bf c}_\mathsf{a}=\mu {\bf v}_\mathsf{a}$ where $0<\mu\leq 1$.  If this can be written as the sum of two vectors that are not proportional to each other then one of these vectors must lie outside the cone. Consider such a vector, ${\bf b}_\mathsf{a}$.  Such vectors give rise to negative probabilities and so cannot represent states.  To see that they give rise to negative probabilities, consider the pure state, ${\bf u}^\mathsf{a}$, that is opposite ${\bf b}_\mathsf{a}$ (by opposite, we mean that if ${\bf b}_\mathsf{a}=b_0{\bf s} + {\bf \bar b}_\mathsf{a}$ then ${\bf u}^\mathsf{a}={\bf s} - \nu{\bf \bar b}_\mathsf{a}$ for some positive $\nu$).  The maximal effect opposite this pure state gives probability 0 and hence is orthogonal to it.  Hence, the vector ${\bf b}_\mathsf{a}$ subtends an angle greater than $90^\circ$ and so gives a negative probability for ${\bf u}^\mathsf{a}$.  This proves {\bf T\ref{gebiteffects}}.

Another result follows from this.
\begin{T}\label{gebitzero}
For a gebit, any effect that gives probability zero for a pure state, $U^a[2]$, is proportional to the maximal effect, $U_a[1]$, which identifies the pure state $U^a[1]$.  Here $U^a[1]$ and $U^a[2]$ form a maximal distinguishable set.
\end{T}
As established in the proof of {\bf T\ref{gebiteffects}}, all effects must lie inside the cone (in the sense that they cannot subtend an angle with ${\bf s}$ that is greater than that subtended by any maximal effect.  The angle at the base of the cone is $90^\circ$ as established in the previous proof. Further, this cone coincides with the cone of states.  Hence, any effect that gives zero probability for a given pure state must be proportional to the effect identifying the opposite pure state.  This proves {\bf T\ref{gebitzero}}.

Similarly,
\begin{T}\label{gebitstatezero}
For a gebit, any state that gives probability zero for a maximal effect, $\mathsf{U_a}[2]$, is proportional to the pure state, $U^a[1]$, that is identified by the maximal effect $\mathsf{U_a}[1]$.  Here, $\{\mathsf{U_a}[1], \mathsf{U_a}[2]\}$ is a maximal measurement.
\end{T}
This follows for the same geometric reasons as {\bf T\ref{gebitzero}}.

\subsection{Theorem concerning non-flattening transformations}

We will prove the following useful theorem\index{non-flattening transformations!a theorem}
\begin{T}\label{nonflatteningnewresult}
Any transformation formed from operations consisting only of pure preparations, reversible transformations, and maximal results is non-flattening.
\end{T}
We will illustrate the proof of this with an example. Consider the transformation
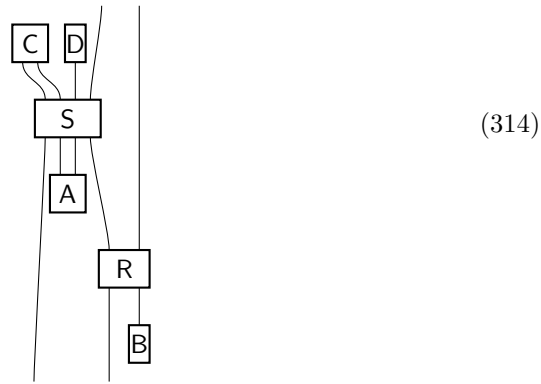
\begin{equation}
\begin{Diagram}{0}{0}
\Opbox[4]{S}{0,0}  \Opbox{R}{3,-8}  \Opbox[2]{A}{0,-4} \Opbox[1]{B}{3.8, -12} \Opbox[2]{C}{-2, 4} \Opbox[1]{D}{0.4, 4}
\thispoint{in1}{-1.8, -14}  \thispoint{in2}{2.2, -14}  \thispoint{out1}{1.8, 6} \thispoint{out2}{3.8, 6}
\wire{in1}{S}{1}{1} \wire{A}{S}{1}{2}  \wire{A}{S}{2}{3} \wire{in2}{R}{1}{1} \wire{B}{R}{1}{3}
\wire{R}{S}{1}{4}
\wire{S}{C}{1}{1}  \wire{S}{C}{2}{2}  \wire{S}{D}{3}{1} \wire{S}{out1}{4}{1}  \wire{R}{out2}{3}{1}
\end{Diagram}
\end{equation}
where $\mathsf A$ and $\mathsf B$ are pure preparations, $\mathsf R$ and $\mathsf S$ are reversible transformations, and $\mathsf C$ and $\mathsf D$ are maximal effects.  We can put this transformation in the form
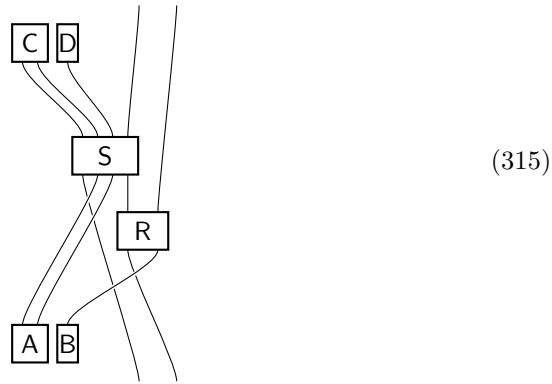
\begin{equation}
\begin{Diagram}{0}{0}
\Opbox[4]{S}{0,0}  \Opbox{R}{2,-4}  \Opbox[2]{A}{-4,-10} \Opbox[1]{B}{-2, -10} \Opbox[2]{C}{-4, 6} \Opbox[1]{D}{-2, 6}
\thispoint{in1}{1.8, -12}  \thispoint{in2}{3.8, -12}  \thispoint{out1}{1.8, 8} \thispoint{out2}{3.8, 8}
\wire{in1}{S}{1}{1} \wire{A}{S}{1}{2}  \wire{A}{S}{2}{3} \wire{in2}{R}{1}{1} \wire{B}{R}{1}{3}
\wire{R}{S}{1}{4}
\wire{S}{C}{1}{1}  \wire{S}{C}{2}{2}  \wire{S}{D}{3}{1} \wire{S}{out1}{4}{1}  \wire{R}{out2}{3}{1}
\end{Diagram}
\end{equation}
Here we have taken all the preparations to the bottom left, all the results to the top left, we have pulled the open inputs to the bottom right and the open outputs out to the top right.  We can put any transformation in this form simply by pulling all the preparations down to the left, all the open inputs down to the right, all the results up to the left, and all the open outputs up to the right.  In the middle we will have a bunch of reversible transformations.  We will now show that, so long as we have only pure preparations, reversible transformations, and maximal results, that any such transformation is equivalent to the following transformation
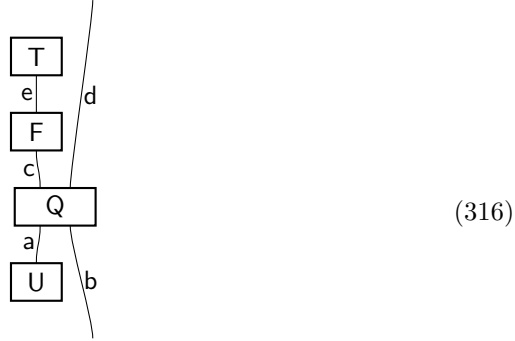
\begin{equation}\label{Qequivtrans}
\begin{Diagram}{0}{0}
\Opbox{U}{-1, -4}
\thispoint{in}{2,-7}
\Opbox[5]{Q}{0,0}
\Opbox{F}{-1, 4}
\Opbox{T}{-1,8}
\thispoint{out}{2,11}
\wire{U}{Q}{2}{2} \opsymbol{a}
\wire{in}{Q}{1}{4}  \otherside \opsymbol{b}
\wire{Q}{F}{2}{2} \opsymbol{c} \wire{F}{T}{2}{2} \opsymbol{e}
\wire{Q}{out}{4}{1} \otherside \opsymbol{d}
\end{Diagram}
\end{equation}
where $\mathsf{U}$ is a pure preparation, $\mathsf Q$ is a reversible transformation, $\mathsf F$ is a filter having capacity equal to one, and $\mathsf T$ is the deterministic result.  To see this first we note that, by {\bf T\ref{comppurestates}}, that if each of the components of a system has a pure preparation then the composite preparation is also pure.  Hence we can replace all the preparations with a single pure preparation for some (generally composite) system of type $\mathsf a$.   We can regard all the input wires on the right as constituting a single system (by {\bf T\ref{compsyst}}) which we represent by a system of type $\mathsf b$.  Similar remarks apply to the output wires (which we represent by a system of type $\mathsf d$).  We know by {\bf P2} that a result on a composite system is maximal if it is comprised of maximal results on each of the components.  Hence, we can represent the effect of all the maximal results on the upper left by a single maximal result (on a system, possibly composite, we take to be of type $\mathsf c$).   Any maximal result is equivalent to a special filter of capacity one followed by the deterministic result (this follows from the properties of special filters and the fact there is, according to {\bf P1}, only one maximal effect identifying a given pure state). The special filter is chosen so that it transmits unchanged only states proportional to the pure state that is identified by the given maximal result.  The system after the filter will be regarded as a system of type $\mathsf e$ (this is the filtered $\mathsf c$ type).  Since the filter has capacity equal to one, $N_\mathsf{e}=1$. Now we come to the bunch of reversible transformations in the middle. These can be regarded as a bunch of reversible transformations in parallel followed by another bunch of transformations in parallel and so on. The wires can be regarded as the identity transformation (which is a reversible transformation).  We know from {\bf T\ref{compositereversible}} that two or more reversible transformations in parallel constitute a reversible transformation. Further, the  sequential composition of reversible transformations gives rise to a reversible transformation itself.  Hence, the overall transformation is reversible.  We represent this by $\mathsf Q$.   If we send a non-flat set of states into the transformation for the system of type $\mathsf b$ shown in (\ref{Qequivtrans}) then it follows from {\bf P1} and {\bf P2} then we have a non-flat set of states for the system $\mathsf{ab}$.  This is because, as $U^{a_1}$ is pure we can, by {\bf P1}, find a maximal measurement for which it only gives rise to a single outcome, and then by {\bf P2}, this maximal measurement on $\mathsf a$ along with the maximal measurement on $\mathsf b$ with respect to which the non-flat set is spanning, constitute a maximal measurement on the composite with respect to which we must have a non-flat set of states.   A non-flat set of states must remain non-flat after a reversible transformation. Hence, a non-flat set emerges from the transformation $\mathsf Q$.  Next the states is subject to a filter on system $\mathsf c$ and the identity transformation on $\mathsf d$.  The identity transformation can be regarded as a filter (the \lq\lq do nothing" filter).  We know from {\bf T\ref{compfiltgeneral}} that two filters in parallel constitute a filter on the composite system.   Hence, we have a non-flat set of states after the filter.
We know that $N_\mathsf{e}=1$.  It follows from {\bf T\ref{KNr}} that $K_\mathsf{e}=1$.  This means that $e_5=1$ only (as the label $e_5$ must run from $1$ to $K_\mathsf{e}$).   Hence we can write the state, $E^{e_5d_4}$ after passing through the filter as $G^{e_5}H^{d_4}$ where $G^{e_5=1}=\text{constant}$ (the only component of $G^{e_5}$ is equal to some constant).  Since $K_\mathsf{e}=1$ a maximal measurement will only have one associated maximal effect which must, then, be equal to $\mathsf T$, the deterministic effect.  Hence, if the set of states for $\mathsf{ed}$ is non-flat, then by {\bf P1} and {\bf P2}, the set of states for $\mathsf d$, taken alone, must be non-flat (for similar reasons as just given for the $\mathsf{ab}$ case).
This proves {\bf T\ref{nonflatteningnewresult}}.

An immediate consequence of this theorem is the following.
\begin{T}\label{purepreparations}
Any preparation formed from operations consisting only of pure preparations, reversible transformations, and maximal results prepares a state proportional to a pure state.
\end{T}
Any preparation of this sort can be formed by sending a system prepared by a pure preparation into the type of transformation considered in proving {\bf T\ref{nonflatteningnewresult}}.  Since all non-flattening transformations are, by {\bf T\ref{nonflatimpliesnonmix}}, also non-mixing this implies that the system emerging must be in a state proportional to a pure state.

An immediate consequence of this theorem is that if we perform a maximal measurement on one component of a bipartite system prepared in a pure state (which may be entangled) then, for each outcome of this measurement, we obtain a pure state (up to normalization) on the other side.

\subsection{Entanglement, teleportation and entanglement swapping}

In this subsection and the following one we adopt the techniques of Chiribella, D'Ariano, and Perinotti to the present situation to show that $K_\mathsf{a}=N^2_\mathsf{a}$.   We will work with gebits of type $\mathsf{a}$.  We will work with a particular set of distinguishable states, $\{U^{a_1}[1], U^{a_1}[2] \}$, which we will think of as corresponding to the computational basis. Corresponding to this is the maximal measurement $\{\mathsf{U_{a_1}}[1], \mathsf{U_{a_1}}[2]\}$.   We define an \emph{equatorial state}\index{equatorial states} to be one that is pure and has probability $\frac{1}{2}$ associated with the two outcomes of the computational basis (if the computational states lie on the poles of the hypersphere then the equatorial states lie on the equator).   We select one particular equatorial state, $U^{a_1}[+]$, which we will use frequently (let $U^{a_1}[-]$ be the opposite state where the two together comprise a maximal distinguishable set).  The corresponding maximal effect is $U_{a_1}[+]$. We have
\begin{equation}
\text{Prob}(\mathsf{U_{a_1}}[1]\mathsf{U^{a_1}}[+]) = \text{Prob}(\mathsf{U_{a_1}}[2]\mathsf{U^{a_1}}[+]) = \frac{1}{2}
\end{equation}
(and similar equations for any other equatorial state).

A \emph{product state}\index{product state} is one that can be written as $A^{a_1}B^{a_2}$.  We will say that any pure state that cannot be written as a product state is an \emph{entangled pure state}\index{entangled states}.  Consider two gebits.  We can define the informational subset $S_{\{11,22\} }$ with respect the maximal measurement,
$\{\mathsf{U_{a_1}}[m] \mathsf{U_{a_2}}[n]: mn=11, 12, 21, 22 \}$ where $O(S_{\{11,22\} })= \{11,22\}$.   A pure state, $E^{a_1a_2}$,  in $S_{\{11,22\} }$ is entangled if
\begin{equation}\label{pEconditions}
p_{11}>0 ~~~\text{and} ~~~ p_{22} >0
\end{equation}
where
\begin{equation}\label{pEdefs}
p_{11}=\text{Prob}(\mathsf{U_{a_1}}[1]\mathsf{U_{a_2}}[1] \mathsf{E^{a_1a_2}}) ~~~\text{and}~~~ p_{22}= \text{Prob}(\mathsf{U_{a_1}}[2]\mathsf{U_{a_2}}[2] \mathsf{E^{a_1a_2}})
\end{equation}
This is equivalent to saying that $E^{a_1a_2}$ is not equal to either $U^{a_1}[1]U^{a_2}[1]$ or $U^{a_1}[2]U^{a_2}[2]$.  These are the only product states in $S_{\{11,22\}}$ since any other product states for these two gebits would have some probability associated with the 12 and/or 21 terms. Hence, this is equivalent to saying that the state is not a product state.

A state, $D^{a_1a_2}$, in $S_{\{11,22\}}$ will be said to be \emph{maximally entangled}\index{maximally entangled state} if it is pure and if
\begin{equation}
\text{Prob}(\mathsf{U_{a_1}}[1]\mathsf{U_{a_2}}[1] \mathsf{D^{a_1a_2}})= \text{Prob}(\mathsf{U_{a_1}}[2]\mathsf{U_{a_2}}[2] \mathsf{D^{a_1a_2}})= \frac{1}{2}
\end{equation}
Maximally entangled states will play an important role in this subsection and the next.  We will define a canonical maximally entangled state, $M^{a_1a_2}$ below.

One way to produce a entangled pure states is to use the permutation transformation $\mathsf{P}_\text{cnot}$ which effects the following permutation
\begin{equation}
\pi_\text{cnot} = \left( \begin{array}{c} 11 \rightarrow 11 \\ 12 \rightarrow 12 \\ 21 \rightarrow 22 \\ 22 \rightarrow 21 \end{array} \right)
\end{equation}
The inverse transformation $\mathsf{\tilde{P}}_\text{cnot}$ effects the same permutation (the inverse of a cnot is a cnot).
We note that
\begin{equation}\label{effectequations}
\begin{Diagram}[1.1]{0}{0}
\opbox{P}{0,0}  \opsymbol{P_\text{\negs\negs\negs cnot}}
\Opbox[1]{1}{0.8,-3}
\wire{1}{P}{1}{3}
\opbox[1]{1'}{-0.8,3} \opsymbol{1}
\opbox[1]{1''}{0.8,3} \opsymbol{1}
\wire{P}{1'}{1}{1}
\wire{P}{1''}{3}{1}
\inwire{P}{1}
\end{Diagram}
\equiv
\begin{Diagram}[1.1]{0}{0}
\Opbox[1]{1}{0,0}
\inwire{1}{1}
\end{Diagram}
~~~~\text{and}~~~~
\begin{Diagram}[1.1]{0}{0}
\opbox{P}{0,0}  \opsymbol{P_\text{\negs\negs\negs cnot}}
\Opbox[1]{1}{0.8,-3}
\wire{1}{P}{1}{3}
\opbox[1]{1'}{-0.8,3} \opsymbol{2}
\opbox[1]{1''}{0.8,3} \opsymbol{2}
\wire{P}{1'}{1}{1}
\wire{P}{1''}{3}{1}
\inwire{P}{1}
\end{Diagram}
\equiv
\begin{Diagram}[1.1]{0}{0}
\Opbox[1]{2}{0,0}
\inwire{2}{1}
\end{Diagram}
\end{equation}
(since we are working in a fixed computational basis, we can simply put \lq\lq 1" and \lq\lq +" rather than $\mathsf{U}[1]$ and $\mathsf{U}[+]$ inside the boxes).  The left equation follows virtue of the choice of permutation and the fact that, according to {\bf P1}, there is only one maximal effect identifying $U^{a_1}[1]$. The right equation follows for similar reasons.

It follows from (\ref{effectequations}), {\bf T\ref{comppurestates}}, and the fact that $\mathsf{P}_\text{cnot}$ is reversible, that
\begin{equation}
\begin{Diagram}[1.1]{0}{0}
\opbox{P}{0,0}  \opsymbol{P_\text{\negs\negs\negs cnot}}
\opbox[1]{+}{-0.8,-3} \opsymbol{A}
\Opbox[1]{1}{0.8,-3}
\wire{+}{P}{1}{1}
\wire{1}{P}{1}{3}
\outwire{P}{1}
\outwire{P}{3}
\end{Diagram}
\end{equation}
produces an entangled pure state as long as $A^{a_1}$ is not equal to either of the computational states.  If $A^{a_1}$ is equatorial then we produce a maximally entangled state. One example of this is the following preparation
\begin{equation}
\begin{Diagram}[1.1]{0}{0}
\opbox{P}{0,0}  \opsymbol{P_\text{\negs\negs\negs cnot}}
\opbox[1]{+}{-0.8,-3} \opsymbol{+}
\Opbox[1]{1}{0.8,-3}
\wire{+}{P}{1}{1}
\wire{1}{P}{1}{3}
\outwire{P}{1}
\outwire{P}{3}
\end{Diagram}
\end{equation}
We will take this to be our canonical maximally entangled\index{maximally entangled state!canonical} state which we will denote $M^{a_1a_2}$:
\begin{equation}\label{cannonicalM}
\begin{Diagram}[1.1]{0}{0}
\Opbox{M}{0,0}
\outwire{M}{1}
\outwire{M}{3}
\end{Diagram}
~~~:=~~~
\begin{Diagram}[1.1]{0}{0}
\opbox{P}{0,0}  \opsymbol{P_\text{\negs\negs\negs cnot}}
\opbox[1]{+}{-0.8,-3} \opsymbol{+}
\Opbox[1]{1}{0.8,-3}
\wire{+}{P}{1}{1}
\wire{1}{P}{1}{3}
\outwire{P}{1}
\outwire{P}{3}
\end{Diagram}
\end{equation}
This entangled state is identified by the maximal result
\begin{equation} \label{cannonicalMeffect}
\begin{Diagram}[1.1]{0}{0}
\Opbox{M}{0,0}
\inwire{M}{1}
\inwire{M}{3}
\end{Diagram}
~~~ := ~~~
\begin{Diagram}[1.1]{0}{0}
\opbox{Pt}{0,0}  \opsymbol{\tilde{P}_{\negs\negs\negs cnot}}
\inwire{Pt}{1} \inwire{Pt}{3}
\Opbox[1]{+}{-0.8,3} \wire{Pt}{+}{1}{1}
\Opbox[1]{1}{0.8,3} \wire{Pt}{1}{3}{1}
\end{Diagram}
\end{equation}
as $\mathsf{\tilde P}_\text{cnot}$ is the inverse of $\mathsf{P}_\text{cnot}$.  That this is a maximal result is clear since the measurement
\begin{equation}
\left\{
\begin{Diagram}[1.1]{0}{-0.3}
\opbox{Pt}{0,0}  \opsymbol{\tilde{P}_{\negs\negs\negs cnot}}
\inwire{Pt}{1} \inwire{Pt}{3}
\Opbox[1]{+}{-0.8,3} \wire{Pt}{+}{1}{1}
\Opbox[1]{1}{0.8,3} \wire{Pt}{1}{3}{1}
\end{Diagram}
,
\begin{Diagram}[1.1]{0}{-0.30}
\opbox{Pt}{0,0}  \opsymbol{\tilde{P}_{\negs\negs\negs cnot}}
\inwire{Pt}{1} \inwire{Pt}{3}
\Opbox[1]{-}{-0.8,3} \wire{Pt}{-}{1}{1}
\Opbox[1]{1}{0.8,3} \wire{Pt}{1}{3}{1}
\end{Diagram}
,
\begin{Diagram}[1.1]{0}{-0.3}
\opbox{Pt}{0,0}  \opsymbol{\tilde{P}_{\negs\negs\negs cnot}}
\inwire{Pt}{1} \inwire{Pt}{3}
\Opbox[1]{+}{-0.8,3} \wire{Pt}{+}{1}{1}
\Opbox[1]{2}{0.8,3} \wire{Pt}{2}{3}{1}
\end{Diagram}
,
\begin{Diagram}[1.1]{0}{-0.3}
\opbox{Pt}{0,0}  \opsymbol{\tilde{P}_{\negs\negs\negs cnot}}
\inwire{Pt}{1} \inwire{Pt}{3}
\Opbox[1]{-}{-0.8,3} \wire{Pt}{-}{1}{1}
\Opbox[1]{2}{0.8,3} \wire{Pt}{2}{3}{1}
\end{Diagram}
\right\}
\end{equation}
distinguishes the four states
\begin{equation}
\left\{
\begin{Diagram}[1.1]{0}{0.3}
\opbox{P}{0,0}  \opsymbol{P_\text{\negs\negs\negs cnot}}
\opbox[1]{+}{-0.8,-3} \opsymbol{+}
\Opbox[1]{1}{0.8,-3}
\wire{+}{P}{1}{1}
\wire{1}{P}{1}{3}
\outwire{P}{1}
\outwire{P}{3}
\end{Diagram}
,
\begin{Diagram}[1.1]{0}{0.3}
\opbox{P}{0,0}  \opsymbol{P_\text{\negs\negs\negs cnot}}
\opbox[1]{+}{-0.8,-3} \opsymbol{-}
\Opbox[1]{1}{0.8,-3}
\wire{+}{P}{1}{1}
\wire{1}{P}{1}{3}
\outwire{P}{1}
\outwire{P}{3}
\end{Diagram}
,
\begin{Diagram}[1.1]{0}{0.3}
\opbox{P}{0,0}  \opsymbol{P_\text{\negs\negs\negs cnot}}
\opbox[1]{+}{-0.8,-3} \opsymbol{+}
\Opbox[1]{2}{0.8,-3}
\wire{+}{P}{1}{1}
\wire{2}{P}{1}{3}
\outwire{P}{1}
\outwire{P}{3}
\end{Diagram}
,
\begin{Diagram}[1.1]{0}{0.3}
\opbox{P}{0,0}  \opsymbol{P_\text{\negs\negs\negs cnot}}
\opbox[1]{+}{-0.8,-3} \opsymbol{-}
\Opbox[1]{2}{0.8,-3}
\wire{+}{P}{1}{1}
\wire{1}{P}{1}{3}
\outwire{P}{1}
\outwire{P}{3}
\end{Diagram}
\right\}
\end{equation}
Hence, the measurement is maximal and $\mathsf M$ is a maximal result (which we will take to be canonical).

We are now in a position to prove the following theorem
\begin{T}\label{teleportationTone}
Consider the gebit preparation
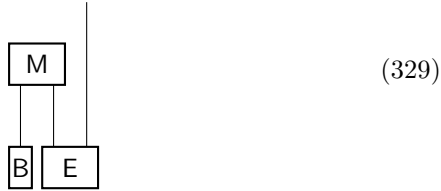
\begin{equation}\label{teleportationone}
\begin{Diagram}[1.1]{0}{0}
\Opbox{M}{0,0}
\Opbox{E}{1.6,-5}
\Opbox[1]{B}{-0.8,-5}
\wire{B}{M}{1}{1}
\wire{E}{M}{1}{3}
\thispoint{out}{2.4,3}
\wire{E}{out}{3}{1}
\end{Diagram}
\end{equation}
where $\mathsf E$ is an entangled pure state in $S_{\{11,22\} }$ and $\mathsf M$ is the canonical maximal result defined in (\ref{cannonicalMeffect}) above.  We can use this preparation to prepare a state proportional to any pure state by making an appropriate choice of preparation $\mathsf B$.
\end{T}
It follows from {\bf T\ref{nonflatteningnewresult}} that, if we send a non-flat set of states in for $B^{a_1}$, then we must get a non-flat set of states out (call the corresponding output states $A^{a_3}$.  Further, it follows from {\bf T\ref{nonflatimpliesnonmix}} that, if $B^{a_1}$ is pure, then $A^{a_3}$ must be proportional to a pure state.   It follows from the fact that $E^{b_2c_3}$ is entangled that the smallest system that can support the output states is a gebit.  To see this we will note that
\begin{equation}\label{Mcondeffects}
\begin{Diagram}[1.1]{0}{0}
\Opbox{M}{0,0}
\Opbox[1]{1}{-0.8,-3}\wire{1}{M}{1}{1}
\inwire{M}{3}
\end{Diagram}
\equiv \frac{1}{2}
\begin{Diagram}[1.1]{0}{0}
\Opbox[1]{1}{0,0}
\inwire{1}{1}
\end{Diagram}
~~~~\text{and}~~~~
\begin{Diagram}[1.1]{0}{0}
\Opbox{M}{0,0}
\Opbox[1]{2}{-0.8,-3}\wire{2}{M}{1}{1}
\inwire{M}{3}
\end{Diagram}
\equiv \frac{1}{2}
\begin{Diagram}[1.1]{0}{0}
\Opbox[1]{2}{0,0}
\inwire{2}{1}
\end{Diagram}
\end{equation}
where the $\frac{1}{2}$ in each equation indicates that the effects corresponding to these results are proportional with constant of proportionality equal to $\frac{1}{2}$. To prove the equation on the left consider sending in $U^{a_1}[1]$ and then $U^{a_1}[2]$.  It follows from the definition of the $\mathsf M$ effect in (\ref{cannonicalMeffect}) that when we send in $U^{a_1}[1]$ the probability is $\frac{1}{2}$ and when we send in $U^{a_1}[2]$ the probability is $0$.  The equation on the left then follows from {\bf T\ref{gebitzero}}.  The equation on the right follows by similar reasoning.  We can also prove
\begin{equation}\label{Econdstates}
\begin{Diagram}[1.1]{0}{-0.5}
\Opbox{E}{0,0}
\opbox[1]{left}{-0.8,3} \opsymbol{1}
\wire{E}{left}{1}{1}
\outwire{E}{3}
\end{Diagram}
~~\equiv~~
p_{11} ~
\begin{Diagram}[1.1]{0}{-0.5}
\opbox[1]{out}{0,0}  \opsymbol{1}
\outwire{out}{1}
\end{Diagram}
~~~~~~~\text{and}~~~~~~~
\begin{Diagram}[1.1]{0}{-0.5}
\Opbox{E}{0,0}
\opbox[1]{left}{-0.8,3} \opsymbol{2}
\wire{E}{left}{1}{1}
\outwire{E}{3}
\end{Diagram}
~~\equiv~~
p_{22} ~
\begin{Diagram}[1.1]{0}{-0.5}
\opbox[1]{out}{0,0}  \opsymbol{2}
\outwire{out}{1}
\end{Diagram}
\end{equation}
where $p_{11}$ and $p_{22}$ are defined in (\ref{pEdefs}).  To prove the equation on the left, we note that if put the result $\mathsf{U_{a_2}}[2]$ on the output of the LHS of this equation, then we must get zero since $E^{a_2a_3}$ is in $S_{\{11,22\} }$.  Hence the equation follows from {\bf T\ref{gebitstatezero}} (and the constant of proportionallity, $p_{11}$, follows from considering the definition in (\ref{pEdefs})).  The equation on the right follows by similar reasoning. It follows from (\ref{Mcondeffects}, \ref{Econdstates}) that if we put $B^{a_1}=U^{a_1}[1]$ in (\ref{teleportationone}) we get $\frac{p_{11}}{2} U^{a_3}[1]$ out, and if we put $B^{a_1}=U^{a_1}[1]$ in we get $\frac{p_{22}}{2} U^{a_3}[2]$ out.  Since the state $E^{a_2a_3}$ is entangled we have, by (\ref{pEdefs}), that $p_{11}$ and $p_{22}$ are non-zero.  Hence the smallest system that can support the preparation in (\ref{teleportationone}) for all pure inputs, $B^{a_1}$, is a gebit. We established that these output states must be non-flat.  The input states correspond to the full set of pure states on a hypersphere.  The output set of states correspond to states that are proportional to pure states. Hence have a linear transformation on a hypersphere of input states, $B^{a_1}$, to a set of output states, $A^{a_3}$, having the same dimension.
Under such a linear transformation, a hypersphere can only transform to a hyper-elipsoid of the same dimension. Hence, there exists an input state $B^{a_1}$ giving rise to any point on this output hyper-elipsoid. These points on the hyper-elipsoid correspond to states that are proportional to pure states.  For every point on the output hyper-elipsoid there must be a corresponding point on the input hypersphere.   This proves {\bf T\ref{teleportationTone}}.

We now prove the following theorem
\begin{T}\label{PPPPprepT}
The preparation
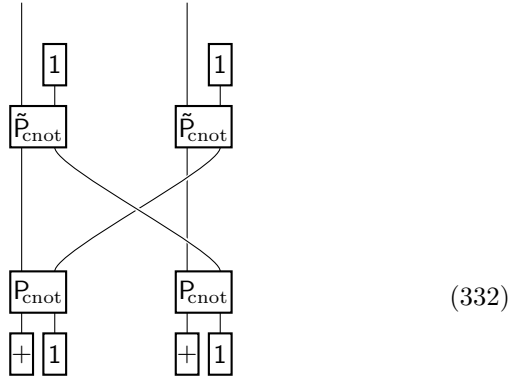
\begin{equation}\label{PPPPprep}
\begin{Diagram}[1.1]{0}{0}
\opbox{PL}{0,0} \opsymbol{P_\text{\negs\negs\negs cnot}}
\opbox{PR}{8,0} \opsymbol{P_\text{\negs\negs\negs cnot}}
\opbox{PtL}{0,8} \opsymbol{\tilde P_\text{\negs\negs\negs cnot}}
\opbox{PtR}{8,8} \opsymbol{\tilde P_\text{\negs\negs\negs cnot}}
\wire{PL}{PtL}{1}{1} \wire{PR}{PtR}{1}{1}
\wire{PL}{PtR}{3}{3} \wire{PR}{PtL}{3}{3}
\opbox[1]{+L}{-0.8,-3} \opsymbol{+} \wire{+L}{PL}{1}{1}
\opbox[1]{1L}{0.8,-3} \opsymbol{1}  \wire{1L}{PL}{1}{3}
\opbox[1]{+R}{7.2,-3} \opsymbol{+} \wire{+R}{PR}{1}{1}
\opbox[1]{1R}{8.8,-3} \opsymbol{1}  \wire{1R}{PR}{1}{3}
\opbox[1]{1Lt}{0.8,11} \opsymbol{1} \wire{PtL}{1Lt}{3}{1}
\opbox[1]{1Rt}{8.8,11} \opsymbol{1} \wire{PtR}{1Rt}{3}{1}
\thispoint{Lout}{-0.8,14} \wire{PtL}{Lout}{1}{1}
\thispoint{Rout}{7.2,14} \wire{PtR}{Rout}{1}{1}
\end{Diagram}
\end{equation}
prepares a state proportional to a maximally entangled state in $S_{\{11,22\} }$ where the constant of proportinality is $\frac{1}{2}$.
\end{T}
To see this note first that the set of results
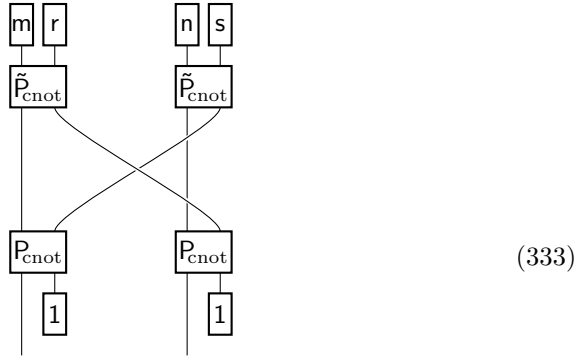
\begin{equation}\label{PPPPmaxmeas}
\begin{Diagram}[1.1]{0}{0}
\opbox{PL}{0,0} \opsymbol{P_\text{\negs\negs\negs cnot}}
\opbox{PR}{8,0} \opsymbol{P_\text{\negs\negs\negs cnot}}
\opbox{PtL}{0,8} \opsymbol{\tilde P_\text{\negs\negs\negs cnot}}
\opbox{PtR}{8,8} \opsymbol{\tilde P_\text{\negs\negs\negs cnot}}
\wire{PL}{PtL}{1}{1} \wire{PR}{PtR}{1}{1}
\wire{PL}{PtR}{3}{3} \wire{PR}{PtL}{3}{3}
\thispoint{+L}{-0.8,-5}  \wire{+L}{PL}{1}{1}
\opbox[1]{1L}{0.8,-3} \opsymbol{1}  \wire{1L}{PL}{1}{3}
\thispoint{+R}{7.2,-5}  \wire{+R}{PR}{1}{1}
\opbox[1]{1R}{8.8,-3} \opsymbol{1}  \wire{1R}{PR}{1}{3}
\opbox[1]{1Lt}{0.8,11} \opsymbol{r} \wire{PtL}{1Lt}{3}{1}
\opbox[1]{1Rt}{8.8,11} \opsymbol{s} \wire{PtR}{1Rt}{3}{1}
\opbox[1]{Lout}{-0.8,11} \opsymbol{m}\wire{PtL}{Lout}{1}{1}
\opbox[1]{Rout}{7.2,11}  \opsymbol{n}\wire{PtR}{Rout}{1}{1}
\end{Diagram}
\end{equation}
constitute a maximal measurement on the input gebits in the computational basis ($11,12,21,22$).  This is clear because each of the input states $U^{a_1}[i]U^{a_2}[j]$ into this measurement must give rise to a different outcome for $mnrs$ by virtue of the fact that we have permutations in the computational basis.  Now consider the transformation
\begin{equation}
\begin{Diagram}[1.1]{0}{0}
\opbox{N}{0,0} \opsymbol{R}
\inwire{N}{1} \inwire{N}{3}
\outwire{N}{1} \outwire{N}{3}
\end{Diagram}
~~~:=~~~
\begin{Diagram}[1.1]{0}{-0.9}
\opbox{PL}{0,0} \opsymbol{P_\text{\negs\negs\negs cnot}}
\opbox{PR}{8,0} \opsymbol{P_\text{\negs\negs\negs cnot}}
\opbox{PtL}{0,8} \opsymbol{\tilde P_\text{\negs\negs\negs cnot}}
\opbox{PtR}{8,8} \opsymbol{\tilde P_\text{\negs\negs\negs cnot}}
\wire{PL}{PtL}{1}{1} \wire{PR}{PtR}{1}{1}
\wire{PL}{PtR}{3}{3} \wire{PR}{PtL}{3}{3}
\thispoint{+L}{-0.8,-5}  \wire{+L}{PL}{1}{1}
\opbox[1]{1L}{0.8,-3} \opsymbol{1}  \wire{1L}{PL}{1}{3}
\thispoint{+R}{7.2,-5}  \wire{+R}{PR}{1}{1}
\opbox[1]{1R}{8.8,-3} \opsymbol{1}  \wire{1R}{PR}{1}{3}
\opbox[1]{1Lt}{0.8,11} \opsymbol{1} \wire{PtL}{1Lt}{3}{1}
\opbox[1]{1Rt}{8.8,11} \opsymbol{1} \wire{PtR}{1Rt}{3}{1}
\thispoint{Lout}{-0.8,13} \wire{PtL}{Lout}{1}{1}
\thispoint{Rout}{7.2,13}\wire{PtR}{Rout}{1}{1}
\end{Diagram}
\end{equation}
By following through the effect of the permutations (recall that $\pi_\text{cnot}$ is equal to its own inverse) we see that
\begin{equation}\label{Rtransprobs}
\text{Prob}\left(
\begin{Diagram}[1.1]{0}{0}
\Opbox{R}{0,0}
\opbox[1]{i}{-0.8,-3} \opsymbol{\mathnormal{i}} \wire{i}{R}{1}{1}
\opbox[1]{j}{0.8,-3}  \opsymbol{\mathnormal{j}} \wire{j}{R}{1}{3}
\opbox[1]{m}{-0.8,3} \opsymbol{\mathnormal{m}} \wire{R}{m}{1}{1}
\opbox[1]{n}{0.8,3} \opsymbol{\mathnormal{n}}  \wire{R}{n}{3}{1}
\end{Diagram}
\right)
~~ = \left( \begin{array}{cc} 1 & \text{if}~~ ij=mn=11 \\ 1 & \text{if}~~ ij=mn=22 \\ 0 & \text{else} \end{array} \right)
\end{equation}
This probability is zero whenever $m$ is not equal to $n$.  Since (\ref{PPPPmaxmeas}) constitute the effects of a maximal measurement, it follows from {\bf T\ref{redundantoutcomes}}  that we must have probability zero for $m$ and $n$ to be different whatever state we send into transformation $\mathsf R$.  And hence it follows that this transformation $\mathsf R$ outputs states in $S_{\{ 11, 22\} }$.  This means we prepare a gebit (as $|O(S_{\{ 11, 22\} }|=2$).  It follows from {\bf P1} and (\ref{Rtransprobs}) that
\begin{equation}
\begin{Diagram}[1.1]{0}{0}
\Opbox{R}{0,0} \inwire{R}{1}\inwire{R}{3}
\opbox[1]{m}{-0.8,3} \opsymbol{\mathnormal{1}} \wire{R}{m}{1}{1}
\opbox[1]{n}{0.8,3} \opsymbol{\mathnormal{1}}  \wire{R}{n}{3}{1}
\end{Diagram}
~\equiv ~
\begin{Diagram}[1.1]{0}{0}
\opbox[1]{m}{0,0} \opsymbol{1} \inwire{m}{1}
\opbox[1]{n}{1.6, 0} \opsymbol{1} \inwire{n}{1}
\end{Diagram}
~~~~~\text{and}~~~~~
\begin{Diagram}[1.1]{0}{0}
\Opbox{R}{0,0} \inwire{R}{1}\inwire{R}{3}
\opbox[1]{m}{-0.8,3} \opsymbol{\mathnormal{2}} \wire{R}{m}{1}{1}
\opbox[1]{n}{0.8,3} \opsymbol{\mathnormal{2}}  \wire{R}{n}{3}{1}
\end{Diagram}
~\equiv ~
\begin{Diagram}[1.1]{0}{0}
\opbox[1]{m}{0,0} \opsymbol{2} \inwire{m}{1}
\opbox[1]{n}{1.6, 0} \opsymbol{2} \inwire{n}{1}
\end{Diagram}
\end{equation}
Hence,
\begin{equation}\label{equalprobsforprep}
\text{Prob}\left(
\begin{Diagram}[1.1]{0}{0}
\Opbox{R}{0,0}
\opbox[1]{i}{-0.8,-3} \opsymbol{\mathnormal{+}} \wire{i}{R}{1}{1}
\opbox[1]{j}{0.8,-3}  \opsymbol{\mathnormal{+}} \wire{j}{R}{1}{3}
\opbox[1]{m}{-0.8,3} \opsymbol{\mathnormal{1}} \wire{R}{m}{1}{1}
\opbox[1]{n}{0.8,3} \opsymbol{\mathnormal{1}}  \wire{R}{n}{3}{1}
\end{Diagram}
\right)
=
\text{Prob}\left(
\begin{Diagram}[1.1]{0}{0}
\Opbox{R}{0,0}
\opbox[1]{i}{-0.8,-3} \opsymbol{\mathnormal{+}} \wire{i}{R}{1}{1}
\opbox[1]{j}{0.8,-3}  \opsymbol{\mathnormal{+}} \wire{j}{R}{1}{3}
\opbox[1]{m}{-0.8,3} \opsymbol{\mathnormal{2}} \wire{R}{m}{1}{1}
\opbox[1]{n}{0.8,3} \opsymbol{\mathnormal{2}}  \wire{R}{n}{3}{1}
\end{Diagram}
\right)
= \frac{1}{4}
\end{equation}
We know by {\bf T\ref{purepreparations}} that the state prepared by (\ref{PPPPprep}) is proportional to a pure state.  Hence it follows from (\ref{equalprobsforprep}) that the state prepared in (\ref{PPPPprep}) is proportional to a maximally entangled state with constant of proportinality $\frac{1}{2}$.

Next we will prove the following:
\begin{T}\label{entangswapT}
{\bf Entanglement swapping.} \index{entanglement swapping}There exists a choice of pure state $B^{a_1}$ such that the following holds
\begin{equation}\label{entangswap}
\begin{Diagram}[1.1]{0}{-2}
\opbox{ML}{0,0} \opsymbol{M}
\opbox{MR}{8,0} \opsymbol{M}
\opbox{Pt}{4,5} \opsymbol{\tilde P_\text{\negs\negs\negs cnot}}
\wire{ML}{Pt}{3}{3} \wire{MR}{Pt}{1}{1}
\thispoint{Lout}{-0.8,14} \wire{ML}{Lout}{1}{1}
\thispoint{Rout}{8.8,14} \wire{MR}{Rout}{3}{1}
\Opbox[1]{1}{4.8,8} \wire{Pt}{1}{3}{1}
\opbox{Ptt}{2.4,8} \opsymbol{M} \wire{Pt}{Ptt}{1}{3}
\Opbox[1]{B}{1.6, 5} \wire{B}{Ptt}{1}{1}
%\Opbox[1]{+}{1.6,11} \wire{Ptt}{+}{1}{1}
%\opbox[1]{one}{3.2,11} \opsymbol{1} \wire{Ptt}{one}{3}{1}
\end{Diagram}
~~~~~\equiv~~~~ \frac{1}{8} ~
\begin{Diagram}[1.1]{0}{-0.2}
\Opbox{M}{0,0}
\outwire{M}{1} \outwire{M}{3}
\end{Diagram}
\end{equation}
Further, $B^{a_1}$, must be equatorial.
\end{T}
The preparation on the LHS of (\ref{entangswap}) must be in $S_{\{11,22\} }$.  To see this we note that
\begin{equation}\label{Monetohalfone}
\begin{Diagram}[1.1]{0}{0}
\Opbox{M}{0,0} \Opbox[1]{1}{-0.8,3} \wire{M}{1}{1}{1} \outwire{M}{3}
\end{Diagram}
~~=~~
\begin{Diagram}[1.1]{0}{0}
\Opbox{M}{0,0} \Opbox[1]{1}{0.8,3} \wire{M}{1}{3}{1} \outwire{M}{1}
\end{Diagram}
~~=~~\frac{1}{2} ~
\begin{Diagram}[1.1]{0}{0}
\Opbox[1]{1}{0,0} \outwire{1}{1}
\end{Diagram}
\end{equation}
and
\begin{equation}\label{Mtwotohalftwo}
\begin{Diagram}[1.1]{0}{0}
\Opbox{M}{0,0} \Opbox[1]{2}{-0.8,3} \wire{M}{2}{1}{1} \outwire{M}{3}
\end{Diagram}
~~=~~
\begin{Diagram}[1.1]{0}{0}
\Opbox{M}{0,0} \Opbox[1]{2}{0.8,3} \wire{M}{2}{3}{1} \outwire{M}{1}
\end{Diagram}
~~=~~\frac{1}{2} ~
\begin{Diagram}[1.1]{0}{0}
\Opbox[1]{2}{0,0} \outwire{2}{1}
\end{Diagram}
\end{equation}
This follows from {\bf T\ref{gebitstatezero}} and the fact that $\text{Prob}(\mathsf{M^{a_1a_2}U_{a_1}}[m]\mathsf{U_{a_2}}[n])=\frac{1}{2}\delta_{mn}$ as $M^{a_1a_2}$ is a maximally entangled state in $S_{\{11,22\} }$. Hence, regardless of what we choose for the state $\mathsf B^a$, we must have
\begin{equation}
\text{Prob}\left(
\begin{Diagram}[1.1]{0}{-2}
\opbox{ML}{0,0} \opsymbol{M}
\opbox{MR}{8,0} \opsymbol{M}
\opbox{Pt}{4,5} \opsymbol{\tilde P_\text{\negs\negs\negs cnot}}
\wire{ML}{Pt}{3}{3} \wire{MR}{Pt}{1}{1}
\opbox[1]{Lout}{-0.8,11} \opsymbol{1} \wire{ML}{Lout}{1}{1}
\opbox[1]{Rout}{8.8,11} \opsymbol {2}\wire{MR}{Rout}{3}{1}
\Opbox[1]{1}{4.8,8} \wire{Pt}{1}{3}{1}
\opbox{Ptt}{2.4,8} \opsymbol{M} \wire{Pt}{Ptt}{1}{3}
\Opbox[1]{B}{1.6, 5} \wire{B}{Ptt}{1}{1}
%\Opbox[1]{+}{1.6,11} \wire{Ptt}{+}{1}{1}
%\opbox[1]{one}{3.2,11} \opsymbol{1} \wire{Ptt}{one}{3}{1}
\end{Diagram}
\right)
~~=~~
\text{Prob}\left(
\begin{Diagram}[1.1]{0}{-2}
\opbox{ML}{0,0} \opsymbol{M}
\opbox{MR}{8,0} \opsymbol{M}
\opbox{Pt}{4,5} \opsymbol{\tilde P_\text{\negs\negs\negs cnot}}
\wire{ML}{Pt}{3}{3} \wire{MR}{Pt}{1}{1}
\opbox[1]{Lout}{-0.8,11} \opsymbol{2} \wire{ML}{Lout}{1}{1}
\opbox[1]{Rout}{8.8,11} \opsymbol {1}\wire{MR}{Rout}{3}{1}
\Opbox[1]{1}{4.8,8} \wire{Pt}{1}{3}{1}
\opbox{Ptt}{2.4,8} \opsymbol{M} \wire{Pt}{Ptt}{1}{3}
\Opbox[1]{B}{1.6, 5} \wire{B}{Ptt}{1}{1}
%\Opbox[1]{+}{1.6,11} \wire{Ptt}{+}{1}{1}
%\opbox[1]{one}{3.2,11} \opsymbol{1} \wire{Ptt}{one}{3}{1}
\end{Diagram}
\right)
~~=~0
\end{equation}
by virtue of the choice of permutation associated with $\mathsf{P}_\text{cnot}$.  This proves that the preparation on the LHS of (\ref{entangswap}) is in $S_{\{11,22\} }$.  It follows from {\bf T\ref{purepreparations}} that this state is proportional to a pure state.
Given that the state on the LHS of (\ref{entangswap}) is pure and in $S_{\{11,22\} }$, it follows from {\bf T\ref{teleportationTone}}, {\bf T\ref{PPPPprepT}}, and the definition of the canonical maximally entangled state, $M^{a_1b_2}$, in (\ref{cannonicalM}) that there exists a state $B^a$ such that
\begin{equation}\label{PPPPprepBE}
\begin{Diagram}[1.1]{0}{-1.6}
\opbox{PL}{0,0} \opsymbol{M}
\opbox{PR}{8,0} \opsymbol{M}
\opbox{PtL}{0,8} \opsymbol{\tilde P_\text{\negs\negs\negs cnot}}
\opbox{PtR}{8,8} \opsymbol{\tilde P_\text{\negs\negs\negs cnot}}
\wire{PL}{PtL}{1}{1} \wire{PR}{PtR}{1}{1}
\wire{PL}{PtR}{3}{3} \wire{PR}{PtL}{3}{3}
%
%\opbox[1]{+L}{-0.8,-3} \opsymbol{+} \wire{+L}{PL}{1}{1}
%\opbox[1]{1L}{0.8,-3} \opsymbol{1}  \wire{1L}{PL}{1}{3}
%
%\opbox[1]{+R}{7.2,-3} \opsymbol{+} \wire{+R}{PR}{1}{1}
%\opbox[1]{1R}{8.8,-3} \opsymbol{1}  \wire{1R}{PR}{1}{3}
%
\opbox[1]{1Lt}{0.8,11} \opsymbol{1} \wire{PtL}{1Lt}{3}{1}
\opbox[1]{1Rt}{8.8,11} \opsymbol{1} \wire{PtR}{1Rt}{3}{1}
\thispoint{Lout}{-0.8,14} \wire{PtL}{Lout}{1}{1}
\opbox{E}{6.4,11} \opsymbol{M} \wire{PtR}{E}{1}{3}
\Opbox[1]{B}{5.6,8}
\wire{B}{E}{1}{1}
\end{Diagram}
~~~\equiv~~~ \gamma ~
\begin{Diagram}[1.1]{0}{-0.1}
\Opbox[1]{+}{0,0} \outwire{+}{1}
\end{Diagram}
\end{equation}
where $\gamma$ is to be determined.  In obtaining (\ref{PPPPprepBE}), the preparation in {\bf T\ref{PPPPprepT}} plays the role of $\mathsf{E^{a_1a_2}}$ in {\bf T\ref{teleportationTone}}.  These circuit diagrams are interpreted graphically.  We can rewrite the above equation as
\begin{equation}\label{stretch}
\begin{Diagram}[1.1]{0}{-1.6}
\opbox{PL}{0,2} \opsymbol{M}
\opbox{PR}{15,2} \opsymbol{M}
\opbox{PtL}{8,18} \opsymbol{\tilde P_\text{\negs\negs\negs cnot}}
\opbox{PtR}{8,8} \opsymbol{\tilde P_\text{\negs\negs\negs cnot}}
\wire{PL}{PtL}{1}{1} \wire{PR}{PtR}{1}{1}
\wire{PL}{PtR}{3}{3} \wire{PR}{PtL}{3}{3}
%
%\opbox[1]{+L}{-0.8,-3} \opsymbol{+} \wire{+L}{PL}{1}{1}
%\opbox[1]{1L}{0.8,-3} \opsymbol{1}  \wire{1L}{PL}{1}{3}
%
%\opbox[1]{+R}{7.2,-3} \opsymbol{+} \wire{+R}{PR}{1}{1}
%\opbox[1]{1R}{8.8,-3} \opsymbol{1}  \wire{1R}{PR}{1}{3}
%
\opbox[1]{1Lt}{8.8,21} \opsymbol{1} \wire{PtL}{1Lt}{3}{1}
\opbox[1]{1Rt}{8.8,11} \opsymbol{1} \wire{PtR}{1Rt}{3}{1}
\thispoint{Lout}{7.2,24} \wire{PtL}{Lout}{1}{1}
\opbox{E}{6.4,11} \opsymbol{M} \wire{PtR}{E}{1}{3}
\Opbox[1]{B}{5.6,8}
\wire{B}{E}{1}{1}
\begin{foliation}{3}{13}
\Startfoliate{PL}{PtL}{1}{1}{0.35} \Finishfoliate{PR}{PtL}{3}{3}{0.35}
\end{foliation}
\end{Diagram}
~~~\equiv~~~ \gamma ~
\begin{Diagram}[1.1]{0}{-0.1}
\Opbox[1]{+}{0,0} \outwire{+}{1}
\end{Diagram}
\end{equation}
We have already established that the preparation up to the dotted line is a pure state in $S_{\{11,22\} }$.  The maximal results,
\begin{equation}\label{plusminusPmaxmeas}
\begin{Diagram}[1.1]{0}{0}
\opbox{Pt}{0,0} \opsymbol{\tilde P_\text{\negs\negs\negs cnot}}
\inwire{Pt}{1} \inwire{Pt}{3}
\opbox[1]{m}{-0.8, 3} \opsymbol{1}\wire{Pt}{m}{1}{1}
\opbox[1]{n}{0.8,3} \opsymbol{1} \wire{Pt}{n}{3}{1}
\end{Diagram}
~~~~~~~
\begin{Diagram}[1.1]{0}{0}
\opbox{Pt}{0,0} \opsymbol{\tilde P_\text{\negs\negs\negs cnot}}
\inwire{Pt}{1} \inwire{Pt}{3}
\opbox[1]{m}{-0.8, 3} \opsymbol{2} \wire{Pt}{m}{1}{1}
\opbox[1]{n}{0.8,3} \opsymbol{1} \wire{Pt}{n}{3}{1}
\end{Diagram}
\end{equation}
form a maximal measurement for the gebit consisting of states in $S_{\{11,22\} }$ as they distinguish the 11 and 22 states.  It follows from {\bf T\ref{deteffect}} that, for gebits in $S_{\{11,22\} }$ inputted into the transformation
\begin{equation}
\begin{Diagram}[1.1]{0}{0}
\opbox{Pt}{0,0} \opsymbol{\tilde P_\text{\negs\negs\negs cnot}}
\inwire{Pt}{1} \inwire{Pt}{3}
\outwire{Pt}{1}
\Opbox[1]{n}{0.8,3} \wire{Pt}{n}{3}{1}
\end{Diagram}
\end{equation}
then we always get $n=1$.  Hence, it follows that the maximal results
\begin{equation}
\begin{Diagram}[1.1]{0}{0}
\opbox{Pt}{0,0} \opsymbol{\tilde P_\text{\negs\negs\negs cnot}}
\inwire{Pt}{1} \inwire{Pt}{3}
\opbox[1]{m}{-0.8, 3} \opsymbol{+}\wire{Pt}{m}{1}{1}
\opbox[1]{n}{0.8,3} \opsymbol{1} \wire{Pt}{n}{3}{1}
\end{Diagram}
~~~~~~~
\begin{Diagram}[1.1]{0}{0}
\opbox{Pt}{0,0} \opsymbol{\tilde P_\text{\negs\negs\negs cnot}}
\inwire{Pt}{1} \inwire{Pt}{3}
\opbox[1]{m}{-0.8, 3} \opsymbol{-} \wire{Pt}{m}{1}{1}
\opbox[1]{n}{0.8,3} \opsymbol{1} \wire{Pt}{n}{3}{1}
\end{Diagram}
\end{equation}
form a maximal measurement in $S_{\{11,22\} }$. If we were to place the maximal effect on the right after the dotted line in (\ref{stretch}) it follows from this equation that we would get probability zero (as $U^{a_1}[+]U_{a_1}[-]=0$).  Consequently (\ref{entangswap}) follows from (\ref{stretch}) and {\bf T\ref{gebitstatezero}} though we have yet to demonstrate that the constant of proportionality is $\frac{1}{8}$.  It is easy to see that
\begin{equation}
\text{Prob}\left(
\begin{Diagram}[1.1]{0}{0}
\Opbox[1]{+}{-0.8,3}  \Opbox[1]{1}{0.8,3}
        \opbox{Pt}{0,0}          \opsymbol{\tilde P_{\negs\negs\negs cnot}}
\Opbox[1]{m}{-0.8,-3}  \Opbox[1]{n}{0.8,-3}
\wire{m}{Pt}{1}{1} \wire{n}{Pt}{1}{3}
\wire{Pt}{+}{1}{1} \wire{Pt}{1}{3}{1}
\end{Diagram}
\right) ~~=~ \left(
\begin{array}{l} \frac{1}{2} ~ \text{if}~m=n \\
                      0      ~ \text{if}~m\not=n \end{array} \right)
\end{equation}
From this and {\bf T\ref{gebiteffects}} it follows that
\begin{equation}\label{Mntohalfn}
\begin{Diagram}[1.1]{0}{0}
\Opbox{M}{0,0}
\Opbox[1]{n}{0.8,-3}
\wire{n}{M}{1}{3}
\inwire{M}{1}
\end{Diagram}
~~=~\frac{1}{2}
\begin{Diagram}[1.1]{0}{0}
\Opbox[1]{n}{0,0}
\inwire{n}{1}
\end{Diagram}
\end{equation}
Using (\ref{Monetohalfone},\ref{Mtwotohalftwo},\ref{Mntohalfn}) and the properties of $\mathsf{\tilde P}_\text{cnot}$ we obtain
\begin{equation}\label{swappy}
\begin{Diagram}[1.1]{0}{-2}
\opbox{ML}{0,0} \opsymbol{M}
\opbox{MR}{8,0} \opsymbol{M}
\opbox{Pt}{4,5} \opsymbol{\tilde P_\text{\negs\negs\negs cnot}}
\wire{ML}{Pt}{3}{3} \wire{MR}{Pt}{1}{1}
\opbox[1]{Lout}{-0.8,11} \opsymbol{n} \wire{ML}{Lout}{1}{1}
\opbox[1]{Rout}{8.8,11} \opsymbol {n}\wire{MR}{Rout}{3}{1}
\Opbox[1]{1}{4.8,8} \wire{Pt}{1}{3}{1}
\opbox{Ptt}{2.4,8} \opsymbol{M} \wire{Pt}{Ptt}{1}{3}
\Opbox[1]{B}{1.6, 5} \wire{B}{Ptt}{1}{1}
%\Opbox[1]{+}{1.6,11} \wire{Ptt}{+}{1}{1}
%\opbox[1]{one}{3.2,11} \opsymbol{1} \wire{Ptt}{one}{3}{1}
\end{Diagram}
~~=~~ \frac{1}{8}
\begin{Diagram}[1.1]{0}{-0.6}
\Opbox[1]{B}{0,0}
\Opbox[1]{n}{0,3}
\wire{B}{n}{1}{1}
\end{Diagram}
\end{equation}
Using (\ref{Monetohalfone},\ref{Mtwotohalftwo}) we obtain
\begin{equation}\label{Mnnhalf}
\begin{Diagram}[1.1]{0}{-0.6}
\Opbox{M}{0,0}
\opbox[1]{nL}{-0.8,3}\opsymbol{n} \Opbox[1]{n}{0.8,3}
\wire{M}{nL}{1}{1} \wire{M}{n}{3}{1}
\end{Diagram}
~~= \frac{1}{2}
\end{equation}
Since we have established that the LHS of (\ref{swappy}) is proportional to the LHS of (\ref{Mnnhalf}), the same must be true for the right hand sides of these equations.  This tells us that $B^a$ is equatorial (since get same $B^{a_1}U_{a_1}[n]$ for $n=1$ and $n=2$) and that the constant of proportionality is $\frac{1}{8}$.  This proves {\bf T\ref{entangswapT}}.

The $\frac{1}{8}$ above can be thought of as $\frac{1}{2}$ times $\frac{1}{4}$.  The $\frac{1}{4}$ is the standard success probability for teleportation (entanglement swapping can be thought of as an application of teleportation).  The $\frac{1}{2}$ comes from the following result.
\begin{T}\label{BMAcircuit}
If the state $B^a$ is equatorial we have
\begin{equation}
\text{Prob}\left(
\begin{Diagram}[1.1]{0}{0}
       \Opbox{M}{-1.6,3}    \Opbox[1]{1}{0.8,3}
\Opbox[1]{B}{-2.4,0}     \opbox{Pt}{0,0} \opsymbol{\tilde P_\text{\negs\negs\negs cnot}}
\wire{B}{M}{1}{1}  \wire{Pt}{M}{1}{3} \wire{Pt}{1}{3}{1}
                         \opbox{A}{0,-5} \opsymbol{H}
\wire{A}{Pt}{1}{1} \wire{A}{Pt}{3}{3}
\end{Diagram}
\right)
~~\leq~~
\frac{1}{2}
\end{equation}
for all $\mathsf{H^{a_1a_2}}$.
\end{T}
To prove this consider
\begin{equation}
\text{Prob}\left(
\begin{Diagram}[1.1]{0}{0}
\Opbox{M}{0,0}
\Opbox[1]{B}{-0.8,-3} \Opbox[1]{C}{0.8,-5}
\wire{B}{M}{1}{1} \wire{C}{M}{1}{3}
\end{Diagram}
\right)
~=~
\text{Prob}\left(
\begin{Diagram}[1.1]{0}{0}
\Opbox[1]{+}{-0.8,3} \Opbox[1]{1}{0.8,3}
\opbox{M}{0,0}  \opsymbol{\tilde P_\text{\negs\negs\negs cnot}}
\Opbox[1]{B}{-0.8,-3} \Opbox[1]{C}{0.8,-5}
\wire{M}{+}{1}{1} \wire{M}{1}{3}{1}
\wire{B}{M}{1}{1} \wire{C}{M}{1}{3}
\end{Diagram}
\right)
~\leq ~
\text{Prob}\left(
\begin{Diagram}[1.1]{0}{0}
\Opbox[1]{T}{-0.8,3} \Opbox[1]{1}{0.8,3}
\opbox{M}{0,0}  \opsymbol{\tilde P_\text{\negs\negs\negs cnot}}
\Opbox[1]{B}{-0.8,-3} \Opbox[1]{C}{0.8,-5}
\wire{M}{T}{1}{1} \wire{M}{1}{3}{1}
\wire{B}{M}{1}{1} \wire{C}{M}{1}{3}
\end{Diagram}
\right)
\end{equation}
where $\mathsf T$ is the deterministic effect and $\mathsf C$ is an arbitrary preparation.  By {\bf T\ref{deteffect}}, we can write $T_{a_1}=U_{a_1}[1]+U_{a_1}[2]$.  Hence the probability on the right in the above equation is equal to
\begin{equation}
\text{Prob}\left(
\begin{Diagram}[1.1]{0}{0}
\Opbox[1]{T}{-0.8,3} \Opbox[1]{1}{0.8,3}
\opbox{M}{0,0}  \opsymbol{\tilde P_\text{\negs\negs\negs cnot}}
\Opbox[1]{B}{-0.8,-3} \Opbox[1]{C}{0.8,-5}
\wire{M}{T}{1}{1} \wire{M}{1}{3}{1}
\wire{B}{M}{1}{1} \wire{C}{M}{1}{3}
\end{Diagram}
\right)
~=~
\text{Prob}\left(
\begin{Diagram}[1.1]{0}{0}
\opbox[1]{T}{-0.8,3} \opsymbol{1}\Opbox[1]{1}{0.8,3}
\opbox{M}{0,0}  \opsymbol{\tilde P_\text{\negs\negs\negs cnot}}
\Opbox[1]{B}{-0.8,-3} \Opbox[1]{C}{0.8,-5}
\wire{M}{T}{1}{1} \wire{M}{1}{3}{1}
\wire{B}{M}{1}{1} \wire{C}{M}{1}{3}
\end{Diagram}
\right)
~+~
\text{Prob}\left(
\begin{Diagram}[1.1]{0}{0}
\opbox[1]{T}{-0.8,3}\opsymbol{2} \Opbox[1]{1}{0.8,3}
\opbox{M}{0,0}  \opsymbol{\tilde P_\text{\negs\negs\negs cnot}}
\Opbox[1]{B}{-0.8,-3} \Opbox[1]{C}{0.8,-5}
\wire{M}{T}{1}{1} \wire{M}{1}{3}{1}
\wire{B}{M}{1}{1} \wire{C}{M}{1}{3}
\end{Diagram}
\right)
\leq \frac{1}{2}
\end{equation}
where the $\frac{1}{2}$ follows from the properties of the permutation transformation (see comments below (\ref{plusminusPmaxmeas})) and the fact that $B^a$ is equatorial.  Hence, {\bf T\ref{BMAcircuit}} follows.

We will now show that
\begin{T}\label{probteleportation}
{\bf Probabilistic teleportation.} \index{probabilistic teleportation}The following holds
\begin{equation}
\begin{Diagram}[1.1]{0}{0}
\Opbox{M}{-1.6,3}    \Opbox[1]{1}{0.8,3}
\Opbox[1]{B}{-2.4,0}     \opbox{Pt}{0,0} \opsymbol{\tilde P_\text{\negs\negs\negs cnot}}
\wire{B}{M}{1}{1}  \wire{Pt}{M}{1}{3} \wire{Pt}{1}{3}{1}
\Opbox[1]{A}{-2.4,-5} \wire{A}{Pt}{1}{1}
\opbox{M1}{1.6,-5} \opsymbol{M}
\wire{M1}{Pt}{1}{3}
\thispoint{out}{2.4,5} \wire{M1}{out}{3}{1}
\end{Diagram}
~~~\equiv~ \frac{1}{8}~
\begin{Diagram}[1.1]{0}{0}
\Opbox[1]{A}{0,0}
\thispoint{out}{0,5}
\wire{A}{out}{1}{1}
\end{Diagram}
\end{equation}
for any state $A^{a_1}$ where $B^{a_1}$ is the equatorial state in {\bf T\ref{entangswapT}}.
\end{T}
We prove this important theorem following the technique of Chiribella, D'Ariano, and Perinotti.  First, we note from {\bf T\ref{teleportationTone}} that there exists a state $D^{a_1}$ such that
\begin{equation}\label{Dteleportation}
\begin{Diagram}[1.1]{0}{1}
\Opbox[1]{D}{-0.8,-5}
\opbox{M3}{0,0} \opsymbol{M}
\opbox{M2}{1.6,-5} \opsymbol{M}
\wire{D}{M3}{1}{1}
\wire{M2}{M3}{1}{3}
\thispoint{out}{2.4,2} \wire{M2}{out}{3}{1}
\end{Diagram}
~~\equiv~ \mu
\begin{Diagram}[1.1]{0}{-0.2}
\Opbox[1]{A}{0,0}
\thispoint{out}{0,5}
\wire{A}{out}{1}{1}
\end{Diagram}
\end{equation}
where $\mu$ is a constant of proportionality ($0<\mu \leq 1$).   Hence,
\begin{equation}
\begin{Diagram}[1.1]{0}{0}
\Opbox{M}{-1.6,3}    \Opbox[1]{1}{0.8,3}
\Opbox[1]{B}{-2.4,0}     \opbox{Pt}{0,0} \opsymbol{\tilde P_\text{\negs\negs\negs cnot}}
\wire{B}{M}{1}{1}  \wire{Pt}{M}{1}{3} \wire{Pt}{1}{3}{1}
\Opbox[1]{A}{-2.4,-5} \wire{A}{Pt}{1}{1}
\opbox{M1}{1.6,-5} \opsymbol{M}
\wire{M1}{Pt}{1}{3}
\thispoint{out}{2.4,5} \wire{M1}{out}{3}{1}
\end{Diagram}
~~\equiv ~~ \mu^{-1}
\begin{Diagram}[1.1]{0}{0}
\Opbox{M}{-1.6,3}    \Opbox[1]{1}{0.8,3}
\Opbox[1]{B}{-2.4,0}     \opbox{Pt}{0,0} \opsymbol{\tilde P_\text{\negs\negs\negs cnot}}
\wire{B}{M}{1}{1}  \wire{Pt}{M}{1}{3} \wire{Pt}{1}{3}{1}
\Opbox[1]{D}{-5.8,-5}
\opbox{M3}{-5,0} \opsymbol{M}
\opbox{M2}{-3.4,-5} \opsymbol{M}
\wire{D}{M3}{1}{1}
\wire{M2}{M3}{1}{3}
\wire{M2}{Pt}{3}{1}
\opbox{M1}{1.6,-5} \opsymbol{M}
\wire{M1}{Pt}{1}{3}
\thispoint{out}{2.4,5} \wire{M1}{out}{3}{1}
\end{Diagram}
\equiv ~ \frac{\mu^{-1}}{8} ~~
\begin{Diagram}[1.1]{0}{0}
\opbox{M3}{0,0} \opsymbol{M}
\Opbox[1]{D}{-0.8,-5}
\opbox{M2}{1.6,-5}  \opsymbol{M}
\wire{D}{M3}{1}{1} \wire{M2}{M3}{1}{3}
\thispoint{out}{2.4,2} \wire{M2}{out}{3}{1}
\end{Diagram}
~~\equiv ~
\frac{1}{8}~
\begin{Diagram}[1.1]{0}{0}
\Opbox[1]{A}{0,0}
\thispoint{out}{0,5}
\wire{A}{out}{1}{1}
\end{Diagram}
\end{equation}
where we have used (\ref{Dteleportation}) in the first step, {\bf T\ref{entangswapT}} in the second step, and (\ref{Dteleportation}) again in the final step.  This proves {\bf T\ref{probteleportation}}.

\subsection{Proving that $K=N^2$}

We have already proven that $K=N^r$ where $r=1, 2, \dots$ (in {\bf T\ref{KNr}}).  We will now prove that, in the non-classical case (where $r>1$) we must have $r=2$.  We note
\begin{T}\label{Nequalstwocase}
If $K_\mathsf{a} \leq 4$ for a gebit then, in the non-classical case, $K=N^2$ for all systems.
\end{T}
This follows immediately since $r=2$ is the only non-classical case consistent with $K=N^r$ if $K\leq 4$ when $N=2$.

We will prove that $K\leq 4$ for a gebit by using the ingenious techniques developed by Chiribella, D'Ariano, and Perinotti (CDP). In particular, see Lemma 22 of \cite{chiribella2010probabilistic} and Sec. IX.C of \cite{chiribella2010informational}.  In the previous subsection we have laid the groundwork for this proof by proving the entanglement swapping result in {\bf T\ref{entangswapT}} and the probabilistic teleportation result {\bf T\ref{probteleportation}}. CDP proved similar results (though using quite different techniques since they have different postulates).  There are a few differences in setting up the theorem below here compared with CDP because we start with a different set of postulates:  first, we are working gebits rather than general systems (the result {\bf T\ref{Nequalstwocase}} bridges this gap in our case); and second, are carrying around an extra factor of $\frac{1}{2}$ whose origin was explained in {\bf T\ref{BMAcircuit}}.  We will now prove the following important result.
\begin{T}\label{simplicityatlast}
{\bf State space dimension.}  \index{K@$K_\mathsf{a}$!relationship with $N_\mathsf{a}$} In the nonclassical case we have $K_\mathsf{a}=N_\mathsf{a}^2 $.
\end{T}
It follows from {\bf T\ref{probteleportation}} and {\bf P3} that
\begin{equation}\label{teleportationidentity}
\begin{Diagram}[1.1]{0}{0}
\Opbox{M}{-1.6,3}    \Opbox[1]{1}{0.8,3}
\Opbox[1]{B}{-2.4,0}     \opbox{Pt}{0,0} \opsymbol{\tilde P_\text{\negs\negs\negs cnot}}
\wire{B}{M}{1}{1}  \wire{Pt}{M}{1}{3} \wire{Pt}{1}{3}{1}
\thispoint{A}{-0.8,-7} \wire{A}{Pt}{1}{1}
\opbox{M1}{1.6,-5} \opsymbol{M}
\wire{M1}{Pt}{1}{3}
\thispoint{out}{2.4,5} \wire{M1}{out}{3}{1}
\end{Diagram}
~~\cong ~\frac{1}{8}
\begin{Diagram}[1.1]{0}{-1.8}
\thispoint{in}{0,0} \thispoint{out}{2,11} \wire[2]{in}{out}{1}{1}
\end{Diagram}
\end{equation}
We use the $\cong$ symbol rather than $\equiv$ (see Sec. \ref{equivalenceoffragments}) because it is possible on the LHS, but not on the RHS, to feed the output into the input.  To prove (\ref{teleportationidentity}) from {\bf T\ref{probteleportation}} we need to invoke {\bf P5}.  We could send one component of a composite system into input on the LHS (or RHS) of (\ref{teleportationidentity}).  In this case, {\bf P5} ensures that product effects are sufficient to characterize transformations. If we have a product effect then we, effectively, are reduced back to the situation in {\bf T\ref{probteleportation}}.   %!!! need a theorem earlier to cover this - I keep having to explain it - it is basically the F locality point.
From {\bf T\ref{BMAcircuit}} we have
\begin{equation}\label{closeloop}
\text{Prob}\left(
\begin{Diagram}[1.1]{0}{0.15}
\Opbox{M}{-1.6,3}    \Opbox[1]{1}{0.8,3}
\Opbox[1]{B}{-2.4,0}     \opbox{Pt}{0,0} \opsymbol{\tilde P_\text{\negs\negs\negs cnot}}
\wire{B}{M}{1}{1}  \wire{Pt}{M}{1}{3} \wire{Pt}{1}{3}{1}
%
%\thispoint{A}{-0.8,-7} \wire{A}{Pt}{1}{1}
%
\opbox{M1}{1.6,-5} \opsymbol{M}
\wire{M1}{Pt}{1}{3}
%\thispoint{out}{2.4,5} \wire{M1}{out}{3}{1}
%
\wire{M1}{Pt}{3}{1}
\end{Diagram}
~\right)
\leq ~ \frac{1}{2}
\end{equation}
The circuit in (\ref{closeloop}) corresponds to taking the output on the LHS of (\ref{teleportationidentity}) and feeding it into the input.  We can do this because the causal structure of the fragment on the LHS of (\ref{teleportationidentity}) allows it.  However, we cannot make this happen on the RHS of (\ref{teleportationidentity}) since the causal structure does not allow it. Nevertheless, we can make this happen mathematically and it corresponds to taking the trace.  We will show how to do this.  For convenience we put
\begin{equation}
\begin{Diagram}[1.1]{0}{0}
\Opbox{N}{0,0}
\inwire{N}{1}\inwire{N}{3}
\end{Diagram}
~~:=~~
\begin{Diagram}[1.1]{0}{0}
\Opbox{M}{-1.6,3}    \Opbox[1]{1}{0.8,3}
\Opbox[1]{B}{-2.4,0}     \opbox{Pt}{0,0} \opsymbol{\tilde P_\text{\negs\negs\negs cnot}}
\wire{B}{M}{1}{1}  \wire{Pt}{M}{1}{3} \wire{Pt}{1}{3}{1}
\inwire{Pt}{1} \inwire{Pt}{3}
\end{Diagram}
\end{equation}
Then, invoking {\bf T\ref{equivequalsrestrictedequiv}} and {\bf T\ref{sanseriftomathnormaltheorem}}, (\ref{teleportationidentity}) gives
\begin{equation}
N_{a_1a_2}M^{a_2a_3}= \frac{1}{8} I_{a_1}^{a_3}
\end{equation}
where $I_{a_1}^{a_3}$ is the identity. Hence,
\begin{equation}
N_{a_1a_2}M^{a_2a_1}= \frac{1}{8} I_{a_1}^{a_1}= \frac{1}{8} K_\mathsf{a}
\end{equation}
since the trace of the identity is equal to the dimension of the space on which it acts.  But (\ref{closeloop}) gives us
\begin{equation}
N_{a_1a_2}M^{a_2a_1} \leq \frac{1}{2}
\end{equation}
It follows that, for a gebit, $K_\mathsf{a}\leq 4$. Hence, by {\bf T\ref{Nequalstwocase}}, {\bf T\ref{simplicityatlast}} follows.

\subsection{The Bloch sphere}

We now see immediately that a gebit is, in fact, a qubit.
\begin{T}\label{blochsphere}
{\bf The Bloch sphere.}  \index{Bloch sphere}The pure states, $U^{a_1}$ for a gebit correspond to the points, ${\bf \overline u}$, on a unit 2-sphere and, likewise, the maximal effects, $V_a$, for a gebit correspond to the points, ${\bf \overline v}$, on a unit 2-sphere such that
\begin{equation}
\text{Prob}(\mathsf{U^{a_1}V_{a_1}}) = \frac{1}{2}(1+ {\bf \overline u}\cdot {\bf \overline v})
\end{equation}
Antipodal points correspond to distinguishable states in the case of states, and to a maximal measurement in the case of effects.
\end{T}
This follows immediately from the results established in Sec.\ \ref{gebitbasicproperties}, {\bf T\ref{hypersphere}}, and {\bf T\ref{simplicityatlast}}.  This is the Bloch sphere associated with the qubit of quantum theory.

\section{Quantum theory reconstructed}\label{partIII}

In this section we will complete the reconstruction of quantum theory.  To do this we will use the machinery of the duotensor framework provided in Part \ref{theduotensorframework}.   We will recover the following two mathematical axioms for quantum theory.
\begin{description}
\item[Axiom 1] \index{axioms!\textbf{Axiom 1}}Operations correspond to operators.
\item[Axiom 2] \index{axioms!\textbf{Axiom 2}}Every complete set of physical operators corresponds to a complete set of operations.
\end{description}
The operators here are understood to act on a complex Hilbert space.
These axioms were explained in Sec.\ \ref{QTaxioms}.

\subsection{Proving Axiom 1}

First, we simply note that the space of Hermitian operators on an $N$ dimensional complex Hilbert space is of dimension $N^2$.  This follows from the fact that we have $N^2$ real parameters for such Hermitian operators.  Motivated by this we choose a set of positive (and therefore Hermitian) operators, $\{\hat X_\mathsf{a_1}^{a_1}: a_1=1 ~\text{to}~N_\mathsf{a}^2\}$, that span the space of Hermitian operators acting on a $N_\mathsf{a}$ dimensional Hilbert space, $\cal{H}_{N_\mathsf{a}}$.  These operators will be associated with the fiducial effects, $\mathsf{X}_\mathsf{a_1}^{a_1}$.  We choose another set, $\{{}_{a_1}\hat X^\mathsf{a_1}:a_1=1~\text{to}~ N_\mathsf{a}^2\}$, which will be associated with the fiducial preparations, ${}_{a_1} X^\mathsf{a_1}$.  Recall that it follows from {\bf T\ref{correspondencetheorem}} that the operation
\begin{equation}\label{operationcorrespondsthree}
\mathsf{A_{a_1b_2\dots c_3}^{d_4e_5\dots f_6}} \,\,\equiv {}^{d_4e_5\dots f_6}\!A_{a_1b_2\dots c_3}\,\, \mathsf{X}_\mathsf{a_1}^{a_1} \mathsf{X}_\mathsf{b_2}^{b_2} \cdots \mathsf{X}_\mathsf{c_3}^{c_3} \,\,{}_{d_4}\!\mathsf{X}^\mathsf{d_4}{}_{e_5}\!\mathsf{X}^\mathsf{e_5}\cdots {}_{f_6}\!\mathsf{X}^\mathsf{f_6}
\end{equation}
corresponds to the operator
\begin{equation}\label{operatorcorrespondsthree}
\hat A_\mathsf{a_1b_2\dots c_3}^\mathsf{d_4e_5\dots f_6} \,\,= {}^{d_4e_5\dots f_6}\!A_{a_1b_2\dots c_3}\,\, \hat {X}_\mathsf{a_1}^{a_1} \hat{X}_\mathsf{b_2}^{b_2} \cdots \hat{X}_\mathsf{c_3}^{c_3} \,\,{}_{d_4}\!\hat{X}^\mathsf{d_4}{}_{e_5}\!\hat{X}^\mathsf{e_5}\cdots {}_{f_6}\!\hat{X}^\mathsf{f_6}
\end{equation}
if we have
\begin{equation}\label{equalhoppingmetricsthree}
{}_{a_1}\! \hat X^\mathsf{a_1}\hat X_\mathsf{a_1}^{a'_1}=\text{Prob}({}_{a_1}\! \mathsf{X}^\mathsf{a_1}\mathsf{X}_\mathsf{a_1}^{a'_1})
\end{equation}
(i.e. equal hopping metrics for operations and operators).  By definition, when we have such a correspondence\index{correspondence} then the probability is equal to the corresponding operator circuit.  For example,
\begin{equation}
\text{Prob}(\mathsf{A}^\mathsf{a_1b_2}\mathsf{B}_\mathsf{b_2}^\mathsf{c_3a_4}\mathsf{C}_\mathsf{a_1c_3a_4})
=\hat{A}^\mathsf{a_1b_2}\hat{B}_\mathsf{b_2}^\mathsf{c_3a_4}\hat{C}_\mathsf{a_1c_3a_4}
\end{equation}
Hence, if we can satisfy (\ref{equalhoppingmetricsthree}) then we will have proven that operations correspond to operators.  This is Axiom 1 of quantum theory as given in Sec.\ \ref{QTaxioms}.

We will show that we can satisfy condition (\ref{equalhoppingmetricsthree}) for a particular choice of fiducials.  First we note that a gebit is associated with each informational subset $S_{\{m,n\} }$ (having $O(S_{\{ m,n\} })=\{ m,n\}$) defined with respect to maximal measurement $\{ \mathsf{U^{a_1}}[n]: n=1~\text{to}~ N_\mathsf{a} \}$. The pure states in a gebit lie on the surface of a sphere (by {\bf T\ref{blochsphere}}).  We will place an axis system in this sphere such that the pure  states $U^{a_1}[m]$ and $U^{a_1}[n]$ correspond to vectors pointing in the $+$ and $-$ directions along the $z$-axis.   Let $\mathsf{U^\mathsf{a_1}}[mnx+]$ and $\mathsf{U^\mathsf{a_1}}[mnx-]$ be preparations corresponding to pure states pointing along the $+$ and $-$ directions along the $x$-axis where $\mathsf{U_\mathsf{a_1}}[mnx+]$ and $\mathsf{U_\mathsf{a_1}}[mnx-]$ are the corresponding maximal results.  Sometimes we drop the \lq\lq +" and simply write $\mathsf{U^\mathsf{a_1}}[mnx]$ and $\mathsf{U_\mathsf{a_1}}[mnx]$.  We use similar notation for the $y$-axis.  We will prove
\begin{T}\label{needthistheorem}
There exists a maximal measurement
\begin{equation}\label{scewedeffects}
\{\mathsf{U_{a_1}}[n]: \text{all} ~n\not= m_1,n_1\}\cup \{ \mathsf{U_{a_1}}[m_1n_1x+], \mathsf{U_{a_1}}[m_1n_1x-] \}
\end{equation}
(where $m_1\not=n_1$) identifying the maximal distinguishuable set of preparations
\begin{equation}\label{scewedstates}
\{\mathsf{U^{a_1}}[n]: \text{all}~ n\not= m_1,n_1\}\cup \{ \mathsf{U^{a_1}}[m_1n_1x+], \mathsf{U^{a_1}}[m_1n_1x-] \}
\end{equation}
where this maximal measurement can be implemented by placing the maximal measurement, $\{ \mathsf{U_{a_1}}[m_1n_1x+], \mathsf{U_{a_1}}[m_1n_1x-] \}$, on the gebit associated with $S_{\{m_1,n_1\} }$ after the special filter $\mathsf{F}$ associated with the maximal measurement
\begin{equation}
\{\mathsf{U_{a_1}}[n]: \text{all}~ n \}
\end{equation}
having informational subset $S_{\{m_1,n_1\} }$ with $O(S_{\{ m_1,n_1\} })=\{ m_1,n_1\}$.  Similar results hold if $x$ is replaced by $y$.
\end{T}
Consider the special filter followed by a maximal measurement on a gebit as described above.  We note that the preparations $\mathsf{U^{a_1}}[n]$ ($n\not=m_1n_1$) in (\ref{scewedstates}) are identified by the special filter while the states corresponding to $\mathsf{U^{a_1}}[mnx\pm]$ pass unchanged through the filter and are then identified by the  maximal measurement on the gebit.  Hence the states (\ref{scewedstates}) can be distinghished and the measurement (\ref{scewedeffects}) is maximal.

Now consider a Hilbert space, ${\cal H}_\mathsf{a}$, spanned by a basis $\{ |n\rangle_\mathsf{a} : n=1 ~\text{to} ~ N_\mathsf{a}\}$.  Define
\begin{equation}
|mnx\rangle_\mathsf{a} := \frac{1}{\sqrt{2}} (|m\rangle_\mathsf{a} + |n\rangle_\mathsf{a}) ~~~ \text{and}~~~
|mny\rangle_\mathsf{a} := \frac{1}{\sqrt{2}} (|m\rangle_\mathsf{a} + i|n\rangle_\mathsf{a})
\end{equation}
Now we can define some effect operators (actually these are rank one projectors) in an obvious notation
\begin{equation}
\hat U_\mathsf{a}[n] := |n\rangle_\mathsf{a}\langle n|, ~~~ \hat U_\mathsf{a}[mnx]:=|mnx\rangle_\mathsf{a} \langle mnx|, ~~\text{and}~~
\hat U_\mathsf{a}[mny]:= |mny\rangle_\mathsf{a} \langle mny|
\end{equation}
The Hilbert space ${\cal H}^\mathsf{a}$ is spanned by a basis $\{ |n\rangle^\mathsf{a} : n=1 ~\text{to} ~ N_\mathsf{a}\}$ we can define a corresponding set of preparation operators
\begin{equation}\label{Uprojectors}
\hat U^\mathsf{a}[n] := |n\rangle^\mathsf{a}\langle n|, ~~~ \hat U^\mathsf{a}[mnx]:=|mnx\rangle^\mathsf{a} \langle mnx|, ~~\text{and}~~
\hat U^\mathsf{a}[mny]:= |mny\rangle^\mathsf{a} \langle mny|
\end{equation}
in an obvious notation.  Here we are using $\mathsf{a}$ rather than $\mathsf{a_1}$ as a label.  We will sometimes include the integers and sometimes omit them as is convenient.  We define the set
\begin{equation}
{\cal F}_\mathsf{a} = \{n: n=1~\text{to}~N_\mathsf{a} \} \cup \{ mnx, mny: \text{all}~m,n=1~\text{to}~N_\mathsf{a}~\text{with}~m<n\}
\end{equation}
We note that $|{\cal F}_\mathsf{a}|= N_\mathsf{a}^2$.
With these definitions we can show
\begin{T}\label{hoppingUs}
If we choose fiducial sets of preparations and results
\begin{equation}
\{ {}_{a_1}\mathsf{X}^\mathsf{a_1} = \mathsf{U^{a_1}}[a_1]: a_1\in {\cal F}_\mathsf{a} \}
~~~\text{and}~~~
\{ \mathsf{X}_\mathsf{a_1}^{a_1} = \mathsf{U_{a_1}}[a_1]: a_1\in  {\cal F}_\mathsf{a} \}
\end{equation}
and we chose fiducial sets of operators
\begin{equation} \label{fidoperators}
\{ {}_{a_1}\hat{X}^\mathsf{a_1} = \hat{U}^\mathsf{a_1}[a_1]: a_1\in {\cal F}_\mathsf{a} \}
~~~\text{and}~~~
\{ \hat{X}_\mathsf{a_1}^{a_1} = \hat{U}_\mathsf{a_1}[a_1]: a_1\in  {\cal F}_\mathsf{a} \}
\end{equation}
then
\begin{equation}\label{hoppingUcondition}
\text{Tr}( {}_{a_1} \hat X^\mathsf{a_1} \hat X_\mathsf{a_1}^{a'_1} ) =
\text{Prob}({}_{a_1} \mathsf{X}^\mathsf{a_1}  \mathsf{X}_\mathsf{a_1}^{a'_1} ):= {}_{a_1} g^{a'_1}
\end{equation}
and, further, the matrix ${}_{a_1} g^{a'_1}$ is invertible (which means that the fiducial states form a spanning linearly independent set as do the fiducial effects).
\end{T}
Consider the case $N_\mathsf{a}=3$.   We can show that
\begin{equation}\label{DNthree}
\text{Prob}({}_{a_1} \mathsf{X}^\mathsf{a_1}  \mathsf{X}_\mathsf{a_1}^{a'_1} ):= {}_{a_1} g^{a'_1}
=\left( \begin{array}{ccccccccc} 1&0&0&h&h&h&h&0&0 \\
                                  0&1&0&h&h&0&0&h&h \\
                                  0&0&1&0&0&h&h&h&h \\
                                  h&h&0&1&h&q&q&q&q \\
                                  h&h&0&h&1&q&q&q&q \\
                                  h&0&h&q&q&1&h&q&q \\
                                  h&0&h&q&q&h&1&q&q \\
                                  0&h&h&q&q&q&q&1&h \\
                                  0&h&h&q&q&q&q&h&1 \end{array} \right)
\end{equation}
where $h=\frac{1}{2}$ and $q=\frac{1}{4}$.  Here we have ordered the rows and columns according to
\begin{equation}
1, 2, 3, 12x, 12y, 13x, 13y, 23x, 23y
\end{equation}
The 1's down the diagonal of (\ref{DNthree}) follow since each fiducial effect identifies the corresponding fiducial preparation.   The $h$ in position $(1,12x)$ follows from {\bf T\ref{blochsphere}} and the fact that the $U^a[1]$ state corresponds to a vector pointing along the $z$-axis whereas the $U^a[12x]$ state corresponds to an effect pointing along the $x$-axis.  All the other $h$'s follow for similar reasons.  The $q$ in position $(12x,23y)$ corresponds to
\begin{equation}\label{forDmatrixone}
\text{Prob}(\mathsf{U^{a_1}}[12x] \mathsf{U_{a_1}}[23y])
\end{equation}
We can think of this as the preparation of a gebit state $U^{a_1}[12x]$ in the informational subspace $S_{\{1,2\} }$.  Now we know that
\begin{equation}\label{forDmatrixtwo}
\text{Prob}(\mathsf{U^{a_1}}[1] \mathsf{U_{a_1}}[23y])=0   ~~~\text{and} ~~~ \text{Prob}(\mathsf{U^{a_1}}[2] \mathsf{U_{a_1}}[23y])=\frac{1}{2}
\end{equation}
The first equation follows from the fact that the maximal effect, $\mathsf{U_{a_1}}[23y]$, pertains to the $S_{\{2,3\} }$ informational subspace and so, by {\bf T\ref{needthistheorem}}, we can put $\mathsf{U_{a_1}}[23y]= \mathsf{F_{a_1}^{a_2}} \mathsf{U_{a_1}}[23y]$ where $\mathsf F$ is a special filter corresponding to $S_{\{2,3\} }$.
The second equation follows from {\bf T\ref{blochsphere}}. It now follows by {\bf T\ref{gebitzero}} that, for states restricted to $S_{\{1,2\} }$, the result
$\mathsf{U_{a_1}}[23y]$ is equivalent to $\frac{1}{2} \mathsf{U_{a_1}}[2]$.  Thus, using (\ref{forDmatrixtwo}) we get that the probability in (\ref{forDmatrixone}) is equal to $q$.
All the $q$'s in (\ref{DNthree}) follow for similar reasons.  The $0$'s in (\ref{DNthree}) follow immediately from {\bf T\ref{needthistheorem}}.  It is a simple matter to see that, with the choices in (\ref{fidoperators}), the matrix
$\text{Tr}( {}_{a_1} \hat X^\mathsf{a_1} \hat X_\mathsf{a_1}^{a'_1} )$ is the same as in (\ref{DNthree}).   For $N_\mathsf{a}>3$ all the entries can be deduced by similar reasoning and (\ref{hoppingUcondition}) is true in general.  The fiducial operators clearly form a spanning set (for Hermitian operators acting on ${\cal H}_\mathsf{a}$).  From this it follows immediately that ${}_{a_1} g^{a'_1}$ is invertible.  This proves {\bf T\ref{hoppingUs}}

This result, together with {\bf T\ref{correspondencetheorem}} proves that Axiom 1 \index{axioms!\textbf{Axiom 1}}above follows from the postulates.

%Note that, according to {\bf \ref{anyfidswork}}, we can change our fiducial sets to any other set so long as the new set of fiducial operators chosen form a spanning set and correspond to the new set of fiducial operations (preparations and results).

\subsection{Operators for pure states and maximal effects}

We can now prove
\begin{T}\label{allprojectors}
Every operator of the form
\begin{equation}\label{psieqn}
\hat A^\mathsf{a} = |\psi\rangle^\mathsf{a} \langle \psi |     ~~~ \text{where} ~~~\langle\psi |\psi\rangle=1
\end{equation}
corresponds to a pure preparation and every operator of the form
\begin{equation}
\hat B_\mathsf{a} = |\varphi\rangle_\mathsf{a}\langle \varphi |  ~~~ \text{where} ~~~\langle\varphi |\varphi\rangle=1
\end{equation}
corresponds to a maximal result.
\end{T}
We will prove this by induction. First we note that this is trivially true for a system having $N_\mathsf{a}=1$.  We prove that, if this is true for a system of type $\mathsf b$ having a particular value of $N_\mathsf{b}=N$, then it is also true for a system of type $\mathsf a$ having $N_\mathsf{a}=N+1$.
Assume we have a maximal set of $N+1$ distinguishable states for $\mathsf a$, labeled by $n=1$ to $N+1$.  According to {\bf T\ref{hoppingUs}}, we can generate a fiducial set of operators from a basis $|n\rangle_\mathsf{a}$ where these $N+1$ distinguishable states correspond to the operators $|n\rangle_\mathsf{a}\langle n|$.  Consider a special filter, $\mathsf F$, associated with the maximal measurement that distinguishes these states having $O(S)=\{n: n=1~\text{to}~ N\}$.  This will produce a system, $\mathsf b$, having $N_\mathsf{b}=N$.   The first $N$ states in the above maximal distinguishable set for $\mathsf a$ will pass through the filter unchanged and constitute a maximal distinguishable set for $\mathsf b$.  Hence, we can generate a fiducial set of operators as in {\bf T\ref{hoppingUs}}, from $|n\rangle_\mathsf{a}$ (for $n=1$ to $N$). In this case, $|n\rangle_\mathsf{a}\langle n|$ (for $n=1$ to $N$) correspond to the $N$ distinguishable states for $\mathsf b$.  For the $|\psi\rangle_\mathsf{a}$ in (\ref{psieqn}), we can write
\begin{equation}\label{psiab}
|\psi\rangle_\mathsf{a} = a |\tilde\psi \rangle_\mathsf{a} + b |N+1\rangle_\mathsf{a} ~~~\text{where}~~~  \langle N+1 | \tilde\psi \rangle  =0, ~~ |a|^2+|b|^2=1
\end{equation}
We can consider the operator $|\tilde\psi \rangle_\mathsf{a}\langle\tilde\psi|$.  This is in the space spanned by the fiducial operators for $\mathsf b$ (since $\text{Tr}(|\tilde\psi \rangle_\mathsf{a}\langle\tilde\psi|N+1\rangle_\mathsf{a}\langle N+1|)=0$).   Therefore by supposition, there exists a pure preparation corresponding to $|\tilde\psi \rangle_\mathsf{a}\langle\tilde\psi|$ since we are proceeding by induction and hence are assuming that {\bf T\ref{allprojectors}} is true for $\mathsf b$. Since this preparation is pure, we know by {\bf T\ref{origP2}} that it belongs to a maximal set of $N$ distinguishable preparations for the system of type $\mathsf b$.  We are free to let the pure preparation corresponding to $|\tilde\psi \rangle_\mathsf{a}\langle\tilde\psi|$ be the $N$th of the distinguishable preparations employed above. Hence $|\tilde \psi\rangle_\mathsf{a}=|N\rangle_\mathsf{a}$.
We can now focus on the informational subset containing the states associated with the operators $|N \rangle_\mathsf{a}\langle N|$ and $|N+1\rangle_\mathsf{a}\langle N+1|$.  This is a gebit.  It is a standard result that, with the trace formula, pure states on the Bloch sphere corresponds to superpositions $a|N\rangle+b|N+1\rangle$.  Since we established the states on the Bloch sphere exist in {\bf T\ref{blochsphere}}, we have proven that the vector  in (\ref{psiab}) corresponds to a pure state.  This proves the first part of {\bf T\ref{allprojectors}} by induction.  To establish the second part we note that since maximal effects must be represented by positive operators since $\text{Tr}(|\psi\rangle_\mathsf{a}\langle \psi| \hat B_\mathsf{a})$ is a probability and so must be positive for all $|\psi\rangle_\mathsf{a}$.  Further, we know by {\bf P1} that there is a unique maximal effect identifying the pure state represented by $|\varphi\rangle^\mathsf{a}\langle \varphi|$ and that this maximal effect does not identify any other pure state.  The only positive operator doing this when we take the trace is $|\varphi\rangle_\mathsf{a}\langle \varphi|$.  The second part of {\bf T\ref{allprojectors}} follows from these facts.

\subsection{The deterministic effect}

First we will show that
\begin{T}\label{TisI}
The deterministic effect\index{deterministic effect}, $\mathsf{T_\mathsf{a_1}}$, corresponds to the identity operator, $\hat I_\mathsf{a_1}$, acting on ${\cal H}_\mathsf{a_1}$.
\end{T}
The deterministic effect is given by course-graining over the outcomes of any measurement (we are implicitly using the uniqueness property in {\bf T\ref{deteffect}}).  For example, we can write
\begin{equation}
T_{a_1}= \sum_{n=1}^{N_\mathsf{a}} U_{a_1}[n]
\end{equation}
We can chose the effects on the RHS to correspond to the maximal measurement used to generate the fiducial set.  In this case we get
\begin{equation}
\hat T_\mathsf{a_1} = \sum_{n=1}^{N_\mathsf{a}} |n\rangle_\mathsf{a}\langle n| = \hat I_\mathsf{a_1}
\end{equation}
This proves {\bf T\ref{TisI}}.

\subsection{Operators are physical}

In Sec.\ \ref{physicaloperatorssection} we considered \emph{operator supersets} which were subsets of all possible operators that might, for example, be induced by taking the subset of operators that correspond to operations.   Recall that an operator superset is defined to be physical if: (1) the operator circuit formed from operators in the superset is between 0 and 1; (2) the operator superset contains preparations and effect operators equal to all rank one projectors for every type; and (3) the operator superset contains result operators corresponding to the identity operator, $\hat I_\mathsf{a_1}$, for every type.
We can now prove that
\begin{T}\label{alloperatorsphysical}
\index{physical operators}All operators corresponding to operations are physical.
\end{T}
The definition of physical operators is given in Sec.\ \ref{physicaloperatorssection}.  The operator superset obtained by taking all operators that correspond to operations must be physical because: (1) the operator circuit is equal to the probability of a circuit by {\bf T\ref{correspondencetheorem}}; (2) all preparation and result operators equal to rank one projectors belong to the superset by {\bf T\ref{allprojectors}}; (3) The identity operator belongs to the superset by {\bf T\ref{TisI}}.  {\bf T\ref{alloperatorsphysical}} then follows from {\bf T\ref{supersettheorem}}.

By theorems {\bf T\ref{physicalimpliespositive}} and {\bf T\ref{CPmaps}} from Sec.\ \ref{physicaloperatorssection} it follows that preparation operators must be physical and have trace less than or equal to one, effect operators must be positive and be less than or equal to the identity, and the transformations associated operations must be completely positive and trace non-increasing.  We have, then, recovered a substantial part of quantum theory.  We need, however, to prove a few more results to obtain Axiom 2 (that every complete set of physical operators corresponds to a complete set of operations).

\subsection{All projection valued measures possible}

We will now prove that
\begin{T}\label{pvmeasures}
\index{projection valued measures}There exists a maximal measurement, $\{\mathsf{U_{a}}[n]: n=1~\text{to}~N_\mathsf{a} \}$ corresponding to  any set $\{|n\rangle_\mathsf{a} \langle n|: n=1~\text{to}~N_\mathsf{a} \}$ where $|n\rangle_\mathsf{a}$ is an orthnormal basis in ${\cal H}_\mathsf{a}$.  Further, the maximal set of distinguishable states, $\{\mathsf{U_{a}}[n]: n=1~\text{to}~N_\mathsf{a} \}$, corresponds to the set $\{|n\rangle^\mathsf{a} \langle n|: n=1~\text{to}~N_\mathsf{a} \}$.
\end{T}
We know from {\bf T\ref{allprojectors}} that there is a pure state corresponding to each of the projectors $|n\rangle^\mathsf{a}\langle n|$.  Consider the projector $|N\rangle^\mathsf{a}\langle N|$.  This state must belong to some  maximal distinguishable set of states (by {\bf T\ref{origP2}}).  We can construct a special filter, $\mathsf F_N$, that picks off just this pure state and allows remaining states in the maximal distinguishable set to pass through unchanged.  From the properties of the special filter we now have a maximal effect corresponding to $|N\rangle_\mathsf{a}\langle N|$. Each of the states associated with $|n\rangle^\mathsf{a}\langle n|$ for $n=1$ to $N-1$ must pass through the filter because $\text{Tr}(|n\rangle^\mathsf{a}\langle n|N\rangle_\mathsf{a}\langle N|)=0$ and so they must belong to the informational subset, $S_N$, associated with $F_N$.  We can iterate this process picking off the state corresponding to $|N-1\rangle^\mathsf{a}\langle N-1|$ with a special filter $F_{N-1}$ and so on and obtain a maximal effect corresponding to $|N-1\rangle_\mathsf{a}\langle N-1|$.  In this way we are able to construct a maximal measurement corresponding to $\{|n\rangle_\mathsf{a} \langle n|: n=1~\text{to}~N_\mathsf{a} \}$  distinguishing the states corresponding to $\{|n\rangle^\mathsf{a} \langle n|: n=1~\text{to}~N_\mathsf{a} \}$.  This proves {\bf T\ref{pvmeasures}}.

\subsection{All unitary transformations possible}

We will now show that we can obtain reversible transformations corresponding to arbitrary unitary operations.  First we prove
\begin{T}\label{revlinearinH}
\index{unitary transformations}If we have a pure state for a system represented by the operator $\hat A^\mathsf{a}=|\psi\rangle^\mathsf{a}\langle \psi|$ then, under a reversible transformation on the system, $|\psi\rangle_\mathsf{a}$ evolves linearly.
\end{T}
Recall from Sec.\ \ref{evolutionsection} (equation (\ref{evolutionequation}) in particular) that the transformation on a state represented by an operator $\hat A^\mathsf{a_1}$ due to an operation $\hat B_\mathsf{a_1}^\mathsf{a_2}$ is given by $\hat A^\mathsf{a_1} \hat B_\mathsf{a_1}^\mathsf{a_2}$.
It follows from {\bf T\ref{alloperatorsphysical}} that any operator, $\hat B_\mathsf{a_1}^\mathsf{a_2}$, must be physical and from {\bf T\ref{completephysicalcompletecpmaps}} that any physical operator corresponds to a completely positive trace non-increasing map, $\$_{B}(\cdot)$.  Further, we can write any such map in Krauss form form \cite{nielsen2000quantum}:
\begin{equation}\label{Kraussform}
\$_B (\hat A^\mathsf{a_1}) =
\hat B_\mathsf{a_1}^\mathsf{a_2} \hat A^\mathsf{a_1} = \sum_i E_\mathsf{a_1}^\mathsf{a_2}[i] \hat A^\mathsf{a_1} E_\mathsf{a_1}^\mathsf{a_2}[i]^\dagger
\end{equation}
where $E_\mathsf{a_1}^\mathsf{a_2}[i]$ are linear operators that act on vectors in ${\cal H}_\mathsf{a}$ and return vectors in ${\cal H}_\mathsf{a}$ having the property
\begin{equation}
\sum_{i} E_\mathsf{a_1}^\mathsf{a_2}[i]^\dagger E_\mathsf{a_1}^\mathsf{a_2}[i] \leq \hat I_\mathsf{a_1}^\mathsf{a_1}
\end{equation}
where $\hat I_\mathsf{a_1}^\mathsf{a_1}$ is the identity operator on ${\cal H}_\mathsf{a}$.  If $\hat B_\mathsf{a_1}^\mathsf{a_2}$, corresponds to a reversible transformation and it is applied to a pure state, then the state afterwards will be pure.   Hence,
\begin{equation}
\hat B_\mathsf{a_1}^\mathsf{a_2} |\psi\rangle^\mathsf{a_1}\langle\psi| = |\psi'\rangle^\mathsf{a_2}\langle\psi'|
\end{equation}
(where, as usual, we are implicitly taking the partial trace over the $\mathsf{a_1}$ space on the LHS).  If we represent the transformation in Krauss form then the only way this is possible is if $i$ only takes one value (so we drop the sum in (\ref{Kraussform})) and hence we see that we have that $|\psi'\rangle^\mathsf{a}= E_\mathsf{a_1}^\mathsf{a_2}|\psi\rangle^\mathsf{a}$ which is a linear transformation.

We now prove
\begin{T}\label{allunitaries}
We can construct a transformation, $\mathsf B$, corresponding to an arbitrary unitary transformation, $U_B$, such that
\begin{equation}
\hat B_\mathsf{a_1}^\mathsf{a_2} \hat A^\mathsf{a_1} = U_B \hat A^\mathsf{a} U_B^\dagger
\end{equation}
for all operators, $\hat A^\mathsf{a_1}$, representing states.
\end{T}
Any unitary transformation on $\cal{H}_\mathsf{a}$ can be specified by giving a linear map from one orthonormal basis set, $\{|u[n]\rangle_\mathsf{a}: n=1~\text{to}~N_\mathsf{a}\}$, to a new one, $\{|v[n]\rangle_\mathsf{a}: n=1~\text{to}~N_\mathsf{a}\}$.  By {\bf T\ref{pvmeasures}} we know that the projectors corresponding to any orthonormal basis set of ${\cal H}_\mathsf{a}$ form a maximal distinguishable set of states.  Hence, by {\bf T\ref{revperm}} we know that there exists a reversible map that takes the the pure states $|u[n]\rangle^\mathsf{a}\langle u[n]|$ and maps them to $|v[n]\rangle^\mathsf{a}\langle v[n]|$.  Using {\bf T\ref{revlinearinH}}, this means we have a linear map, $f_B$, which performs the transformation
\begin{equation}\label{fthetauv}
f_B |u[n]\rangle^\mathsf{a} = e^{i\theta[n]} |v[n]\rangle^\mathsf{a}  ~~~\text{for}~ n=1 ~\text{to} ~ N_\mathsf{a}
\end{equation}
We must have the $\text{exp}(i\theta[n])$ phase factors since they cancel when we form the projectors $|v[n]\rangle_\mathsf{a}\langle v[n]|$.  {\bf T\ref{revperm}} does not fix the values of the $\theta[n]$ since it only guarantees the existance of some reversible transformation mapping between these sets of states.  The $\theta[n]$ terms are important since [as we know from quantum theory] they give rise to interference when the transformation is applied to a general state.  However, whatever values $\theta[n]$ take, we will show that we can construct a transformation that maps maps these phases to zero (note that the vectors $|v[n]\rangle^\mathsf{a}$ can already have an arbitrary phase absorbed into them).  Hence, we can perform an arbitrary unitary.  To construct this, first we note that we can construct an arbitrary unitary for a gebit (a qubit).  We know that, in an appropriate representation, pure states are represented by the points on a 2-sphere.  By {\bf T\ref{revt}} we know that there exists a reversible transformation taking any point on the sphere to any other point.  In other words, the group of reversible transformations must be transitive on the 2-sphere.  There is only one such group \cite{montgomery1943transformation, borel1949some} which also corresponds to a completely positive map (we know by {\bf T\ref{alloperatorsphysical}} and {\bf T\ref{CPmaps}} that this must be a completely positive map) and this is $SO(3)$ (i.e. the group of orthogonal transformations). Note that we need $SO(3)$ rather than $O(3)$ because because the map must be completely positive.
It is well known that special orthogonal transformations on the Bloch sphere correspond to unitary transformations in the Hilbert space for a qubit.  Hence, we can obtain an arbitrary unitary for a qubit.  One particular such transformation is the phase gate:
\begin{equation}
|0\rangle^\mathsf{b} \rightarrow |0\rangle^\mathsf{b} ~~~~~~ |1\rangle^\mathsf{b} \rightarrow e^{i\phi}|1\rangle^\mathsf{b}
\end{equation}
(this is a rotation in about the $z$ axis by an angle $\phi$).
If this is applied to $N_\mathsf{a}$ qubits then we have
\begin{flalign}\label{phaseonNqubits}
 |100\dots 0\rangle &\rightarrow e^{i\phi[1]} |100\dots 0\rangle  \nonumber\\
 |010\dots 0\rangle &\rightarrow e^{i\phi[2]} |010\dots 0\rangle \nonumber\\
 |001\dots 0\rangle &\rightarrow e^{i\phi[3]} |001\dots 0\rangle \\
&~~\vdots  \nonumber\\
 |000\dots 1\rangle &\rightarrow e^{i\phi[N_\mathsf{a}]} |000\dots 1\rangle \nonumber
\end{flalign}
where $\phi[n]$ is the phase applied to the $n$th qubit. Note that we are implicitly invoking the fact that it follows from correspondence that a product preparation of qubits corresponds to a product of preparation operators. For example,
\begin{equation}
|1\rangle\langle 1| \otimes |0\rangle\langle 0| \otimes |0\rangle\langle 0| \otimes \dots \otimes |0\rangle\langle 0| :=|100\dots 0\rangle\langle 100\dots 0|
\end{equation}
In (\ref{phaseonNqubits}) we have just considered the cases where one qubit is prepared in the 1 state and the remainder are prepared in the 0 states.  Of course there are other terms such as $|110\dots 1\rangle$ that accumulate a more complicated phase which we will not make particular use of. There are $N_\mathsf{a}$ qubits and, hence, a total of $2^{N_\mathsf{a}}$ states in the maximal distinguishable set associated with these qubits.  We are particularly interested in the $N_\mathsf{a}$ of these apearing above, namely $\{|100\dots 0\rangle$, $|010\dots 0\rangle$, $|001\dots 0\rangle$, $\dots$,$|000\dots 1\rangle\}$ which we will label by $m=1$ to $N_\mathsf{a}$ and represent as $|m\rangle^\mathsf{q}$. We will label the $|000 \dots 0\rangle$ state by $|0\rangle^\mathsf{q}$.  The remaining states in the computational basis for the qubits can be labled by $m=N_\mathsf{a}+1$ to $2^{N_\mathsf{a}}-1$.  Here $\mathsf{q}$ stands for the system type comprised of $N_\mathsf{a}$ qubits. We will show that we can perform an arbitrary unitary on a system of type $\mathsf a$ by applying the transformation
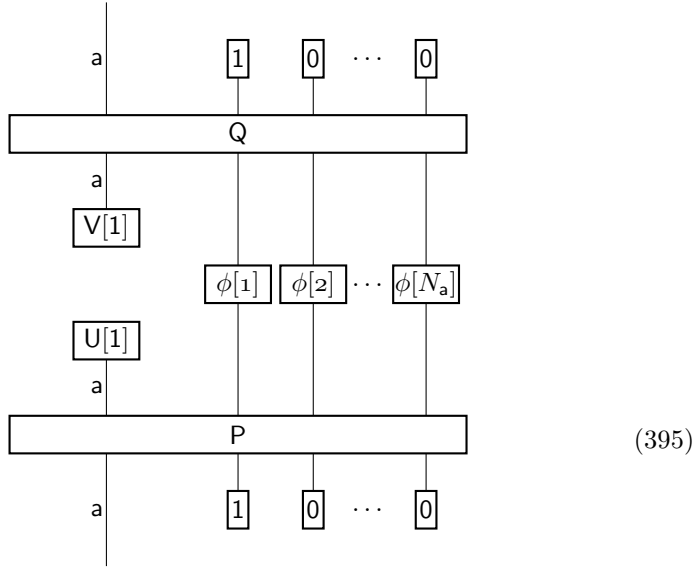
\begin{equation}\label{unitarydiagram}
\begin{Diagram}{0}{0}
\opbox[1]{zero1}{0,-4} \opsymbol{1} \opbox[1]{zero2}{4,-4} \opsymbol{0} \opbox[1]{zero3}{10,-4} \opsymbol{0}
\Opbox[30]{P}{0,0}
\opbox[4]{U1}{-7,5} \opsymbol{U[1]}
\opbox[4]{phi1}{0,8} \opsymbol{\mathnormal{\phi[1]}}
\opbox[4]{phi2}{4,8} \opsymbol{\mathnormal{\phi[2]}}
\opbox[4]{phi3}{10,8} \opsymbol{\mathnormal{\phi[N_\mathsf{a}]}}
\opbox[4]{V1}{-7,11} \opsymbol{V[1]}
\Opbox[30]{Q}{0,16}
\opbox[1]{zero1t}{0,20} \opsymbol{1} \opbox[1]{zero2t}{4,20} \opsymbol{0} \opbox[1]{zero3t}{10,20} \opsymbol{0}
\thispoint{in}{-7,-7} \thispoint{out}{-7,23}
\wire{in}{P}{1}{6.75} \opsymbol{a}
\wire{Q}{out}{6.75}{1}\opsymbol{a}
\wire{P}{U1}{6.75}{2.5} \opsymbol{a}
\wire{V1}{Q}{2.5}{6.75}  \opsymbol{a}
\wire{zero1}{P}{1}{15.5} \wire{zero2}{P}{1}{20.5} \wire{zero3}{P}{1}{28}
\wire{Q}{zero1t}{15.5}{1} \wire{Q}{zero2t}{20.5}{1} \wire{Q}{zero3t}{28}{1}
\wire{P}{phi1}{15.5}{2.5} \wire{P}{phi2}{20.5}{2.5} \wire{P}{phi3}{28}{2.5}
\wire{phi1}{Q}{2.5}{15.5} \wire{phi2}{Q}{2.5}{20.5} \wire{phi3}{Q}{2.5}{28}
\placelatex{7,-4}{\dots} \placelatex{7,8}{\dots}\placelatex{7,20}{\dots}
\end{Diagram}
\end{equation}
The unlabeled wires are qubits.  Applying {\bf P4$'$}, we choose the transformation $\mathsf{P}$ to implement a permutation, $\pi_\mathsf{P}$, with respect to the maximal set of distinguishable states formed by $|u[n]\rangle^\mathsf{a}|m\rangle^\mathsf{q}$  where
\begin{equation}
\pi_\mathsf{P} = \left( \begin{array}{l} nm \leftrightarrow mn ~~\text{for}~~ n,m=1~~\text{to}~~ N_\mathsf{a} \\
                           \text{other}~~ nm ~~ \text{not permuted}  \end{array} \right)
\end{equation}
This basically swaps the incoming state onto the space spanned by the first $N_\mathsf{a}$ states of the qubits.  Applying {\bf P4$'$}, we choose $\mathsf Q$ to implement a permutation, $\pi_\mathsf{Q}$, of the maximal distinguishable set of states formed by $|v[n]\rangle^\mathsf{a}|m\rangle^\mathsf{q}$ where
\begin{equation}
\pi_\mathsf{Q} = \left( \begin{array}{l} nm \leftrightarrow mn  ~~\text{for}~~ n,m=1~~\text{to}~~ N_\mathsf{a} \\
                           \text{other}~~ mn ~~ \text{not permuted}  \end{array} \right)
\end{equation}
This basically swaps the state back onto a system of type $\mathsf{a}$.   The phase rotations, $\phi[n]$, do not effect the distinguishable set of states $|m\rangle_\mathsf{b}\langle m|$ since the phase cancels.  Further, the phase rotations are all reversible transformations.  Hence we can apply exactly the same reasoning as in proving {\bf T\ref{revperm}} (with a slight but obvious elaboration to deal with the reversible transformation due to phase gates) to prove the transformation in (\ref{unitarydiagram}) is reversible and maps the states $\{|u[n]\rangle^\mathsf{a}\langle u[n]|: n=1~\text{to}~N_\mathsf{a}\}$, to the states, $\{|v[n]\rangle^\mathsf{a}\langle v[n]|: n=1~\text{to}~N_\mathsf{a}\}$.  Since the transformation is reversible it follows from {\bf T\ref{revlinearinH}} that it corresponds to the linear evolution in $\mathcal{H}^\mathsf{a}$ given in (\ref{fthetauv}).  We can chose the $\phi[n]$ in (\ref{unitarydiagram}) to cancel the $\theta[n]$ and hence achieve an arbitrary unitary transformation.  This proves {\bf T\ref{allunitaries}}.

\subsection{Proving Axiom 2}

We now prove
\begin{T}\label{magicoperation}
\index{complete set!of physical operators}For any complete set of physical operators, $\{\hat{ B}_\mathsf{a_1}^\mathsf{b_2}[l]: l=1~\text{to}~L\}$, a corresponding complete set of operations $\{\mathsf{ B}_\mathsf{a_1}^\mathsf{b_2}[l]: l=1~\text{to}~L\}$ are given by
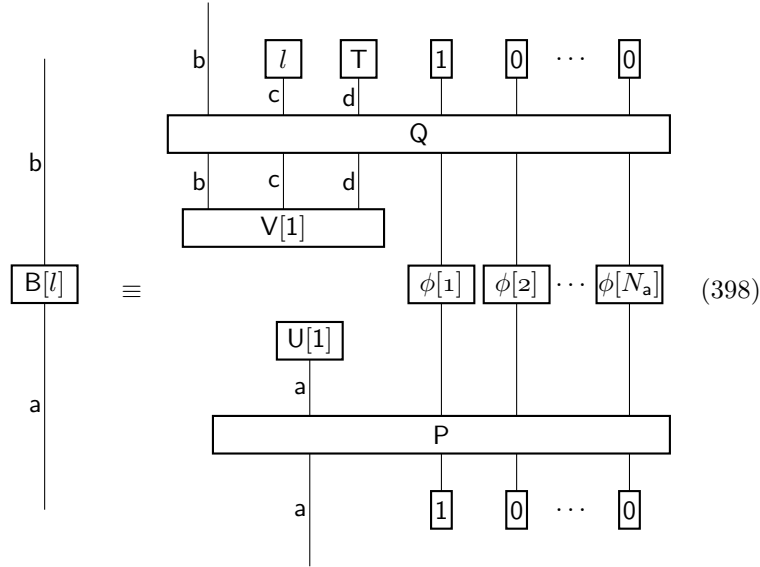
\begin{equation}\label{magic}
\begin{Diagram}{0}{0}
\opbox[4]{B}{0,0} \opsymbol{B\mathnormal{[l]}}
\thispoint{in}{0,-12} \thispoint{out}{0,12}
\wire{in}{B}{1}{2.5} \opsymbol{a}
\wire{B}{out}{2.5}{1} \opsymbol{b}
\end{Diagram}
~~~~\equiv ~~
\begin{Diagram}{0}{-2}
\opbox[1]{zero1}{0,-4} \opsymbol{1} \opbox[1]{zero2}{4,-4} \opsymbol{0} \opbox[1]{zero3}{10,-4} \opsymbol{0}
\Opbox[30]{P}{0,0}
\opbox[4]{U1}{-7,5} \opsymbol{U[1]}
\opbox[4]{phi1}{0,8} \opsymbol{\mathnormal{\phi[1]}}
\opbox[4]{phi2}{4,8} \opsymbol{\mathnormal{\phi[2]}}
\opbox[4]{phi3}{10,8} \opsymbol{\mathnormal{\phi[N_\mathsf{a}]}}
\opbox[13]{V1}{-8.4,11} \opsymbol{V[1]}
\opbox[2]{aux1t}{-8.4,20} \opsymbol{\mathnormal{l}}
\opbox[2]{aux2t}{-4.4,20} \opsymbol{T}
\Opbox[33]{Q}{-1.2,16}
\opbox[1]{zero1t}{0,20} \opsymbol{1} \opbox[1]{zero2t}{4,20} \opsymbol{0} \opbox[1]{zero3t}{10,20} \opsymbol{0}
\thispoint{in}{-7,-7} \thispoint{out}{-12.4,23}
\wire{in}{P}{1}{6.75} \opsymbol{a}
\wire{Q}{out}{3}{1}\opsymbol{b}
\wire{P}{U1}{6.75}{2.5} \opsymbol{a}
\wire{V1}{Q}{2}{3} \opsymbol{b}
\wire{V1}{Q}{7}{8} \opsymbol{c}
\wire{V1}{Q}{12}{13} \opsymbol{d}
\wire{Q}{aux1t}{8}{1.5} \opsymbol{c}
\wire{Q}{aux2t}{13}{1.5} \opsymbol{d}
\wire{zero1}{P}{1}{15.5} \wire{zero2}{P}{1}{20.5} \wire{zero3}{P}{1}{28}
\wire{Q}{zero1t}{18.5}{1} \wire{Q}{zero2t}{23.5}{1} \wire{Q}{zero3t}{31}{1}
\wire{P}{phi1}{15.5}{2.5} \wire{P}{phi2}{20.5}{2.5} \wire{P}{phi3}{28}{2.5}
\wire{phi1}{Q}{2.5}{18.5} \wire{phi2}{Q}{2.5}{23.5} \wire{phi3}{Q}{2.5}{31}
\placelatex{7,-4}{\dots}  \placelatex{7,8}{\dots}\placelatex{7,20}{\dots}
\end{Diagram}
\end{equation}
where $\mathsf P$ and $\mathsf Q$ are appropriately chosen reversible permutation transformations, $\{\phi[n]: n=1~\text{to}~N_\mathsf{a}\}$ are appropriately chosen phases, $\mathsf c$ and $\mathsf d$ are ancillary systems having appropriate $N_\mathsf{c}$ and $N_\mathsf{d}$, and $\mathsf{V^{b_2c_3d_4}}[1]$ is an appropriately chosen preparation (for a pure state).  The unlabeled wires represent qubits. $\mathsf T$ is the deterministic effect.
\end{T}
We know from {\bf T\ref{completephysicalcompletecpmaps}} that a complete set of physical operators, $\{B_\mathsf{a_1}^\mathsf{b_2}[l]: l=1 ~\text{t}~L\}$, can be associated with a set of superoperators, $\{ \$_B^l(\cdot): l=1 ~\text{to}~ L\}$ that are completely positive and whose sum is trace preserving.   We can write these superoperators in Krauss form \cite{nielsen2000quantum}:
\begin{equation}\label{Choidecomp}
\$_B^l (\hat A^\mathsf{a_1}) =
\hat B_\mathsf{a_1}^\mathsf{b_2}[l] \hat A^\mathsf{a_1} = \sum_i E_\mathsf{a_1}^\mathsf{b_2}[li] \hat A^\mathsf{a_1} E_\mathsf{a_1}^\mathsf{b_2}[li]^\dagger
\end{equation}
where $E_\mathsf{a_1}^\mathsf{b_2}[li]$ is any set of operators that act on vectors in ${\cal H}_\mathsf{a}$ to return vectors in ${\cal H}_\mathsf{b}$ having the property
\begin{equation}\label{EEIdent}
\sum_{li} E_\mathsf{a_1}^\mathsf{b_2}[li]^\dagger E_\mathsf{a_1}^\mathsf{b_2}[li] = \hat I_\mathsf{a_1}^\mathsf{a_1}
\end{equation}
where $\hat I_\mathsf{a_1}^\mathsf{a_1}$ is the identity operator on ${\cal H}_\mathsf{a}$.  We define
\begin{equation}
|v[n]\rangle^\mathsf{b_2c_3d_4} := \sum_{il} E_\mathsf{a_1}^\mathsf{b_2}[il] |u[n]\rangle^\mathsf{a_1}|l\rangle^\mathsf{c_3}|i\rangle^\mathsf{d_4}
\end{equation}
We see that these states are orthonormal:
\begin{equation}
\langle v[n']|v[n]\rangle=
\sum_{li} \langle u[n']| E_\mathsf{a_1}^\mathsf{b_2}[li]^\dagger  E_\mathsf{a_1}^\mathsf{b_2}[li] | u[n] \rangle
=\delta_{n'n}
\end{equation}
where we have used (\ref{EEIdent}).   We can complete this set of $N_\mathsf{a}$ orthonormal states for $\mathsf{bcd}$ into an orthonormal basis set,
$\{ |v[n]\rangle^\mathsf{bcd}: n=1~\text{to}~N_\mathsf{bcd} \}$, by adding a further $N_\mathsf{bcd} - N_\mathsf{a}$ basis states (any set that gives a basis will do).  By {\bf T\ref{pvmeasures}}, we know that projectors onto these states comprise a maximal distiguishable set of states.  We will use the same notation for the $N_\mathsf{a}$ qubits as in the proof of {\bf T\ref{allunitaries}}.  By {\bf P4}$'$ we can choose the transformation $\mathsf P$ to be a reversible permutation that permutes the maximal distinguishable set of states corresponding to the operators $|u[n]\rangle^\mathsf{a}\langle u[n]|\otimes |m\rangle^\mathsf{q}\langle m|$ according to the permutation
\begin{equation}
\pi_\mathsf{P} = \left( \begin{array}{l} nm \leftrightarrow mn  ~~\text{for}~~ n,m=1~~\text{to}~~ N_\mathsf{a} \\
                           \text{other}~~ nm ~~ \text{not permuted}  \end{array} \right)
\end{equation}
This basically swaps the state of $\mathsf a$ onto the qubits.  Since we have a reversible transformation we have, by {\bf T\ref{revlinearinH}},
\begin{equation}
|u[n]\rangle^\mathsf{a} |1\rangle^\mathsf{q} \rightarrow  \exp{i\theta_P[n]} |u[1]\rangle^\mathsf{a} |n\rangle^\mathsf{q}
\end{equation}
after the $\mathsf P$ transformation.  Next, the $\mathsf a$ system is detected with certainty at the $U_\mathsf{a}[1]$ effect (as its state is $|u[1]\rangle^\mathsf{a}\langle u[1]|$).  If we take the circuit trace we can eliminate the space associated with $\mathsf a$.  The qubits each pass through a phase gate and the state becomes $\exp i(\theta_P[n]+\phi[n]) |n\rangle^\mathsf{q}$ (see the proof of {\bf T\ref{allunitaries}}).  Hence, we have
\begin{equation}
\exp i(\theta_P+\phi) |v[1]\rangle^\mathsf{bcd} |n\rangle^\mathsf{q}
\end{equation}
impinging on the transformation $\mathsf Q$.   Invoking {\bf P4}$'$ we can choose the transformation $\mathsf Q$ to be a reversible permutation that permutes the maximal distinguishable set of states associated with
$|v[n]\rangle^\mathsf{bcd}\langle v[n]|\otimes |m\rangle^\mathsf{q}\langle m|$ according to the permutation
\begin{equation}
\pi_\mathsf{Q} = \left( \begin{array}{l} nm \leftrightarrow mn  ~~\text{for}~~ n, m=1~~\text{to}~~ N_\mathsf{a} \\
                           \text{other}~~ nm ~~ \text{not permuted}  \end{array} \right)
\end{equation}
The state after this transformation is, hence,
\begin{equation}
\exp i(\theta_P[n]+\phi[n]+\theta_Q[n]) |v[n]\rangle^\mathsf{bcd} |1\rangle^\mathsf{q}
\end{equation}
Since the outgoing state of the qubits is that associated with $|100\dots 0\rangle$, the effects for the qubits after $Q$ all fire with certainty. If we put the state in projector form we can take the circuit trace we eliminate the $\mathsf{q}$ part of the state.   We are left with the state, $\exp i(\theta_P[n]+\phi[n]+\theta_Q[n]) |v[n]\rangle^\mathsf{bcd}$ for $\mathsf{bcd}$.  We choose $\phi[n]=-(\theta_P[n]+\theta_Q[n])$.  This means that if we send $|u[n]\rangle^\mathsf{a}$ in, we get $|v[n]\rangle^\mathsf{bcd}$ out for $\mathsf{bcd}$.  That is,
\begin{equation}
|u[n]\rangle^\mathsf{a_1} \rightarrow \sum_{il} E_\mathsf{a_1}^\mathsf{b_2}[il] |u[n]\rangle^\mathsf{a_1}|l\rangle^\mathsf{c_3}|i\rangle^\mathsf{d_4}
\end{equation}
The transformation up to this stage is reversible by the same reasoning as we used in proving {\bf T\ref{revperm}} (we can build a transformation that does each of the steps so far discussed in reverse).  Hence by {\bf T\ref{revlinearinH}},
\begin{equation}
|\psi\rangle^\mathsf{a_1} := \sum_n c_n |u[n]\rangle^\mathsf{a_1} \rightarrow
\sum_{il} E_\mathsf{a_1}^\mathsf{b_2}[il] |\psi\rangle^\mathsf{a_1}|l\rangle^\mathsf{c_3}|i\rangle^\mathsf{d_4}
\end{equation}
A general positive operator can be written as
\begin{equation}
\hat A^\mathsf{a} = \sum_\mu a_\mu |\psi_\mu\rangle^\mathsf{a}\langle\psi_\mu|
\end{equation}
(for some real coefficients $a_\mu$)
and will, therefore, evolve according to
\begin{equation}
\hat A^\mathsf{a} \rightarrow  \sum_{li,l'i'} E_\mathsf{a_1}^\mathsf{b_2}[li] \hat A^\mathsf{a} E_\mathsf{a_1}^\mathsf{b_2}[l'i']^\dagger \otimes
|l\rangle^\mathsf{c}\langle l'| \otimes |i\rangle^\mathsf{d} \langle i'|
\end{equation}
Finally, we have effects in the $\mathsf c$ and $\mathsf d$ paths.  The operator associated with the effect in path $\mathsf c$ is, by {\bf T\ref{pvmeasures}}, equal to $|l\rangle_\mathsf{c}\langle l|$.  The deterministic effect, $\mathsf{T_d}$, is assocaited with the identity operator, $\hat I_\mathsf{c}$ (by {\bf T\ref{TisI}}). If we take the circuit trace after these effects have been included then we obtain (when we have outcome $l$)
\begin{equation}
\hat A^\mathsf{a_1} \rightarrow \sum_i E_\mathsf{a_1}^\mathsf{b_2}[li] \hat A^\mathsf{a_1} E_\mathsf{a_1}^\mathsf{b_2}[li]^\dagger
\end{equation}
which is what we had in (\ref{Choidecomp}). Hence the complete set of operations in ({\ref{magic}}) can have any corresponding complete set of physical operators with appropriate choices of $\mathsf P$, $\mathsf Q$, $\mathsf c$, $\mathsf d$, ${\mathsf V^{bcd}}[1]$ and $\{\phi[n]\}$. This proves {\bf T\ref{magicoperation}}.

The complete set of operations in {\bf T\ref{magicoperation}} can simulate any complete set of operations whether they constitute a set of preparations, a set of transformations, or a set of results (a measurement).  In the case that we are simulating a preparation we should think of $\mathsf a$ as the trivial system (having $N_\mathsf{a}=1$) and we should send in a pure state (that is normalized). In the case we are simulating an effect we should, likewise, think of $\mathsf b$, as being the trivial system (having $N_\mathsf{b}=1$) and we should place a maximal effect in the output.  Interestingly, we can also consider the case where both $\mathsf a$ and $\mathsf b$ are trivial systems and where we send in a pure state and have a maximal effect in the output.  In this case, we obtain a complete set of circuits.  Each circuit will have a probability associated with it.   {\bf T\ref{magicoperation}} applied to this special case proves that we can obtain any set of probabilities, $p[l]$, that add up to one.

Axiom 2 \index{axioms!\textbf{Axiom 2}}follows from {\bf T\ref{magicoperation}}.  Hence we have derived quantum theory (for finite dimensional Hilbert space) from the five postulates {\bf P1-5}.  Further classical probability theory and quantum theory are the only two theories consistent with {\bf P1, P2, P3, P4}$'$, and {\bf P5}.

\newpage

\part{Discussion}

\section{The nature of the reconstruction}

The reconstruction given in this paper is in the context of an operational framework.   This raises questions about the role of operationalism in fundamental physics which we will discuss below (in Sec.\ \ref{ontology}).   Many other reconstructions are also cast in operational frameworks (see Sec.\ \ref{previouswork}).  What is it that distinguishes the reconstruction here?  The most notable feature is the special role played by maximal sets of distinguishable states.  They are mentioned explicitly or implicitly (by reference to maximal effects and measurements) in three of the five postulates.  Concepts that are defined with respect to maximal sets of distinguishable states play an important role both in setting up the basic ideas (such as filters and systems) and in constructing the proofs in this paper.  Such maximal sets of distinguishable states are a good concept for a reconstruction because they are very basic in our classical conception of the world.  Further, without distinguishable states in the world we would not be able to do very much.  For example, this paper could not be written as we would not have an alphabet by which to form words.  It is possible (even likely) that, in some deeper theory, distinguishable states are an approximate concept.   Distinguishable states allow us to carry information forward in time.  If we have indefinite causal structure (as in a theory of quantum gravity) then there will not be a fundamental notion of \lq\lq forward in time" and so we may not have distinguishable states at the deepest level.

\section{Previous work on reconstruction}\label{previouswork}

In this section we will consider the relationship of operational postulates presented here with some previous papers that take a similar point of view (such as adopting tomographic locality as a postulate).

In 2001 the author provided a set of operationally motivated axioms for quantum theory \cite{hardy2001quantum}.  In modern form (see \cite{hardy2009foliable}), these are
\begin{description}
\item[Information] Systems having, or constrained to have, a given information carrying
capacity have the same properties.
\item[Information locality] Same as {\bf P2}.
\item[Tomographic locality] Same as {\bf P3}.
\item[Continuity] There exists a continuous reversible transformation between any
pair of pure states.
\item[Simplicity] States are specified by the smallest number of probabilities consistent
with the other axioms.
\end{description}
Causality was implicit as a background assumption in \cite{hardy2001quantum} whereas it follows as a theorem ({\bf T\ref{deteffect}}) in the present work.  The biggest improvement in the present work is that we get rid of the simplicity axiom.  We do this using the technique discovered by Chiribella, D'Ariano, and Perinotti \cite{chiribella2010probabilistic, chiribella2010informational} in which the output of a teleportation transformation is fed into the input. This technique has to be adapted to the present work since we start with rather different postulates.  The information axiom of \cite{hardy2001quantum} is a rather strong constraint.  In the present work it follows very naturally from more basic postulates.  The idea is that we can swap the state onto a second system by choosing an appropriate permutation of a set of distinguishable states for the composite system.   The continuity axiom of \cite{hardy2001quantum} is no longer necessary.  The continuity axiom has two parts - a transitivity part and a continuity part.  The transitivity part in the continuity axiom (that there exists a reversible transformation between any pair of pure states) now follows as a theorem ({\bf T\ref{revt}}) from {\bf P1, P2, P3} and {\bf P4}$'$.  The continuity property of this axiom does not follow at a low level in the reconstruction.  If we use {\bf P4} (so we assume compound permutatability) then we immediately deduce that there is at least one pure state \lq\lq between" any pair of distinguishable states for a gebit.  This takes us a little way in the direction of the continuity property. It is striking that this is enough to construct the continuum of pure states (as done finally in {\bf T\ref{allprojectors}}).

In 2009 Daki\'c and Brukner \cite{dakic2009quantum} attempted a reconstruction from the following axioms: (1) Information, (2) Information locality, (3) Tomographic locality, (4) Transitivity (that there exists a reversible transformation between any pair of pure states), and (5) Gebit state spectrality (any state for a gebit can be written as a convex combination of a pair of maximally distinguishable states).  Causality is taken as a background assumption. Information locality is not explicitly stated as an axiom but is explicitly used in the reconstruction.  They claim that the only two theories consistent with these axioms are classical probability theory and quantum theory and suggest that if transitivity is replaced by the continuity axiom then quantum theory is singled out. With this substitution the axioms of Daki\'c and Brukner are the same those of \cite{hardy2001quantum} except that gebit state spectality replaces simplicity.  The authors show how it follows that all points on the hypersphere must correspond to pure states.  First they use transitivity to to show that pure states must correspond to points on a hypersphere.  It then follows almost immediately from gebit spectrality that all points on this hypersphere correspond to pure states.  This is an important step in \cite{dakic2009quantum} on route to getting rid of the simplicity axiom.  To actually get rid of the need for simplicity axiom (i.e. to show that the hypersphere is a regular 2-sphere) Daki\'c and Brukner employ a quite remarkable technique involving two gebits.  There is a significant technical difficulty with the proof provided in their paper as the authors assume that any group of transformations which are transitive on a $(d-1)$-sphere contains, at least, $SO(d)$.  This is not true \cite{montgomery1943transformation, borel1949some}.  There are counterexamples for an infinite number of cases with odd $(d-1)$.  There is only one counterexample for the case of even $(d-1)$.  For $(d-1)=6$ the exceptional Lie group $G_2$ (which is a proper subgroup of $SO(7)$) is transitive on the sphere. Now $K=N^r$, $N=2$ (for a gebit), and $d=K-1$ (the minus one is for normalisation of states).   Hence we actually only need to consider even spheres of dimension $2^r-2$.  Unfortunately the 6-sphere is one such case.  Hence there exists one pertinent counterexample relevant to Daki\'c and Brukner's work that needs to be addressed.  Work by Masanes and M\"uller \cite{masanes2010derivation} suggests this counterexample can be eliminated.   Ignoring this technical gap in the proof of \cite{dakic2009quantum}, the main advantage of the present work over the work of Daki\'c and Brukner is that deeper reasons are given for all the points on the hypersphere representing pure states.  Daki\'c and Brukner use gebit state spectrality to do this.   It is better if axioms apply to all types of system rather than being restricted to special cases (such as gebits).  This is particularly true in quantum theory because qubits are rather special when compared with general quantum systems.  A more general statement that Daki\'c and Brukner could have used would simply be \emph{state spectrality}: any state can be written as a convex combination of the states in some maximal set of distinguishable states \cite{hardynature}.  It follows from the fact that density matrices can be diaganolised that this is true in quantum theory.  This property is rather suprising.  It means that any state can be written as a convex sum of $N_\mathsf{a}$ extremal points.  If we want to include the unnormalised states then we need to include the null state in our convex combinations and so we need $N_\mathsf{a}+1$ extremal points.  According to Carath\'eodory's theorem, any point in a any convex set of dimension $K_\mathsf{a}$ can be written as a convex combination of $K_\mathsf{a}+1$ extremal points.  Since $K_\mathsf{a}=N_\mathsf{a}^2$ in quantum theory, this raises the question of why states in quantum theory can be written in terms of far fewer extremal states than we would expect in the generic case.  This is the sort of thing that should be explained by axioms for quantum theory rather than assumed.  Postulate {\bf P5} that filters are non-mixing and non-flattening is, we argue here, more natural than state spectrality in view of the special role that filters play in allowing us to preparing systems for experiments.

In 2010 Masanes and M\"uller \cite{masanes2010derivation} considered the following axioms: (1) Information, (2) Tomographic locality, (3) Transitivity, and (4) All measurements (for a gebit, all mathematically well-defined measurements are allowed by the theory). Additionally, they have an axiom that gebits are characterized by a finite number of probabilities (in the present work a similar role is played by the background assumption {\bf Assump2}).  Causality is taken as a background assumption.  It is shown that classical probability theory and quantum theory are the only two theories consistent with these axioms.  They suggest substituting the continuity axiom for the reversibility axiom to single out quantum theory.  Note that information locality is not assumed.  Rather, Masanes and M\"uller derive it from their axioms. They use their axiom that all mathematically well-defined measurements are allowed to prove that all points on the hypersphere for a gebit correspond to pure states.   They use group theoretic methods to show the hypersphere must be a 2-sphere (employing a trick used by Daki\'c and Brukner along the way).  They address the issue of the exception for the 6-sphere mentioned above by considering and eliminating the exceptional Lie group $G_2$ as a possible space for the gebit.  Their 4th axiom appears to be related to gebit state spectality (as used by Daki\'c and Brukner) in that it immediately gives the property that all points on the hypersphere correspond to pure states in the same environment of assumptions.  Masanes and M\"uller also show how to achieve this step in their proof using \emph{perfect distinguishability} (this is one of the axioms of Chiribella, D'Ariano, and Perinotti given below).

The present work was, in part, motivated by \cite{dakic2009quantum} and \cite{masanes2010derivation}.  Daki\'c and Brukner showed that there is a route to getting rid of the simplicity axiom.  Then Masanes and M\"uller dealt with the technical issue mentioned above.  Additionally, they showed that information locality follows from other axioms.  This suggests the following thought: if we turn this round,  maybe information locality can be used to derive some of those other axioms.   Indeed, it turns out that information locality plays a central role in deriving transitivity ({\bf T\ref{revt}}) and the fact that systems having the same $N_\mathsf{a}$ are equivalent (the information axiom above).  Information locality is a far more natural assumption than transitivity and the information axiom.

In 2006 D'Ariano initiated his own research program aimed at reconstructing quantum theory from operational axioms (see \cite{d2008probabilistic} and references therein).   This program culminated in 2010 with the beautiful work of Chiribella, D'Ariano, and Perinotti (CDP) \cite{chiribella2010informational} (see also \cite{chiribella2010probabilistic}) who give the following axioms:
\begin{description}
\item[Causality] The probability of preparations is independent of the choice of observations.
\item[Perfect distinguishability] Every state that is not completely mixed can be perfectly distinguished from some other state.
\item[Ideal compression] For every state there exists an ideal compression scheme.
\item[Local distinguishability] If two bipartite states are different, then they give different probabilities for at least one product experiment.
\item[Pure conditioning] If a bipartite system is in a pure state, then each outcome of an atomic measurement on one side induces a pure state on the other.
\item[Purification] Every state has a purification. For fixed purifying system, every two purifications of the same state are connected by a reversible transformation on the purifying system.
\end{description}
The local distinguishability axiom is identical to the assumption of tomographic locality.  CDP explicitly state causality as one of their axioms.  They regard the first five axioms as being standard in that they define a broad class of information processing theories.  The purification postulate is then regarded as the assumption that singles out quantum theory within this broad class. (We could take  a similar attitude to the five postulates presented in this paper.  {\bf P1}, {\bf P2}, {\bf P3}, and {\bf P4}$'$ define a broad class of physical theories developed in Sec.\ \ref{filtersandsystemssection} and {\bf P5} singles out classical probability theory and quantum theory.)  A longer version of the third axiom is provided \lq\lq Ideal compression axiom: every source of information can be encoded in a suitable physical system in a lossless and maximally efficient fashion. Here lossless means that the information can be decoded without error and maximally efficient means that every state of the encoding system represents a state in the information source."  CDP arrived independently at a similar derivation of information locality ({\bf P3} in the present work) to Masanes and M\"uller.   CDP use avoid the need for a simplicity axiom by a novel technique involving teleportation.   This technique has has been adopted in the present work.

\section{Infinite dimensional Hilbert spaces?}

Our objective here was simply to deal with finite dimensional Hilbert spaces.  Many applications of quantum theory make use of an infinite dimensional Hilbert space.  The postulates given here make sense when $N_\mathsf{a}$ is countably infinite.  Hence, the postulates are applicable when we have countably infinite Hilbert space dimension.  Further, in any restriction of such a theory to the finite $N_\mathsf{a}$ case we would obtain finite dimensional quantum theory from the postulates.  However, the following questions are open: (a) do postulates {\bf P1-5} give a unique theory in the case where $N_\mathsf{a}$ can be countably infinite and, if so, (b) is this this theory standard quantum theory for countably infinite dimensional Hilbert space?  Experimentally we would never be able to distinguish between the finite and infinite dimensional cases so the interest in answering these questions has more to do with which theories we can formulate with respect to these postulates (in particular, it is possible to impose various continuous symmetries in the case of infinite dimensional Hilbert spaces that cannot be easily accommodated in finite dimensional Hilbert spaces).

\section{Quantum field theory}

\index{quantum field theory}The duotensor framework approach to quantum theory outlined in Part \ref{theduotensorframework} may offer a route to reformulating quantum field theory.
In fact, first we could consider the problem of formulating probabilistic field theories (for continuous fields) in general.  The approach in this paper was for finite systems only. By this we mean that (i) any operation or fragment has a finite number of inputs and outputs,
and (ii) each system type is associated with a finite $K_\mathsf{a}$. To do field theory for continuous fields we would need to relax these requirements. One possible way to proceed is the following. Work with a fixed Minkowski background. Each fragment would be associated with a spacetime region having a boundary. Fragments have settings and outcome sets. The setting might be imposed by some classical apparatus that is part of the experiment.  The outcomes would be read off detectors or other output devices.  We are able to wire together two fragments if some part of their boundaries fit together (this may require a boost). We can consider infinitesimal areas on the boundary. Associated with each infinitesimal area with outward normal pointing to the future would be an output. Associated with each infinitesimal
area with outward normal pointing to the past would be an input. Associated with any part of the boundary with normal pointing in a spacelike direction
would be both an input and an output. The type associated with the input and output would be determined by (a) the type of field and (b) the invariant area of
the infinitesimal. Under system composition the areas would add. By imposing full decomposability of operations we could set up the duotensor framework so long as we are able to consistently replace sums with integrals as required.  If this worked it would allow us to do both classical field theory (for a probabilistic version of electromagnetism for example) and quantum field theory. To do quantum field theory we could use the operator duotensor approach.  Associated with each region of spacetime having a specified setting and outcome set would be a positive operator.  These operators could be combined by the circuit trace operation to form an operator for a composite region.

We could not do general relativity this way since we have assumed a fixed background metric.

This approach to field theory would enforce a different attitudes than normally taken by field theorists in two respects.  First, the need to think about settings and outcomes makes this an operational rather than ontological approach.  Second, the approach is intrinsically probabilistic.  In the case of quantum theory it is fundamentally based on objects (operators) which are generalizations of density matrices and completely positive maps rather than pure states and unitary evolution.  In particular, the full decomposability property of operators makes most sense in this setting.

\section{Quantum Gravity}\label{quantumgravity}

\index{quantum gravity}When Newton accounted for Kepler's laws of plannetary motion in terms of his three laws of motion and his universal law of gravitation he actually did more.  He also accounted for the motion of the bodies in arbitrary gravitational systems.  By finding a deeper explanation of the adhoc laws of Kepler he was able to go beyond Kepler's physics to new physics.  Einstein's derivation of the Lorentz transformations from two simple axioms paved the way for the development of general relativity (by Einstein), and once again this approach led to new physics.   The great open problem in fundamental physics today is the problem of quantum gravity.  This is to find a theory that reduces, in appropriate limits, to the physics of quantum theory on the one hand, and the physics of general relativity on the other.  The new theory may be as different mathematically from either quantum theory or general relativity as these theories are from the physics that preceded them.   To obtain such a theory it is most likely that we need to understand the present theories at a deep conceptual level.  The approach of reformulating them in mathematical and operational terms is likely to be helpful here.

General relativity and quantum theory are both conservative and radical compared with the physics that preceded them, but in complementary respects. General relativity is conservative in that it is deterministic.  It is radical in that it has non-fixed causal structure.  Quantum theory is conservative in that it has fixed causal structure.  It is radical in that it is intrinsically probabilistic (it cannot be formulated in its standard form without resort to probabilities). It seems likely that a theory of quantum gravity will inherit the radical features of both theories.  In fact, we can expect it to be a little more radical still.  In general relativity causal structure is non-fixed. However, once we have a solution to the field equations, the metric is everywhere given.  Hence the causal structure is definite.  However, in quantum theory, any quantity that can vary is subject to quantum superposition.  This leads to a \lq\lq no-matter-of-factness" or \lq\lq fundamental indefiniteness" about the value of the quantity.  While we may not expect the mathematics that leads to linear superpositions to survive in quantum gravity (the theory could be formulated in terms of very different mathematics) it is likely that the qualitative feature of \lq\lq no-matter-of-factness" will survive.  Since causal structure varies in general relativity this suggests that we will have fundamentally indefinite causal structure in quantum gravity. If so then there will be no-matter-of-the-fact about whether a particular interval is spacelike or timelike.  The circuit model considered here has wires.   The obvious interpretation of these wires is that they allow systems to pass from one apparatus use to another.  In other words, they are the circuit analogues of timelike intervals.  This interpretation is enforced by the property of causality.  All this suggests that we need to work out how to formulate operational theories without using the notion of timelike intervals at a fundamental level.  The causaloid framework
\cite{hardy2005probability,hardy2007towards} is a preliminary effort in this direction.

In an operational approach to physics we need to start by considering how we go about describing the world operationally.  Typically this consists of considering small parts of the world and specifying different ways in which they can be connected up.  The operational framework established in Part \ref{thecircuitframework} is, most likely, insufficient for the task of accommodating quantum gravity for two reasons.  First, it started with the idea of aligning the apertures on apparatus uses (such that we can imagine systems passing through).  However, at the operational level we can imagine experiments that cannot be described in these terms.  Indeed, for the reasons given above, this aspect of the circuit model may be particularly problematic for quantum gravity.  Second, the circuit model deals with the case where we calculate the probability for outcomes on the operations given a fixed wiring.  In general relativity, however, we are interested in predicting coincidences between bodies.  In some probabilistic version of general relativity (where we have probabilistic ignorance as to the value of observable quantities) we would be interested in calculating probabilities for different configurations of coincidences. Graphically this would correspond to calculating probabilities for different graphs.  Preliminary ideas in this direction were outlined in \cite{hardy2009operational}.

\section{Operational methodology and ontology}\label{ontology}

Our ultimate objective in fundamental physics must be to gain the deepest understanding of the world that is possible.   This should give an account of what is real at the deepest level. That is, it should provide an ontology.   We have adopted an operational approach in this paper.  It might be thought that this is inconsistent with an attempt to discover the correct ontology\index{ontology} of the world.  This might be true if we were pursuing operationalism as a fundamental philosophy (in which it is asserted that there is no reality beyond instrument settings and readings as described at the macroscopic level).  However, we do not need to take this attitude.  Rather we can view the approach taken in this paper as an \emph{operational methodology}\index{operational methodology} aimed at gaining a deeper insight into certain structural properties of quantum theory.  The point of adopting a certain methodology is that it may  help us along the road in constructing the next fundamental physical theory (such as a theory of quantum gravity).   Operationalism as a methodology played an essential role in Einstein's approach to special relativity.  By thinking in operational terms, Einstein was able to see that the idea of absolute simultaneity need not be a feature of the fundamental ontology.  This was a pretty dramatic success for the operational methodology.  It is difficult to imagine how he could have had this insight had he not thought in operational terms (about how to synchronize distant clocks and so on).  Likewise, we can hope that the operational methodology will help us to divest ourselves of unnecessary ontological notions we currently take for granted that should play no role in quantum theory and, possibly, beyond.  An operational methodology enables us to proceed in a conceptual manner in the absence of a deeper fundamental picture of the world.  It is better that physics is driven by conceptual ideas than purely mathematical ones.  In particular, if we can formulate some essential idea in operational terms then it is more mobile as we move between mathematical frameworks.

It is, in some respects, deeply shocking how successful operationalism is for the purposes of reconstructing quantum theory.  The five postulates given in this paper can all be understood in operational terms.  The entire framework of quantum theory can be derived from operational ideas. Any ontology must account for the success of operationalism.  Of course, there are things that appear in applications of quantum theory that have not been derived such as particular Hamiltonians along with the constants that appear in them.  But surely quantum theory itself is more fundamental than the values of particular constants that appear in various applications.

While our ultimate aim ought to be to come up with an ontology, we should perhaps slightly temper rush to get there.  It is likely that quantum gravity (or whatever more fundamental theory supersedes quantum theory) will look quite different from quantum theory (as suggested in Sec.\ \ref{quantumgravity}).  In such a case there is a danger that the exercise of finding the best ontology for quantum theory will be entirely academic.  It could turn out that none of this ontology passes over to the more fundamental theory.  Then the world would not actually be as suggested by this ontology.  On the other hand, it is possible that that the path to the next fundamental theory will be via an incorrect ontological understanding of the world.   Indeed, there are many examples in the history of physics where something like this has happened.  In other words, we might usefully pursue an ontological methodology along side, or instead of, an operational one.   Of course, most physicists are not so dispassionate.   In the end we are driven to search for what we hope will turn out to be the correct ontology of the world.  After all, it is the desire to understand what reality is like that burns deepest in the soul of any true physicist.

\newpage

\section*{Postlude: a conjecture}\label{conjecturesection}
\addcontentsline{toc}{section}{Postlude: a conjecture}

We conjecture\index{conjecture} that {\bf P5} can be replaced with the following postulate
\begin{quote}
{\bf P5}$'$ Filters are non-mixing.
\end{quote}
To motivate this conjecture, consider the way in which the non-flattening assumption for filters was used in the reconstruction.  In Sec.\ \ref{hyperspherepopulated} we considered a set of states for a getrit that had support on a particular gebit space (call this $V_1$). We then sent them through a filter acting on the getrit that filtered down to a different gebit space (call this $V_2$).  It was clear that any state in the original gebit space would pass through the filter with some probability and that the smallest system that would support the outgoing states was a gebit (associated with $V_2$).  Assume all the incoming states lie on the surface of the convex cone of states associated with the incoming gebit (this means they are proportional to pure states).  If the incoming set of states is non-flat for $V_1$ then they would span space of this convex cone.  The outgoing set of states must lie on the surface of the cone associated with the outgoing gebit as we are still assuming that filters are non-mixing.  However, if the filtering transformation were flattening in this case then these outgoing states would not span the space of states of this cone - they would be flattened into a lower dimensional space.  We know from {\bf T\ref{filtersproject}} that the filtering transformation is a projection transformation into $V_2$.  It is very difficult to take a set of points on the surface a cone and flatten them into a lower dimensional space by a projective map while keeping them on the surface of the cone.  Certainly, if all points on the cone are in our set then the can be no such transformation.  However, we were using the no flattening assumption exactly to show that there must be all points on the surface of the cone (which has a hypersphere base) so this does not help us prove the conjecture.  It is possible to imagine sets of points on the initial cone that could be flattened.  One strategy to prove the conjecture would be to show that it follows from the postulates that the set of points representing states on the cone is such that it cannot be flattened.

Non-flattening property of filters was also used in proving {\bf T\ref{teleportationTone}}.  There we also had an input gebit space and an output gebit space so the above strategy should cover this case also.

Even if the conjecture is proven, the non-flattening property remains a deeply interesting property of quantum theory and, in this paper, it drives two key parts of the reconstruction: showing that all points on the hypersphere correspond to states and showing that we can prepare any state with a teleportation-type transformation.  These two proofs are important in showing that $K_\mathsf{a}=N_\mathsf{a}^2$.

% To he or she that solves this conjecture in maths I can understand I offer the prize of five Canadian dollars.  Solving it means either proving it false or true.

\newpage

\section*{Acknowledgements}
\addcontentsline{toc}{section}{Acknowledgements}

I am especially grateful to Chris Fuchs for getting me thinking about the issue of reconstructing quantum theory in the first place. I thank Gilles Brassard who, along with Chris Fuchs, invited me to a workshop in Montreal in 2000 that, in the event, I was unable to attend \cite{fuchs2009coming}.  The talk I began preparing for that workshop has matured into the present work.  I am very grateful to Giullio Chiribella, for explaining elements of the quantum combs approach \cite{chiribella2009theoretical} and, additionally, to Mauro D'Ariano, Paulo Perinotti, and Tony Short for additional remarks leading to the characterization of the operators considered in Sec.\ \ref{positivityunderinputtranspose} as having positive input transpose.  I am grateful to Markus Mueller for comments  that led led me to improve the section on probabilities.  I would like to thank Prakash Panangaden for inviting me to a workshop in Bellairs.  I am grateful to Bill Edwards for an awkward question at that workshop that led me to the duotensor way of thinking.  I would like to thank Bob Coecke for providing funding for a research visit to ComLab in the University of Oxford in the summer of 2010 where some of the ground work this project was laid and the summer of 2011 where some of the final improvements were made.   I would also like to thank Bob Coecke for many discussions on quantum picturalism which has been embraced in this paper and Samson Abramsky for long discussions on the content of this paper.

I am very grateful to everyone at the Casa Mia Cafe in Waterloo where most of this work was done.

I am grateful to Vivienne for teaching me much more about operationalism than I had previously known and to Vanessa for numerous useful discussions that helped shape the form of this rather long paper.

Research at Perimeter Institute for Theoretical Physics is supported in part by the Government of Canada through NSERC and by the Province of Ontario
through MRI.

\newpage

\part*{Appendices}
\addcontentsline{toc}{part}{Appendices}

\appendix

\section{Filters non-flattening in quantum theory}\label{appendixflattening}

\index{non-flattening transformations!in quantum theory}In this appendix we prove that filters are non-flattening in quantum theory. In fact we will prove that all non-mixing transformations are non-flattening in quantum theory. First, though, consider filters.  Assume we have a filter which acts on a Hilbert space ${\cal H}_\mathsf{a}$.  Consider states corresponding to density operators acting on ${\cal H}_\mathsf{a}$.   Assume we have a set of such states which have support on subspace, ${\cal H}_\mathsf{b}$ (and there exists no smaller subspace supporting these states).   Such a set is non-flat if it spans the space of positive operators acting on ${\cal H}_\mathsf{b}$ (because we can have a subset of outcomes for a maximal measurement associated with this subspace).   Now consider sending this set through a filter, $\mathsf F$ (this could be any filter).   Let the set of states which emerges from the filter have support on the subspace ${\cal H}_\mathsf{c}$ (where there exists no smaller subspace supporting these states).  This set of output states is non-flat if it spans the space of positive operators acting on ${\cal H}_\mathsf{c}$ (because we can have a filter that projects onto this subspace).   We will say the transformation is non-flattening if a non-flat set of input states gives rise to a non-flat set of output states.  In quantum theory the map on density operators due to any transformation is linear. By linearity, if a transformation is non-flattening for some (spanning) input set having support on a given ${\cal H}_\mathsf{b}$, then it is non-flattening for any other (spanning) input set having support on ${\cal H}_\mathsf{b}$.
%And by linearity again, if it is flattening for any input set having support on ${\cal H}_\mathsf{b}$, then it is flattening for any other input set having support on ${\cal H}_\mathsf{b}$.
Hence, just need to prove that the filter is non-flattening for some non-flat input set for every ${\cal H}_\mathsf{b}$.  In fact, it is sufficient to prove that there exists some set of input states having support on ${\cal H}_\mathsf{b}$ which give rise to a non-flat set of output states having support on ${\cal H}_\mathsf{c}$ (where there exists no smaller subspace supporting these output states).  If this is the case then we can, clearly, complete the input set into a non-flat set without effecting the property that the output set is non-flat also.

Now consider sending a non-flat set of states having support on ${\cal H}_\mathsf{b}$ onto the filter.  We will assume that, after the filter, the smallest subspace that the states have support in is of dimension $N_\mathsf{c}$.  We will refer to this as system $\mathsf c$.  Let the Hilbert space for $\mathsf c$ be spanned by the orthonormal set $\{ |n\rangle: n=1~\text{to}~N_c\}$.  Hence,

Filters are non-mixing.  If we send in a state $|\psi\rangle$ we get out a state $\hat F|\psi\rangle$ (where $\hat F$ is the projector associated with the filter). Since the output states have support on ${\cal H}_\mathsf{c}$ there must exist $N_\mathsf{c}$ linearly independent pure states, $\{|u'_n\rangle: n=1~\text{to}~ N_\mathsf{c}\}$, in the output set (linearly independent when represented as vectors in $\mathcal H_{\mathsf{c}}$) since this is what it means for $\mathsf c$ to be the smallest system. Let the input state that leads to $|u'_n\rangle$ be $|u_n\rangle$.  Define
\begin{equation}
|u_{mnx}\rangle=\frac{1}{\sqrt{2}} (|u_m\rangle+|u_n\rangle)~~~~\text{and}~~~~ |u_{mny}\rangle=\frac{1}{\sqrt{2}} (|u_m\rangle+i|u_n\rangle)
\end{equation}
Consider the states
\begin{equation}
A=\{\sigma_n: n=1~\text{to}~N_\mathsf{c} \}\cup\{\sigma_{mnx},\sigma_{mny}: m,n=1~\text{to}~N_\mathsf{c}, m<n\}
\end{equation}
where
\begin{equation}
\sigma_n=|u_n\rangle\langle u_n|, ~~~~\sigma_{mnx}=|u_{mnx}\rangle\langle u_{mnx}|, ~~~~\sigma_{mny}=|u_{mny}\rangle\langle u_{mny}|
\end{equation}
If we send the set of states $A$ onto the filter then we will get the states
\begin{equation}
A'=\{\sigma'_n: n=1~\text{to}~N_\mathsf{c} \}\cup\{\sigma'_{mnx},\sigma'_{mny}: m,n=1~\text{to}~N_\mathsf{c}, m<n\}
\end{equation}
out (which are defined as above but with primes on the $u$'s).  Now this set of states is non flat.  To prove this we use the result, due to Duan and Guo \cite{duan1998probabilistic}, that there exists a linear quantum operation which converts any linearly independent set of states such as $|u'_n\rangle$ to the same number of orthogonal states $\sqrt{\gamma_n}|n\rangle$.  This transformation is probabilistic and the success probability is greater than zero for each $n$. Hence, $\gamma_n>0$ for all $n$.  Since the transformation is linear, the state $a|u'_m\rangle+b|u'_n\rangle$ gets converted to $a\sqrt{\gamma_m}|m\rangle + b\sqrt{\gamma_n}|n\rangle$.  Hence, the states in $A'$ get converted to the states
\begin{equation}
A''=\{\rho_n: n=1~\text{to}~N_\mathsf{c} \}\cup\{\rho_{mnx},\rho_{mny}: m,n=1~\text{to}~N_\mathsf{c}, m<n\}
\end{equation}
Here we define
\begin{equation}
\sigma_n=\gamma_n|n\rangle\langle n|, ~~~~\sigma_{mnx}=|{mnx}\rangle\langle {mnx}|, ~~~~\sigma_{mny}=|{mny}\rangle {mny}|
\end{equation}
with
\begin{equation}
|{mnx}\rangle=\frac{1}{\sqrt{2}} (\sqrt{\gamma_m}|m\rangle+\sqrt{\gamma_n}|n\rangle)~~~~\text{and}~~~~ |{mny}\rangle=\frac{1}{\sqrt{2}} (\sqrt{\gamma_m}|m\rangle+i\sqrt{\gamma_n}|n\rangle)
\end{equation}
The set of states in $A''$ certainly constitute a spanning set for system $\mathsf c$.  Since it is impossible for a linear transformation to turn a flat set into a non-flat set for a system of the same dimension, the states in set $A'$ (immediately after the filter) must be a non-flat also.  Hence, filters are non-flattening.

Although we presented this proof for filters, the only property of filters it depends on is that they enact a transformation of the form $|\psi\rangle\rightarrow \hat E|\psi\rangle$ for some operator $\hat E$.  This captures the non-mixing aspect of filters.  Indeed, in quantum theory, all non-mixing transformations are of the form $\rho\rightarrow \hat G \rho \hat G^\dagger$.  Hence, in general, in quantum theory all non-mixing transformations are also non-flattening.

\section{Proof that linearity follows from mixing}\label{linearityfrommixing}

\index{linearity}In Sec.\ \ref{associatingstateswithpreparations} we introduced the idea of representing a state by a list of probabilities, $A^{a_1}$, associated with minimal set of fiducial results, $\{\mathsf{X}_\mathsf{a_1}^{a_1}: a_1=1~\text{to}~K_\mathsf{a}\}$, in such a way that
\begin{equation}
\text{Prob}(\mathsf{A^{a_1}B_{a_1}})= A^{a_1}B_{a_1}
\end{equation}
for any effect, $\mathsf{B_{a_1}}$.  In particular, we chose $K_\mathsf{a}$ big enough that probability is given by a linear function of the probabilities in $A^{a_1}$. We can always do this since we can, if necessary, choose $A^{a_1}$ to be a list of probabilities for all results (i.e.\ the set of fiducial results could consist of all results for this type of system).   We might, however, imagine that there exists a smaller set of fiducial results such that the probability $\text{Prob}(\mathsf{A^{a_1}B_{a_1}})$ is given by a nonlinear function of the probabilities in this list.   We will call the process of going from a list of all probabilities to a minimal fiducial list of probabilities (sufficient to specify the state) \emph{physical compression}.  This name is appropriate because the  physical laws of the theory allow us to decompress this compressed list and calculate a general probability.   We will prove that
\begin{T}
If we allow arbitrary probabilistic mixtures of preparations then (1) linear physical compression is optimal and (2) optimal physical compression is necessarily linear.
\end{T}
This proof was previously given in (\cite{hardy2009foliable}).  To prove the first point we observe that, under linear compression, there must exist $K_\mathsf{a}$ linearly independent states, $\{A^{a_1}[k]:k=1~\text{to}~K_\mathsf{a}\}$, since otherwise we could compress further.  We can take an arbitrary probabilistic mixture of these linearly independent states with weightings $\lambda_k\geq 0$ where $\sum_{k}^{K_\mathsf{a}} \lambda_k \leq 1$.   Note we need not have the $\lambda_k$'s sum to one because we can include an extra weighting, $\lambda_0$, for the null state.  With this extra weighting they would sum to one.  The $\lambda_k$'s (for $k=1$ to $K_\mathsf{a}^\text{min}$) can all be varied independently in the convex sum
\begin{equation}
\sum_{k}^{K_\mathsf{a}} \lambda_k A^{a_1}[k]
\end{equation}
giving rise to a volume of dimension $K_\mathsf{a}$ in the state space.  Hence, we need $K_\mathsf{a}$ probabilities to specify the state and so linear compression is optimal.  To prove the second point consider a representing the state by a list of $K_\mathsf{a}$ probabilities, $\breve{A}[k]$ ($k=1$ to $K_\mathsf{a}$), where we do not demand that the general probability is given by a linear function.  The entries in $\breve{A}[k]$ correspond to some set of fiducial results, $\mathsf{Y}_\mathsf{a_1}[k]$, such that
\begin{equation}
\breve{A}^{a_1}:= \text{Prob}(\mathsf{A^{a_1}}\mathsf{Y}_\mathsf{a_1}[k])
\end{equation}
Now, since linear compression is optimal, there must by (1) exist a set of $K_\mathsf{a}$ fiducial results, $\mathsf{X}_\mathsf{a_1}^{a_1}$, for which the physical compression is linear. Let $A^{a_1}$ be the list of probabilities with respect to this set of fiducial results.  We can write
\begin{equation}
\breve{A}[k]= \text{Prob}(\mathsf{A^{a_1}}\mathsf{Y}_\mathsf{a_1}[k])= A^{a_1}Y_{a_1}[k]
\end{equation}
We can think of $Y^{a_1}[k]$ as a matrix with entries labeled by $a_1 k$.   This matrix must be invertible since otherwise we could specify $\breve{A}[k]$ with fewer than $K_\mathsf{a}$ probabilities.  Hence, we can write
\begin{equation}
\text{Prob}(\mathsf{A^{a_1}B_{a_1}})=  A^{a_1}B_{a_1} =  (Y^{a_1}[k])^{-1} \breve{A}[k]) B_{a_1}
\end{equation}
but this equation is linear $\breve{A}[k]$.  Hence, optimal compression is linear.

This technique for proving linearity follows from allowing arbitrary mixtures is much simpler than the approach adopted in \cite{hardy2001quantum}.

\section{Compactness}\label{compactness}

In this appendix we consider the vector formed from fiducial probabilities characterizing fragments with a given input-output structure.  We consider the case where only a finite number of fiducial probabilities are required.  We show that {\bf Assump 3} implies that the space of such vectors is compact (i.e.\ bounded and closed).  This means that the sets of allowed states, effects, and transformations, are compact for the finite dimensional cases we are interested in.  It also implies that the sets of allowed duotensors associated with fragments having a given input-output structure are compact (so long as they are specified by a finite number of fiducial probabilities).

We consider the cases where we can characterize fragments having a given input-output structure by a finite set of probabilities defined
with respect to a finite set of $K$ fiducial fragments, $\{\mathsf{X}[k]: k=1~\text{to}~K \}$, each of which can complete any fragment in having this input-output structure into a circuit (i.e.\ they have the mirror input-output structure).  Any two fragments that are equivalent have the same set of probabilities
\begin{equation}
{\bf p}_\mathsf{A}:=\{ \text{Prob}(\mathsf{X}[k]\mathsf{A}): k=1~\text{to}~K\}
\end{equation}
(this is understood to be a vector with the given components).  We use slightly antiquated notation here - in Part \ref{theduotensorframework} we see that this object can correspond to a duotensor with all black dots.  If we assume full decomposability (or, equivalently, tomographic locality) then $K$ is equal to the product of the $K_\mathsf{a}$'s associated with each input and output.   In any case, we choose $K$ to be just sufficient that we can write
\begin{equation}\label{rdotpagain}
\text{Prob}(\mathsf{AC})= {\bf r}_\mathsf{C}\cdot {\bf p}_\mathsf{A}
\end{equation}
The argument for being able to do this is the same as in Sec.\ \ref{associatingstateswithpreparations}.  The ${\bf r}$ vectors belong to some set, $R$, associated with this mirror input-output structure.

We will now show that it follows from {\bf Assump 3} that the set of allowed ${\bf p}_\mathsf{A}$ is compact\index{compactness}.  We know that probabilities are bounded by 0 and 1.  Further, if we put $\delta=a/l$ where $a$ is a constant and $l$ is a positive integer, then we have a convergent series, $\{\mathsf{A}[a/l]: l=1, 2, \dots\}$. Since, by {\bf Assump 3}, the limit point exists, the set of allowed ${\bf p}_\mathsf{A}$ must be closed with respect an appropriately chosen norm.  One norm that will do this is just $||\cdot ||$ defined by
\begin{equation}
||{\bf p}||=|{\bf p}|
\end{equation}
since, if $\mathsf A$ and $\mathsf B$ are operationally indiscernible to accuracy $\delta$ then it follows from the linearity of (\ref{rdotpagain}) that there exists a constant $c$ such that $|{\bf p}_\mathsf{A}-{\bf p}_\mathsf{B}|\leq c\delta$. To see this we note that
\begin{equation}
|{\bf p}_\mathsf{A} - {\bf p}_\mathsf{B}| = \max_{\bf \hat x} {\bf \hat x} \cdot ({\bf p}_\mathsf{A} - {\bf p}_\mathsf{B}) ~~~\text{where}~~|{\bf \hat x}|=1
\end{equation}
We can expand any unit vector, ${\bf \hat x}$, in terms of a (not necessarily orthonormal) basis set of vectors chosen from $R$,
\begin{equation}
{\bf \hat x}= \sum_{k=1}^K x_k {\bf r}_k
\end{equation}
Then we can write
\begin{equation}
{\bf \hat x} \cdot ({\bf p}_\mathsf{A} - {\bf p}_\mathsf{B}) = \sum_{k=1}^K x_k {\bf r_k} \cdot ({\bf p}_\mathsf{A} - {\bf p}_\mathsf{B})
\end{equation}
There must exist some finite $\alpha$ such that $\alpha\geq |x_k|$ for all $k$ and for all unit vectors ${\bf \hat x}$.  Then we have
\begin{equation}
{\bf \hat x} \cdot ({\bf p}_\mathsf{A} - {\bf p}_\mathsf{B}) \leq K \alpha \delta
\end{equation}
Hence,
\begin{equation}
|{\bf p}_\mathsf{A} - {\bf p}_\mathsf{B}| \leq K\alpha \delta
\end{equation}
as required.

We have shown that the ${\bf p}$ vectors belong to compact sets.  In duotensor language this means that the duotensors with all black dots belong to compact sets. However, we can use the hopping metric (and its inverse) to change the form of the duotensor (so it does not necessarily have all black dots).  The hopping metric is invertible and hence duotensors of a given form always belong to compact sets.

\section{Transforming duotensors}\label{transformingduotensors}

\index{duotensors!transforming}For an object to be a duotensor it must transform appropriately under transformation of the fiducial preparations and results. We will indicate the original fiducial preparations and results by $\mathsf X$ and the new set by $\tilde{\mathsf X}$.  Then we can write the old in terms of the new. For results we have
\begin{equation}
\mathsf{X}_\mathsf{a_1}^{a_1} =   \mathcal{E}^{a_1}_{\tilde{a}_1} \,\, \mathsf{\tilde{X}}_\mathsf{a_1}^{\tilde{a}_1}
\end{equation}
where $\mathcal{E}^{a_1}_{\tilde{a}_1}$ is the transformation matrix between fiducial sets of results.  For preparations we have
\begin{equation}
{}_{a_1}\!\mathsf{X}^\mathsf{a_1} =   {}^{\tilde{a}_1}_{a_1}\! \mathcal{P}  \,\, {}_{\tilde{a}_1}\!\mathsf{\tilde{X}}^\mathsf{a_1}
\end{equation}
where ${}_{\tilde{a}_1}^{a_1}\!\mathcal{P}$ is the transformation matrix between fiducial sets of preparations.  Consider
\begin{eqnarray}
\mathsf{A_{a_1b_2}^{c_3d_4}} \,\,& \equiv & {}^{c_3d_4}\!A_{a_1b_2}\,\, \mathsf{X}_\mathsf{a_1}^{a_1} \mathsf{X}_\mathsf{b_2}^{b_2}  \,\,{}_{c_3}\!\mathsf{X}^\mathsf{c_3}{}_{d_4}\!\mathsf{X}^\mathsf{d_4}   \nonumber\\
& {} & \nonumber \\
& \equiv & {}^{\tilde{c}_3\tilde{d}_4}\!\tilde{A}_{\tilde{a}_1\tilde{b}_2}\,\, \mathsf{\tilde{X}}_\mathsf{{a}_1}^{\tilde{a}_1} \mathsf{\tilde{X}}_\mathsf{b_2}^{\tilde{b}_2}  \,\,{}_{\tilde{c}_3}\!\mathsf{\tilde{X}}^\mathsf{c_3}{}_{\tilde{d}_4}\!\mathsf{\tilde{X}}^\mathsf{d_4}
\nonumber
\end{eqnarray}
Clearly
\begin{equation}\label{presupsubtrans}
{}^{\tilde{c}_3\tilde{d}_4}\!\tilde{A}_{\tilde{a}_1\tilde{b}_2} =      {}^{c_3d_4}\!A_{a_1b_2} \,\,  \mathcal{E}^{a_1}_{\tilde{a_1}}\,\, \mathcal{E}^{b_2}_{\tilde{b_2}}   \,\, {}_{c_3}^{\tilde{c}_3} \mathcal{P} \,\,{}_{d_4}^{\tilde{d}_4} \mathcal{P}
\end{equation}
This equation shows how a duotensor transforms if it has only pre-superscripts and subscripts.  To see how it transforms if we have indices in other positions we note
\begin{equation}
\text{Prob}(\mathsf{A^{a_1} B_{a_1}}) = A^{a_1} B_{a_1} = \tilde{A}^{\tilde{a}_1} \tilde{B}_{\tilde{b}_1} = \tilde{A}^{\tilde{a}_1}\,\, \mathcal{E}^{a_1}_{\tilde{a_1}}\, B_{a_1}
\end{equation}
where we transform the subscript as in (\ref{presupsubtrans}).  This equation must hold for any $B_{a_1}$.  Hence we must have
\begin{equation}
 A^{a_1} = \tilde{A}^{\tilde{a}_1}\,\, \mathcal{E}^{a_1}_{\tilde{a_1}}
\end{equation}
or
\begin{equation}
\tilde{A}^{\tilde{a}_1} =  A^{a_1} \,\, \mathcal{E}_{a_1}^{\tilde{a_1}}
\end{equation}
where $\mathcal{E}_{a_1}^{\tilde{a_1}}$ is the inverse of $\mathcal{E}^{a_1}_{\tilde{a_1}}$ such that
\begin{equation}
\mathcal{E}^{a_1}_{\tilde{a_1}} \, \mathcal{E}_{a_1}^{\tilde{a'_1}} = \delta_{\tilde{a}_1}^{\tilde{a}'_1}
\end{equation}
Hence superscripts on duotensors transform with $\mathcal{E}_{a_1}^{\tilde{a_1}}$.

By considering
\begin{equation}
\text{Prob}(\mathsf{A^{a_1} B_{a_1}}) = {}^{a_1}\!A \,\,\, {}_{a_1}\!B = {}^{\tilde{a}_1}\!\!\tilde{A}\,\,\, {}_{\tilde{b}_1}\!\!\tilde{B}
\end{equation}
and employing similar reasoning to that above, we can easily prove that pre-subscripts transform with  ${}^{a_1}_{\tilde{a}_1} \mathcal{P}$, this being is the inverse of ${}_{a_1}^{\tilde{a}_1} \mathcal{P}$, i.e.\
\begin{equation}
{}^{a_1}_{\tilde{a}'_1} \mathcal{P}\,\, {}_{a_1}^{\tilde{a}_1}\! \mathcal{P} = {}^{\tilde{a}_1}_{\tilde{a}'} \delta
\end{equation}
Hence, the transformation rule for a duotensor with indices in all positions is illustrated by
\begin{equation}\label{duotensortrans}
{}^{\tilde{c}_3}_{\tilde{d}_4}\!\tilde{A}_{\tilde{a}_1}^{\tilde{b}_2} =      {}^{c_3}_{d_4}\!A_{a_1}^{b_2} \,\,\,\,  \mathcal{E}^{a_1}_{\tilde{a_1}}\,\, \, \mathcal{E}_{b_2}^{\tilde{b_2}}   \,\,\, {}_{c_3}^{\tilde{c}_3} \mathcal{P} \,\,\,{}^{d_4}_{\tilde{d}_4} \mathcal{P}
\end{equation}
We see that a duotensor that has only subscripts and superscripts (i.e.\ is in standard form) transforms as a tensor with respect to the transformation matrix for results.  A duotensor that has only pre-superscripts and pre-subscripts transforms as a tensor with respect to the transformation matrix for preparations.  However, a duotensor with indices in all positions behaves like a new object that transforms with transformation matrices for the results and the preparations.  Further, there exists a hopping metric which can take indices from the left to the right and vice-versa.  The duotensor is a generalization of the idea of a tensor.  It has particular application to operational probabilistic theories. We should note that we have a choice of fiducial results and fiducial preparations for each type.  In general we do not expect $K_{\mathsf a}$ and $K_{\mathsf b}$ to be equal.  Hence the indices for different types will, in general, run over different numbers of values.

\section{Proof that $K=N^r$}\label{AppendixC}

\index{K@$K_\mathsf{a}$!relationship with $N_\mathsf{a}$}We will prove that if we have a integer valued function, $K(N)$, satisfying (i) $K(N+1)>K(N)$, (ii) $K(N_AN_B)=K(N_A)K(N_B)$, for $N=1, 2, 3, \dots$ then $K=N^r$ for $r=1, 2, 3, \dots$. This was first proven (for the purpose of reconstructing quantum theory) in \cite{hardy2001quantum}.  Here we give the simpler proof in \cite{hardy2010limited}.  We can expand $N$ as a product of primes.  Thus, we write $N=\prod_i (p_i)^{n_i}$ where $p_i$ is the $i$th prime and $n_i$ is the power of this prime in the expansion.  It follows from (ii) that
\begin{equation}\label{Kexpand}
K(N) = \prod_i [K(p_i)]^{n_i}
\end{equation}
We will prove that $K(p_i)=(p_i)^r$ for a fixed value of $r$ that is independent of $p_i$.  To prove this we will assume the converse and obtain a contradiction. Thus, assume that there exist two primes, $p$ and $q$ such that $K(p)=p^{r_p}$ and $K(q)=q^{r_q}$ where $r_p\not= r_q$. Let $N_A=p^a$ and $N_B=q^b$ where $a$ and $b$ are positive integers.  Using (\ref{Kexpand}) we obtain
\begin{equation}
\frac{\log K_B  }{\log K_A} = \frac{br_q \log q}{ar_q \log p}
\end{equation}
We also have
\begin{equation}
\frac{\log N_B  }{\log N_A} = \frac{b \log q}{a \log p}
\end{equation}
The two real numbers $\log q/\log p$ and $(r_q/r_p)(\log q/\log p)$ are distinct since we assume $r_p\not= r_q$.  Hence we can choose a value of $a/b$ that is strictly in between these two real numbers.  If we to this then it follows that the ratios $\log K_B /\log K_A$ and $\log N_B / \log N_A$ lie on opposite sides of $1$.  This contradicts (i). Hence we must have $K(p_i)=p_i^r$ for all $p_i$.  Using (\ref{Kexpand}) we immediately obtain $K=N^r$. Since $K$ must be an integer for all $N$, $r$ must take non-negative integer values.  We cannot have $r=0$ by (i).  Hence $r=1, 2, \dots$.

\section{The duotenzor drawing package}\label{duotenzor}

\index{duotenzor drawing package}All figures in this work were drawn using version 1.1 of the duotenzor package \cite{hardyduotenzor}. Version 1.1 has additional commands to draw operator boxes.  The \verb+duotenzor+ drawing package (spelled with a \verb+z+) is a purpose built package for drawing circuits and duotensor diagrams \cite{hardy2010formalism}.  It consists of about eighty commands (defined using the LaTeX \verb+\newcommand+ command) that call on the TikZ package written by Till Tantau \cite{tantautikz}.  Here is a simple example.  The code on the left produces the example on the right.
\begin{verbatim}
\begin{diagram}
\Opbox{A}{0,0}
\Opbox{B}{2,4}
\wire{A}{B}{1}{2}
\end{diagram}
\end{verbatim}
\makebox[3cm][l]{
\begin{diagram}
\boundingbox{0,0}{0,0}
\begin{move}{30,5}
\Opbox{A}{0,0}
\Opbox{B}{2,4}
\wire{A}{B}{1}{2}
\end{move}
\end{diagram}
}

\vspace{-13pt}
\noindent
Here \verb+\Opbox{B}{2,4}+ puts a box at coordinate $(2,4)$ with the symbol $\mathsf B$ in it.   The \verb+\wire{A}{B}{1}{2}+ command draws a wire from output 1 of box \verb+A+ to input 2 of box \verb+B+.  A comprehensive tutorial for the package is provided with this package \cite{hardyduotenzor} (see also the appendix to \cite{hardy2010formalism}).  For the drawing the kind of circuits used in quantum computing papers the Q-circuit package written by Brian Eastin and Steve Flammia \cite{eastinqcircuit} (which is powered by the XY-pic LaTeX package) is more suitable than the duotenzor package.

\bibliography{quantum}
\bibliographystyle{plain}

\printindex

\end{document}